\documentclass[manuscript]{aastex}

\usepackage{color}

\slugcomment{Revised version accepted for publication in Astrophys. J.}

\shorttitle{3D Neutrino transfer in supernovae}
\shortauthors{Sumiyoshi et al.}

\begin{document}


\title{Multi-dimensional Features of Neutrino Transfer \\
in Core-Collapse Supernovae}


\author{K. Sumiyoshi}
\affil{Numazu College of Technology, 
       Ooka 3600, Numazu, Shizuoka 410-8501, Japan}
\email{sumi@numazu-ct.ac.jp}

\author{T. Takiwaki\footnote{
Current address: Institute of Physical and Chemical Research (RIKEN), 
2-1 Hirosawa, Wako, Saitama 351-0198, Japan}}
\affil{National Astronomical Observatory of Japan, 
       2-21-1 Osawa, Mitaka, Tokyo 181-8588, Japan}
\email{takiwaki.tomoya@nao.ac.jp}

\author{H. Matsufuru}
\affil{Computing Research Center,
       High Energy Accelerator Research Organization
       1-1 Oho, Tsukuba, Ibaraki 305-0801, Japan}
\email{hideo.matsufuru@kek.jp}

\and

\author{S. Yamada}
\affil{Science and Engineering
       \& 
       Advanced Research Institute for Science and Engineering, \\
       Waseda University, 
       Okubo, 3-4-1, Shinjuku, Tokyo 169-8555, Japan}
\email{shoichi@heap.phys.waseda.ac.jp}



\begin{abstract}
We study the multi-dimensional properties of neutrino transfer inside supernova cores 
by solving the Boltzmann equations 
for neutrino distribution functions in genuinely six dimensional (6D) phase space.  
Adopting representative snapshots of the post-bounce core from other supernova simulations 
in three dimensions, 
we solve the temporal evolutions to stationary states of neutrino distribution functions 
by our Boltzmann solver.  
Taking advantage of the multi-angle and multi-energy feature realized by the S$_n$ method in our code, 
we reveal the genuine characteristics of spatially three dimensional (3D) neutrino transfer 
such as non-radial fluxes and non-diagonal Eddington tensors.  
In addition, we assess the ray-by-ray approximation, 
turning off the lateral-transport terms in our code.  
We demonstrate that the ray-by-ray approximation tends to 
propagate fluctuations in thermodynamical states around the neutrino-sphere 
along each radial ray and overestimate the variations 
between the neutrino distributions on different radial rays.  
We find that the difference in the densities and fluxes of neutrinos 
between the ray-by-ray approximation and the full Boltzmann transport becomes 
$\sim20\%$, which is also the case for the local heating rate, 
whereas the volume-integrated heating rate in the Boltzmann transport 
is found to be only slightly larger ($\sim2\%$) than the counterpart 
in the ray-by-ray approximation due to cancellation among different rays.  
These results suggest that 
we had better assess carefully the possible influences 
of various approximations in the neutrino transfer employed in the current simulations 
on supernova dynamics.  
Detailed information on the angle and energy moments 
of neutrino distribution functions will be 
profitable for the future development of numerical methods 
in neutrino-radiation hydrodynamics.  
\end{abstract}




\section{Introduction}

Neutrinos play an essential role 
in the mechanism of core-collapse supernovae 
as a driving force in the dynamics 
starting from the gravitational collapse of massive stars.  
Transport of energy and lepton number via neutrinos 
is particularly crucial to determine their outcomes, i.e. explosions 
with an energy of 10$^{51}$ erg \citep{bet90,kot06,jan07}.  
Electron-type 
neutrinos are produced by electron captures 
in the collapsing phase 
and are trapped inside the central core 
to give pressure support.  
Pairs of neutrinos are created after bounce as 
a part of the thermal 
energy converted from the gravitational energy.  
Emissions of the trapped neutrinos result in 
an energy release of $\sim$10$^{53}$ erg 
in the form of supernova neutrinos.  
The release of this huge energy in neutrinos was indeed vindicated 
in the observed events of SN 1987A \citep{hir87}.  
The emitted neutrinos carry 
valuable information of the mechanism of 
explosion as well as the properties of 
dense matter inside compact stars 
and progenitors 
(see \citet{nak13}, for example, and references therein).  

Interactions of neutrinos with material 
in the supernova core contribute to the energy transfer 
in a significant size.  
The shock wave generated by core bounce 
stalls soon after the launch 
and fizzles later in the case of strictly spherical geometry.  
Without such an artificial restriction in geometry, 
multi-dimensional hydrodynamical instabilities occur and 
push the stalled shock so that it could hover at larger radii.  
A portion of the emitted neutrinos is absorbed by the matter 
behind the stagnant shock wave and may lead eventually to 
its outward propagation again.  
This so-called neutrino heating mechanism \citep{bet85} 
combined with the multi-dimensional hydrodynamical instabilities 
is one of the most promising ways to trigger the explosion 
particularly under a marginal condition.  
In fact, energy gain by the neutrino heating 
amounts to $\sim$10$^{51}$ erg,  
which is comparable to the observed explosion energy of 
supernovae \citep{jan96}.  
Whether the neutrino heating mechanism is 
the essential cause of explosion or not 
remains an unsolved issue, though \citep{jan12,bur13}.  

In order to advance further 
our understanding of the role of neutrinos 
in the intrinsically multi-dimensional dynamics, 
the evaluation of neutrino heating and cooling 
should be as accurate as possible.  
Neutrinos interact with matter very frequently 
and diffuse outward only gradually inside the proto-neutron star 
just born after the bounce.  
Since the temperatures are high in the proto-neutron star, 
neutrinos are coolant there.  
On the other hand, a partial absorption of the emitted neutrinos 
takes place outside the neutrino-sphere 
below the stalled shock wave before they stream out of the star freely, 
which heats the matter there.  

Quantitatively accurate treatment of neutrino transfer is hence mandatory 
to provide correct rates of cooling and heating since 
the energy transport proceeds across the region, 
where neither diffusion nor free-streaming approximation is applied.  
The emergent fluxes and energy spectra of neutrinos 
are determined by 
the neutrino-matter interactions 
around the energy-dependent neutrino-spheres.  
We need to do 
multi-energy group computations, 
since the interactions 
strongly depend on the neutrino energy.  
In the evaluation of the heating rate, 
on the other hand, 
the local number density of neutrinos are important 
in addition to the energy spectra.  
It is hence 
the angular distributions of neutrinos 
in their momentum space that should be given precisely.  
Although they are isotropic near the center and 
become forward-peaked at large radii, 
the angular distributions are highly non-trivial 
in the transitional region, which includes the heating region.  

The treatment of neutrino transfer should, therefore, 
be as accurate as possible, particularly in the description of 
the energy and angle distributions.  
In most of multi-dimensional numerical simulations of supernovae, 
however, some approximations are employed at the moment.  
This is simply because the full description of neutrino distributions 
is a six-dimensional 
(three spatial coordinates and three momenta of neutrinos) problem, 
which is formidable even for modern supercomputing resources.  
While computations with the multi-energy groups are 
becoming a standard nowadays, 
angular distributions are still commonly approximated one way or another.  
Since such approximations can be eliminated 
in spherically symmetric simulations, 
the influences of the approximations have been extensively studied 
\citep{jan92,mes98,yam99,lie05}.  
It is well known that the diffusion approximation 
with a flux-limiter tends to overestimate the degree of forward-peak 
and thus underestimate the heating rate \citep{jan92,mes98}.  
Although the difference is not so drastic, 
such a small change may be crucial, tipping the balance 
for explosion particularly in the marginal conditions \citep{jan96}.  

Validation of the approximate methods 
in multi-dimensions has growing importance, 
since the post-shock conditions obtained in multi-dimensional simulations 
appear more marginal for successful explosion.  
The diffusion approximations (including its variants such as IDSA) 
and/or the ray-by-ray methods are frequently 
used in the currently state-of-the-art 
2D/3D simulations \citep{suw10,mul12b,bru13,tak12,han13,knak14a,suw14b,bru14}.  
Although they have reported a considerable number of successful explosions 
produced by the neutrino heating combined 
with the neutrino-driven convection 
and/or the standing accretion shock instability (SASI), 
their approximate treatments of neutrino transfer introduce 
a certain level of uncertainty.  
The influences of these approximations should be examined carefully 
before drawing any conclusion.  

In the past years 
we have indeed seen 
the multi-dimensional treatment of neutrino transfer 
progress substantially, 
keeping pace with the rapid increase of supercomputing resources.  
In two dimensions, multi-angle and multi-energy groups 
neutrino radiation-hydrodynamics have been performed 
to follow the post-bounce evolution of supernova cores 
\citep{liv04,ott08,bra11}.  
They adopted the discrete-ordinate (S$_n$) method 
to fully describe angle distributions in 2D space \citep{liv04}.  
\citet{ott08} reported the comparisons 
with the diffusion approximation.  
They demonstrated that the multi-angle transport is 
needed to describe the neutrino distributions 
in globally aspherical supernova cores.  
\citet{bra11} analyzed further the neutrino heating 
and showed that the multi-angle transfer 
leads to less lateral variations in the neutrino distributions 
owing to an averaging effect.  
Such a detailed study has been limited to 2D.  

In three dimensions, there has been a remarkable progress 
in the multi-energy group transport, 
but not 
in the multi-angle group transport \citep{tak12,han13,tak14,tam14,mez14,hor14}.  
These simulations commonly adopted the ray-by-ray approximations 
although the radial transport was treated differently:  
the isotropic diffusion source approximation (IDSA) 
is employed by \citet{tak12,tak14} 
whereas the moment equations are solved together 
with a variable Eddington factor obtained from the solution of 
the 1D model Boltzmann equations in \citet{han13,tam14}.  
Results obtained by the two groups differ in many respects, 
most notably on the success or failure of explosion 
(see e.g. their 11.2M$_{\odot}$ model).  
Note, however, that the origin of the differences 
might be the employed neutrino reactions \citep{tak14b}.  
Provided that the number of reported models is still small 
and the employed physics and numerics are considerably different among them, 
it is still premature to draw a firm conclusion 
from these 3D simulations and 
further studies should be done systematically 
with the assessment of the influences 
from the approximations in 3D neutrino transfer.  

The purpose of this study is hence 
to explore the basic features of 
3D neutrino transfer in realistic supernova cores 
with the multi-angle and multi-energy group Boltzmann solver 
for the first time.  
Adopting a set of profiles from 3D numerical simulations 
by \citet{tak12,hor14,tak14b}, 
we run our newly developed code \citep{sum12} 
to obtain stationary neutrino distribution functions 
for those fixed matter configurations.  
Taking advantage of the multi-angle treatment of our code, 
we explore the non-radial (polar and azimuthal) transport 
in the globally and locally aspherical matter distributions.  
Note that in our previous paper \citep{sum12} 
the background models were constructed 
by artificially deforming spherically symmetric models.  
We pay particular attention to 
the behavior of the angle moments of neutrino distribution functions 
in the transitional regime, 
which will be useful for development of 
numerical codes using the moment formalism.  

Employing our Boltzmann code, we can study 
the quality of the ray-by-ray approximation, 
which is routinely used in many simulations 
in 2D and 3D\footnote{Some simulations implemented 
the ray-by-ray plus approximation, in which 
the lateral coupling is partially taken into account.  }.  
As a matter of fact, simply dropping 
the polar and azimuthal advection terms, 
we can emulate the ray-by-ray approximation 
without changing other settings.  
We examine the non-radial transport, 
which will not be fully described in the ray-by-ray approximation.  
As we will show, the contributions of polar and azimuthal 
fluxes are significant even outside the optically thick region.  
It is also found that 
the ray-by-ray approximation tends to exaggerate 
the contrast between different radial rays, which 
may affect local neutrino-heating rates.  

The paper is organized as follows.  
We describe the methodology in \S \ref{section:3D-nu}.  
The realization of the ray-by-ray approximation in our code 
is also explained there.  
The 3D profiles taken 
from \citet{tak12,hor14,tak14b} 
are presented in \S \ref{section:3D-hyd}.  
We report in \S \ref{section:d-boltz} the basic features 
of 3D neutrino transfer, 
showing density, radial and non-radial fluxes 
obtained by the 6D Boltzmann solver.  
The corresponding results by the ray-by-ray approximation 
are presented in \S \ref{section:d-ray} for comparison.  
We investigate the heating rates obtained 
by both the 6D Boltzmann solver and the ray-by-ray approximation 
in \S \ref{section:h-boltz-ray}.  
In \S \ref{section:h-boltz-ray.others}, we present 
the results for other profiles and demonstrate 
that our findings are generic.  
The volume-integrated heating rates are examined 
in \S \ref{section:h-total} to discuss the global 
influences of 3D neutrino transfer on the supernova dynamics.  
In \S \ref{section:moments}, 
we show the angle and energy moments of the neutrino 
distribution functions.  
Discussions and implications of our study 
are given in \S \ref{section:discuss} 
followed by the summary in \S \ref{section:summary}.  

\section{Numerical settings}

\subsection{Neutrino transfer in 3D space}\label{section:3D-nu}

We adopt the numerical code 
to solve the Boltzmann equation in 6D \citep{sum12}.  
The code solves the time evolution of 
the 6D neutrino distribution functions 
in the spherical coordinate system 
by using neutrino energy and two angles 
to designate the neutrino momenta.  
The 6D Boltzmann equation in the conservative form is 
discretized in a finite-differenced method 
on a spatial grid as wells as 
multi-energy and multi-angle grids.  
We employ the discrete-ordinate (S$_n$) method 
to describe the angular distributions.  
The implicit differencing is adopted 
for the advance of time step 
to solve stiff equations and 
to ensure the stability for the equilibrium solution.  
Further details of the numerical scheme 
can be found in \citet{sum12}.  

A basic set of the neutrino reactions is 
implemented in the collision term \citep{bru85,sum05}.  
The set includes the emission, absorption and scattering 
of neutrinos via nucleons and nuclei.  
The pair processes via electron-positron annihilation, 
creation and the nucleon-nucleon bremsstrahlung 
are also taken into account.  
We treat the three neutrino species, 
$\nu_e$, $\bar{\nu}_e$ and $\nu_\mu$.  
The last one is actually a representative
of four species, 
$\nu_{\mu}$, $\bar{\nu}_{\mu}$, $\nu_{\tau}$ and $\bar{\nu}_{\tau}$\footnote{This group is 
often denoted by $\nu_x$ or $\nu_{\mu/\tau}$ and is 
sometimes used to show the summed contributions of four species.  }.   
It is to be noted that 
the densities, fluxes and angle moments of $\nu_{\mu}$ is 
for one species whereas the contributions of the four species 
are summed for the cooling and heating rates.  
The same setting has been used in our previous paper \citep{sum12} 
for detailed comparisons 
with the 1D results of core-collapse supernovae from \citet{sum05}.  
These neutrino reactions are 
the most important but certainly not sufficient. 
The electron-neutrino scattering, for example, is omitted 
in the current study \citep{tho03,bur06a}.  
This is mainly due to the limitation of 
numerical resources and modernization will be definitely needed \citep{len12a}.  
The equation of state of \citet{lat91} are used 
for all calculations in the current study 
so that better consistency would be obtained 
with the profiles 
we adopt from the supernova simulations by \cite{tak12,hor14,tak14b}.  

We follow the time evolution of the neutrino distribution functions 
by the 6D Boltzmann solver for the fixed background profiles 
until stationary states are reached (typically in $\sim$20 ms) 
from arbitrarily set initial conditions, in which essentially no neutrino exists.  
We ignore fluid motions although the original numerical data include such information.  
We neglect all the velocity-dependent terms in the current study 
and do not distinguish the inertial frame from the fluid-rest frame, 
since 
our aim here is to understand the basic 3D features of the neutrino transfer.  
It should be emphasized, however, that these 
velocity-dependent terms in the Boltzmann equation 
cannot be neglected in more realistic simulations 
as demonstrated by \citet{len12b}.  
As a matter of fact, the implementation 
of all the velocity-dependent effects 
to the current Boltzmann solver 
has been already done and its application to 
2D neutrino-radiation hydrodynamics simulations 
is currently under way \citep{nag14}.  

In addition, 
we perform approximate calculations by a ray-by-ray prescription 
in our framework 
by dropping off the advection terms 
for spatial $\theta$- and $\phi$-derivatives 
as well as neutrino-angle $\phi_{\nu}$-derivative 
in the Boltzmann equation.  
We compute the stationary states of the 3D neutrino transfer 
with the same setting for profiles and reactions.  
We compare the numerical results by the two methods 
and examine the differences due to the approximation.  

The numbers of spatial grid points employed 
for neutrino distribution functions are 
$N_r=256$, $N_{\theta}=64$, $N_{\phi}=32$ 
for our baseline 3D models.  
The numbers of neutrino angle ($\theta_{\nu}$, $\phi_{\nu}$) 
and energy ($\varepsilon_{\nu}$) grids are 
$N_{\theta_{\nu}}=6$, $N_{\phi_{\nu}}=12$ and 
$N_{\varepsilon_{\nu}}=14$, respectively.  
The energy grid points are placed logarithmically from 0.9 MeV 
to 300 MeV with additional grid points for high-energy tails.  
Computations with this mesh using 1.6TB memory 
are the largest scale available on 8 nodes of HITACHI SR16000 at KEK.  
In order to obtain one configuration of the model 
by the time evolution in several tens of milliseconds, 
we need a run of $\sim$4$\times10^{4}$ steps 
using the MPI-parallel code of the 6D Boltzmann solver.  
It typically takes $\sim$200~hours of wall-clock time 
on 8 nodes of SR16000 by 128 MPI processes.  

Although 
the current setting of 
the angle and energy grids is sufficient to obtain 
the reasonable results 
as validated by the convergence test in \citet{sum12}, 
we have additionally studied the convergence 
by increasing the number of grid in some sample models.  
For this purpose, we perform the numerical simulations 
for the inner part of the supernova core ($\le$~360~km) 
using $N_r=128$ and increase the angle resolution.  
When we increase $N_{\theta_{\nu}}$ from 6 to 8, 
we find typical 
differences within 2~$\%$ in the neutrino densities, 
fluxes and heating rates.  
In the case of increasing $N_{\phi_{\nu}}$ from 12 to 16, 
the differences are within $\sim$0.8~$\%$.  
Appearance of large differences of the heating rates 
is limited near the gain radius 
due to very small values close to zero.  

\subsection{3D supernova cores}\label{section:3D-hyd}

We utilize the post-bounce matter profiles taken from the 3D numerical simulations 
of core-collapse supernovae by \citet{tak12,hor14,tak14b}, who followed 
the evolutions of 11.2M$_{\odot}$ and 27.0M$_{\odot}$ stars of \citet{woo02}.  
Their computations employed 
the extended ZEUS-MP code for hydrodynamics \citep{hay06,iwa08} 
and the IDSA scheme \citep{lie09} for neutrino transport 
with the ray-by-ray approximation.  
%
%
We use the 11.2M$_{\odot}$ model (11M) as the baseline model 
for detailed analysis.  
The profiles adopted from this model 
represent the typical situations, in which the revived 
shock wave is expanding toward an explosion.  
On the other hand, the 27.0M$_{\odot}$ model (27M) \citep{hor14,tak14b}, 
which were computed with the improved code 
and details will be reported elsewhere, 
does not lead to shock revival 
and represents the situations in such prolonged stagnations.  
It is to be noted that 
since the profiles we use in this study 
are obtained in the simulations with the ray-by-ray approximation, 
our results may be somewhat influenced indirectly by the approximation.  
%

We adopt the 3D profiles of density, temperature and 
electron fraction (but not velocity) from the dynamical simulations 
and remap them 
onto our numerical mesh.  
We use the same radial (logarithmically spaced) grid 
up to 2600 km 
as those employed in the simulation of 11.2M$_{\odot}$.  
To cover the whole solid angle in space, 
we deploy the same number of uniform azimuthal 
grid points as in the dynamical simulation 
while a different (non-uniform) polar grid is used 
so that it would fit our Boltzmann solver.  
We take the same spacial mesh also for 27M, 
interpolating the density, temperature and electron fraction.  
Other thermodynamical quantities are derived from 
the equations of state by \citet{lat91} 
with the incompressibility of 180 MeV 
for 11M and 220 MeV for 27M, respectively, 
just as in the original simulations.  
%

\begin{figure}
\epsscale{0.40}
\plotone{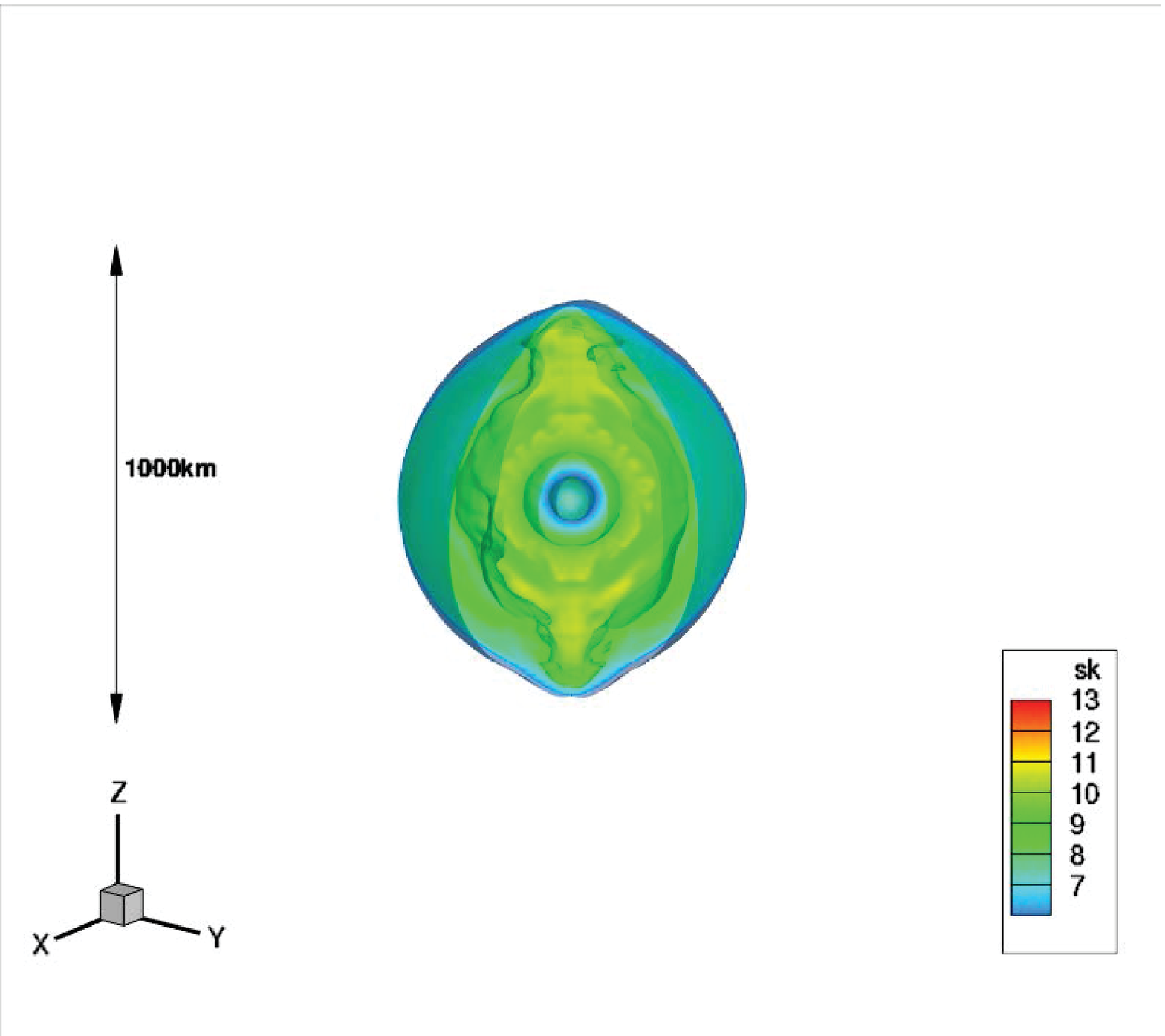}\\
\plotone{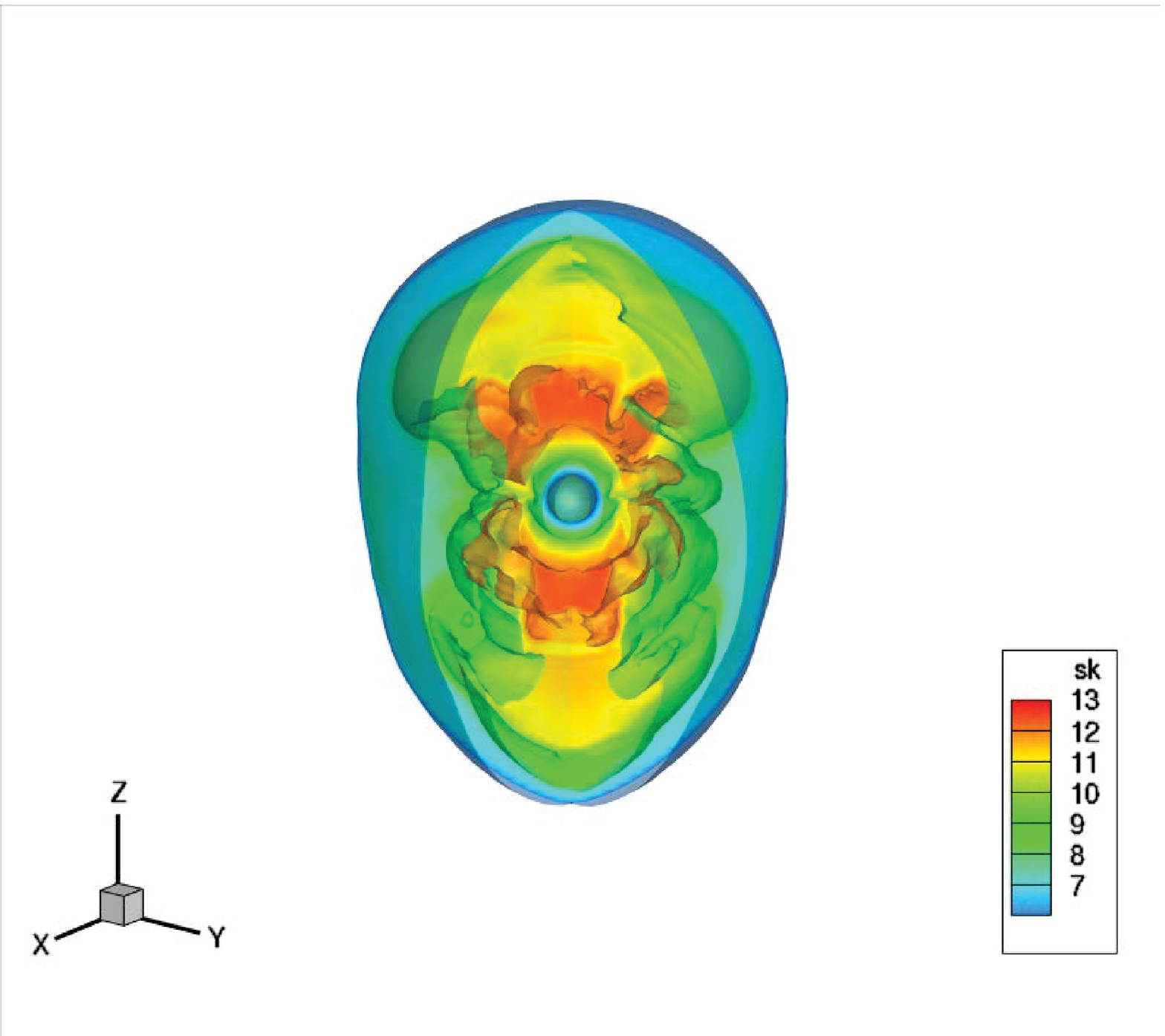}\\
\plotone{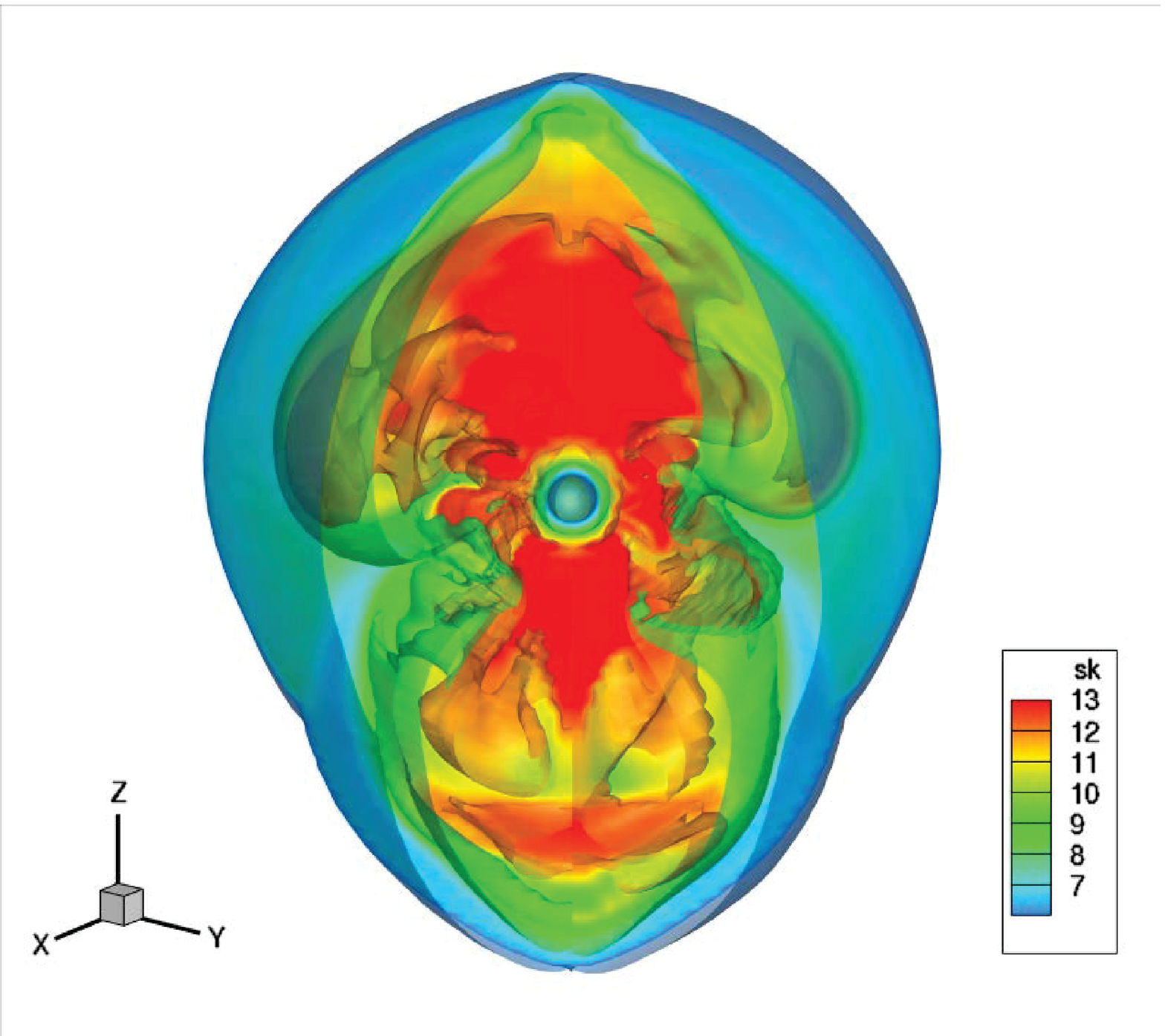}
\caption{Profiles of the 3D supernova core 
from the simulation of 11.2M$_{\odot}$ star \citep{tak12}.  
Iso-surfaces of the entropy per baryon 
in the unit of the Boltzmann constant, $k_B$, 
are shown for the snapshots 
at 100, 150 and 200 ms after the bounce 
in the top, middle and bottom panels, respectively.  }
\label{fig:3d-entropy}
\end{figure}

In Figure \ref{fig:3d-entropy}, 
we show the 3D profiles of supernova core 
for the 11.2M$_{\odot}$ star 
at 100, 150 and 200 ms after the bounce.  
They correspond to 
the phase, in which the shock wave expands gradually 
from $\sim$300 km to $\sim$600 km 
in a non-spherical manner.  
The revival of shock wave occurs earlier (at $\sim$50 ms) 
due to the neutrino heating 
and its outward propagation is assisted by 
the enhanced heating 
through the hydrodynamical instabilities.  
This 3D model leads to a successful explosion.  

\begin{figure}
\epsscale{0.9}
\begin{center}
\plottwo{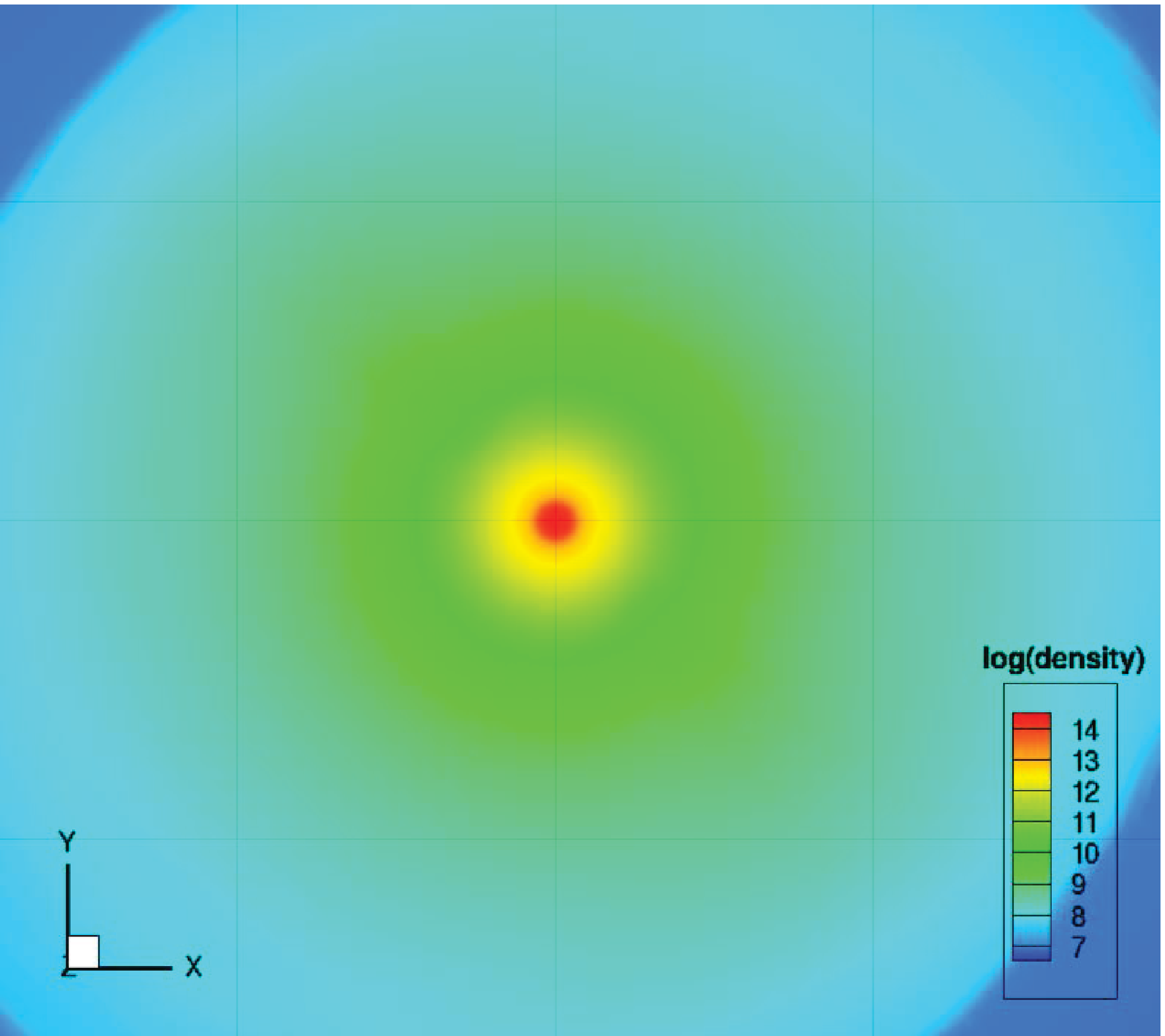}{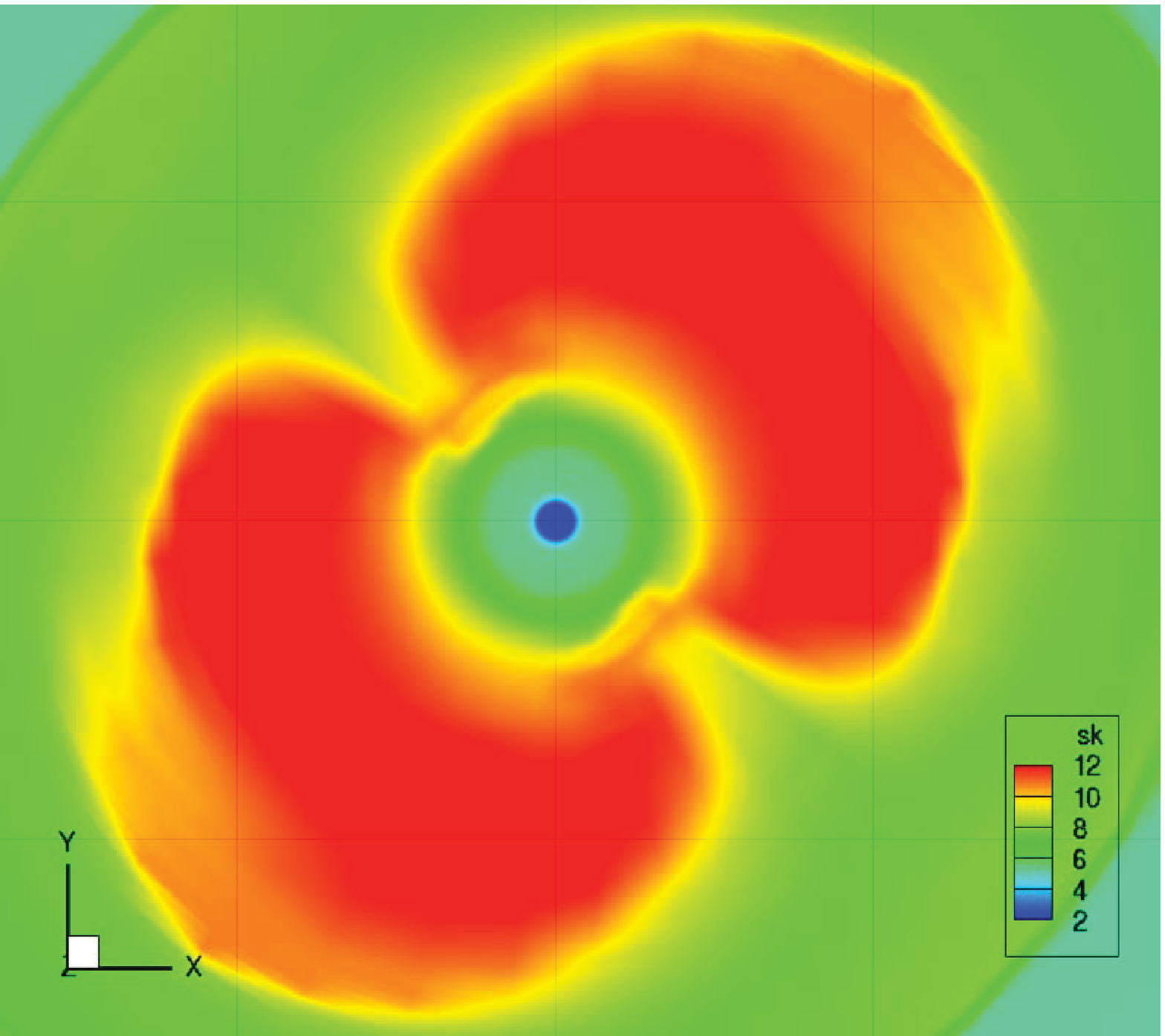}\\
\plottwo{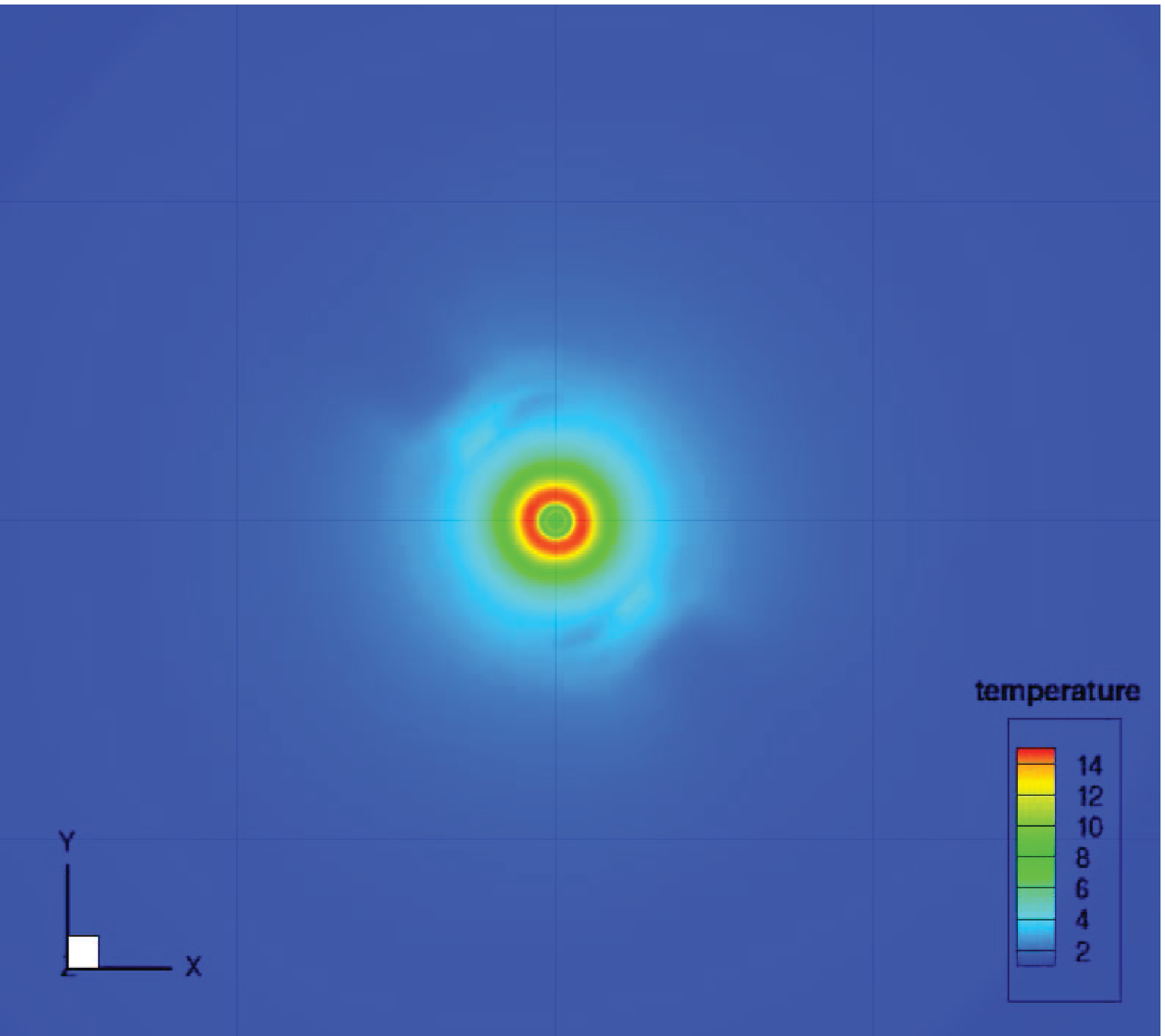}{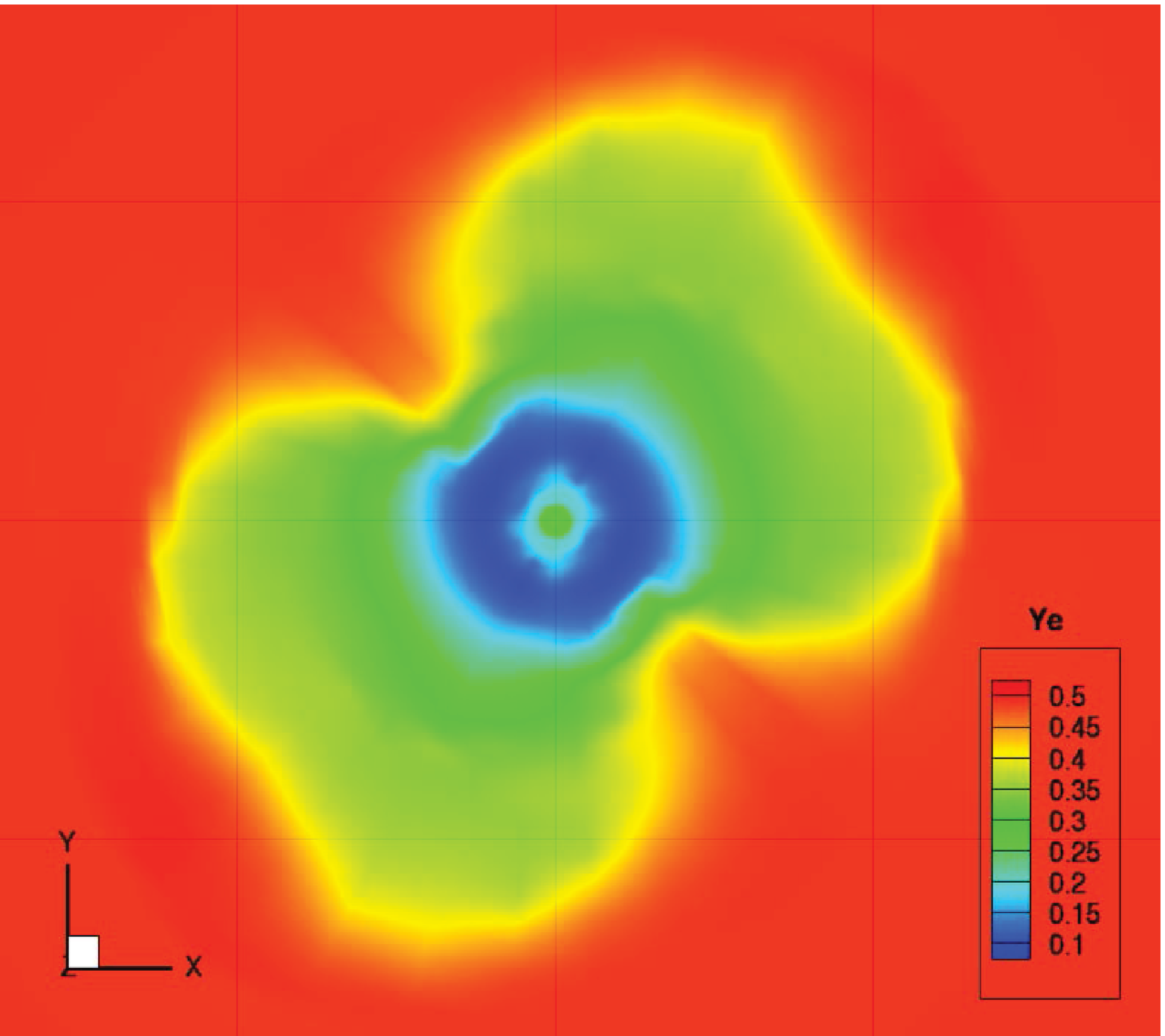}\\
\end{center}
\caption{Color maps of the 3D supernova core (11M) on the xy-plane (z=0) 
are shown at 150 ms after the bounce.  
Density [g/cm$^3$], entropy per baryon [$k_B$], 
temperature [MeV] and electron fraction 
are displayed  
in the left-top, right-top, left-bottom and right-bottom panels, 
respectively.  
Grid lines with 200 km spacing are shown in the background.  
}
\label{fig:3db-xyslice}
\end{figure}

We display in Fig. \ref{fig:3db-xyslice} 
the color maps of the core (11M) on the xy-plane (z=0) 
at 150 ms after the bounce.  
The shock wave is located around 400--500 km 
with an elongated shape 
at $\phi \sim45^\circ$ and $\phi \sim225^\circ$.  
The low entropy region prevails 
along $\phi \sim135^\circ$ and $\phi \sim315^\circ$, 
which correspond to 
down flows of proton-rich matter.  
This nearly dipolar configurations may have arisen 
from the weak $\pi/2$ periodicity in $\phi$, 
which was planted accidentally in the perturbation 
imposed initially in the dynamical simulation\footnote{
\citet{tak12} indeed added a perturbation pointwise-randomly. 
The resultant perturbation, however, had a $\pi/2$ periodicity 
weakly in the azimuthal direction by accident, 
which was not mentioned in their paper.  }.  
The density and temperature distributions 
of the central object (proto-neutron star) 
are almost spherical 
with only the outer layer being deformed.  
The configurations at other times 
are qualitatively similar but with elongations at different $\phi$-angles.  

\begin{figure}
\epsscale{0.9}
\plotone{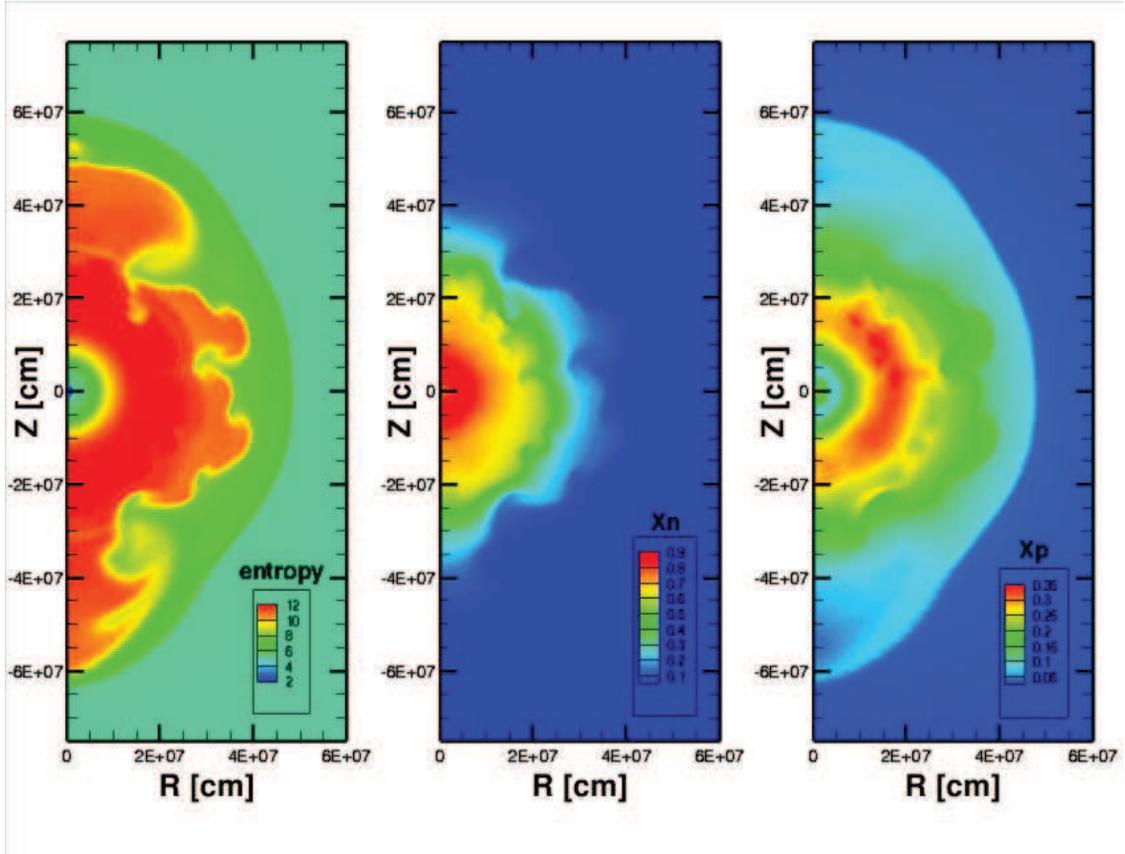}
\caption{Color maps of the 3D supernova core (11M) 
on the meridian slice at $\phi$=51$^\circ$ 
are shown at 150 ms after the bounce.  
Entropy per baryon [$k_B$] 
and mass fractions of free neutrons and protons are displayed 
in the left, middle and right panels, respectively.  }
\label{fig:3db-phi.slice.iph05}
\end{figure}

\begin{figure}
\epsscale{0.9}
\plotone{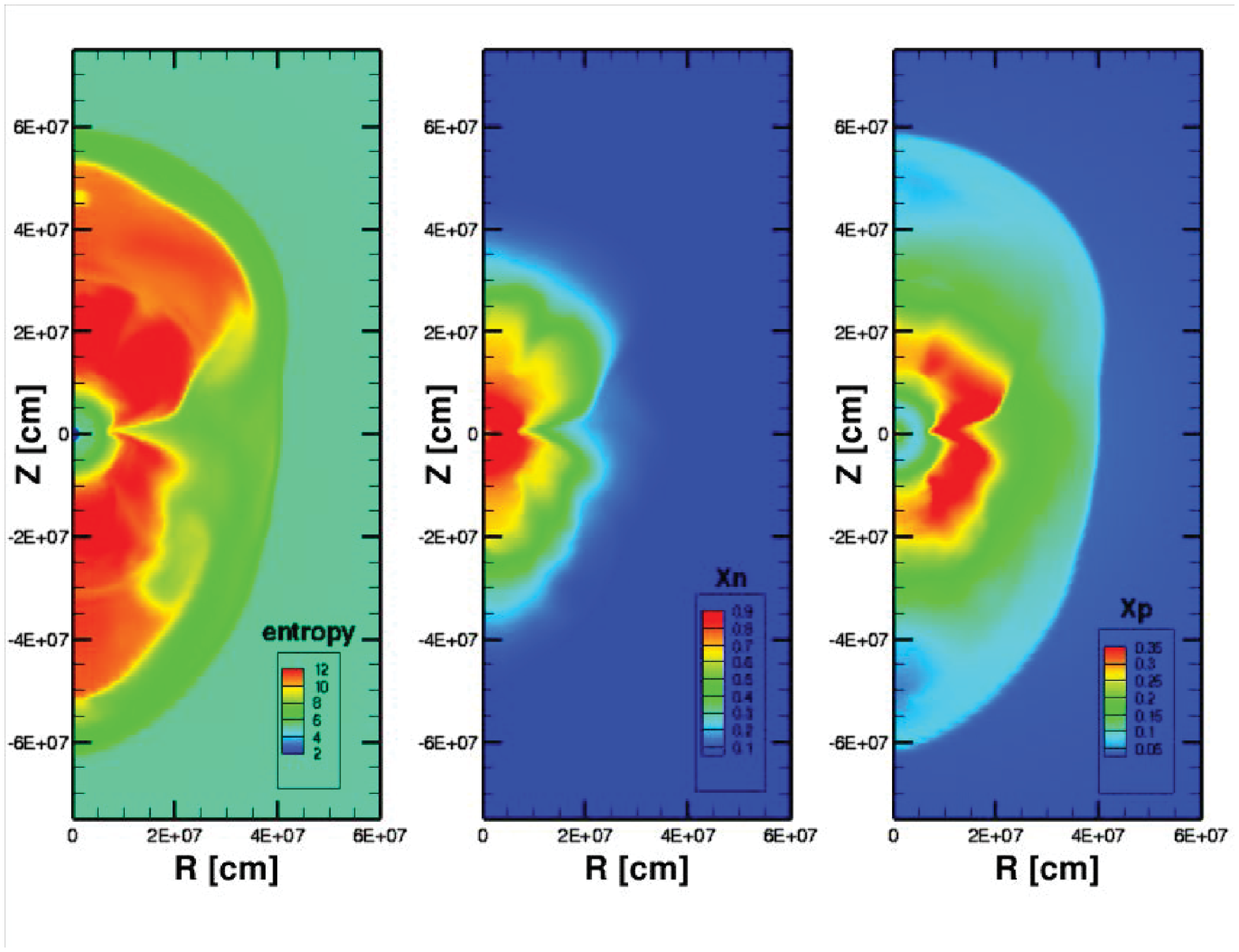}
\caption{Same as Figure \ref{fig:3db-phi.slice.iph05}, 
but on the meridian slice at $\phi$=141$^\circ$.  }
\label{fig:3db-phi.slice.iph13}
\end{figure}

Figures \ref{fig:3db-phi.slice.iph05} 
and \ref{fig:3db-phi.slice.iph13} show 
the color maps of entropy and mass fractions of nucleons 
on the meridian slices at $\phi$=51$^\circ$ and 141$^\circ$, respectively.  
Mushroom-like shapes of high entropy regions 
reflect convective motion and the associated shock dynamics.  
On the equator, the high entropy region 
extends up to the shock front at $\phi$=51$^\circ$ 
and there is a narrow channel of low entropy matter 
at $\phi$=141$^\circ$.  
Free neutrons are abundant close to and 
inside the proto-neutron star 
whereas free protons populate rather abundantly 
at larger radii where the entropy is high.  
Nucleons behind the shock wave 
act as chief absorbing material 
in the neutrino heating.  

\begin{figure}
\epsscale{0.48}
\plotone{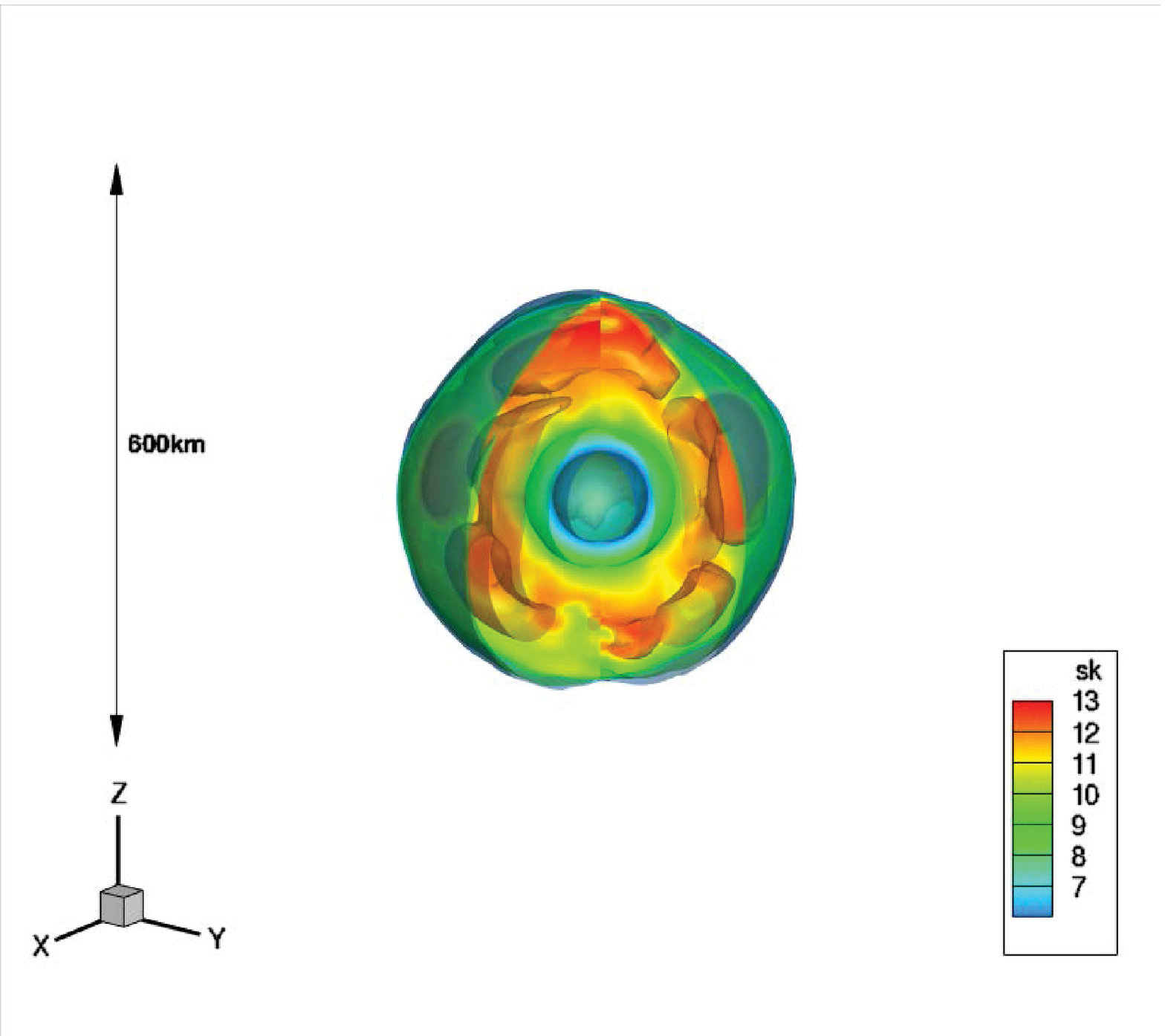}
\plotone{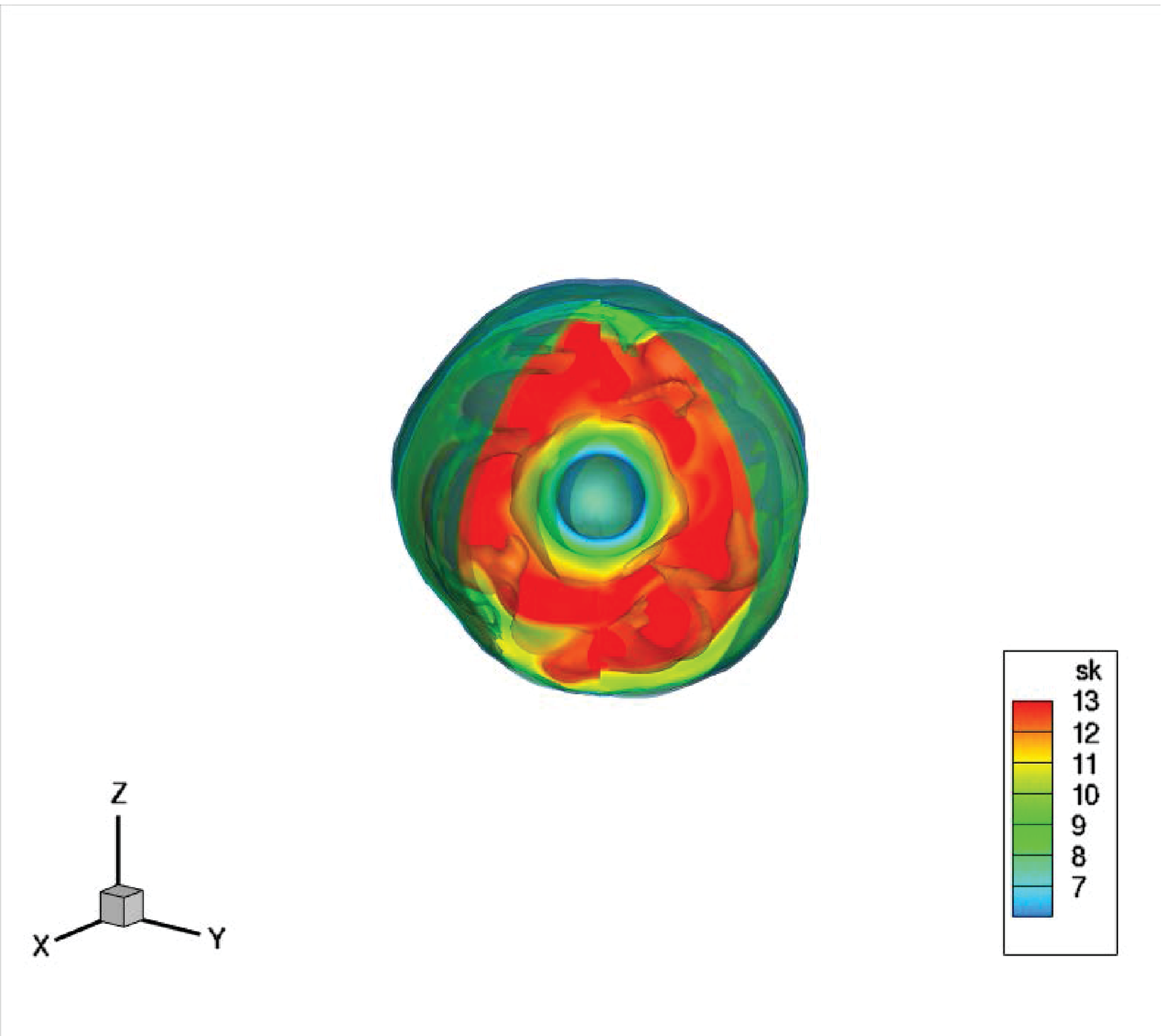}
\caption{
Profiles of the 3D supernova core from the simulation of 
27.0M$_{\odot}$ star \citep{hor14,tak14b}.  
Iso-surfaces of the entropy per baryon 
in the unit of the Boltzmann constant, $k_B$, 
are shown for the snapshots at 150 and 200 ms after the bounce 
in the left and right panels, respectively.  
}
\label{fig:3d-entropy.3d.s27}
\end{figure}

\begin{figure}
\epsscale{0.48}
\plotone{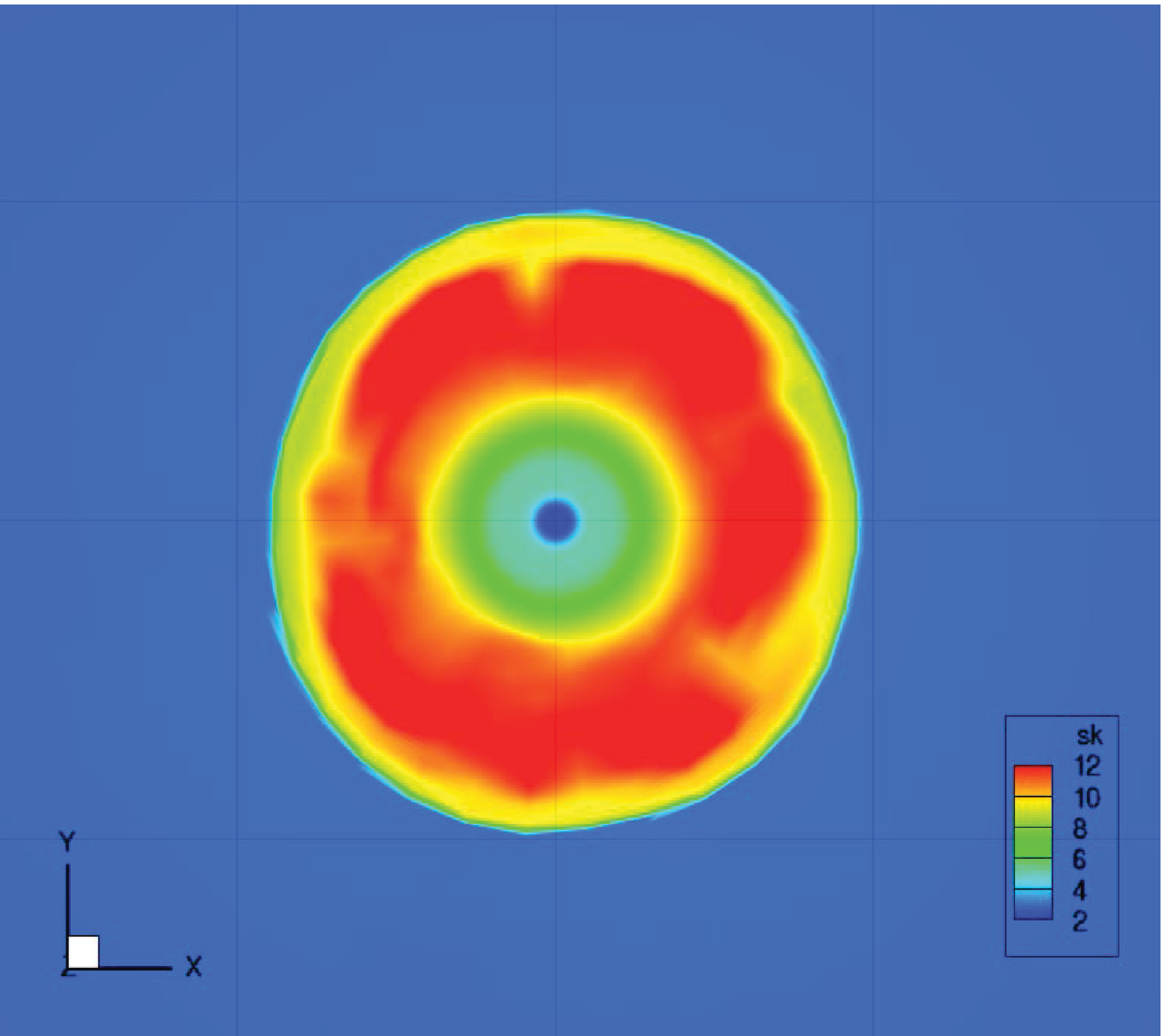}
\plotone{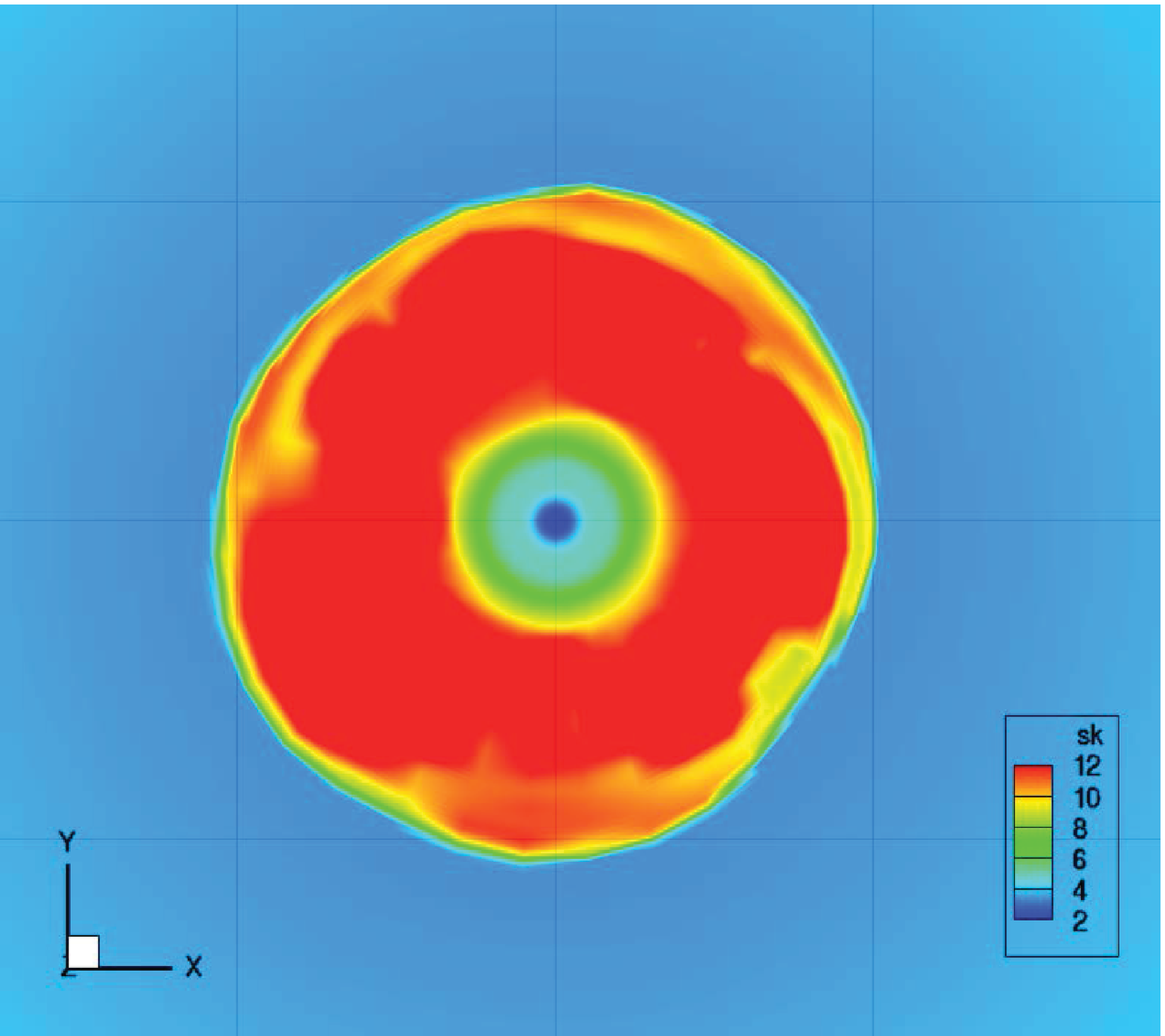}
\caption{
Profiles of the 3D supernova core (27M) are shown 
for the snapshots at 150 and 200 ms after the bounce 
in the left and right panels, respectively.
Entropy per baryon [$k_B$] is plotted by color maps 
on the xy-plane (z=0).  
Grid lines with 200 km spacing are shown in the background.  
}
\label{fig:3d.s27-xyslice-entropy}
\end{figure}

\begin{figure}
\epsscale{0.4}
\plotone{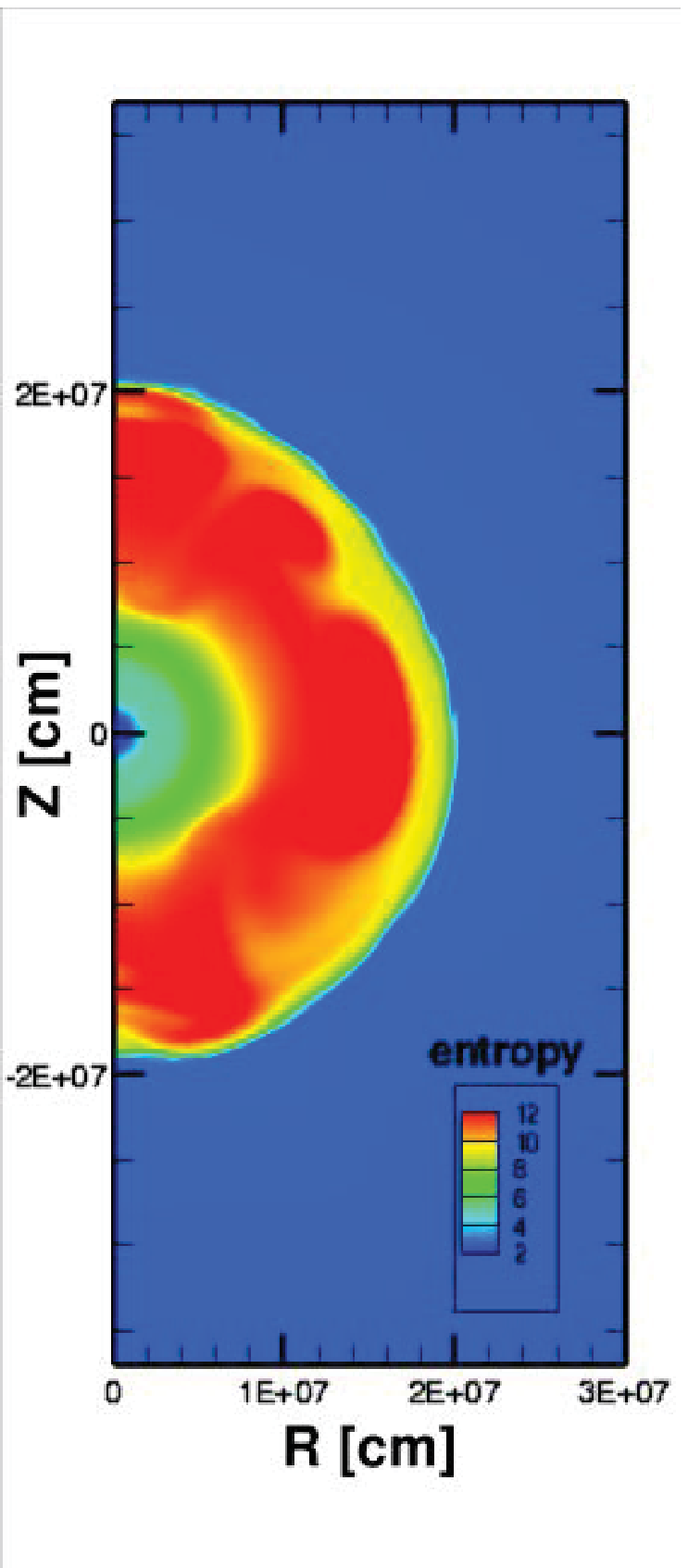}
\plotone{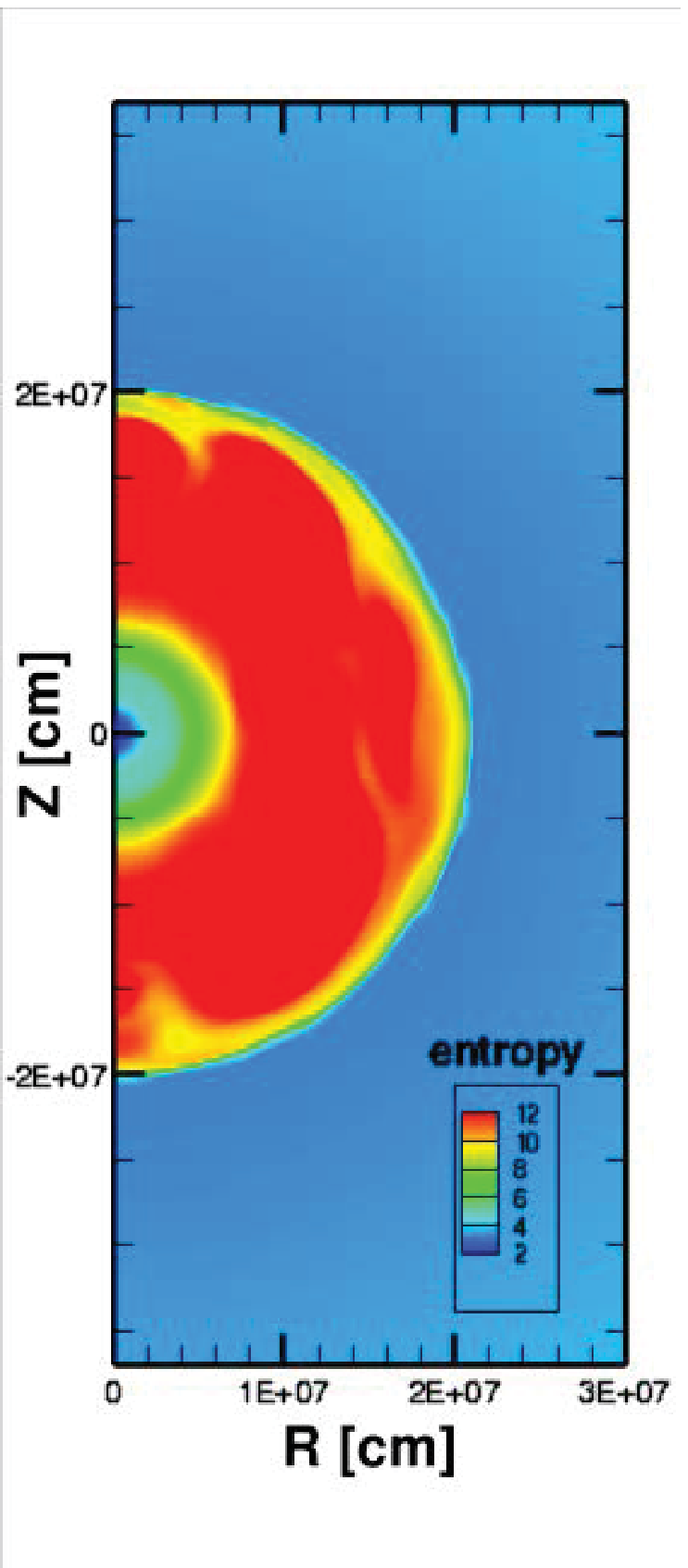}
\caption{
Profiles of the 3D supernova core (27M) at 150 and 200 ms after the bounce 
are shown on the meridian slice at $\phi$=51$^\circ$ 
in the left and right panels, respectively.  
Entropy per baryon [$k_B$] is shown by color maps.  
}
\label{fig:3d.s27-phi.slice.iph05}
\end{figure}

In Figure \ref{fig:3d-entropy.3d.s27}, we show the 
3D snapshots for the 27.0M$_{\odot}$ model 
at 150 and 200 ms after the bounce.  
In the supernova simulation, the shock wave 
stalls and is hovering 
around 200 km up to 700 ms.  
There is no sign of shock revival.  
Figures \ref{fig:3d.s27-xyslice-entropy} 
and \ref{fig:3d.s27-phi.slice.iph05} 
present the color maps of entropy 
at 150 and 200 ms, respectively, on the xy-plane (z=0) and 
the meridian slice at $\phi$=51$^\circ$.  
The shape of the stalled shock wave is nearly spherical 
with a slight deformation due to the convection in the heating region.  
High-entropy blobs grow and move upward and 
low-entropy matter falls downward 
in channels formed between the mushroom-like shapes.  
The convective motion with the down flows leads to 
the non-uniform composition 
in the central core as in 
Figs. \ref{fig:3db-phi.slice.iph05} 
and \ref{fig:3db-phi.slice.iph13} for the 11M model.  
We note that there is no periodic pattern in the azimuthal 
direction in this model.  

\section{Neutrino distributions}

\subsection{Density and flux by Boltzmann evaluation}\label{section:d-boltz}

We show the 3D features of the neutrino distributions and 
its associated moments obtained 
for the snapshots of the 3D supernova core.  
We mainly display the numerical results 
for the snapshot of 11M at 150 ms after the bounce.  
The 3D characteristics of the neutrino transfer shown below 
are common also for the other snapshots.  

\begin{figure}
\epsscale{0.9}
\plotone{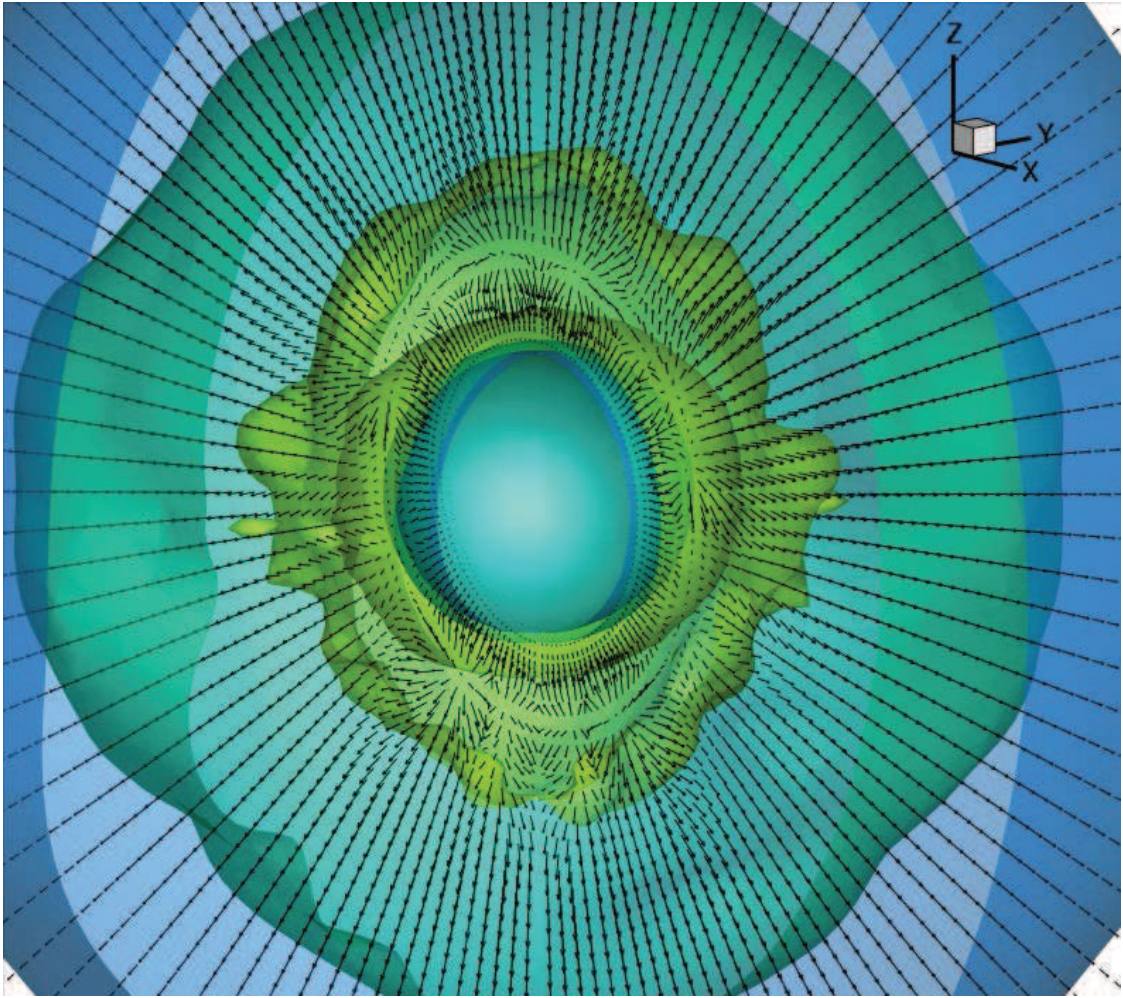}
\caption{Iso-surfaces of the density of 
electron-type anti-neutrinos ($\bar{\nu}_e$) 
for the 11M model at 150 ms after the bounce.  
Arrows represent the flux vector of neutrinos.  }
\label{fig:3db-density.3d.in2}
\end{figure}

We show in Fig. \ref{fig:3db-density.3d.in2} 
the iso-surfaces of the density of 
electron-type anti-neutrinos ($\bar{\nu}_e$)
with arrows of its flux vector.  
The distribution of neutrinos is spherical at center, 
but deformed in the surrounding region, 
reflecting the convective high entropy region.  
The corresponding fluxes are non-radial 
according to the gradients of the deformed distributions.  
The distribution further out is again spherical 
due to the outward neutrino fluxes at large radii.  
These 3D features are seen also for the densities 
and fluxes for electron-type neutrinos ($\nu_e$) and 
$\mu$-type neutrinos ($\nu_{\mu}$).  

\begin{figure}
\epsscale{0.45}
\plotone{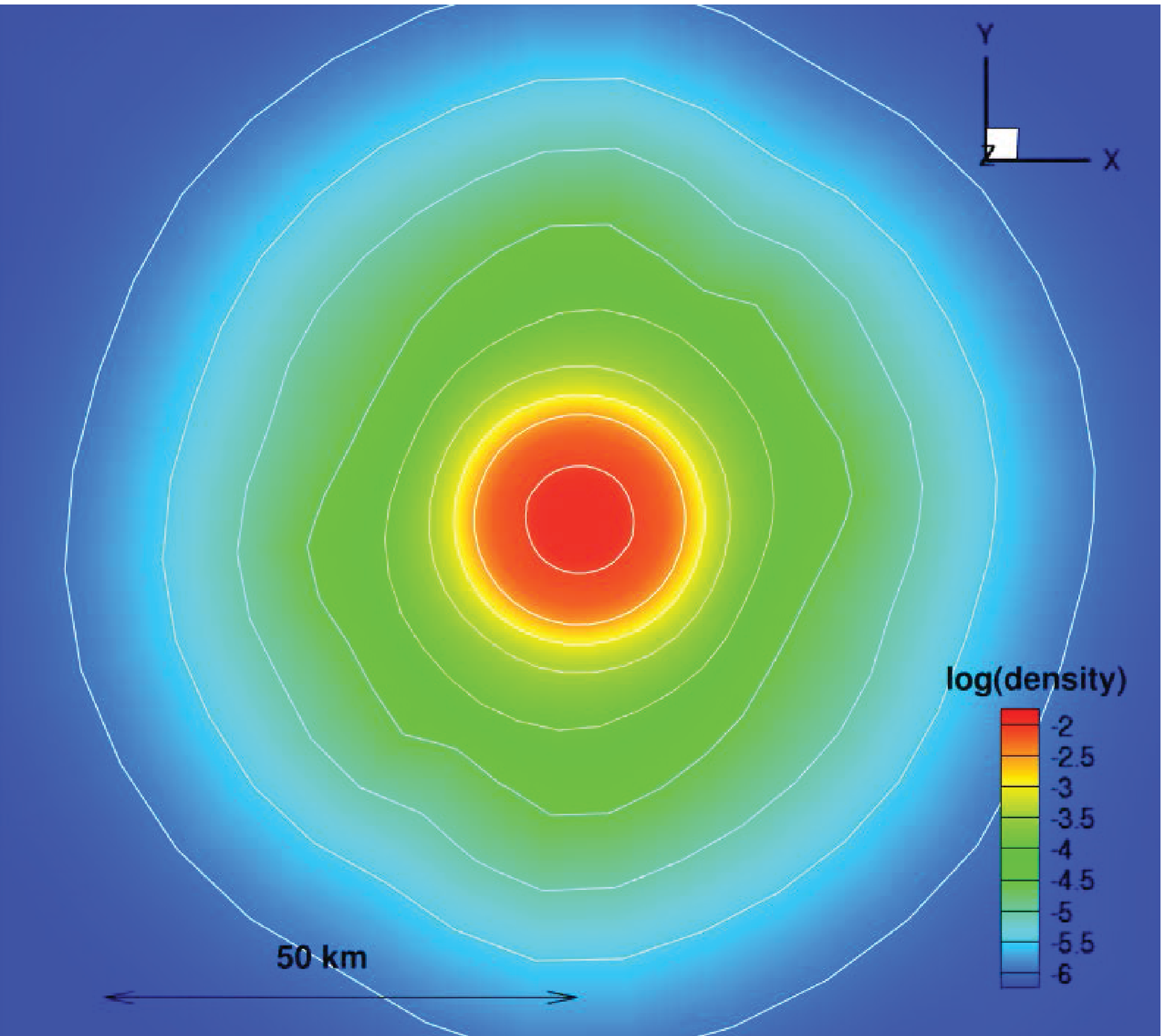}\\
\plotone{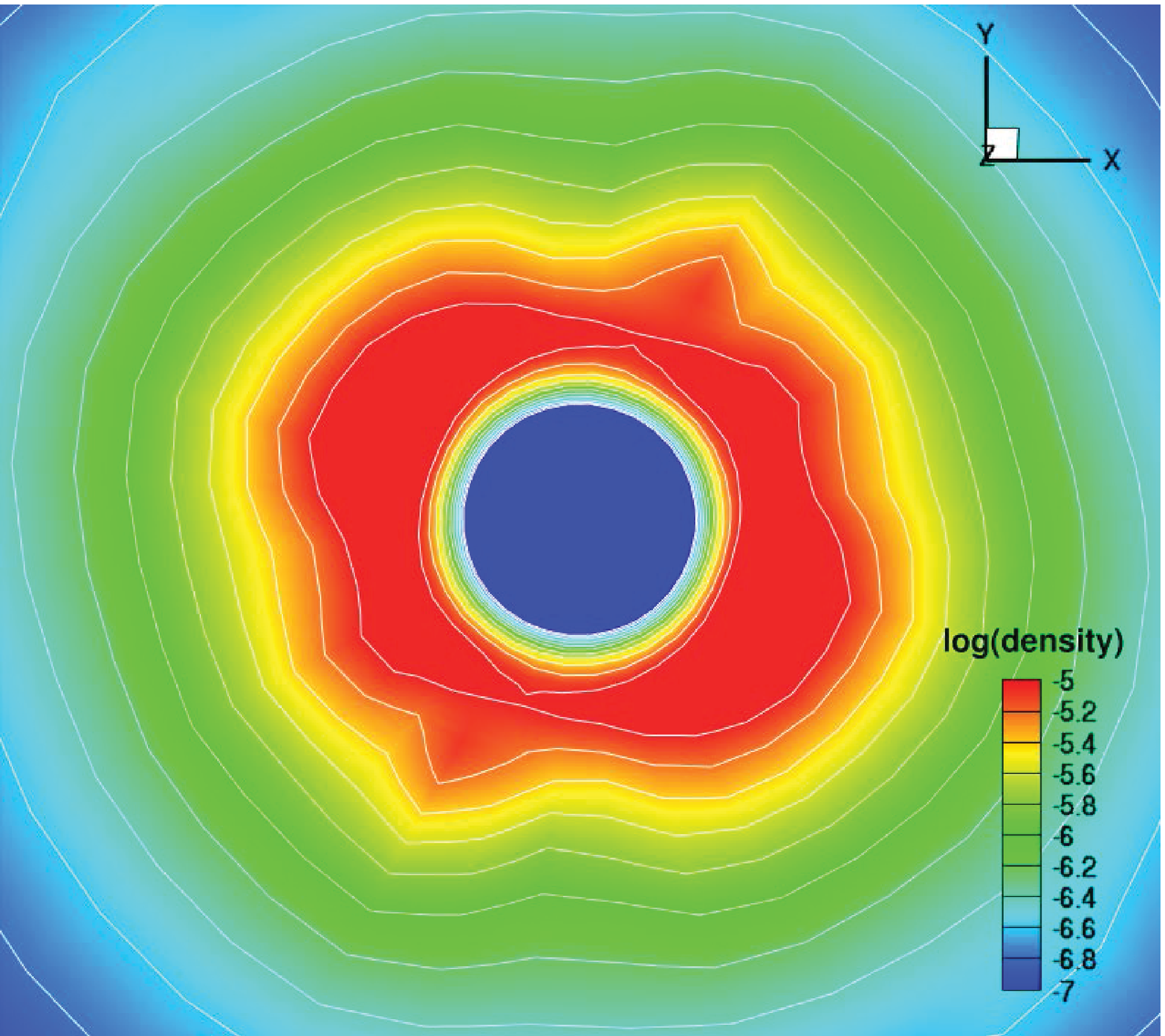}\\
\plotone{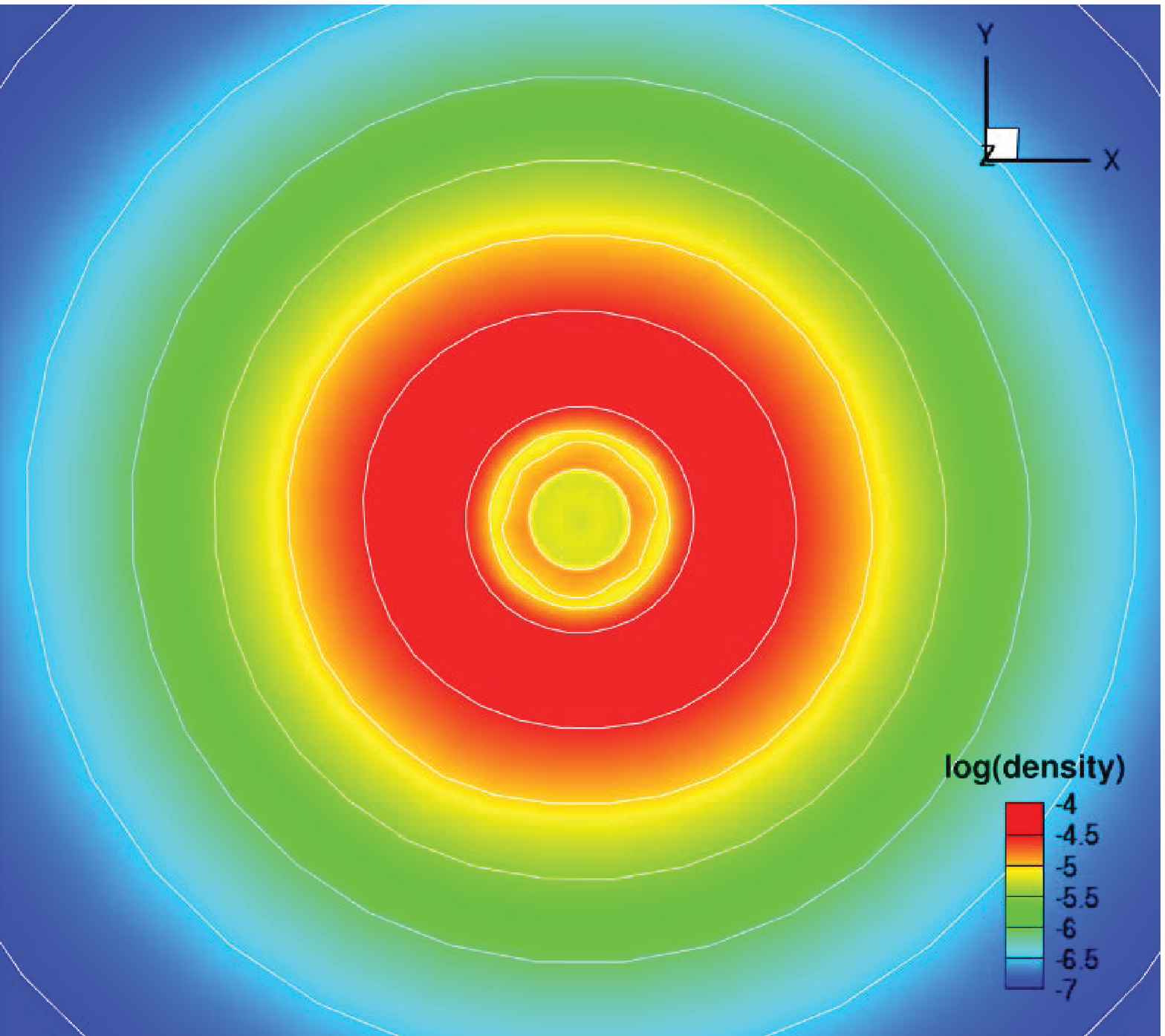}
\caption{Color maps of the neutrino density 
in log scale in the unit of fm$^{-3}$ 
for three species ($\nu_e$, $\bar{\nu}_e$ and $\nu_{\mu}$) 
on the xy-plane (z=0) are shown 
for the 11M model at 150 ms after the bounce 
in the top, middle and bottom panels, respectively.  }
\label{fig:3db-density.xyslice}
\end{figure}

Figure \ref{fig:3db-density.xyslice} displays 
the color maps of 
the density of neutrinos for three species 
($\nu_e$, $\bar{\nu}_e$ and $\nu_{\mu}$) 
on the xy-plane (z=0).  
The $\nu_e$-distribution is focused 
inside the proto-neutron star 
with high chemical potential of neutrinos.  
The $\bar{\nu}_e$- and $\nu_{\mu}$-distributions 
are concentrated in the off-center region, 
where the degeneracy is low due to high entropies.  
Among the three species, 
the $\bar{\nu}_e$-distribution exhibits 
the most non-spherical shape.  
The $\nu_e$-distribution is spherical at center 
and slightly deformed in the surrounding region.  
They are affected by the non-spherical shape of 
the degeneracy of neutrinos ($\eta_{\nu}=\mu_{\nu}/T$).  
The $\nu_{\mu}$-distribution is entirely spherical, 
reflecting the spherical shape of temperature distribution 
as seen in Fig. \ref{fig:3db-xyslice}.

\begin{figure}
\epsscale{0.48}
\plotone{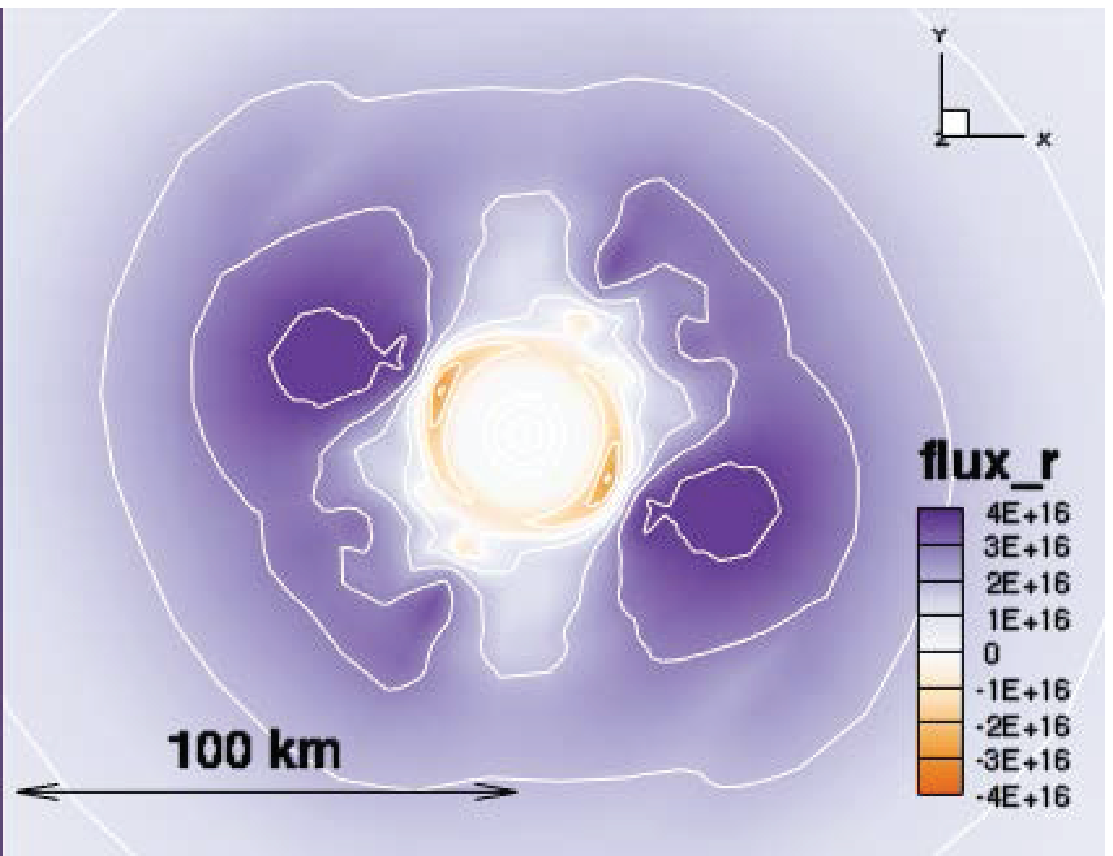}\\
\plotone{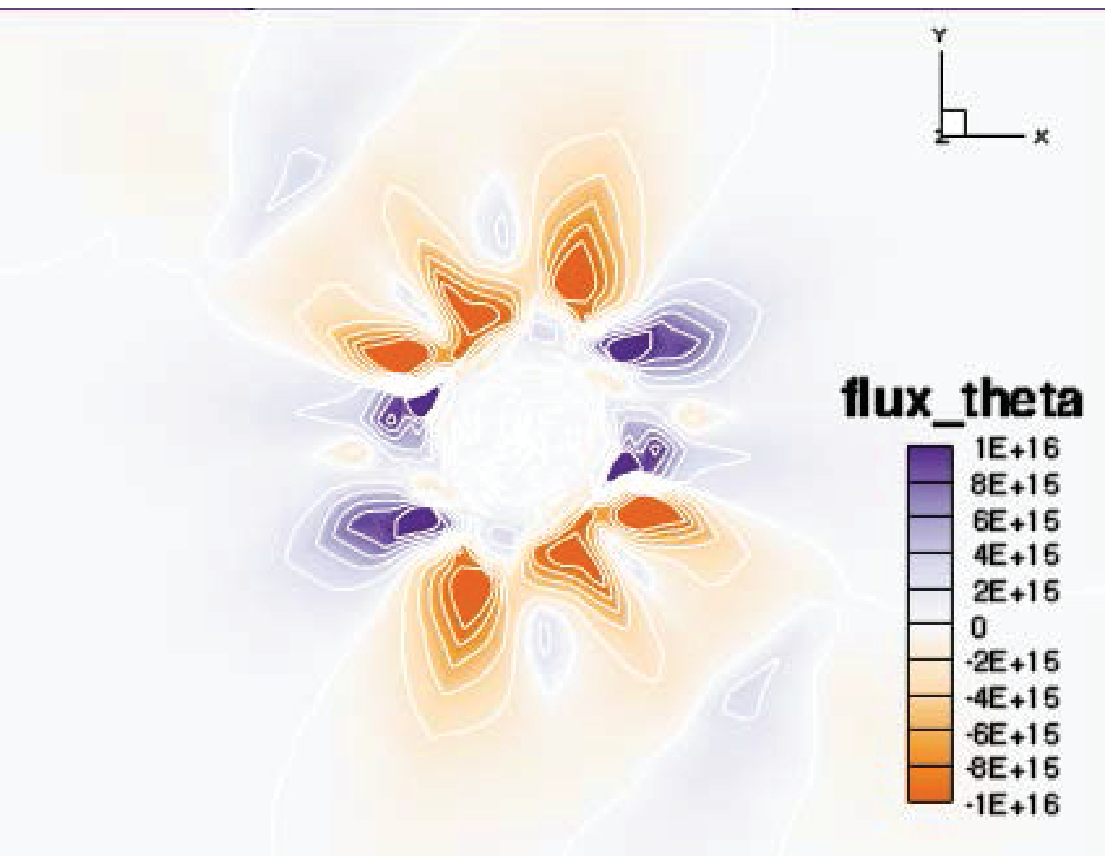}\\
\plotone{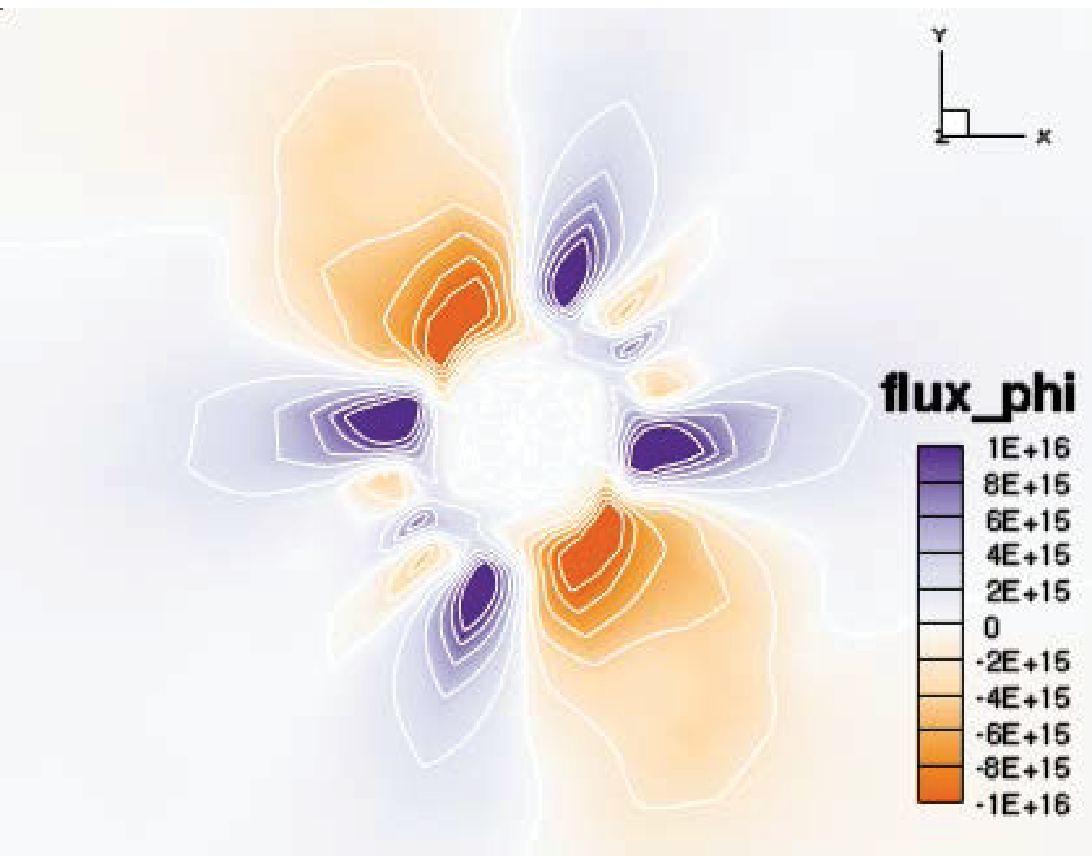}
\caption{Color maps of the number flux of $\bar{\nu}_e$ 
in the unit of fm$^{-2}$s$^{-1}$ 
on the xy-plane (z=0) are shown 
for the 11M model at 150 ms after the bounce.  
The radial, polar and azimuthal fluxes are shown 
in the top, middle and bottom panels, respectively.  }
\label{fig:3db-flux.xyslice}
\end{figure}

The 6D Boltzmann solver handles the advection in 3D space 
and, therefore, describes the non-radial advection 
as well as the radial advection.  
Figure \ref{fig:3db-flux.xyslice} reveals 
the three components of the flux of $\bar{\nu}_e$ 
on the xy-plane (z=0) 
corresponding to Fig. \ref{fig:3db-density.xyslice}.  
There are significant contributions of polar ($\theta$) 
and azimuthal ($\phi$) fluxes 
due to the deformed $\bar{\nu}_e$-distribution.  
The direction of the flux follows roughly 
the gradient of neutrino distribution in the central region.  
The magnitude of the non-radial component of the flux 
reaches more than 50 $\%$ of the radial component 
around 20--50 km inside the proto-neutron star.  

The 6D Boltzmann solver enables 
the seamless description of non-radial fluxes 
in the wide region 
while the diffusion approximation is applicable 
only under the optically thick condition.  
The non-radial fluxes remain significant 
up to large radii, going out the optically thick region.  
The neutrino transport is diffusive well inside a radius of 50 km 
and transitional to the free-streaming around 100 km 
as we will see that 
the second angle moment, $\langle \mu_{\nu}^2 \rangle$, 
is less than 0.35 inside 50 km 
in Figs. \ref{fig:3db-mom.mu.iph05.slice} and 
\ref{fig:3db-mom.mu.iph13.slice}.  

\begin{figure}
\epsscale{0.75}
\plotone{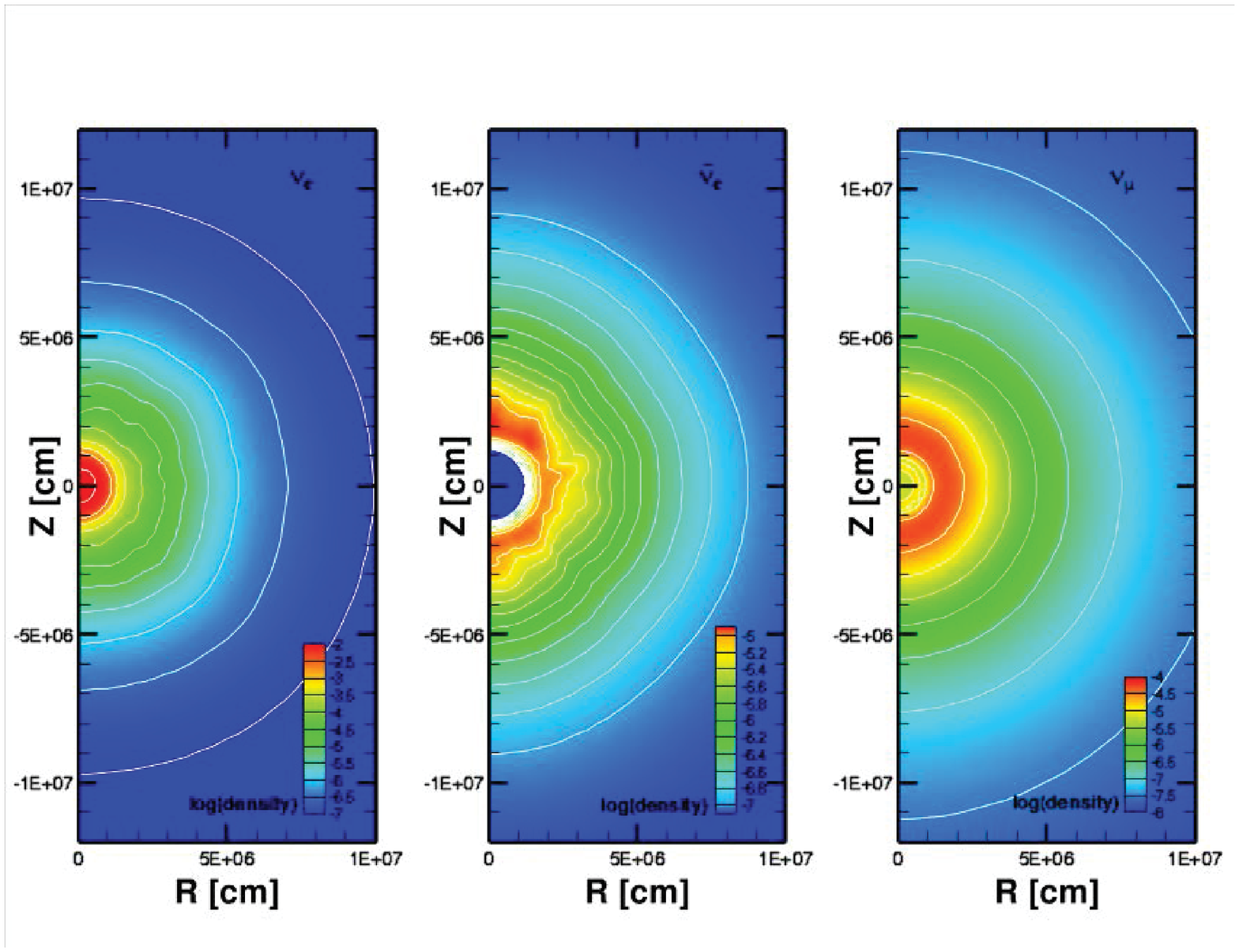}\\
\plotone{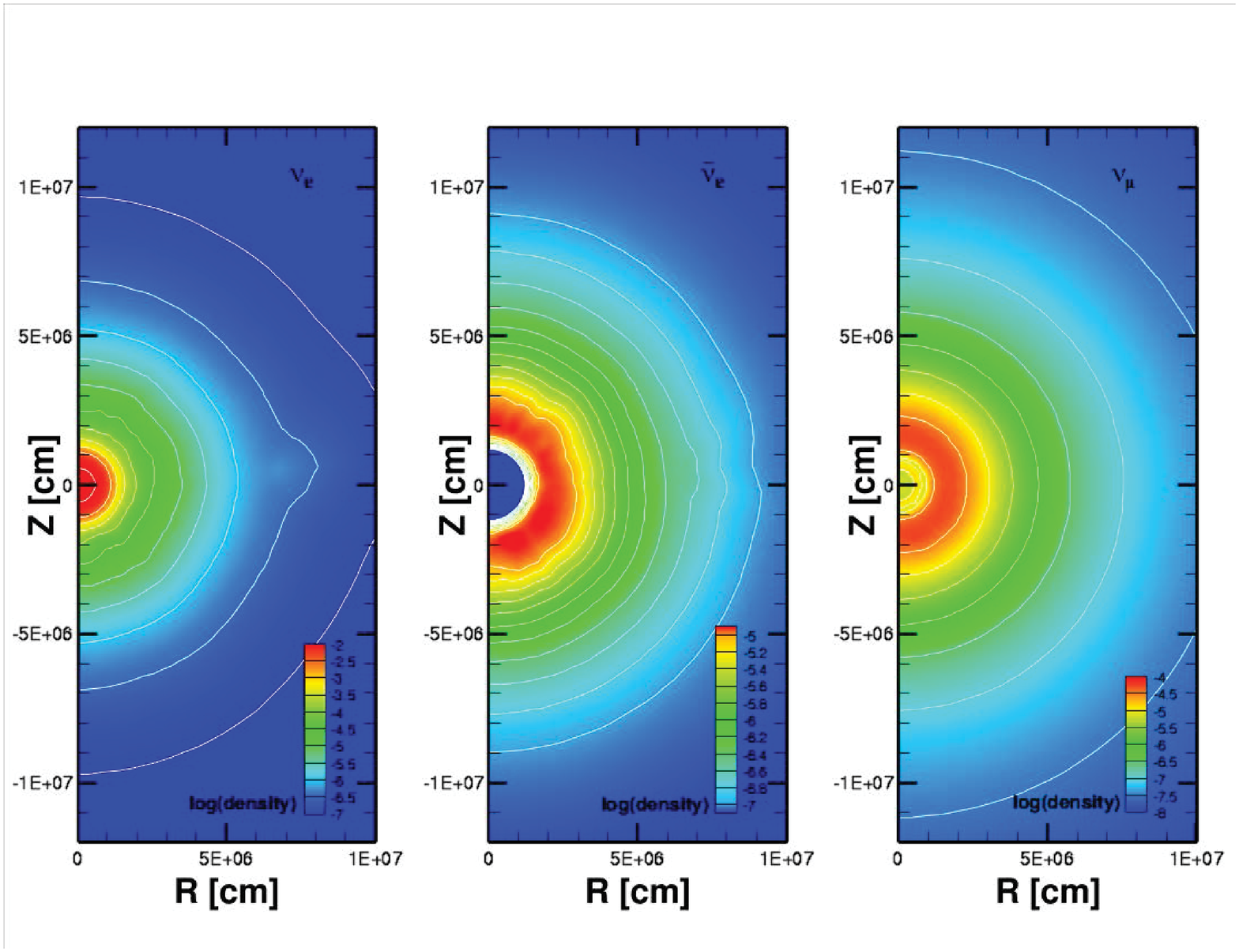}
\caption{Profiles of the neutrino density 
on the meridian slices at $\phi$=51$^{\circ}$ (top) 
and 141$^{\circ}$ (bottom) are shown 
for the 11M model at 150 ms after the bounce.  
The neutrino densities 
in log scale in the unit of fm$^{-3}$ 
for three species ($\nu_e$, $\bar{\nu}_e$ and $\nu_{\mu}$) 
are plotted by color maps 
in the left, middle and right panels, respectively.  }
\label{fig:3db-density.iphxx.slice}
\end{figure}

Figure \ref{fig:3db-density.iphxx.slice} shows 
the neutrino densities for three species 
in the central region 
on the meridian slices at $\phi$=51$^\circ$ and 141$^\circ$.  
The deformed shapes of density distributions 
are seen for $\nu_e$ and $\bar{\nu}_e$ 
while the shape is spherical for $\nu_{\mu}$ 
as we saw in Fig. \ref{fig:3db-density.xyslice}.  
The $\bar{\nu}_e$ and $\nu_{\mu}$ are abundant 
in the off-center region due to their origin 
of thermal production and 
the neutrino degeneracy for the former.  
There is a bump around the equator 
in the case of $\phi$=141$^\circ$ 
for $\nu_e$ and $\bar{\nu}_e$.  
This corresponds to the narrow channel 
with the proton-rich and low entropy condition 
seen in Fig. \ref{fig:3db-phi.slice.iph13}.  
The production of neutrinos is amplified 
by local enhancement of thermodynamical quantities, 
which we discuss later.  

\begin{figure}
\epsscale{0.75}
\plotone{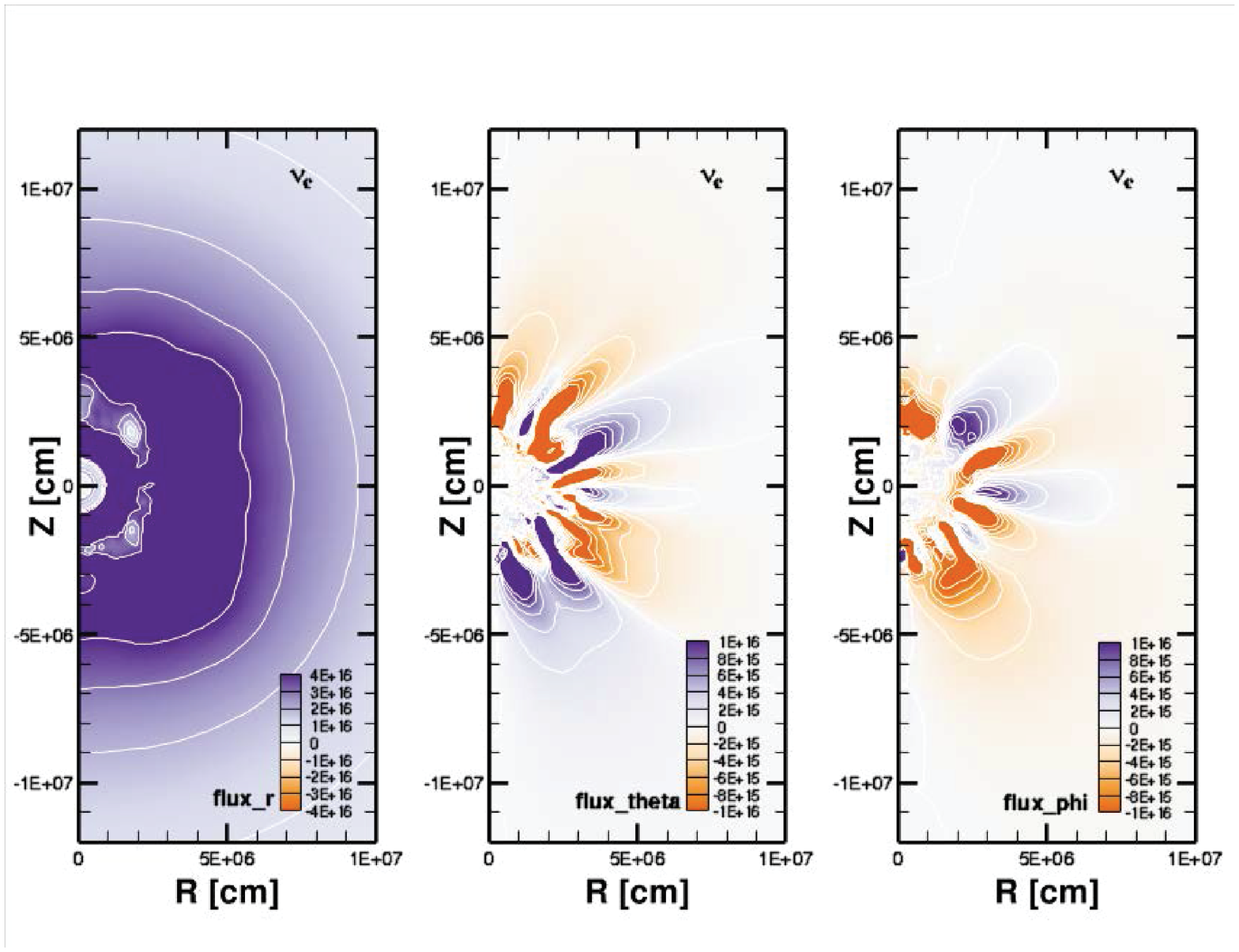}\\
\plotone{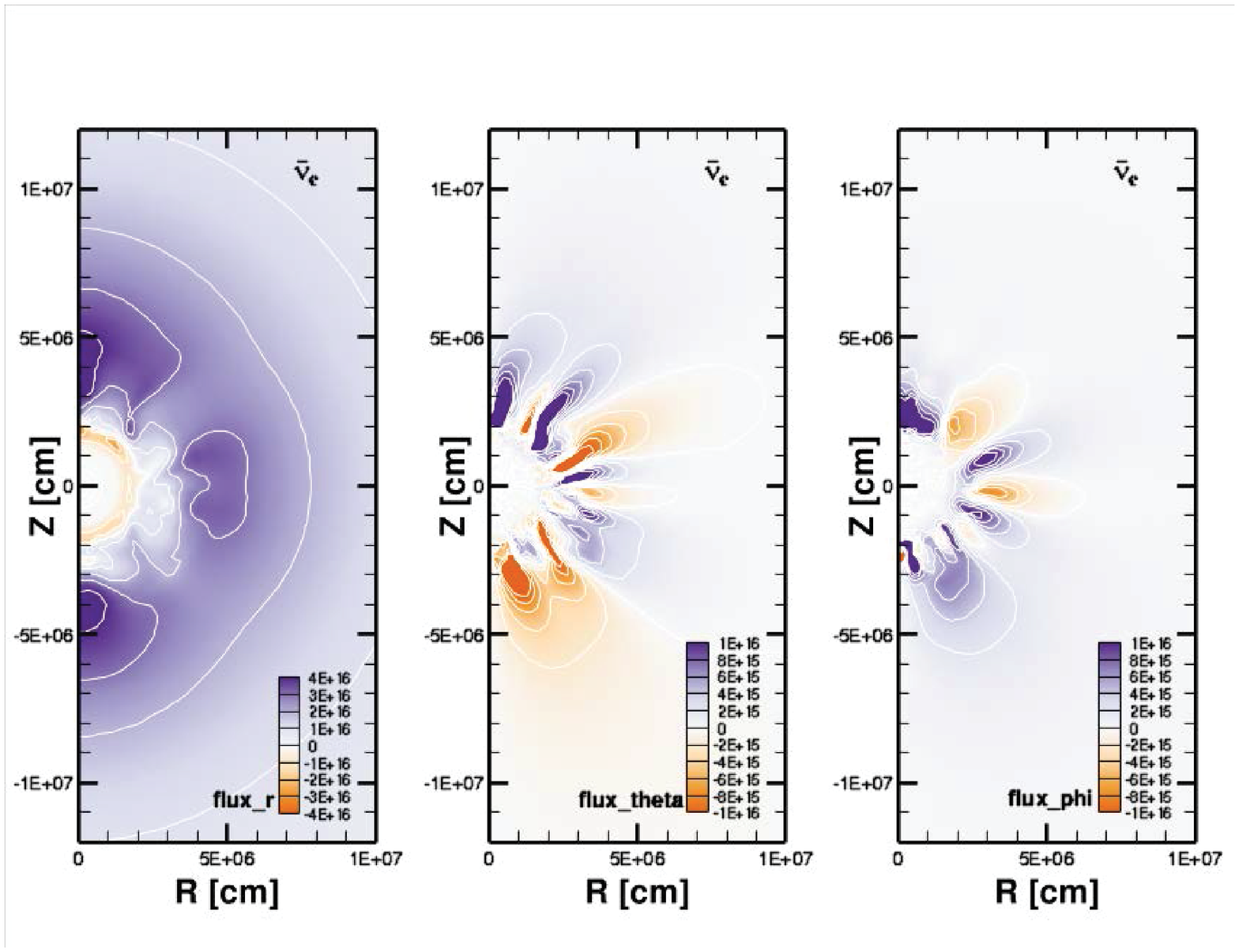}
\caption{Profiles of the number flux of $\nu_e$ (top) 
and $\bar{\nu}_e$ (bottom) 
in the unit of fm$^{-2}$s$^{-1}$ 
on the meridian slice at $\phi$=51$^{\circ}$ are shown 
for the 11M model at 150 ms after the bounce.  
The radial, polar and azimuthal fluxes are plotted by color maps 
in the left, middle and right panels, respectively.  }
\label{fig:3db-flux.iph05.slice.inx}
\end{figure}

\begin{figure}
\epsscale{0.75}
\plotone{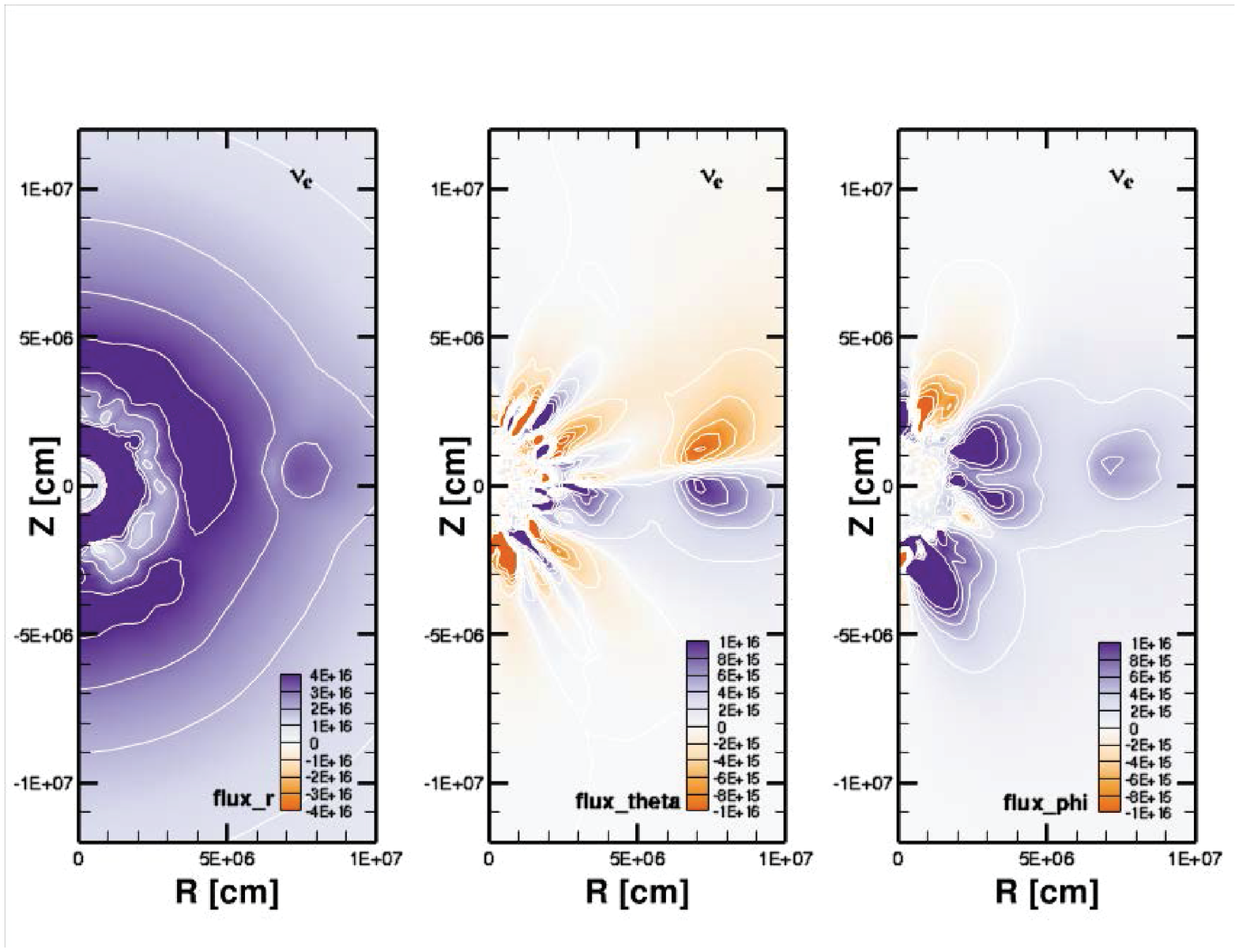}
\plotone{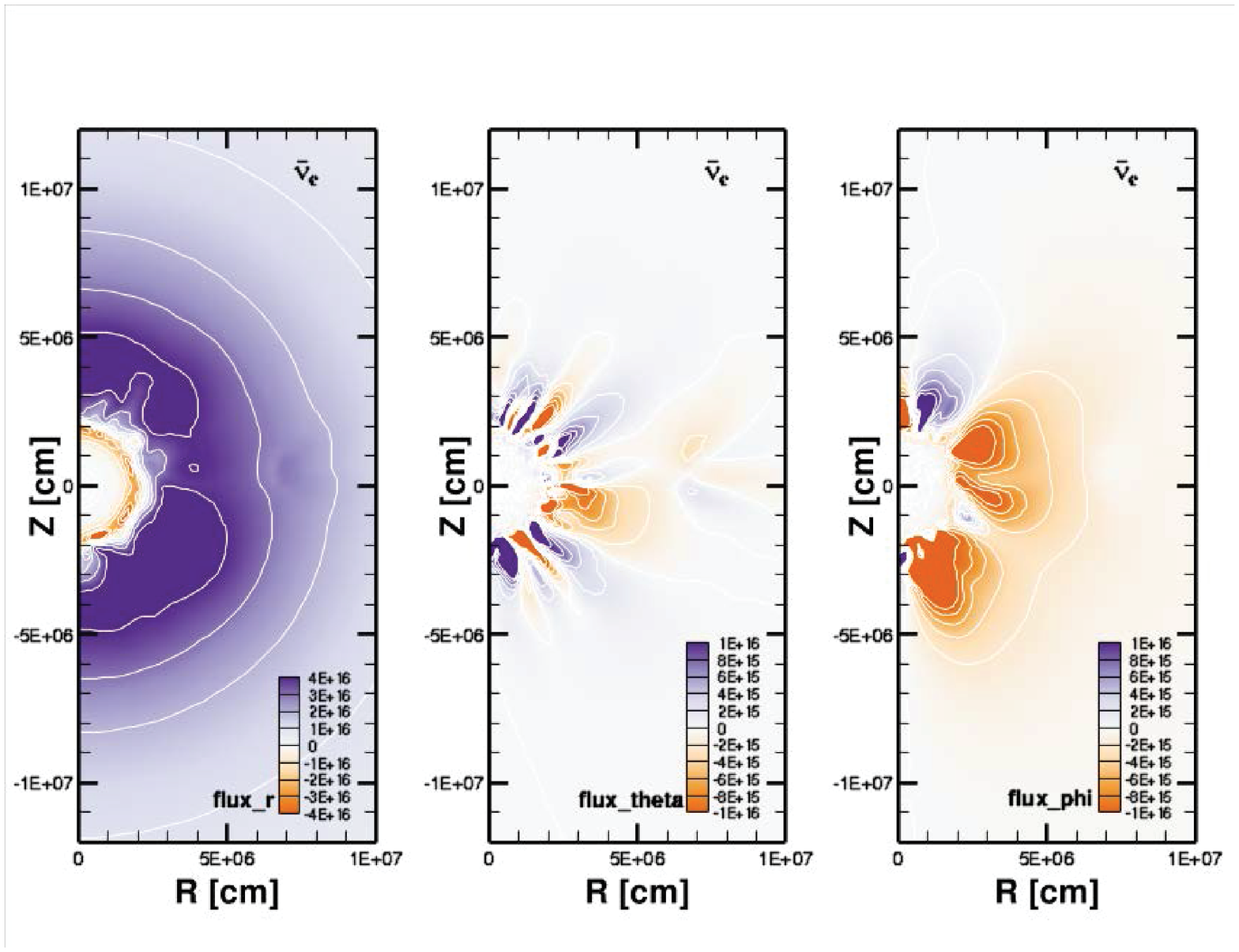}
\caption{Same as Figure \ref{fig:3db-flux.iph05.slice.inx} 
but on the meridian slice at $\phi$=141$^{\circ}$.  }
\label{fig:3db-flux.iph13.slice.inx}
\end{figure}

\begin{figure}
\epsscale{0.3}
\plotone{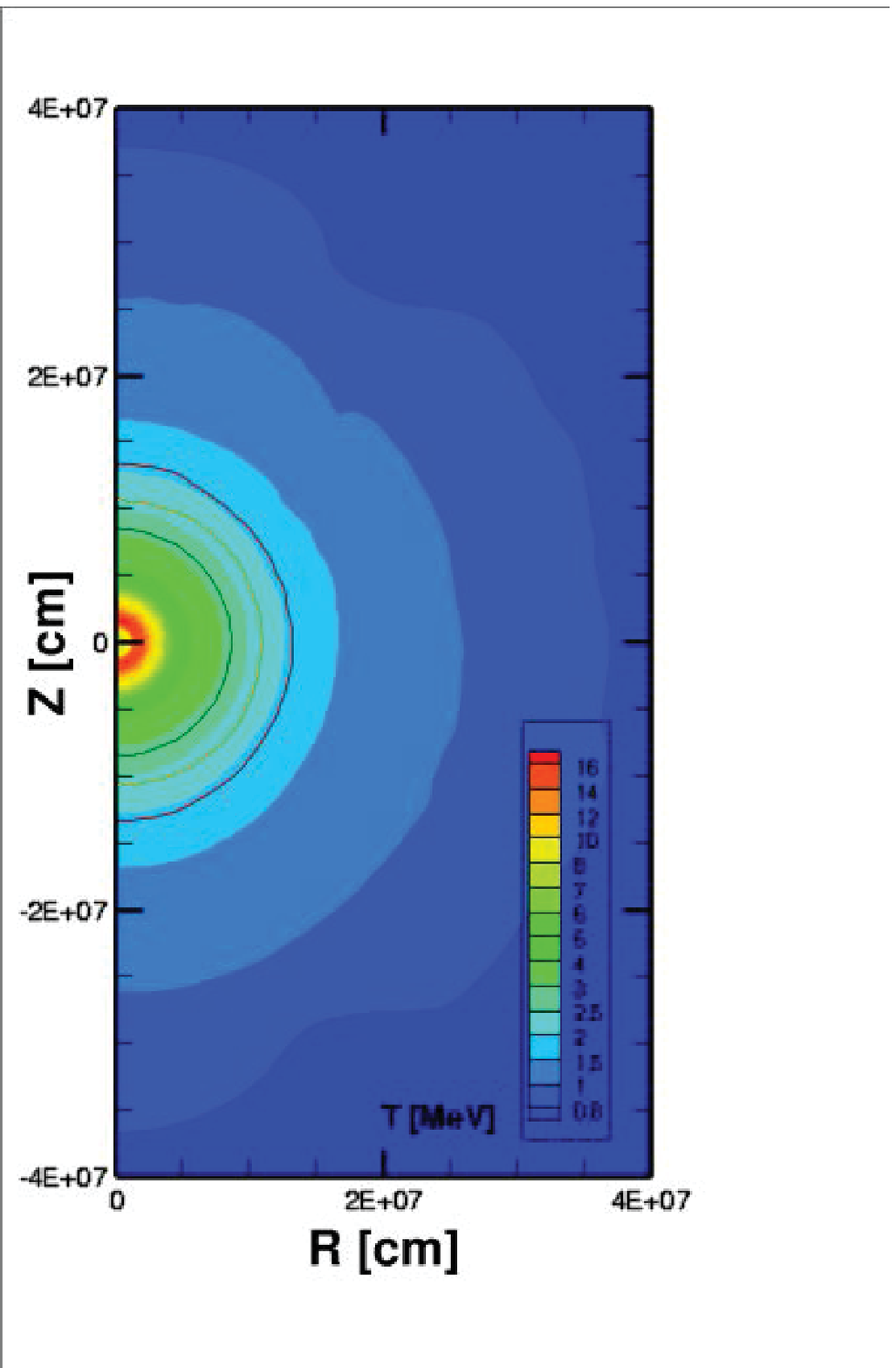}
\plotone{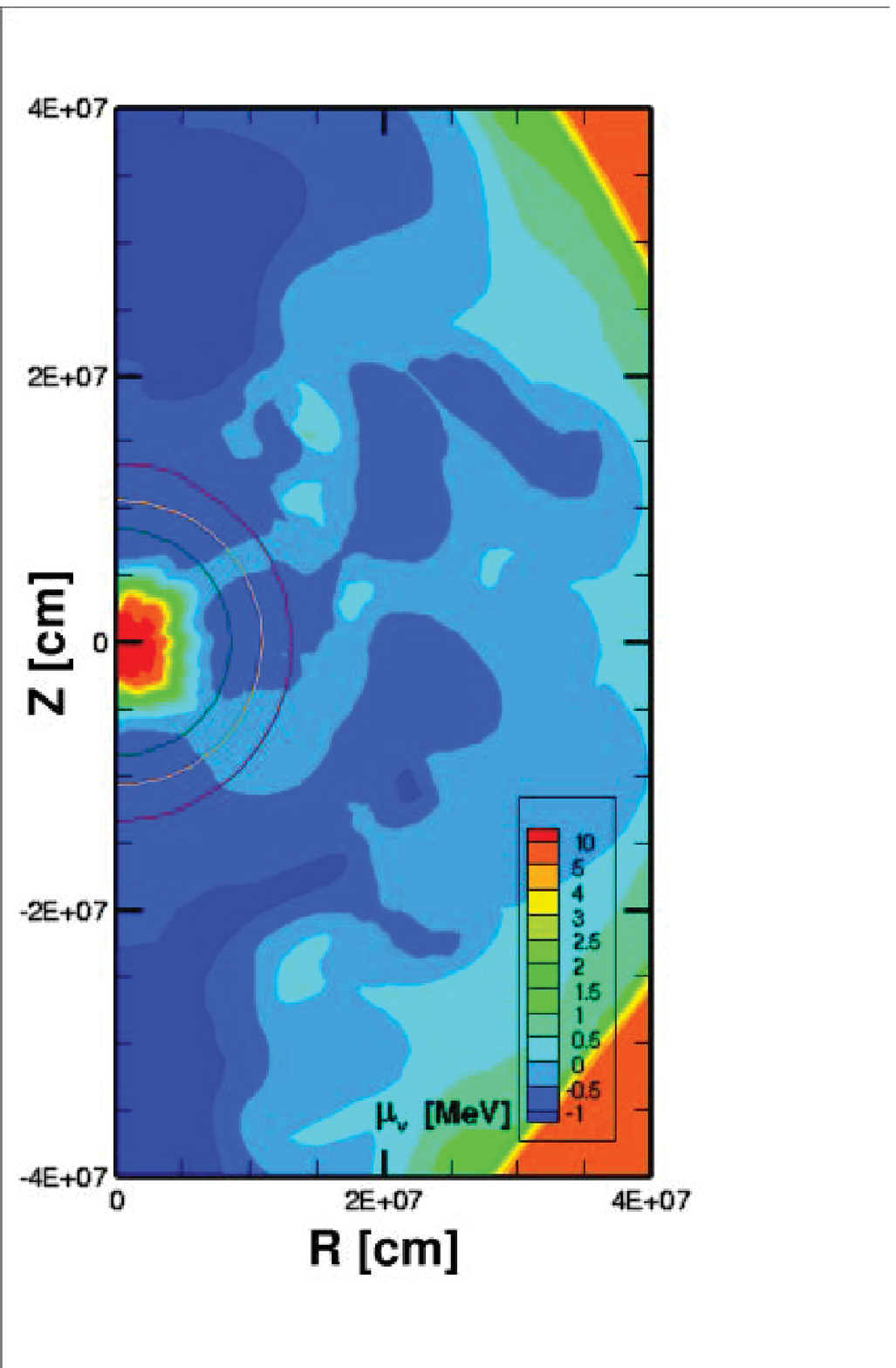}
\plotone{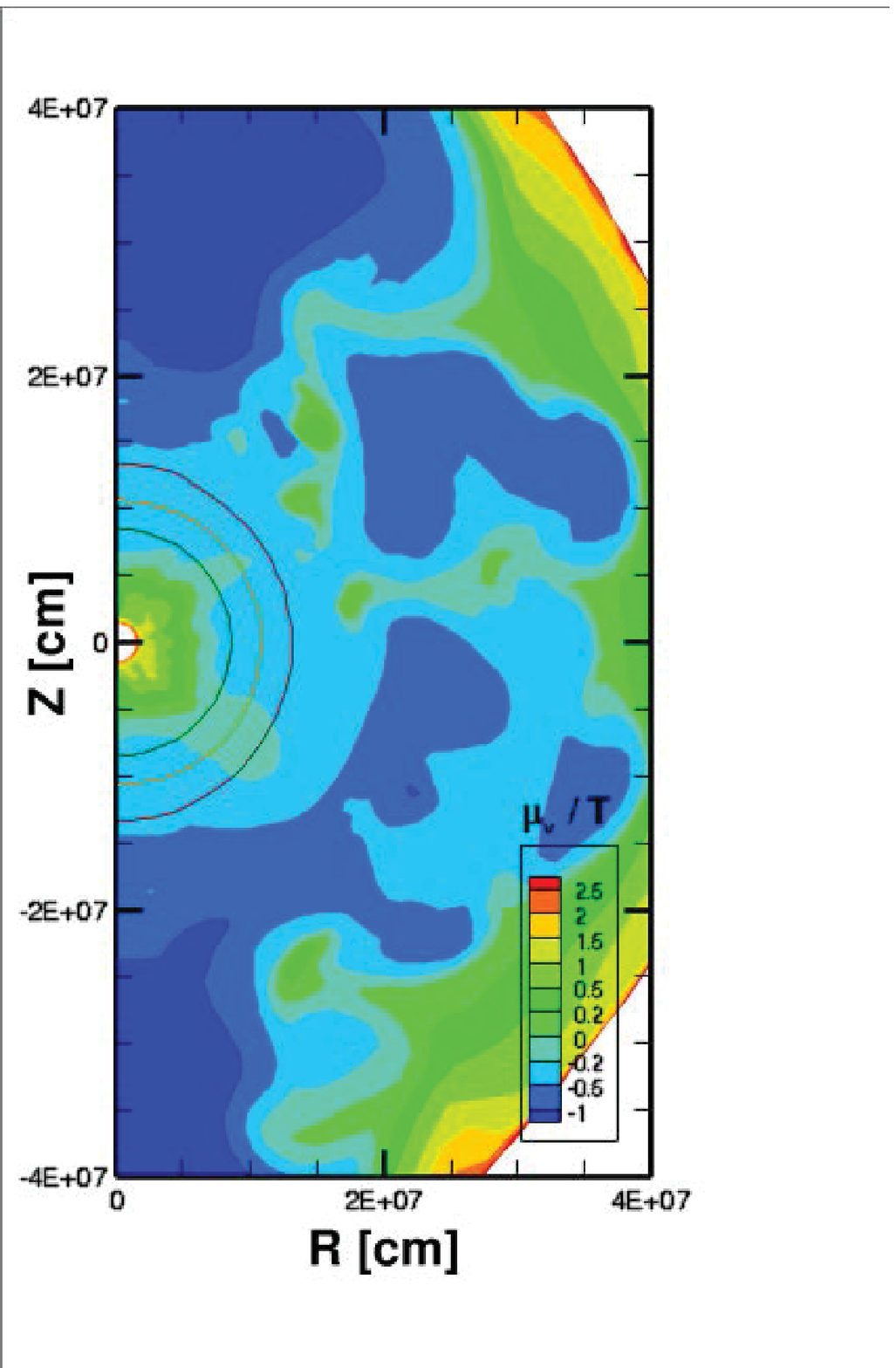}\\
\plotone{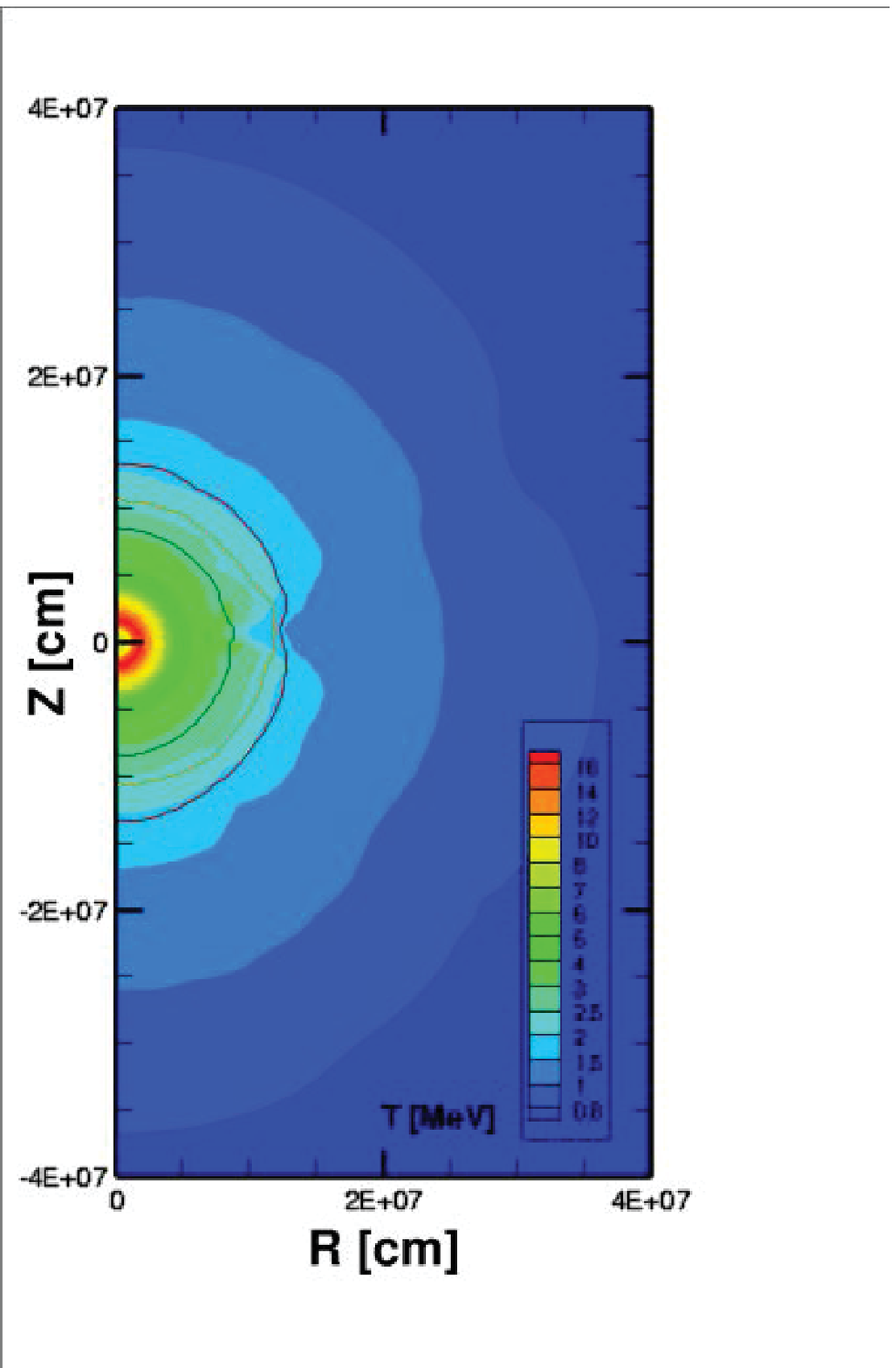}
\plotone{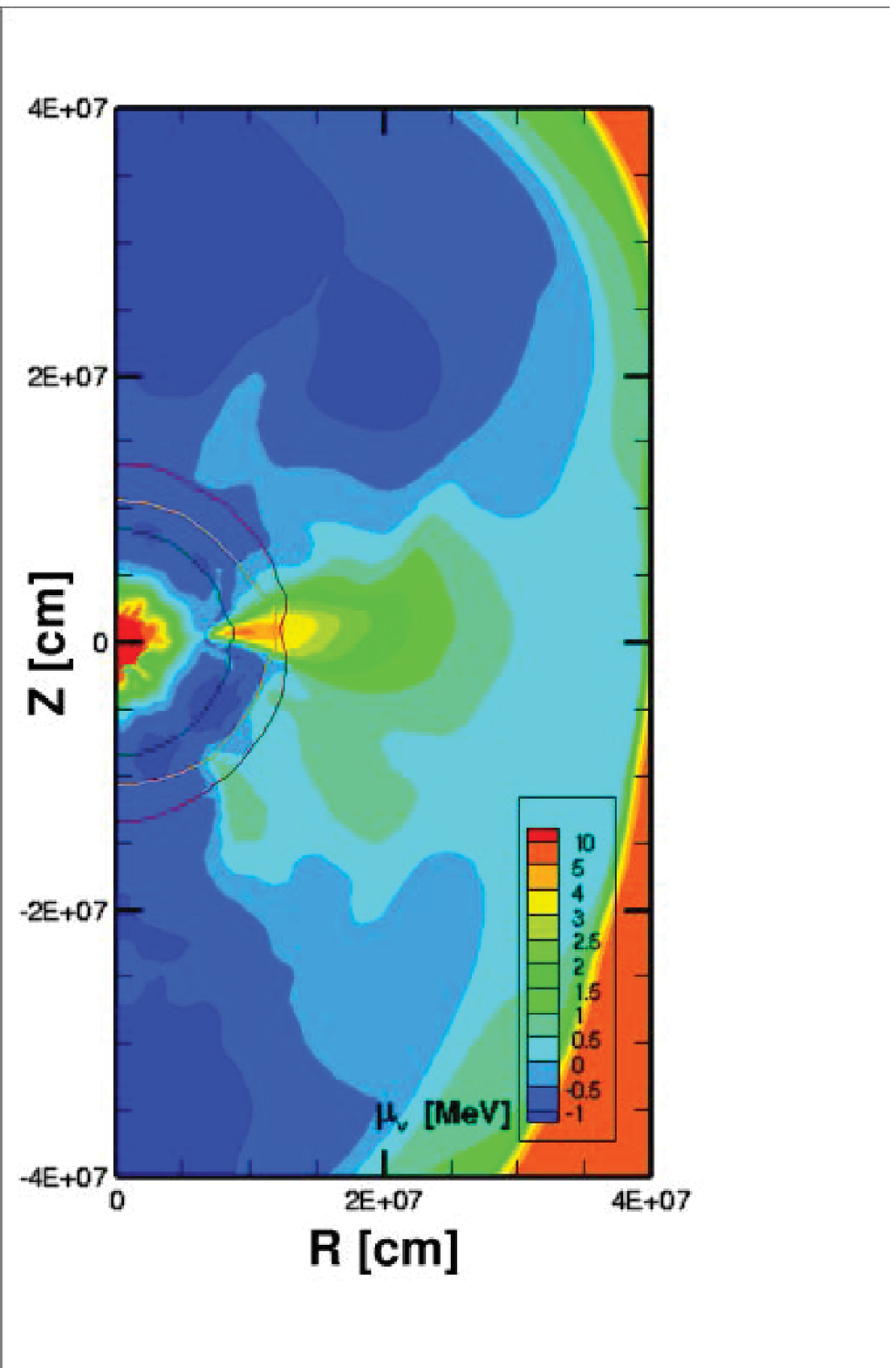}
\plotone{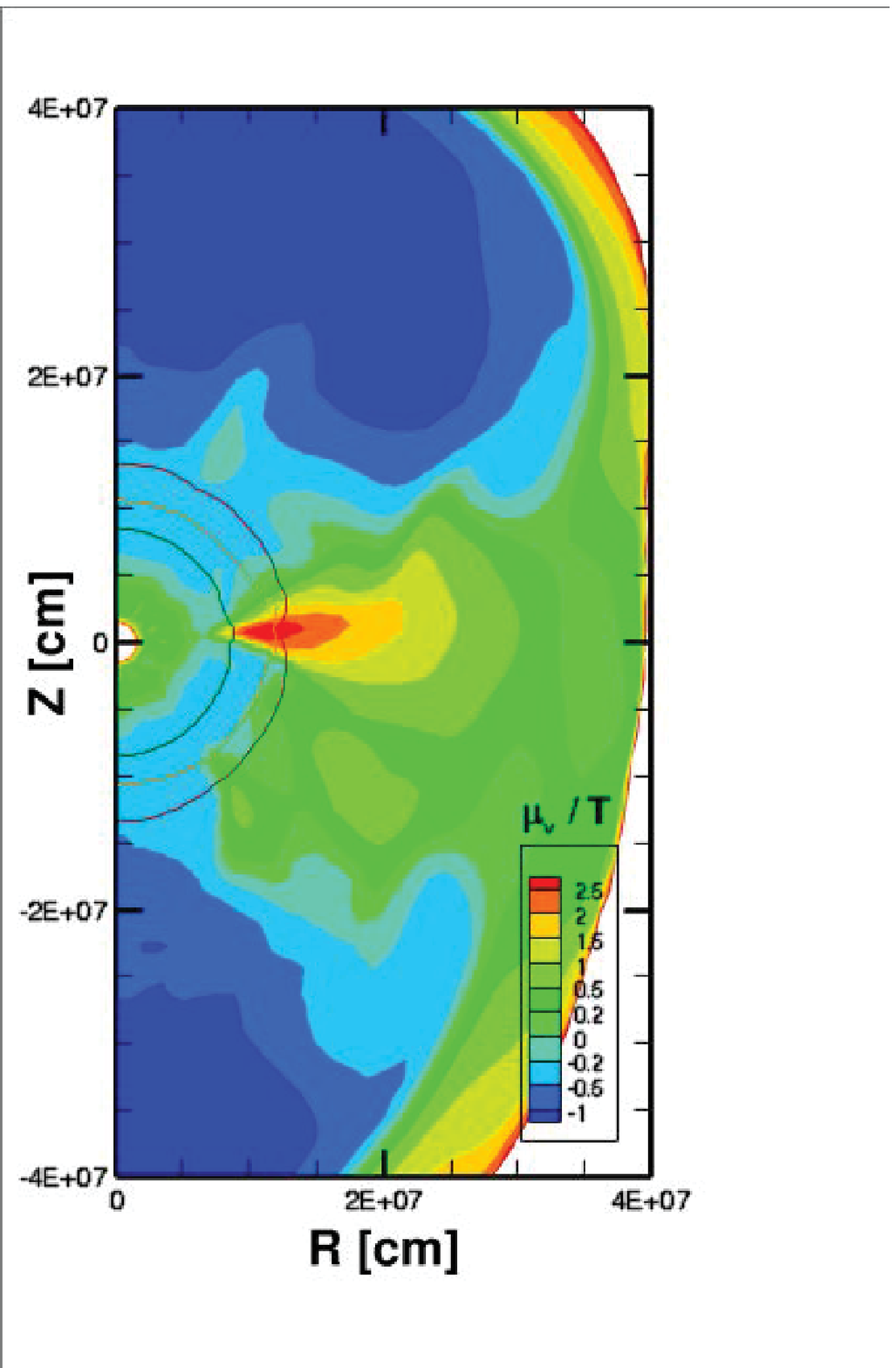}
\caption{Profiles of 
temperature [MeV] (left), chemical potential [MeV] (center) 
and degeneracy of neutrinos (right) 
with the location of neutrino-sphere 
on the meridian slices at $\phi$=51$^{\circ}$ (top) 
and 141$^{\circ}$ (bottom) 
are shown 
for the 11M model at 150 ms after the bounce.  
The neutrino-spheres for $\nu_e$, $\bar{\nu}_e$ and $\nu_{\mu}$ 
are shown in the order from outside to the center.  }
\label{fig:2d.phslice-etanu.3db.iph13}
\end{figure}

Prominent non-radial fluxes of $\nu_e$ and $\bar{\nu}_e$ 
can be seen in Figs. \ref{fig:3db-flux.iph05.slice.inx} 
and \ref{fig:3db-flux.iph13.slice.inx}, 
where the three components of flux are shown on the meridian slices.  
The polar component exhibits fine structure 
with alternating directions 
and the azimuthal component widely extends.  
It is remarkable to see that 
the directions of flux for $\nu_e$ and $\bar{\nu}_e$ 
are opposite each other.  
This is because 
the neutrino chemical potential and temperature 
(and their gradient) affect the distributions of two species 
in an opposite way.  
(See Fig. \ref{fig:2d.phslice-etanu.3db.iph13} and discussion below) 
The flux for $\nu_{\mu}$ is almost 
spherically symmetric (not shown here) 
and its radial flux is concentrated 
around its off-center density distribution.  
The magnitude of the non-radial fluxes 
is comparable to that of radial flux in some regions.  
Its ratio to the magnitude of the radial flux reaches more than 50 \% 
around the central object up to $r \sim40$ km.  
The non-radial fluxes extend up to 100 km for $\nu_e$ in the case
of $\phi$=141$^\circ$.  
This corresponds to the narrow region 
near the equator noted above.  

As we have seen, the neutrino distributions are 
affected by the neutrino degeneracy.  
While the temperature distribution is almost spherical, 
the chemical potential (i.e. degeneracy) shows 
non-uniform features in the central region inside the neutrino-sphere 
and the surrounding region up to the location of the shockwave.  
We show in Fig. \ref{fig:2d.phslice-etanu.3db.iph13} the temperature, 
neutrino chemical potential and neutrino degeneracy 
on the meridian slices corresponding to 
Figs. \ref{fig:3db-density.iphxx.slice}, 
\ref{fig:3db-flux.iph05.slice.inx} and
\ref{fig:3db-flux.iph13.slice.inx}.  
Having a deformed shape in the profile of electron fraction, 
the distribution of the neutrino chemical potential is non-spherical.  
We overlay the location of the neutrino-sphere 
for the neutrinos with an energy bin of $E_{\nu}=$ 34 MeV 
for three species.  
Here the neutrino-sphere is defined by the location 
where the optical depth is $2/3$ along the radial coordinate.  
For this purpose, 
we evaluated the optical depth by using the opacities 
for the neutrinos in the forward angle bin.  

In the outer layer, the neutrino distributions become rather 
spherical due to an averaging effect by the integration of neutrino fluxes 
from various angles 
while the neutrino distributions are inhomogeneous in the central region.  
This is the characteristics of the 6D Boltzmann solver 
to describe the advection in 3D space.  
This finding is in accord with the result found in 2D space by \citet{ott08}.  
We will demonstrate characteristics of the ray-by-ray approximation in 3D 
through comparisons with the 6D Boltzmann results in the following sections.  

The deviations from the spherical shape even in the outer layer occur 
in the case of $\phi$=141$^\circ$ due to a spot in the narrow channel 
near the equator.  
In this region, the neutrino-spheres of $\nu_e$ and $\bar{\nu}_e$ 
pass through the degenerate regime, 
i.e. low temperature and high neutrino chemical potential.  
This spot effect is evident in densities and fluxes 
for $\nu_e$, modest for $\bar{\nu}_e$ 
and not discernible for $\nu_{\mu}$.  
This order corresponds to the positions of neutrino-sphere 
for three species from outside toward the center.  
In the case of $\phi$=51$^\circ$, variations of neutrino degeneracy 
around the neutrino-spheres are not drastic, therefore, 
leading to rather spherical distributions in the outer layer 
besides non-uniform distributions at center.  
We see these two types of behavior in non-uniform situations 
depending on positions and timing in the models.  
The spot effect may cause enhanced/reduced neutrino distributions 
and associated changes of heating rates as we will examine below.  

\subsection{Density and flux by ray-by-ray evaluation}\label{section:d-ray}

In this section, we demonstrate 
by comparison between the full Boltzmann transport and the ray-by-ray approximation 
that the latter tends to exaggerate asymmetries outside the neutrino-sphere.
The relative difference of a quantity $Q$ between the two methods 
is hereafter denoted by 
$\delta_Q = ( Q_{RbR} - Q_{Boltz} ) / Q_{Boltz} $, 
where $Q_{RbR}$ and $Q_{Boltz}$ are the values of $Q$ 
obtained by the ray-by-ray and Boltzmann evaluations, respectively.  
As already mentioned, 
the background models employed in the current study 
were obtained with the ray-by-ray approximation and 
the following results may be somewhat affected indirectly.  

\begin{figure}
\epsscale{0.4}
\plotone{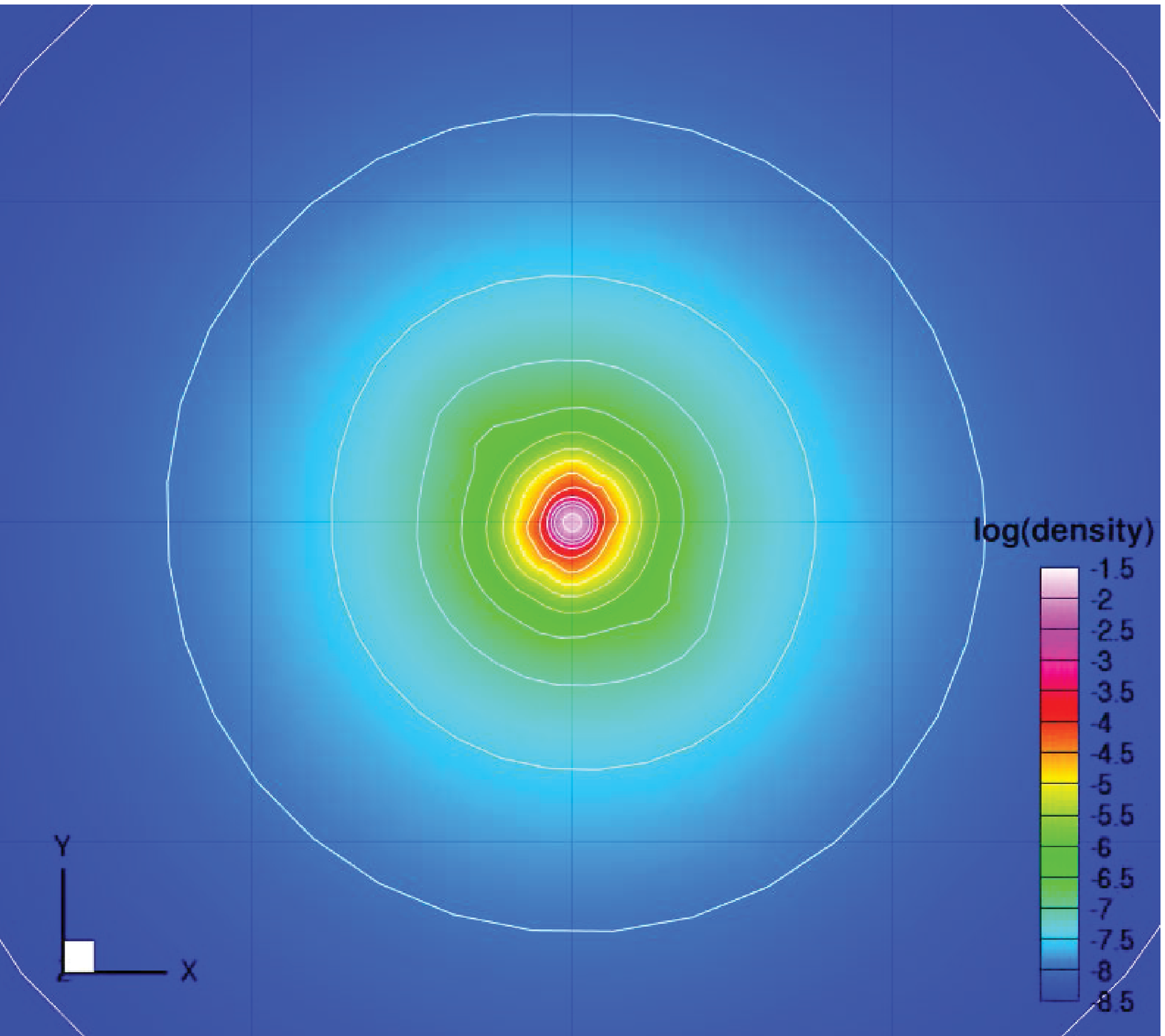}
\plotone{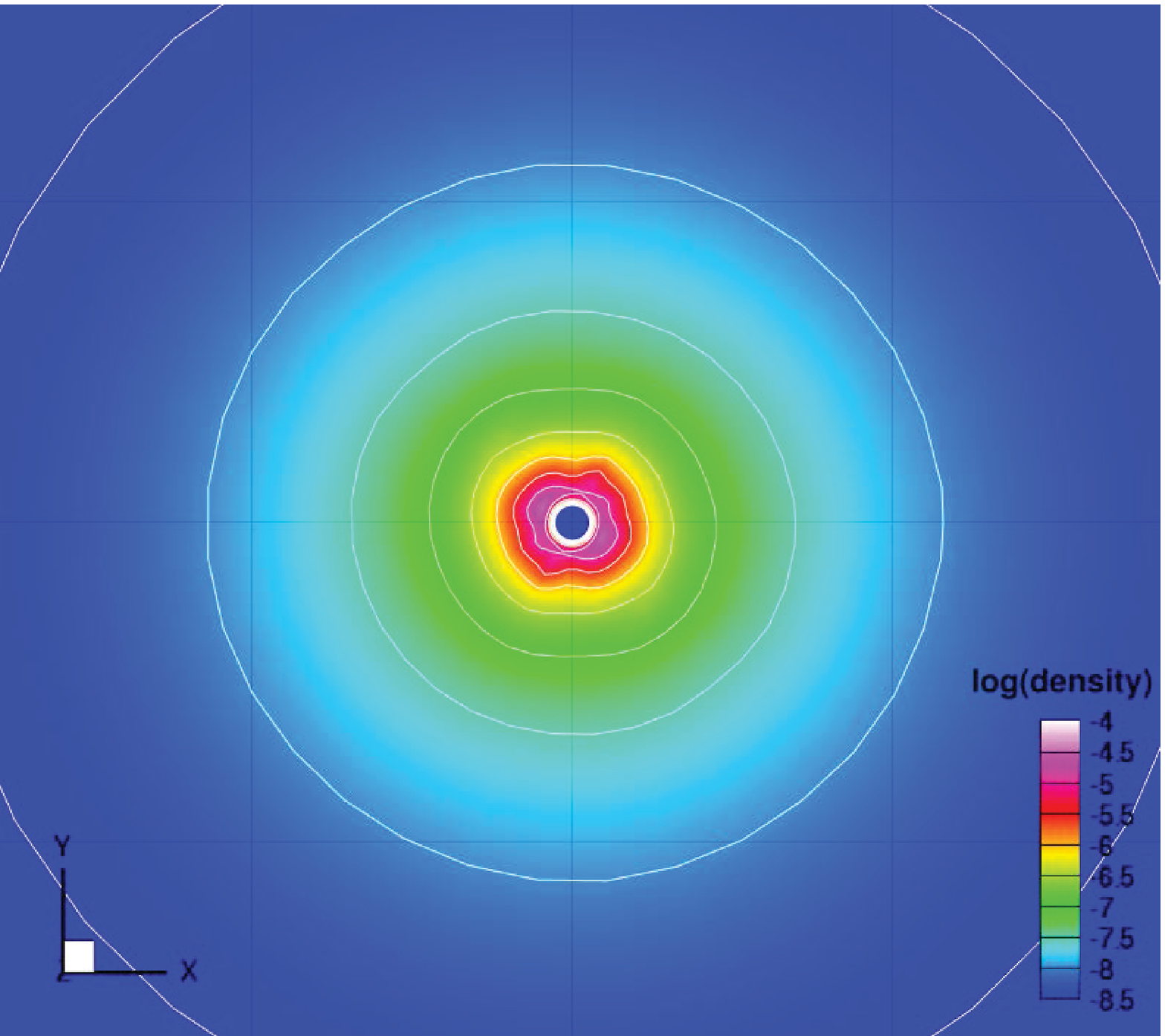}\\
\plotone{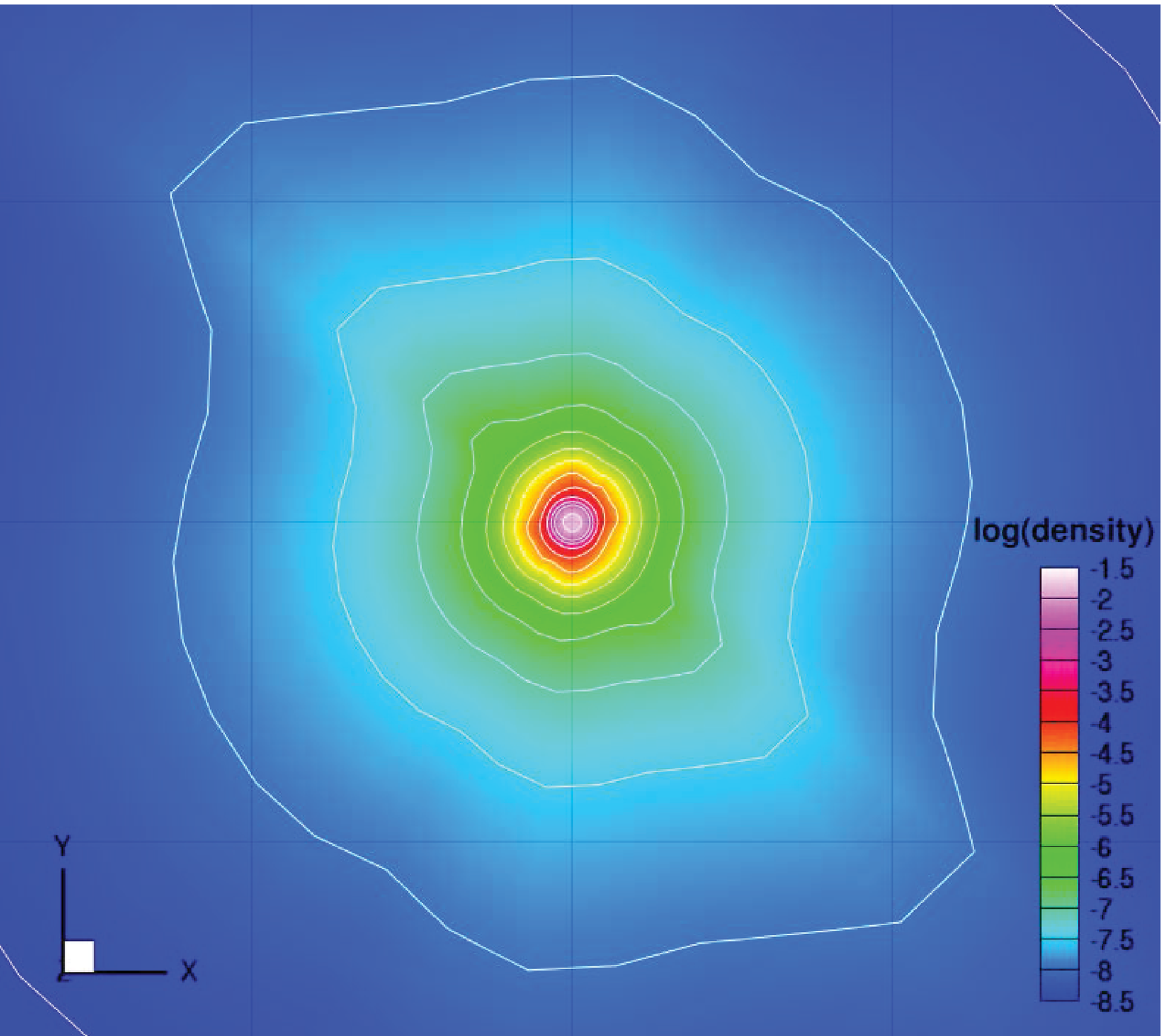}
\plotone{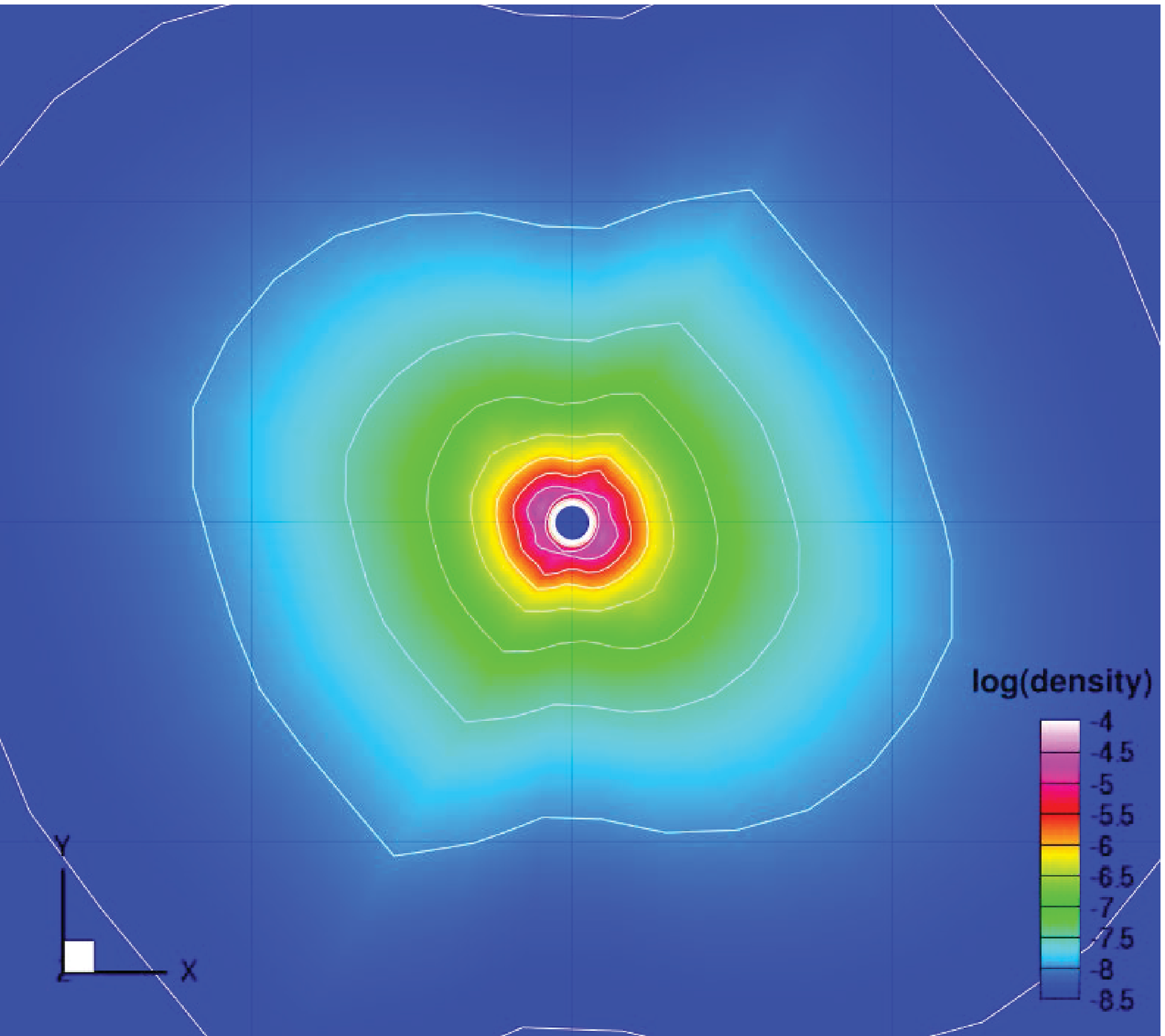}\\
\plotone{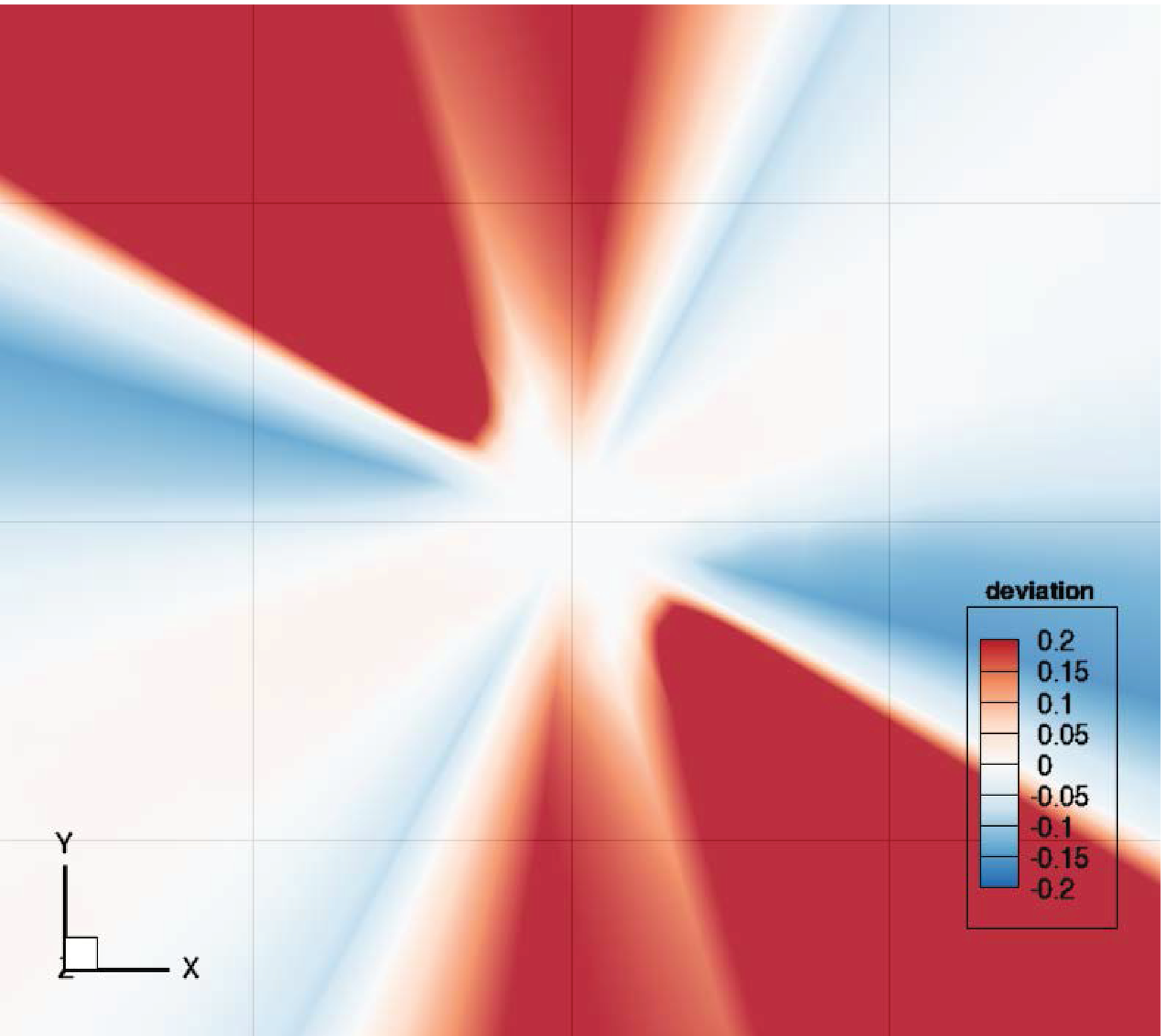}
\plotone{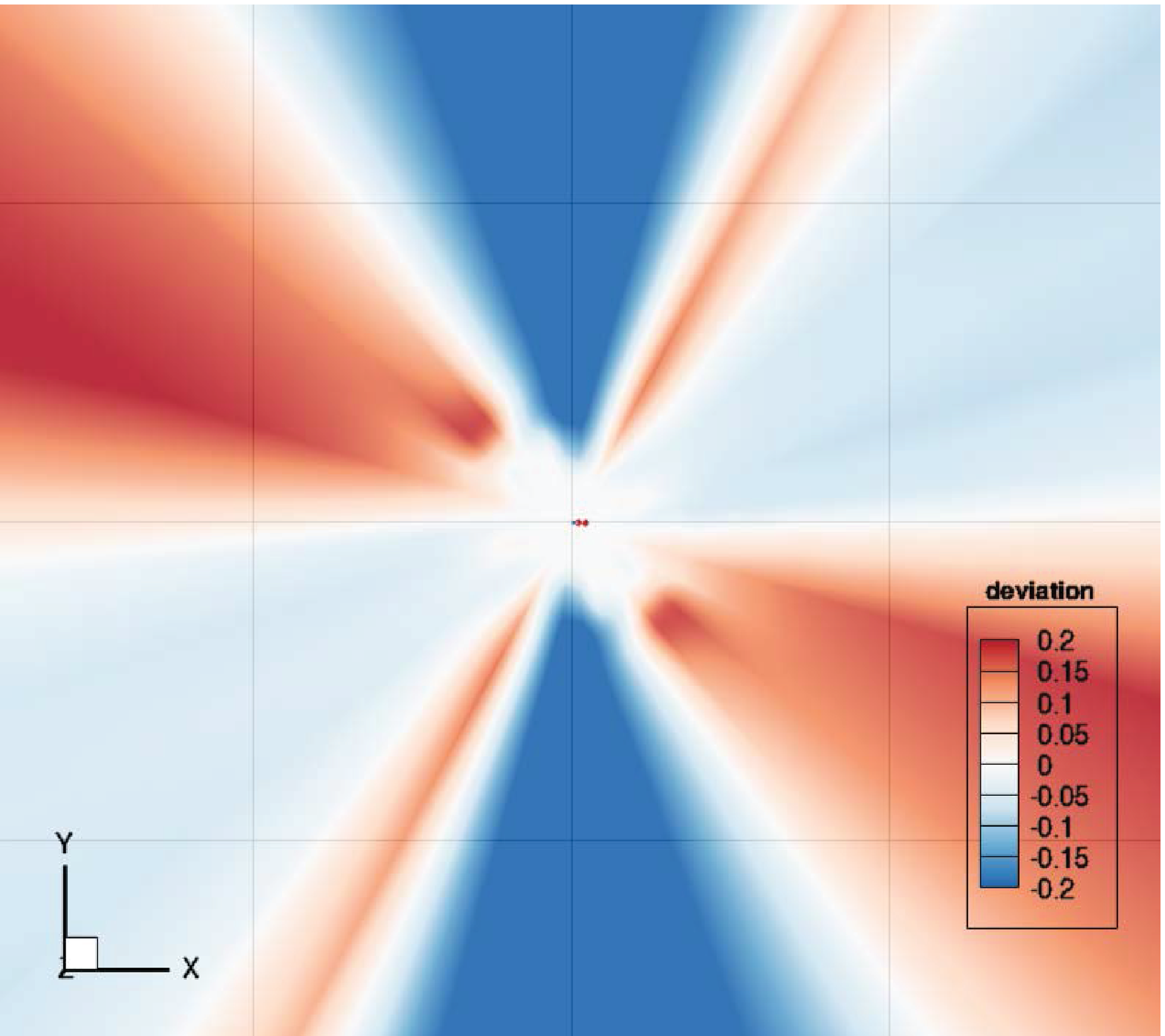}
\caption{Comparison of the neutrino densities 
for two species ($\nu_e$ and $\bar{\nu}_e$) on the xy-plane (z=0) 
is displayed by color maps 
in the left and right panels, respectively.  
The setting of the 3D supernova core (11M) at 150 ms 
is the same as in Fig. \ref{fig:3db-density.xyslice}.  
The neutrino densities evaluated 
by the 6D Boltzmann solver (top) 
and the ray-by-ray approximation (middle) 
are displayed together with the relative differences 
between the two evaluations (bottom).  
Grid lines with 200 km spacing are shown in the background.  }
\label{fig:rbr.3db-density.xyslice}
\end{figure}

In Fig. \ref{fig:rbr.3db-density.xyslice}, we display 
the densities of $\nu_e$ and $\bar{\nu}_e$ on the xy-plane 
obtained for the 11M model at 150 ms by the two methods.  
The relative difference between them 
is also shown in the bottom panels.  
The distribution is not spherical even in the outer region 
($>$ 100 km) for the ray-by-ray evaluation.  
This is the artifact of the ray-by-ray approximation.  
The neutrino fluxes are mainly determined 
by the environment around the neutrino-sphere.  
In the ray-by-ray approximation, 
the local enhancement (or suppression) of neutrino fluxes 
at a particular point on the neutrino-sphere 
due to variations of temperature and/or chemical potential 
are propagated along the radial ray 
that emanates from the point alone. 
In reality, however, neutrinos move also in other directions 
and affect the densities and fluxes on different radial rays. 
In the absence of such a mixing among different radial rays, 
the ray-by-ray approximation exaggerates the contrast 
among different rays
as shown in the bottom panels.  

\begin{figure}
\epsscale{0.4}
\plotone{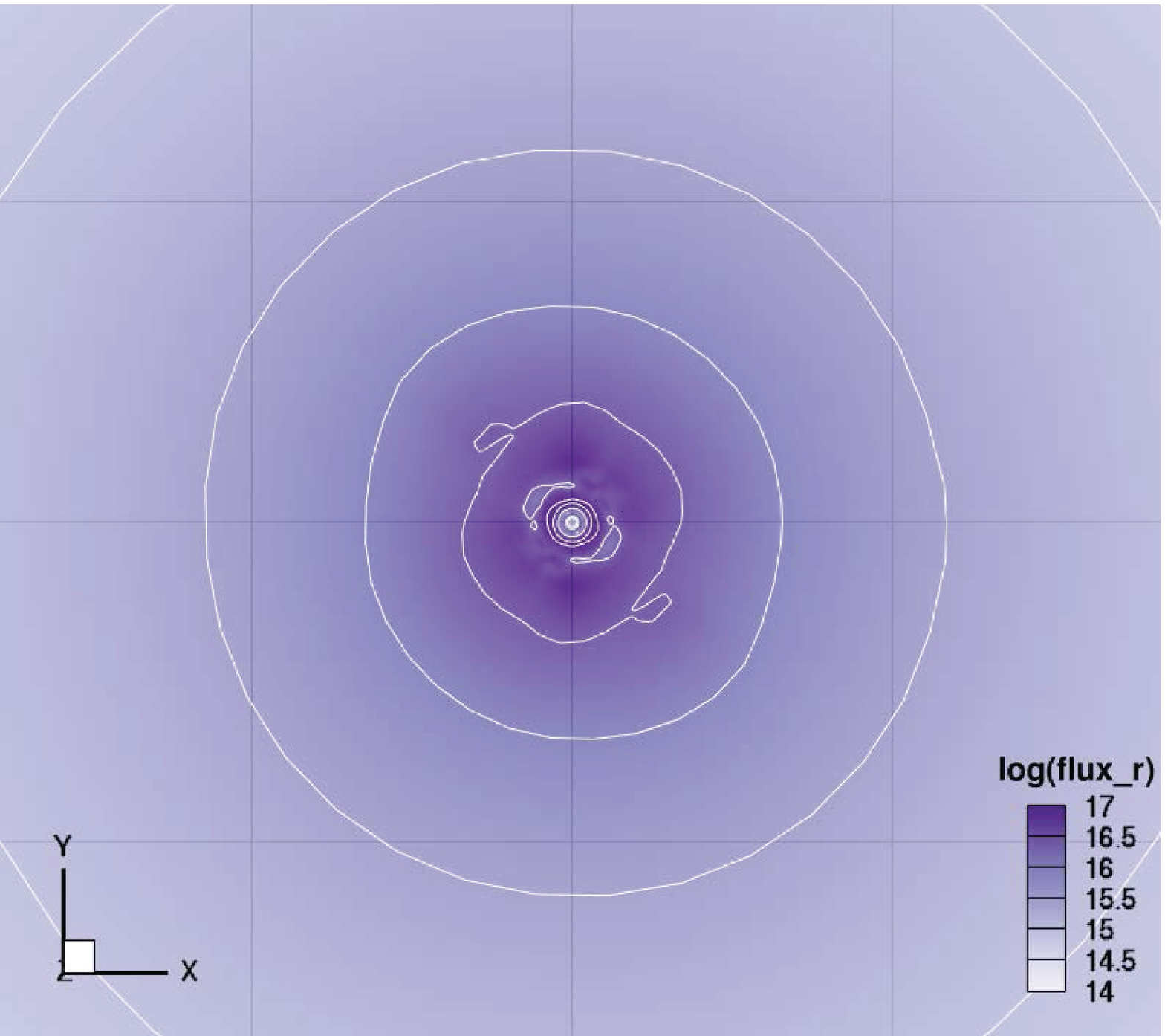}
\plotone{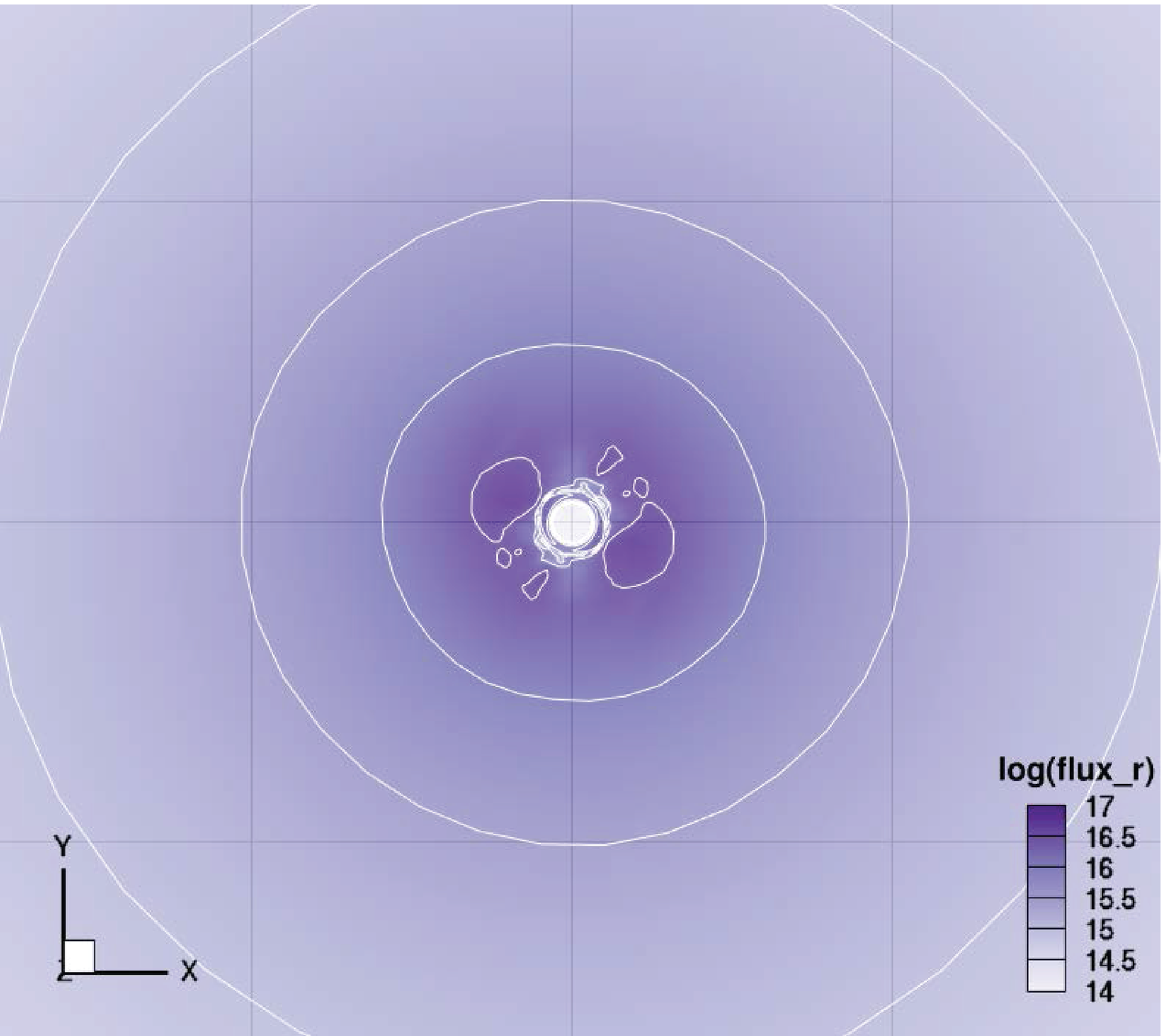}\\
\plotone{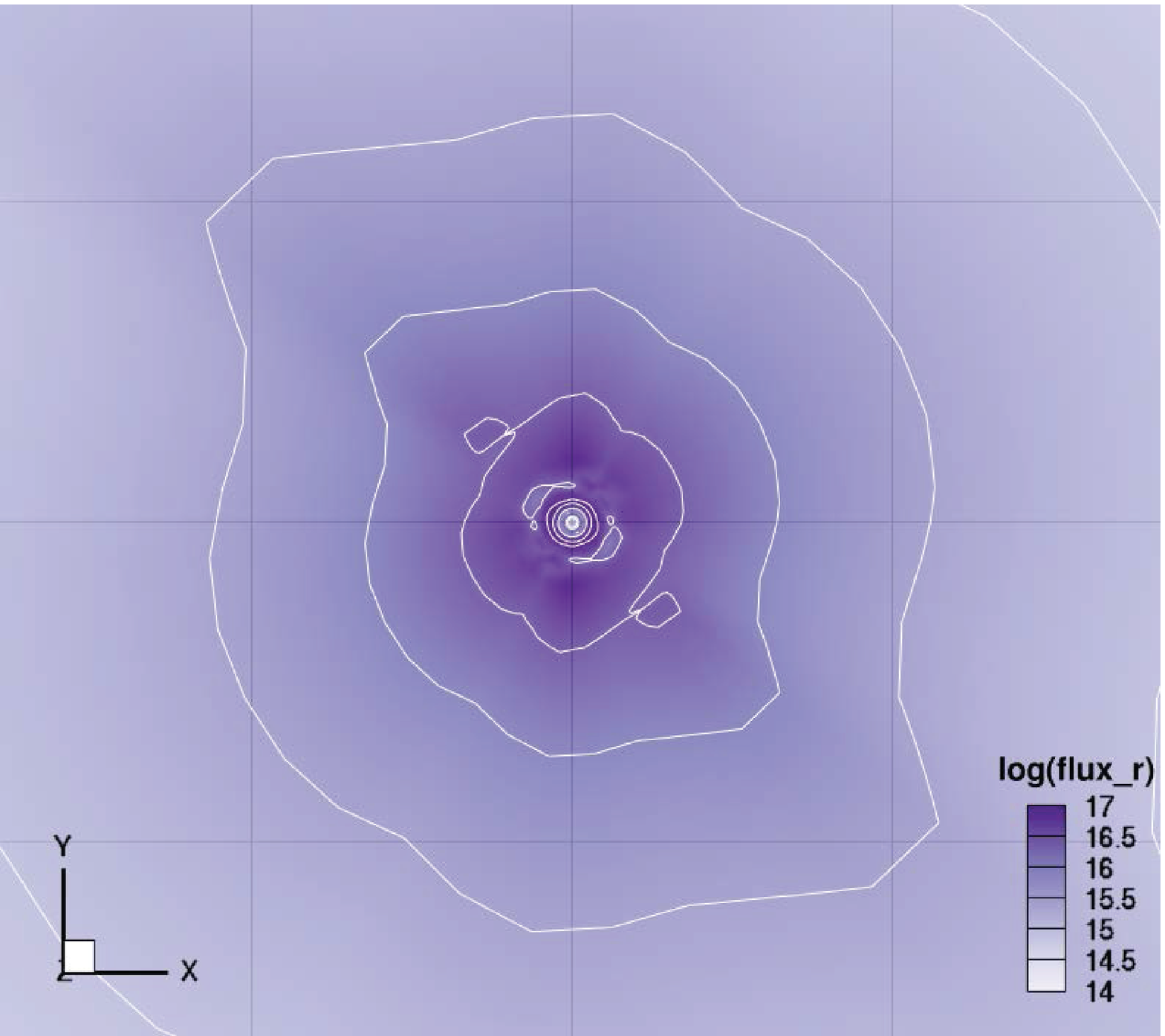}
\plotone{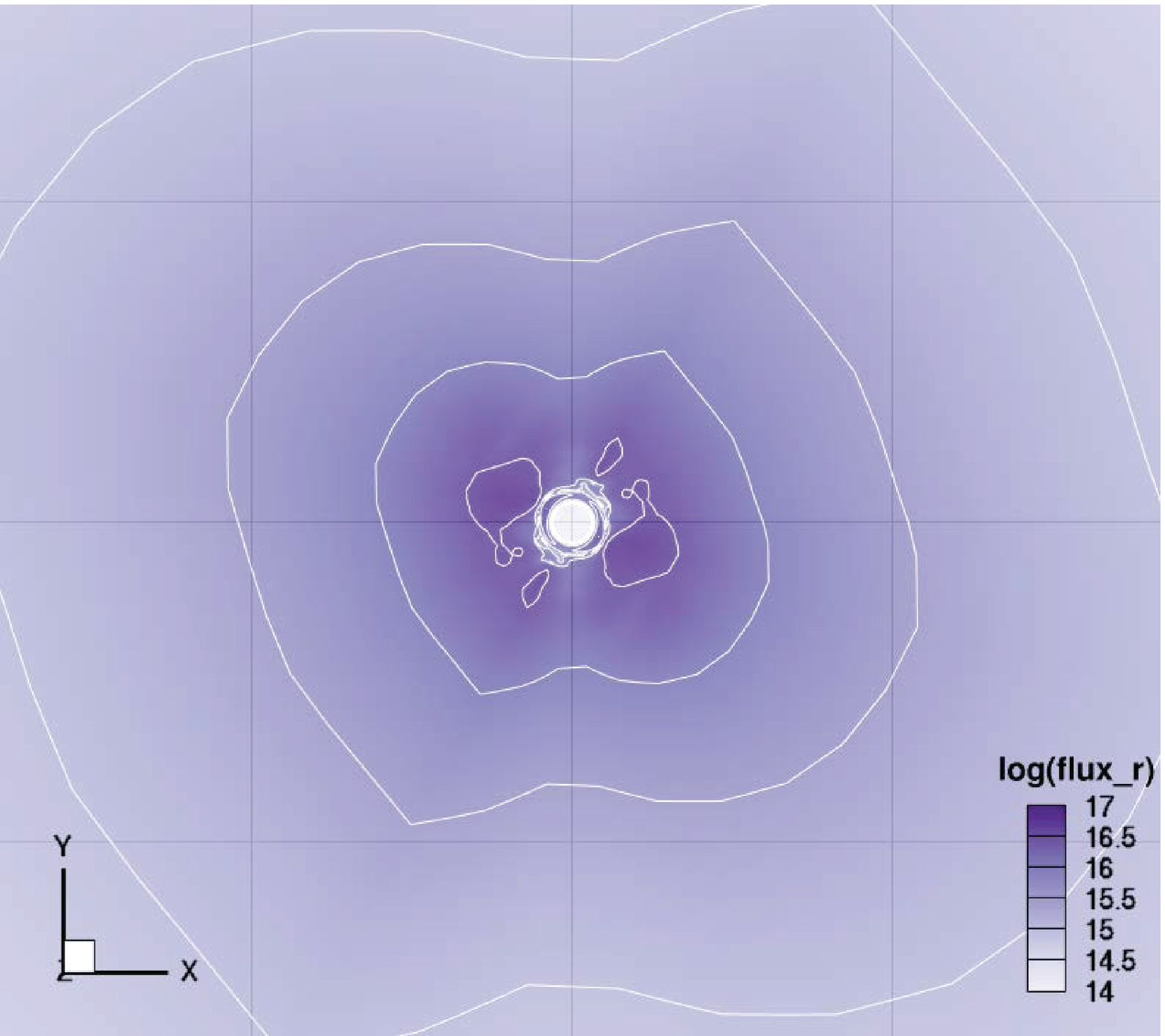}\\
\plotone{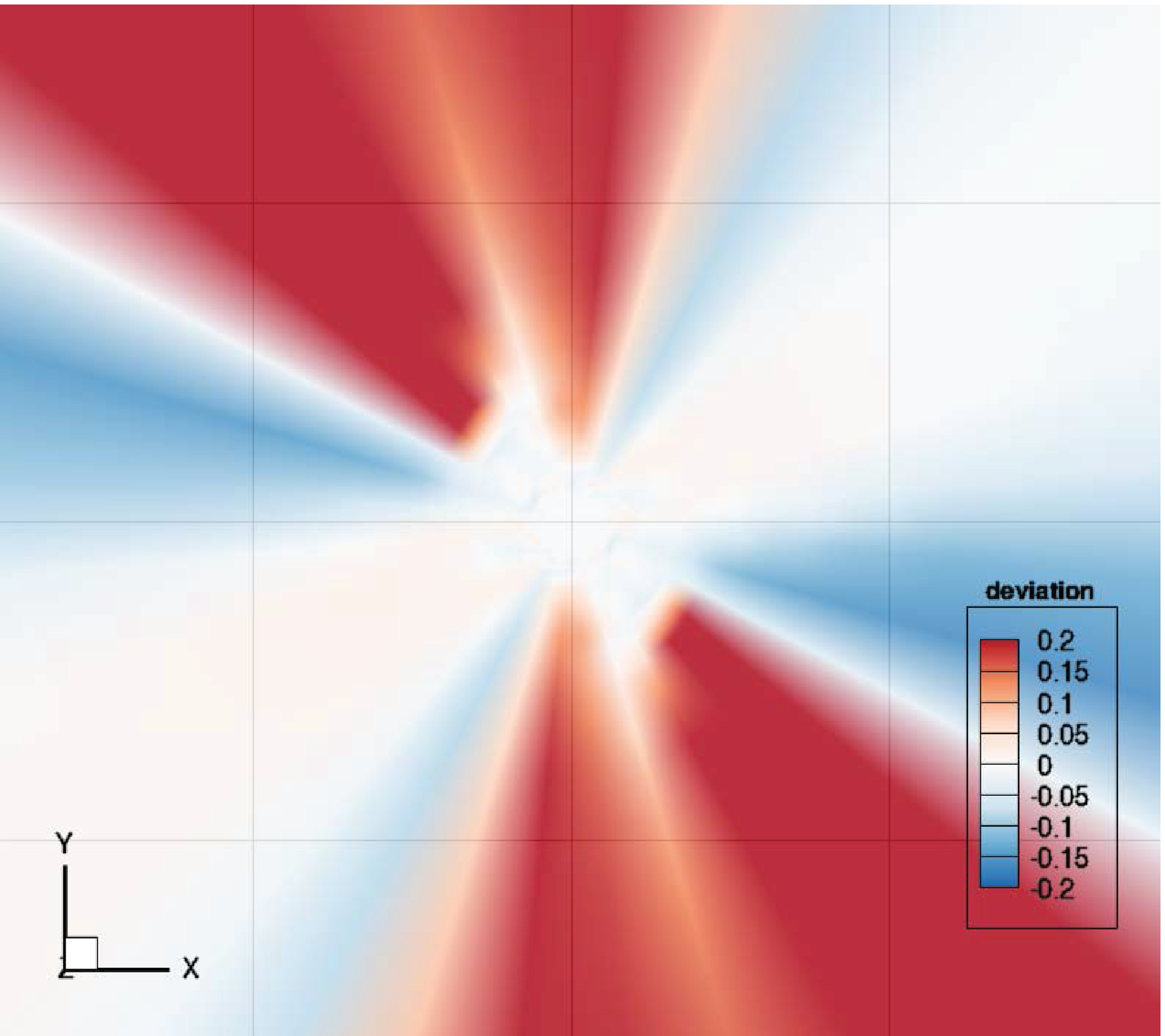}
\plotone{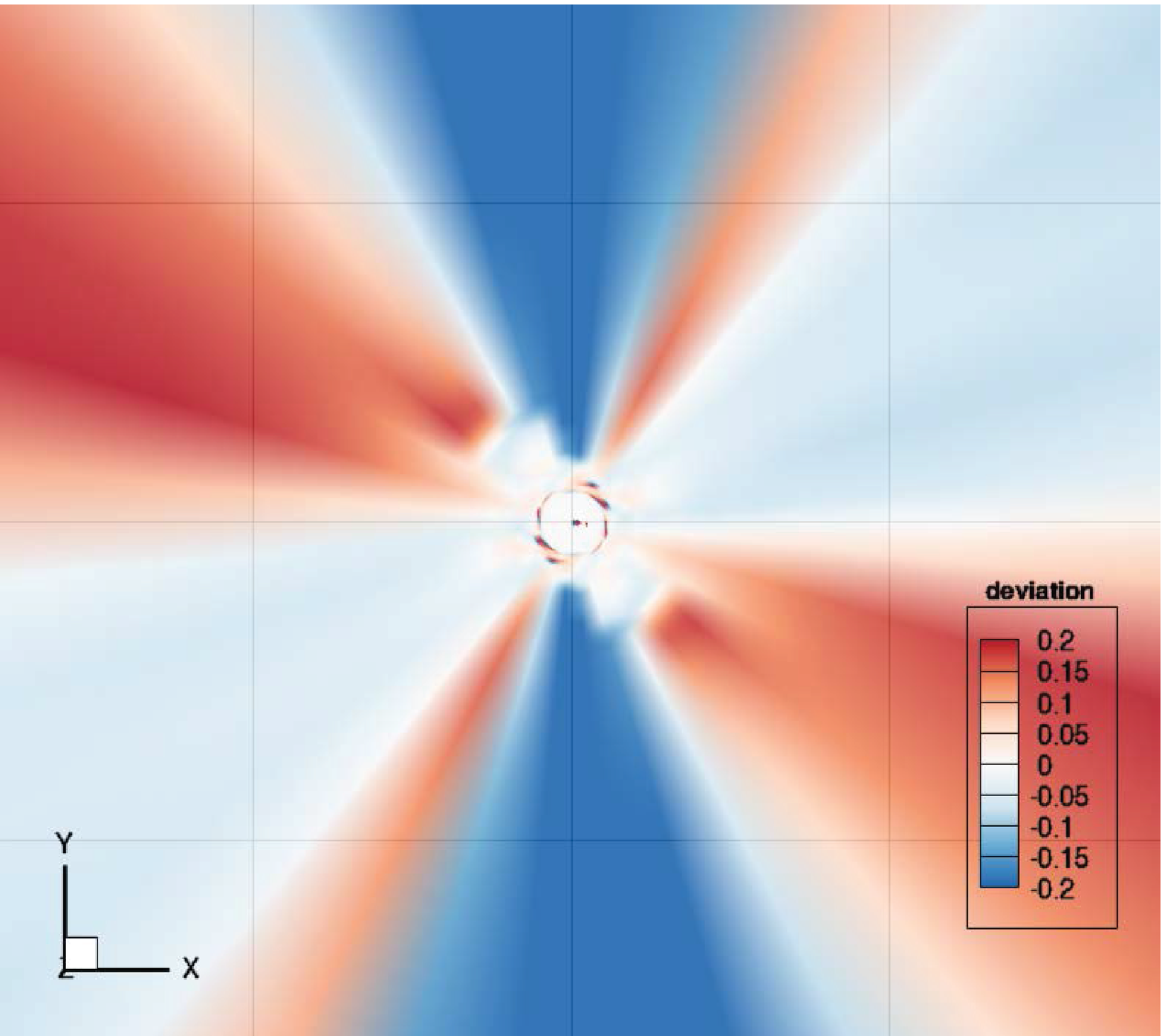}
\caption{Comparison of the number fluxes 
in log scale in the unit of fm$^{-2}$s$^{-1}$ 
for two species ($\nu_e$ and $\bar{\nu}_e$) on the xy-plane (z=0) 
is displayed by color maps 
in the left and right panels, respectively.  
The setting of the 3D supernova core (11M) at 150 ms 
is the same as in Fig. \ref{fig:3db-flux.xyslice}.  
The radial fluxes evaluated 
by the 6D Boltzmann solver (top) 
and the ray-by-ray approximation (middle) 
are displayed together with the relative differences 
between the two evaluations (bottom).  
Grid lines with 200 km spacing are shown in the background.  }
\label{fig:rbr.3db-flux.xyslice}
\end{figure}

The same features can be consistently seen 
in the behavior of fluxes 
shown in Fig. \ref{fig:rbr.3db-flux.xyslice}.  
The distributions of fluxes in the ray-by-ray approximation 
have non-spherical features in the outer region 
due to enhancement or reduction along the radial rays.  
In some cases, the difference of $\nu_e$ flux 
has anti-correlation with that of $\bar{\nu}_e$ flux.  
When the neutrino chemical potential is high 
at the neutrino-sphere in one ray, 
the $\nu_e$ flux is enhanced along the ray 
but $\bar{\nu}_e$ flux is suppressed.  
Although the neutrino fluxes for two species may depend 
on the location of neutrino-spheres, 
the environment and other factors, 
we find that these features often appear in many profiles 
of our models as we will show.  

\begin{figure}
\epsscale{0.75}
\plotone{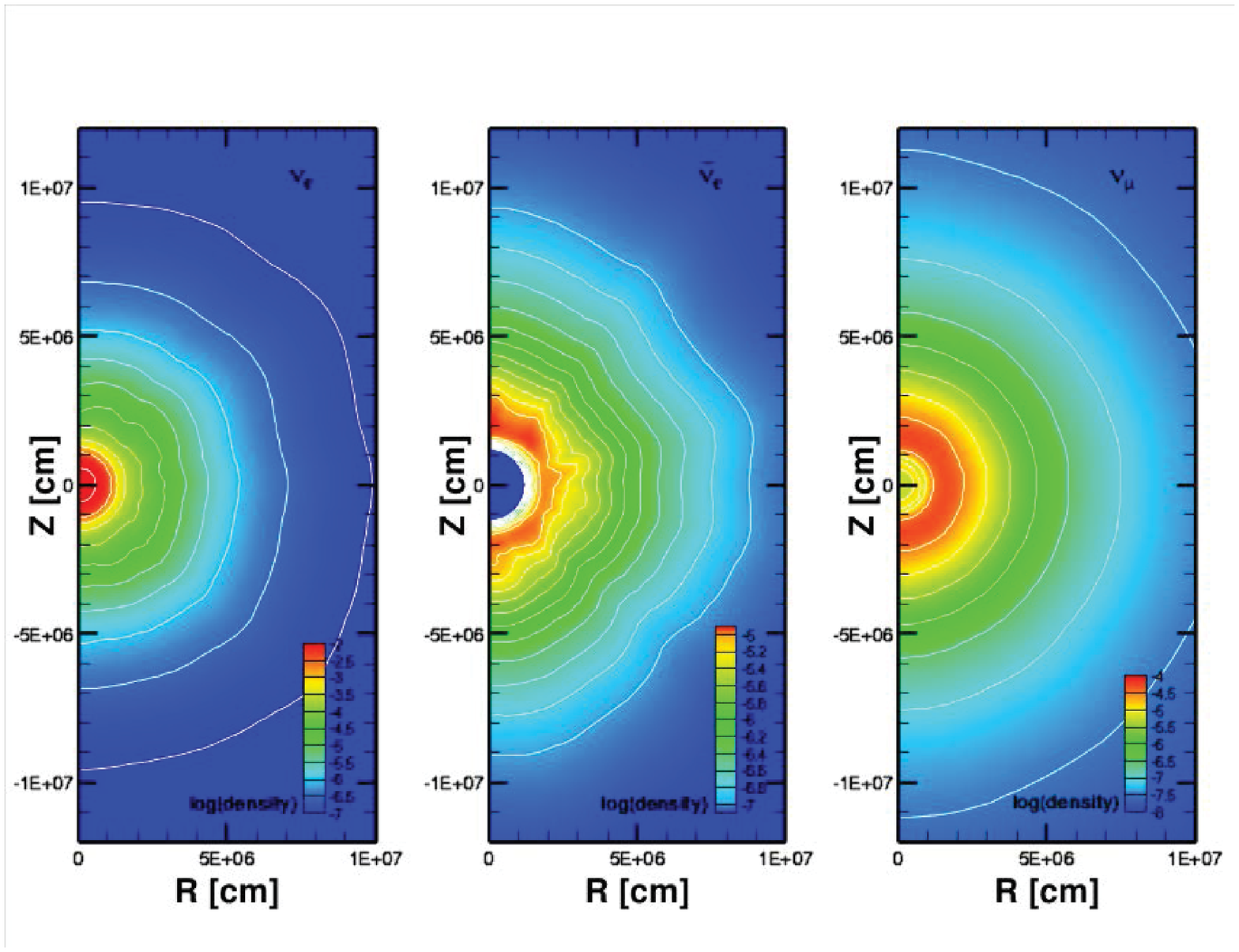}
\plotone{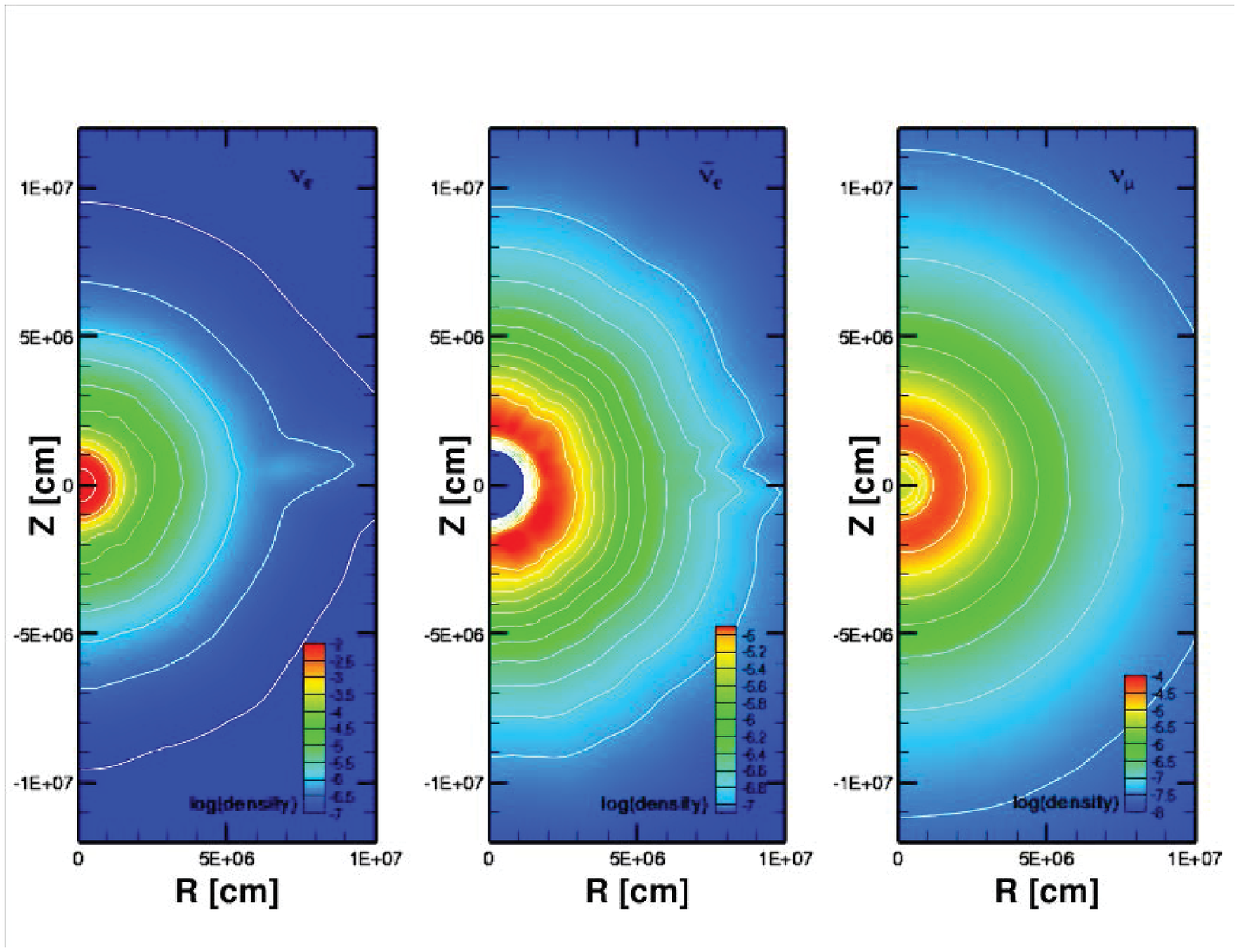}
\caption{Profiles of the neutrino density 
evaluated by the ray-by-ray approximation 
for the 11M model at 150 ms.  
The locations of meridian slice and the notation 
are the same as in Fig. \ref{fig:3db-density.iphxx.slice}.  }
\label{fig:rbr.3db-density.iphxx.slice}
\end{figure}

\begin{figure}
\epsscale{0.75}
\plotone{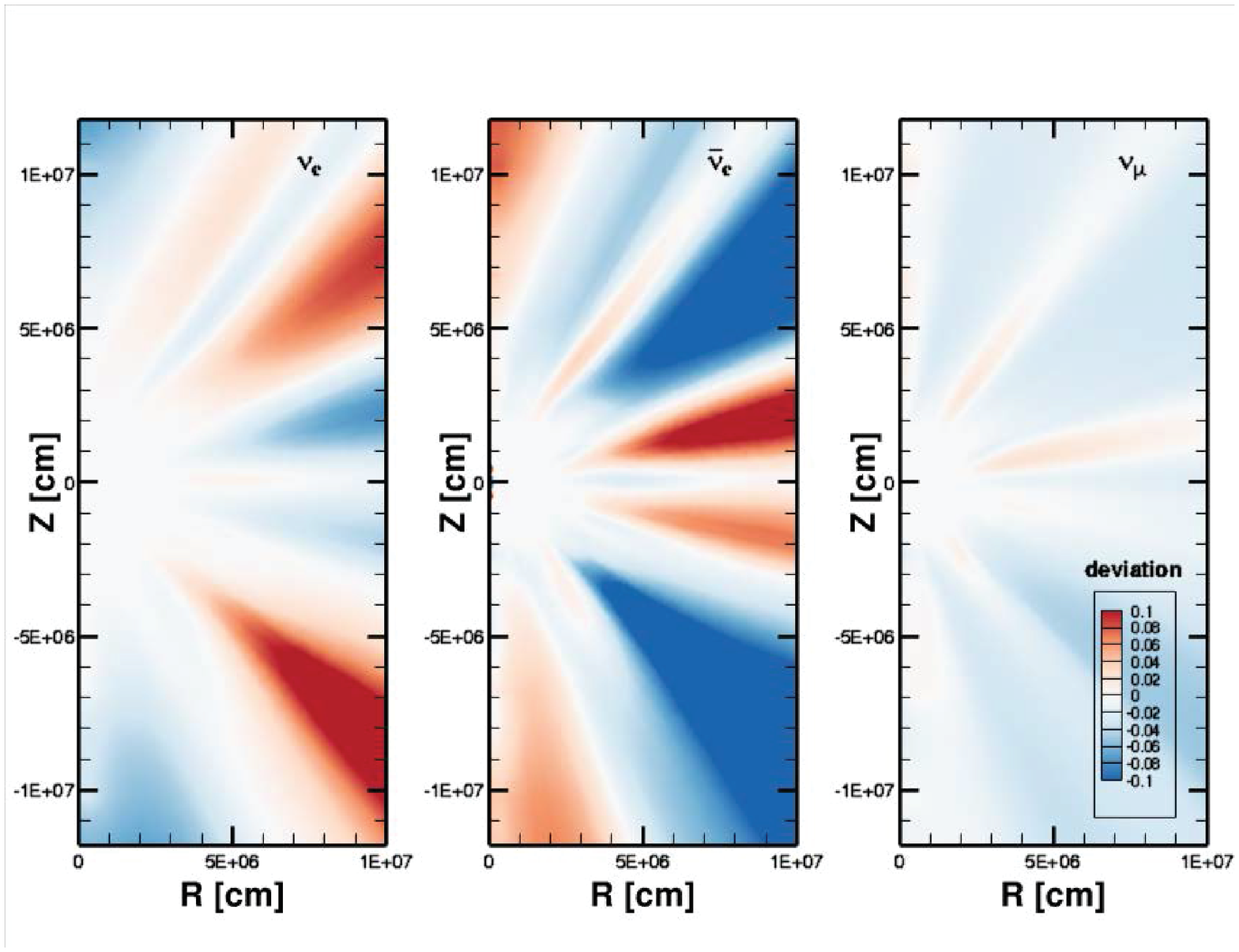}
\plotone{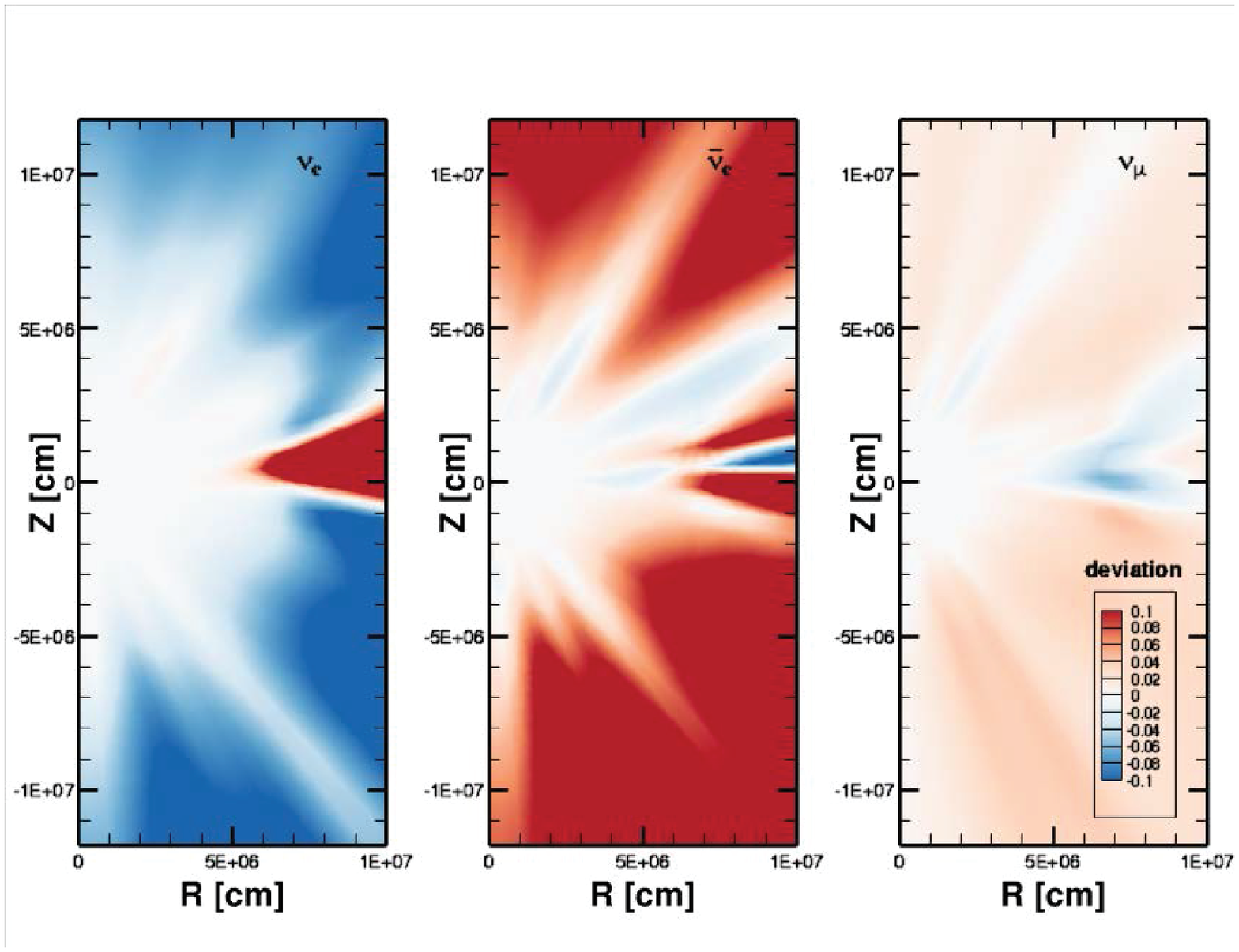}
\caption{Color maps for 
the relative differences of the neutrino density 
by the ray-by-ray evaluation 
with respect to the 6D Boltzmann evaluation 
for the 11M model at 150 ms 
corresponding to Figs. 
\ref{fig:3db-density.iphxx.slice} and 
\ref{fig:rbr.3db-density.iphxx.slice}.  }
\label{fig:ratio.3db-density.iphxx.slice}
\end{figure}

We demonstrate that the anisotropy in the ray-by-ray approximation 
can be large enough to affect the densities and fluxes at large distances, 
depending on the position.  
Fig. \ref{fig:rbr.3db-density.iphxx.slice} shows the neutrino densities 
obtained by the ray-by-ray approximation on the meridian slices, 
which corresponds to the ones in Fig. \ref{fig:3db-density.iphxx.slice}.  
The neutrino densities of $\nu_e$ and $\bar{\nu}_e$ are not spherical 
at $\sim$100~km in contrast to the spherical shape found in the 6D Boltzmann evaluation.  
In the case of $\phi$=141$^\circ$, the anisotropy around the equator 
is enhanced more than that in the 6D Boltzmann evaluation.  
We display the relative difference between the two methods 
in Fig. \ref{fig:ratio.3db-density.iphxx.slice}.   
Depending on the polar direction, there are regions of overestimation 
or underestimation of densities for $\nu_e$ and $\bar{\nu}_e$ 
in the outer region.  
The enhancement (or reduction) at the emission region persists along the ray 
due to the averaged radial propagation of neutrinos.  
There is anti-correlation between the two species ($\nu_e$ and $\bar{\nu}_e$) 
in the differences along the ray.  
This is again because the emission is determined by the neutrino degeneracy, 
which enhances and reduces the flux of $\nu_e$ and $\bar{\nu}_e$, respectively.  

The anisotropy of densities arises from angular variations of 
the fluxes of neutrinos emitted from the region of neutrino spheres.  
We show the behavior of neutrino fluxes in Figs. \ref{fig:rbr.3db-flux.iph05.slice.inx} 
and \ref{fig:rbr.3db-flux.iph13.slice.inx} for the two meridian slices 
at $\phi$=51$^\circ$ and 141$^\circ$, respectively.  
The relative differences of the ray-by-ray evaluation from the 6D Boltzmann evaluation 
are also displayed for each profile in the right panels.  
One can clearly see the overestimation or underestimation of the fluxes 
along the radial ray, depending on the polar angle.  
The anti-correlation between the two species appears apparently 
on the two slices as seen in the profiles on the xy-plane 
in Fig. \ref{fig:rbr.3db-flux.xyslice}.  
The pattern of differences in the flux corresponds to that in the densities 
shown in Fig. \ref{fig:ratio.3db-density.iphxx.slice}.  
In the case of $\phi$=141$^\circ$, the differences are large 
for both species in a wide region.  
The region around the equator has a significant deviation 
due to the narrow channel.  

\begin{figure}
\epsscale{0.414}
\hspace*{0.8cm}
\plotone{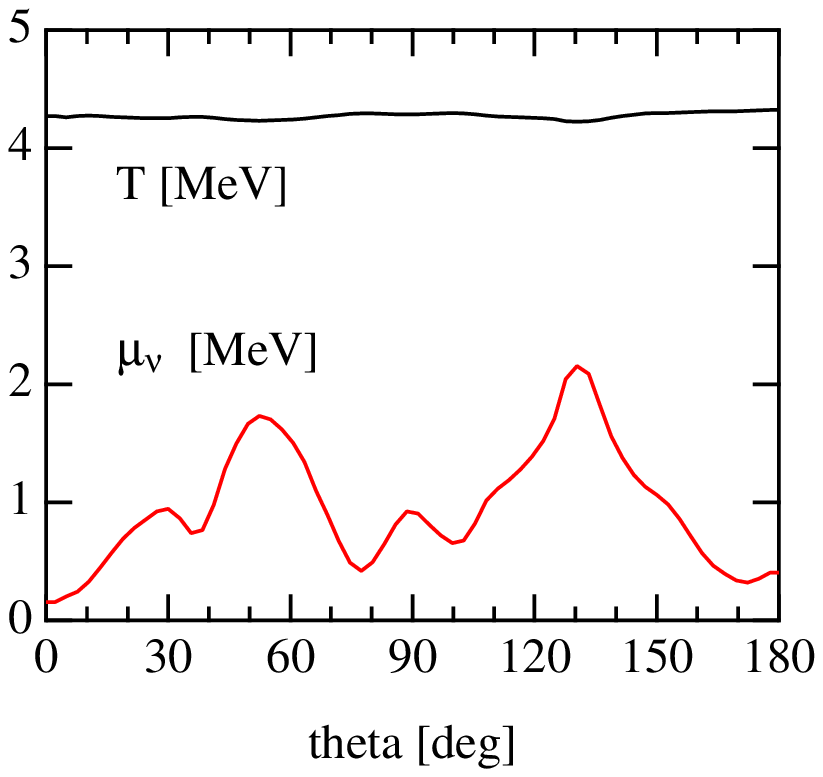}\\
\epsscale{0.486}
\plotone{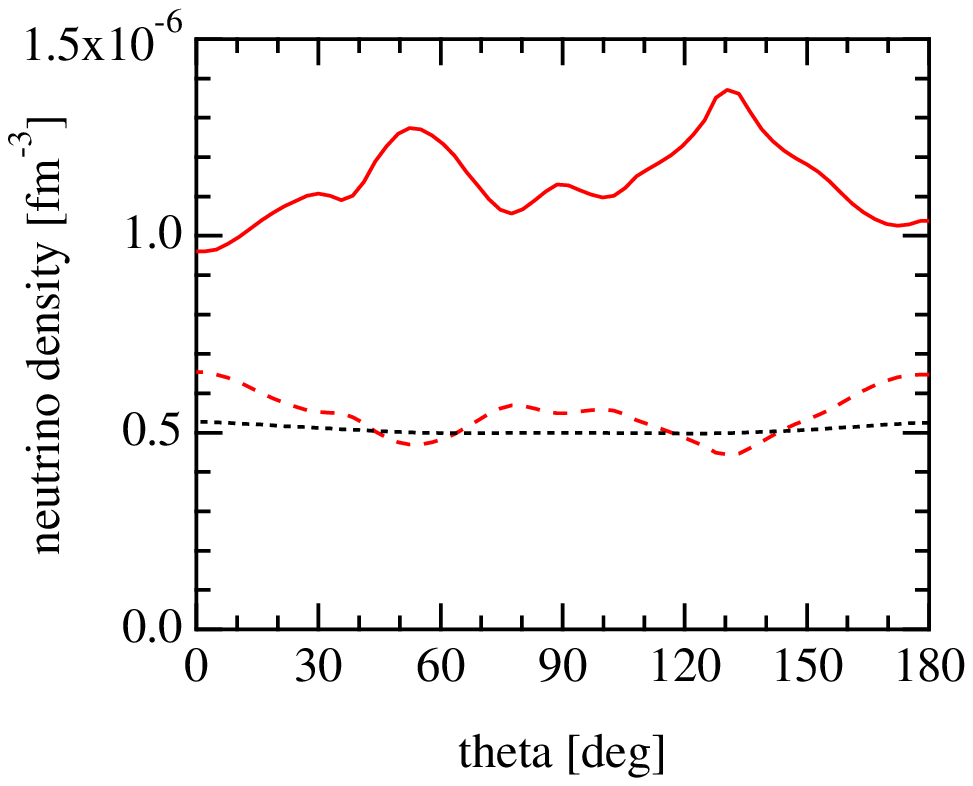}\\
\plotone{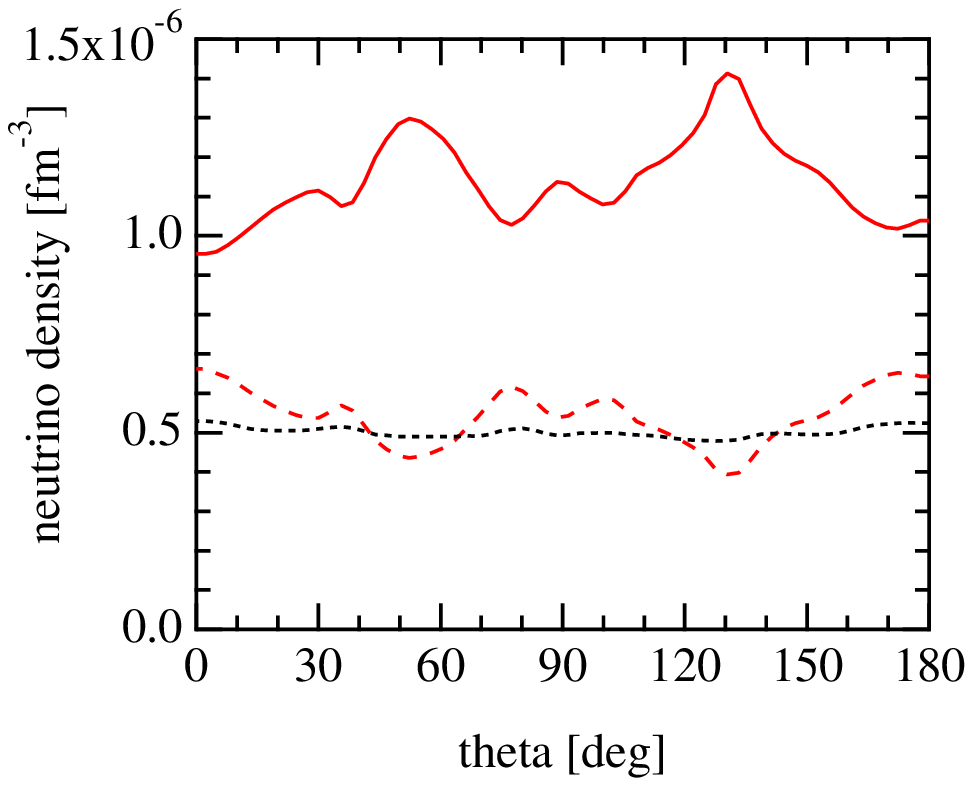}
\vspace*{1cm}
\caption{Distributions of temperature and neutrino chemical potential (top) 
and the neutrino densities 
by 6D Boltzmann (middle) and ray-by-ray evaluations (bottom) 
are plotted 
as functions of the polar angle at the radius of 54~km 
on the meridian slice at $\phi$=51$^{\circ}$ 
for the 11M model at 150 ms.  
In the middle and bottom panels, 
red solid, red dashed and black dotted lines 
denote 
densities for $\nu_e$, $\bar{\nu}_e$ and $\nu_{\mu}$, 
respectively.  }
\label{fig:1d.polar.ir040.3db.iph05}
\end{figure}

\begin{figure}
\epsscale{0.423}
\hspace*{0.3cm}
\plotone{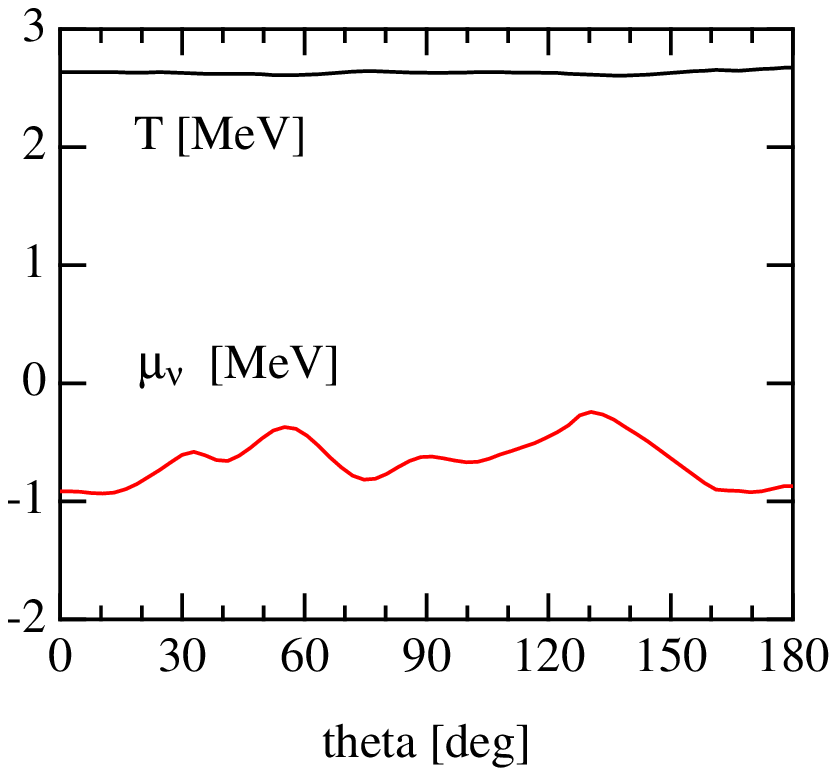}\\
\epsscale{0.486}
\plotone{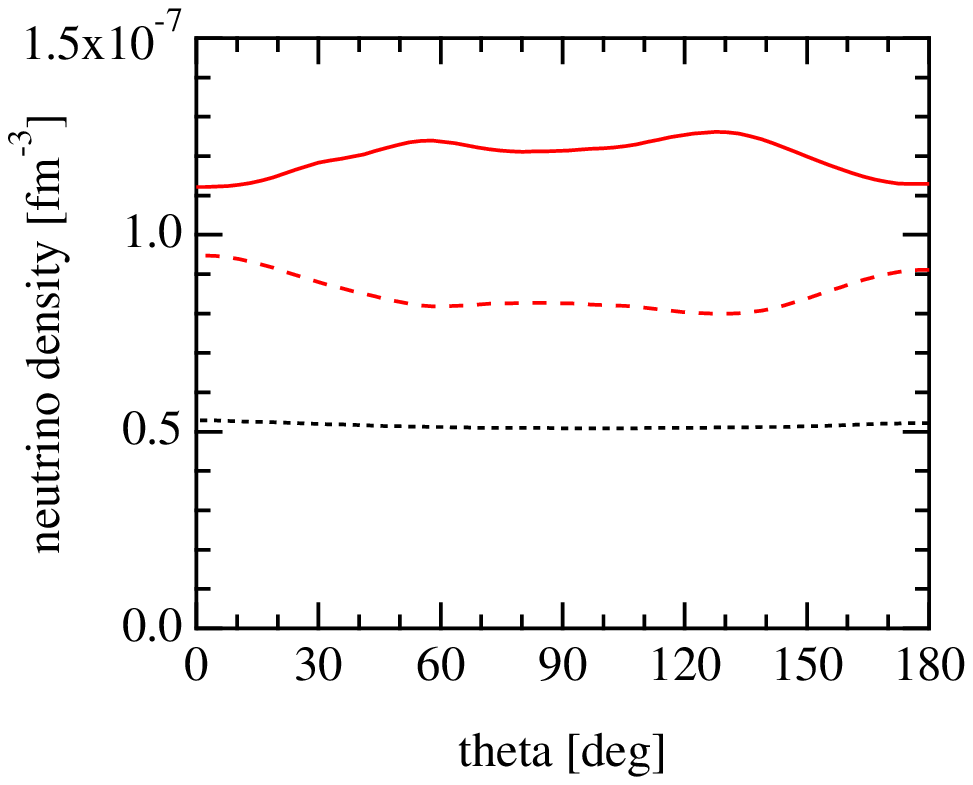}\\
\plotone{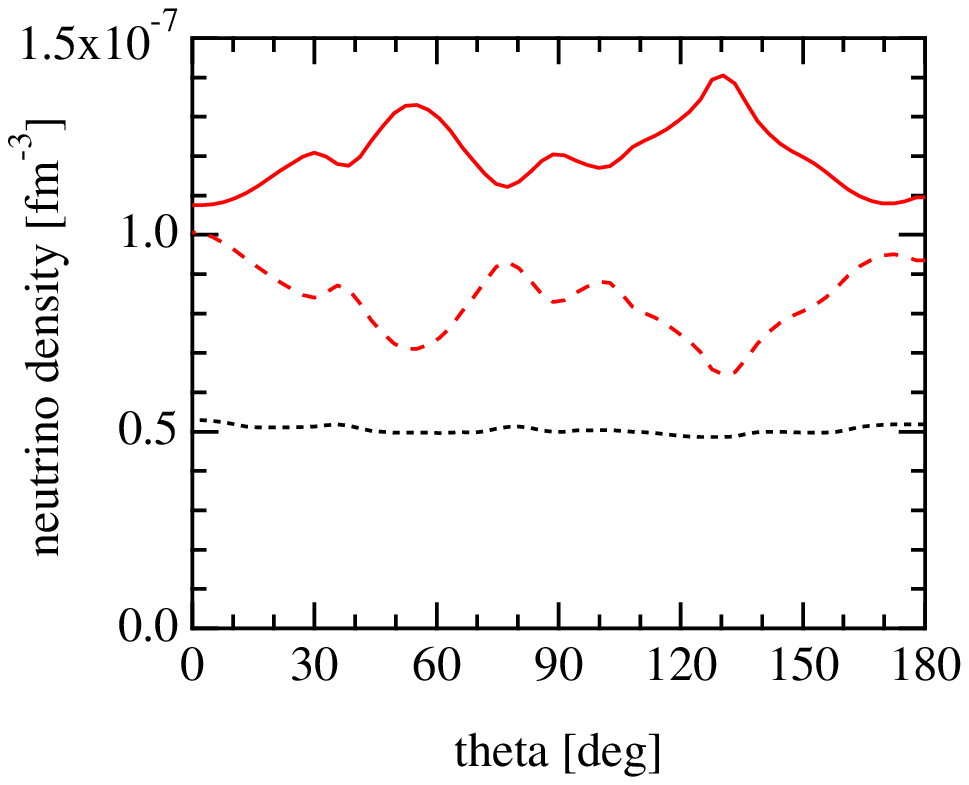}
\vspace*{1cm}
\caption{Same as 
Fig. \ref{fig:1d.polar.ir040.3db.iph05}, 
but at the radius of 94~km.  }
\label{fig:1d.polar.ir060.3db.iph05}
\end{figure}

We further examined the variations of the neutrino distributions 
and the anti-correlation between the two species.  
We show the polar angle dependence of quantities at the fixed radius 
on the meridian slice at $\phi$=51$^\circ$ 
in Figs. \ref{fig:1d.polar.ir040.3db.iph05} 
and \ref{fig:1d.polar.ir060.3db.iph05}.  
The neutrino densities are well determined by the local environment 
(temperature and neutrino chemical potential) in the optically thick region 
at radius of 54~km in Fig. \ref{fig:1d.polar.ir040.3db.iph05}.  
We see the neutrino densities follow 
the variation of the neutrino chemical potential 
while the anti-neutrino densities have the anti-correlated variation.  
$\mu$-type neutrinos have a constant distribution 
since the temperature stays almost constant.  
Both the two evaluations 
provide the neutrino distributions simply reflecting the environment.  
The ray-by-ray evaluation shows slightly stronger variations 
than the 6D Boltzmann evaluation.  

Neutrino distributions become different from the variation of local environment 
as we leave the the optically thick region and go further away 
from the region of the neutrino-sphere.  
At radius of 94 km in Fig. \ref{fig:1d.polar.ir060.3db.iph05}, 
the neutrino densities by the 6D Boltzmann evaluation become 
more flat than the variations in the neutrino chemical potential.  
This is due to the averaging effect by the non-radial fluxes.  
In contrast, the neutrino densities by the ray-by-ray evaluation 
show clear variations due to the degeneracy, 
keeping the anisotropy produced deep inside the core.  
The anti-correlation between the two species remains similar 
to the situation at the inner position.  
As shown in these examples, 
the non-uniform distribution of neutrino chemical potential 
inside the core is important to determine the neutrino distributions 
in the outer region.  
Note that the electron fraction is a key factor 
to determine the neutrino chemical potential 
even when the distributions of density and temperature are 
spherical inside the core.  

\section{Evaluation of energy transfer rates}\label{section:heating}

In this section, 
we examine the heating rates in the supernova core 
by the 6D Boltzmann solver.  
Since the heating rate is a crucial quantity 
to explore the neutrino heating mechanism, 
we investigate also the heating rates obtained by the ray-by-ray approximation 
so that we can assess possible differences due to the approximation 
popularly used in multi-dimensional simulations.  
We show that differences of the heating rate occur along the radial ray 
due to the ray-by-ray approximation 
in the same way as those found above.  
The size of differences depends on the environment around the neutrino-sphere, 
which provides the differences in the densities and fluxes.  

\subsection{Heating rates by Boltzmann and ray-by-ray evaluations}\label{section:h-boltz-ray}

\begin{figure}
\epsscale{0.4}
\plotone{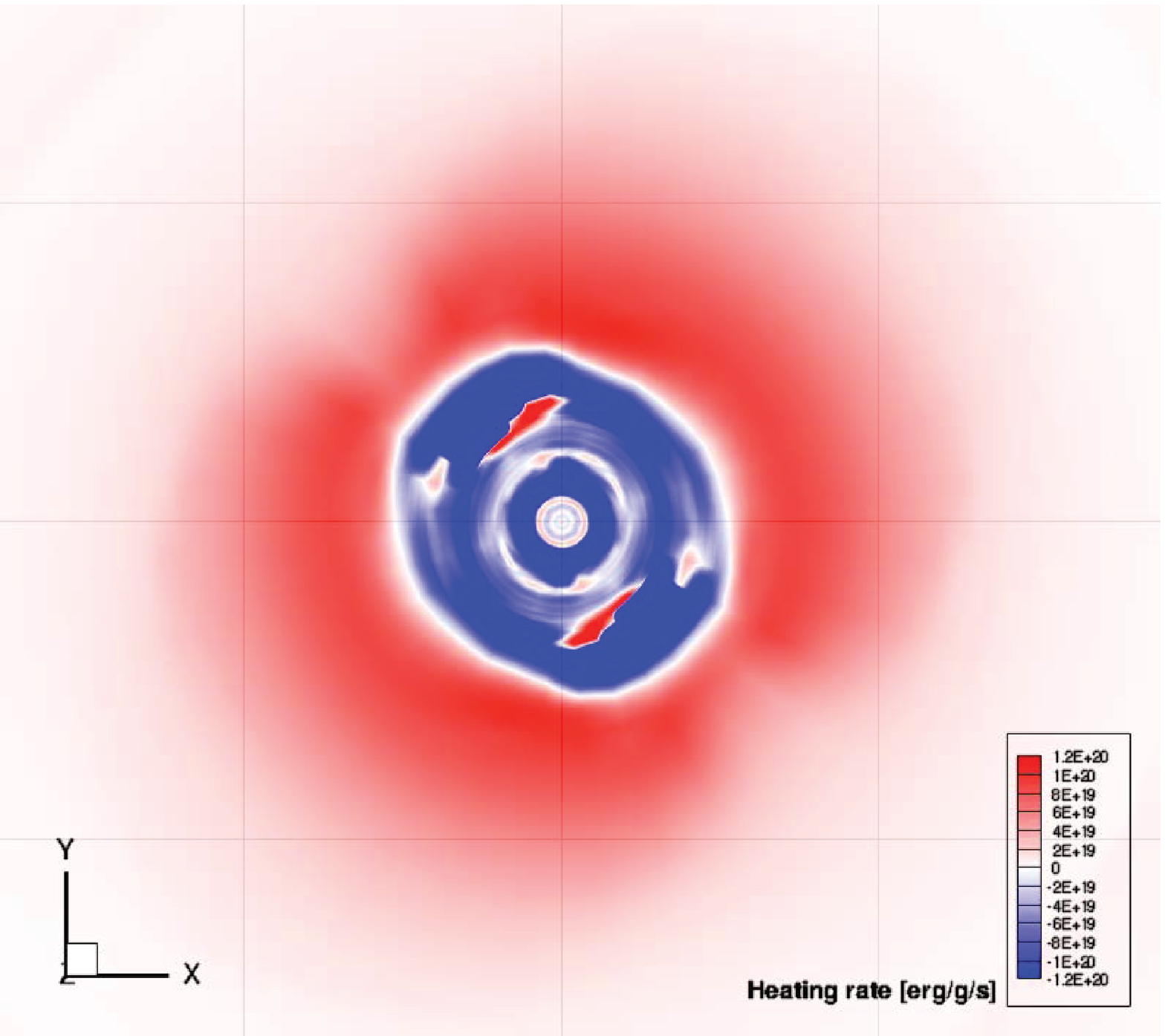}\\
\plotone{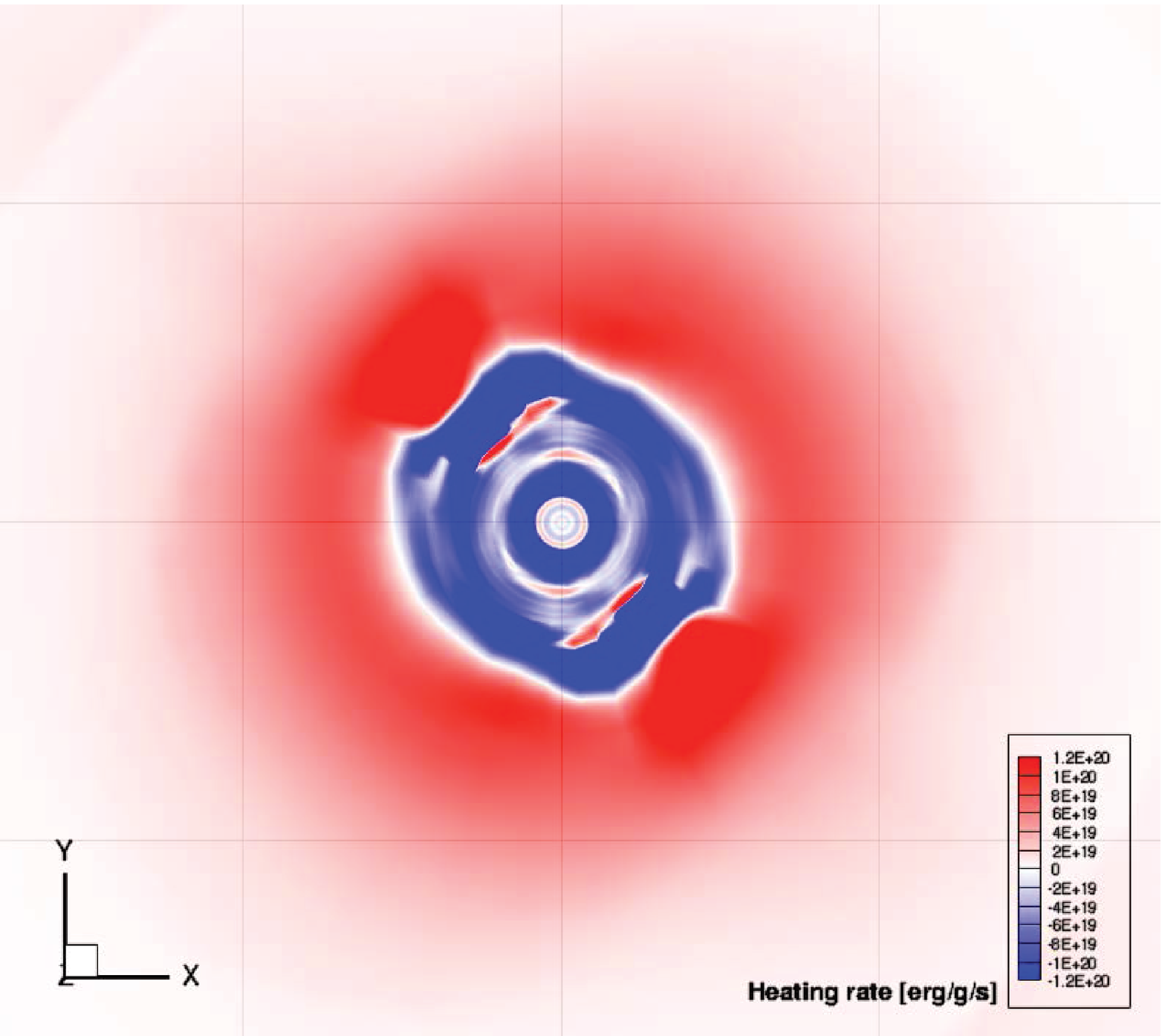}\\
\plotone{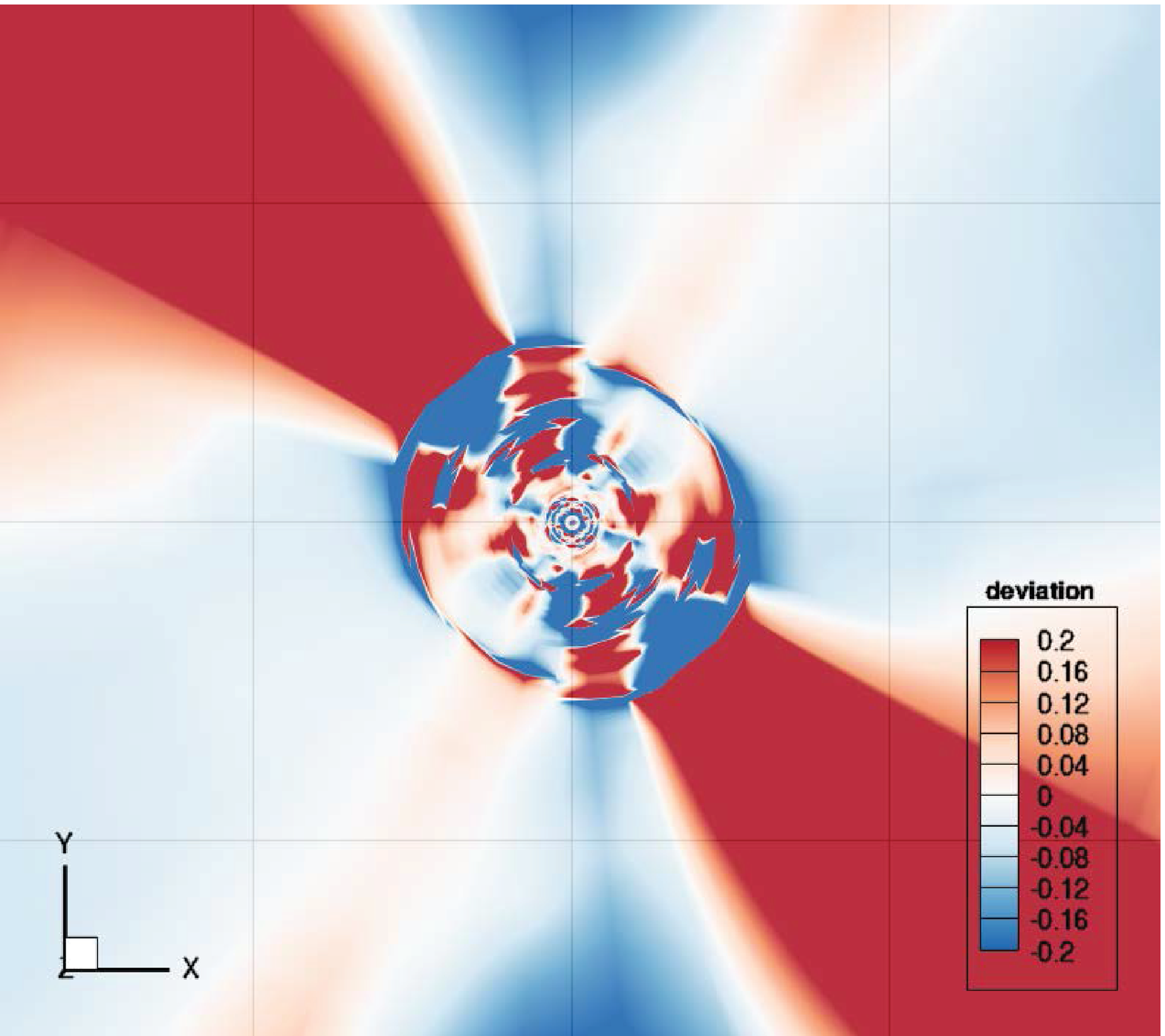}
\caption{Color maps for the neutrino-heating rate on the xy-plane 
are shown for the 11M model at 150 ms after the bounce.  
The top and middle panels display 
the energy transfer rates via neutrinos 
(heating by red and cooling by blue) 
evaluated by the 6D Boltzmann solver 
and the ray-by-ray approximation, respectively.  
The relative difference of the ray-by-ray evaluation 
with respect to the 6D Boltzmann evaluation 
is shown in the bottom panel.  
Grid lines with 200 km spacing are shown in the background.  }
\label{fig:rbr.2d.xyslice-heating.3db}
\end{figure}

We describe here the behavior of the heating rates by the two methods
in the supernova core of 11M at 150 ms after the bounce.  
We display the heating rates on the xy-plane in Fig. \ref{fig:rbr.2d.xyslice-heating.3db} 
by the 6D Boltzmann (top) and ray-by-ray (middle) evaluations.  
The relative differences of the ray-by-ray evaluation 
with respect to the 6D Boltzmann evaluation 
are also shown in the bottom panel.  
While the profiles of heating region in red are similar each other, 
there are two spots with enhanced heating rates in the ray-by-ray evaluation.  
Along the two directions, the ray-by-ray approximation leads to 
the overestimation of $\sim20\%$.  
This enhancement corresponds to the enhanced densities and fluxes 
shown in Figs. \ref{fig:rbr.3db-density.xyslice} and \ref{fig:rbr.3db-flux.xyslice} 
and is caused by 
the inhomogenous distributions of neutrino degeneracy 
with the {\it hot spot} for neutrino emissions as discussed above.  
There are also regions of 
minor differences along some radial directions.  
Note that large differences near the gain radius in the central part 
appear since the energy transfer rate is very small there.  

\begin{figure}
\epsscale{0.32}
\plotone{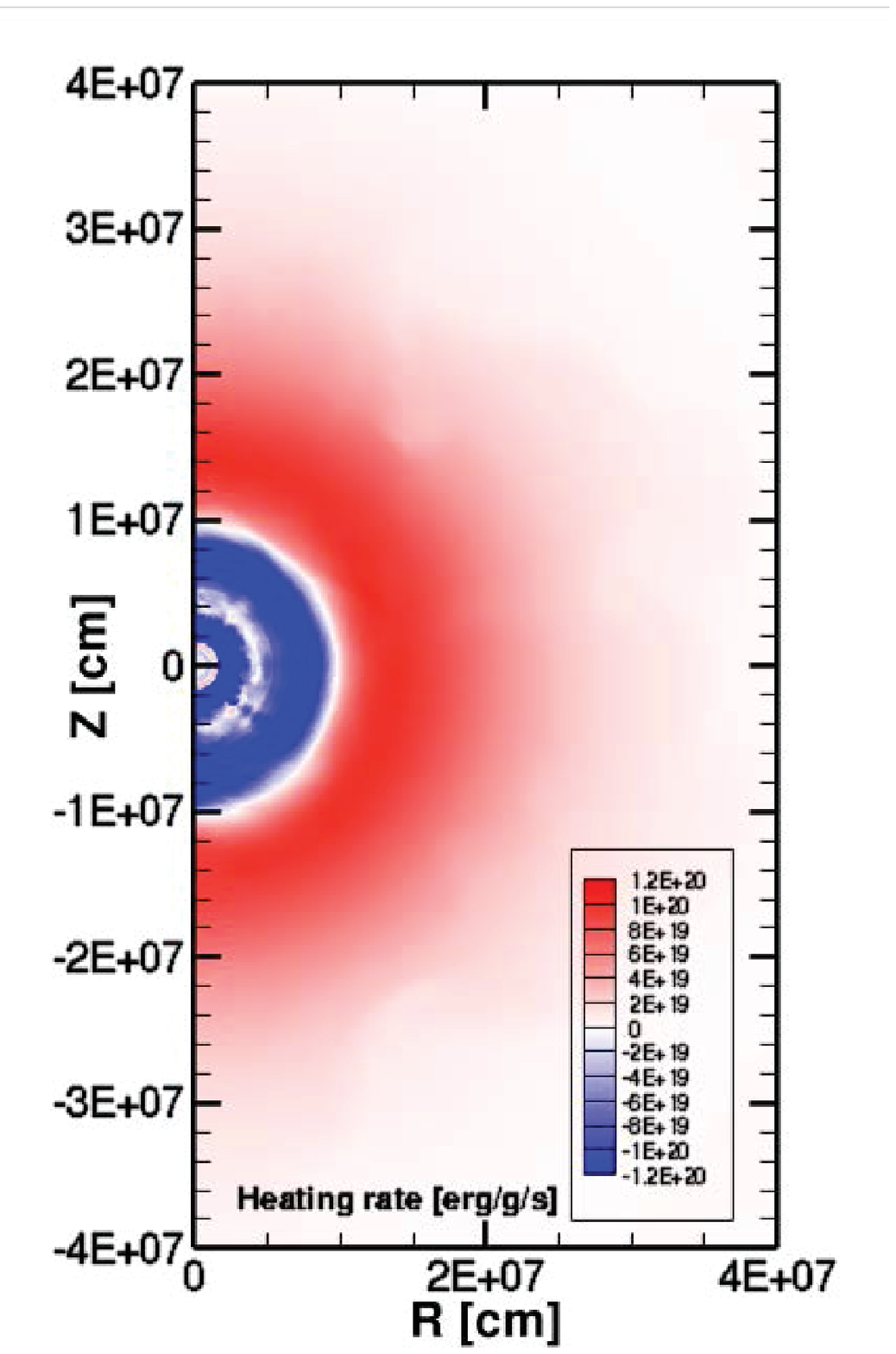}
\plotone{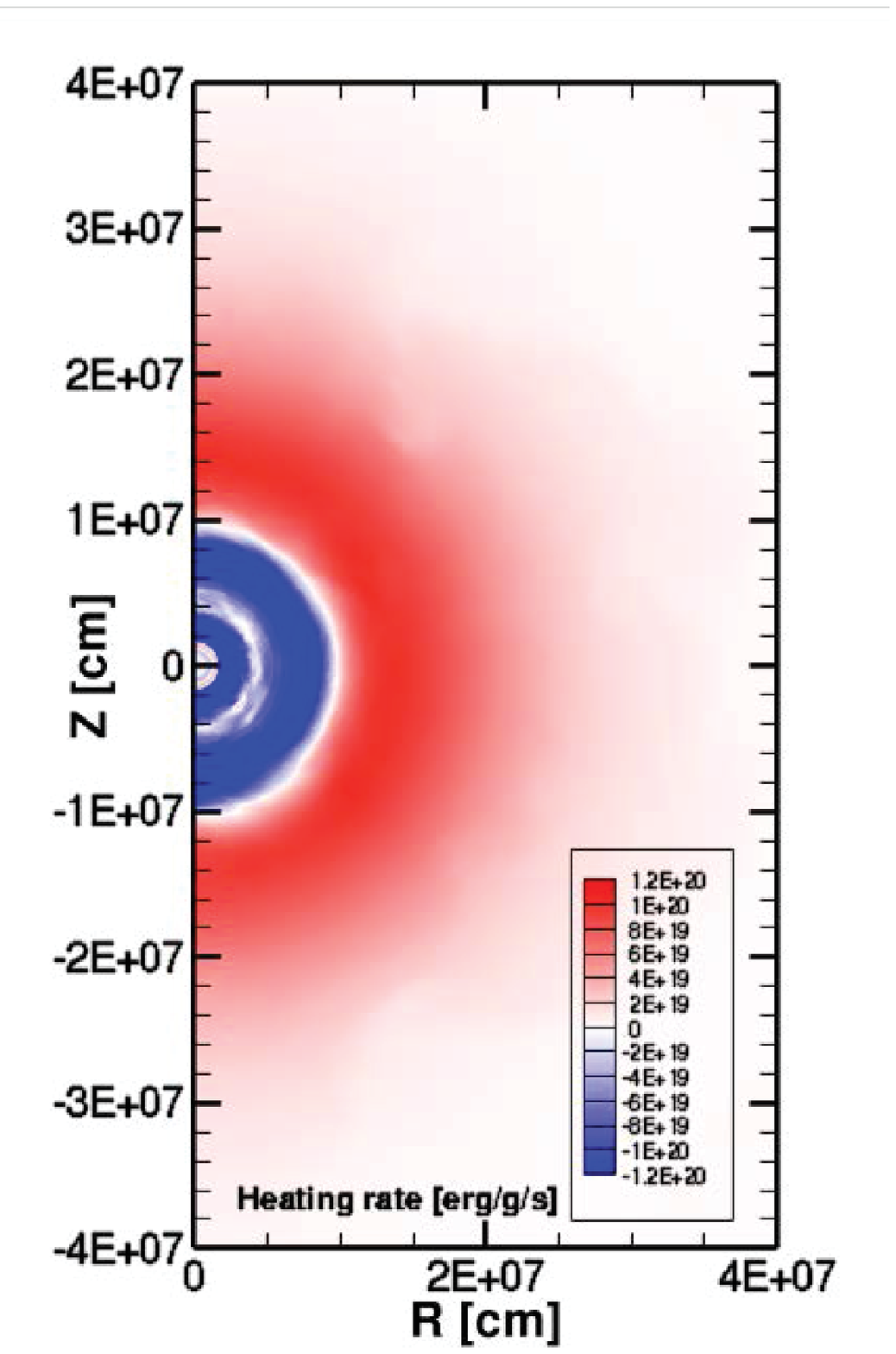}
\plotone{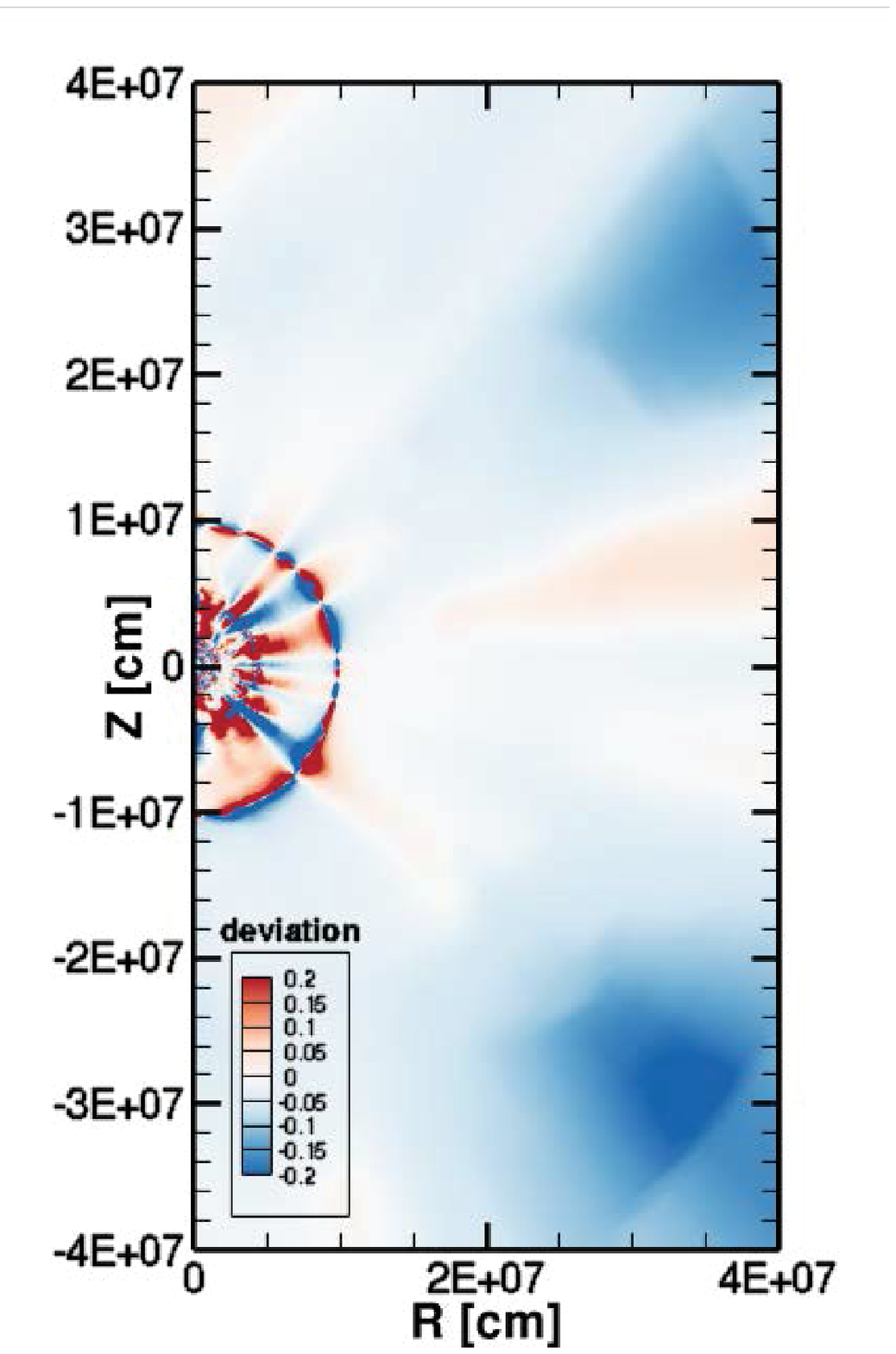}
\plotone{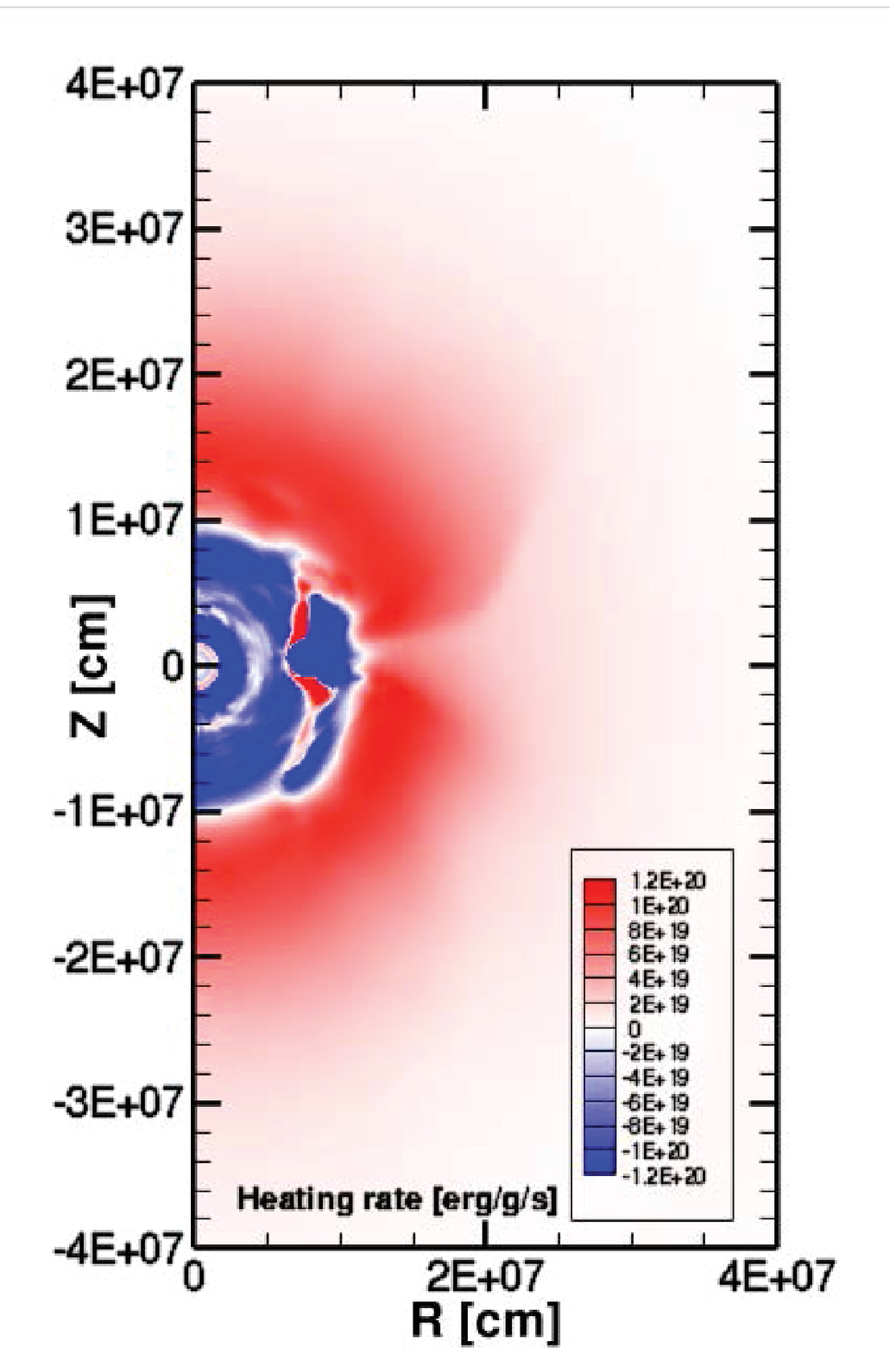}
\plotone{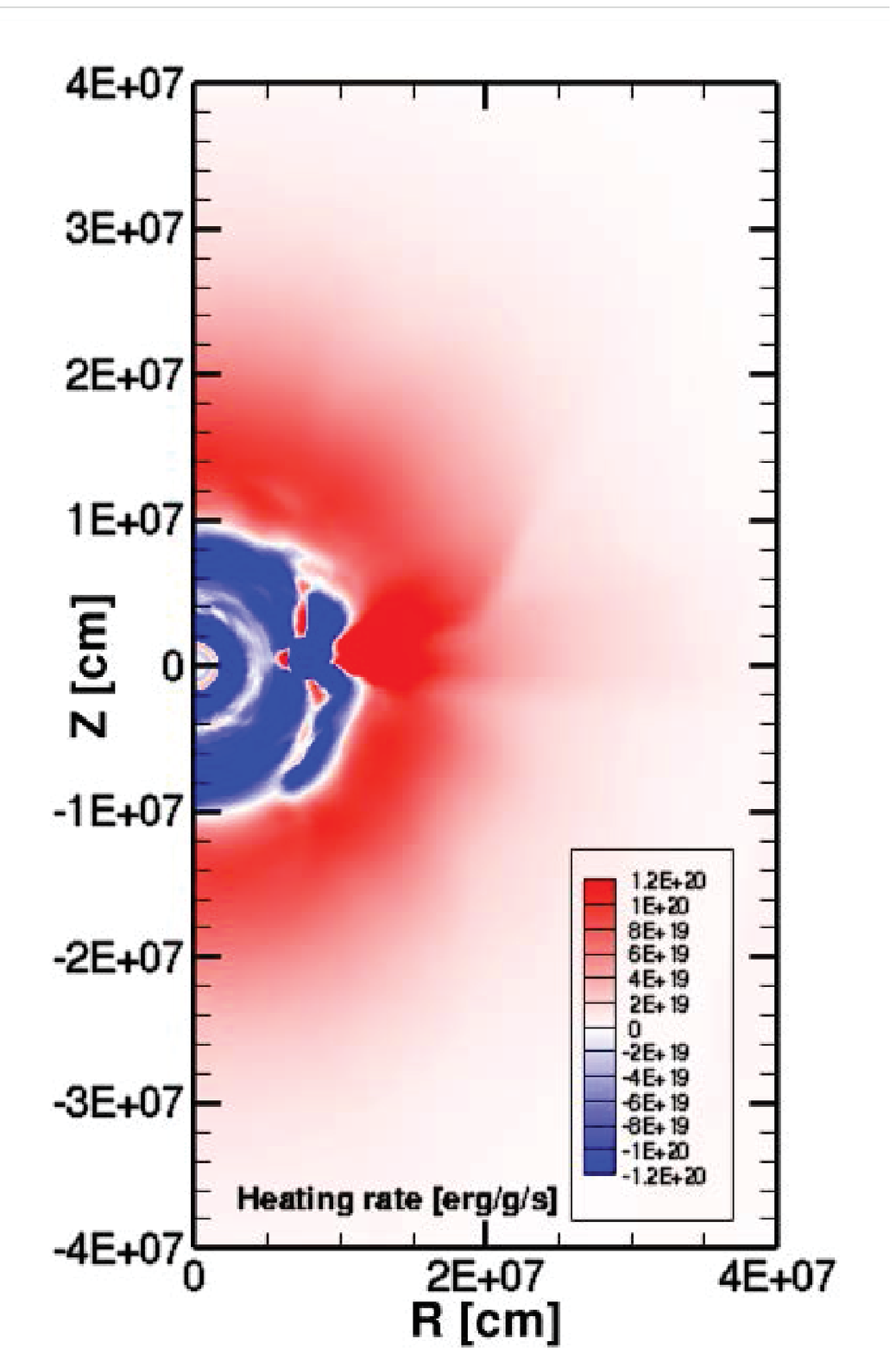}
\plotone{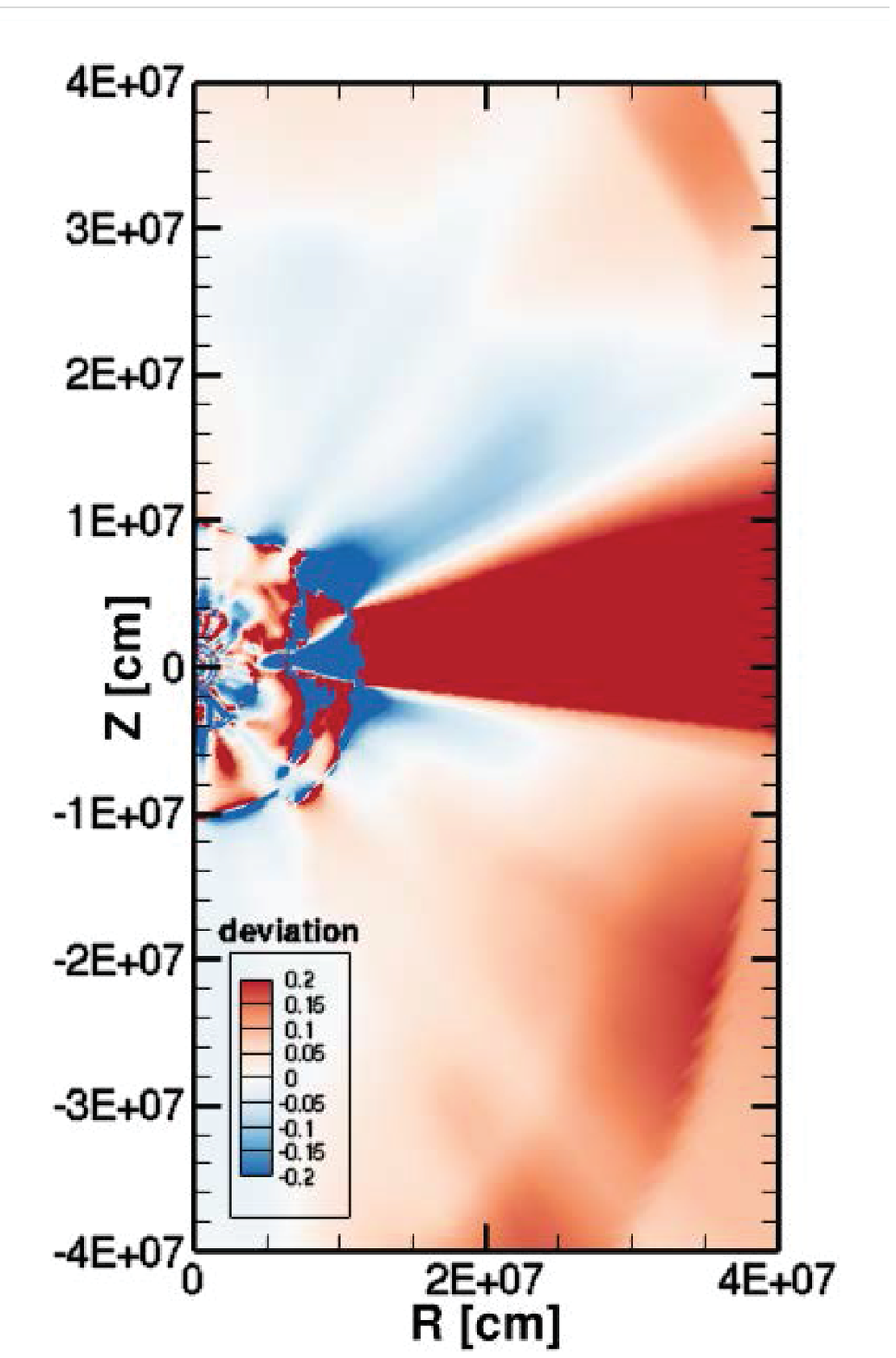}
\caption{Color maps for the neutrino-heating rate 
on the meridian slices at $\phi$=51$^\circ$ and 141$^\circ$ 
are shown for the 11M model at 150 ms after the bounce 
in the upper and lower panels, respectively.  
The left and middle panels display 
the energy transfer rates via neutrinos 
evaluated by the 6D Boltzmann solver 
and the ray-by-ray approximation, respectively.  
The relative difference of the ray-by-ray evaluation 
with respect to the 6D Boltzmann evaluation 
is shown in the right panel.  }
\label{fig:rbr.2d.phslice-heating.3db.iphx}
\end{figure}

Figure \ref{fig:rbr.2d.phslice-heating.3db.iphx} displays 
the profiles for the neutrino-heating rate 
on the meridian slices at $\phi$=51$^\circ$ and 141$^\circ$ 
corresponding to Fig. \ref{fig:rbr.2d.xyslice-heating.3db}.  
The heating rates by 
the 6D Boltzmann (left) and ray-by-ray (middle) evaluations 
are shown along with the relative differences of the ray-by-ray evaluation (right).  
In the case of $\phi$=51$^\circ$, there is no noticeable difference 
in the heating rates between the two methods 
except for a slight underestimation along two directions.  
A significant overestimation 
is seen around the equator in the case of $\phi$=141$^\circ$.  
This corresponds to the one of the azimuthal directions with overestimation 
seen in Fig. \ref{fig:rbr.2d.xyslice-heating.3db}.  
The overestimation in the ray-by-ray approximation occurs 
due to the enhanced neutrino fluxes in the narrow channel.  
There are also wide areas with overestimation 
and a radial direction with underestimation in the heating region.  

\begin{figure}
\epsscale{0.27}
\plotone{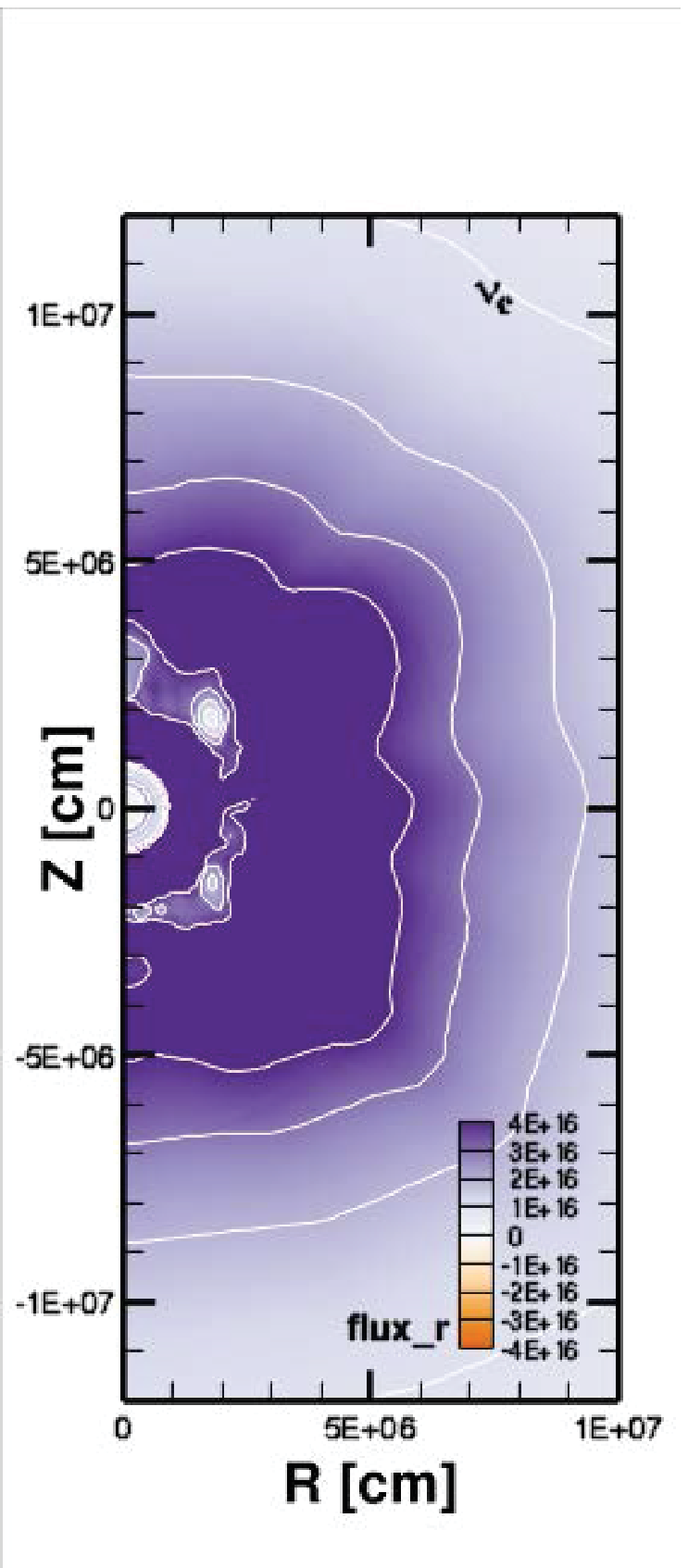}
\plotone{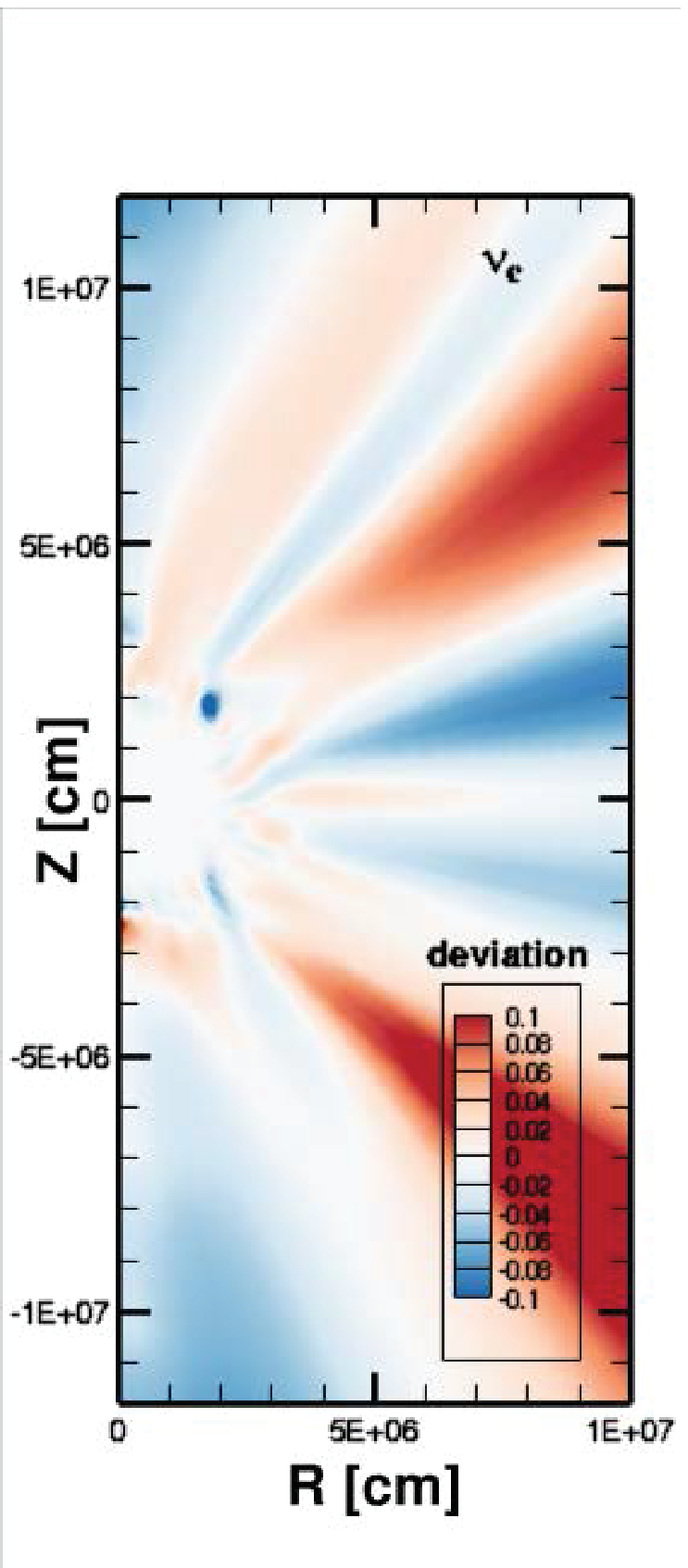}\\
\plotone{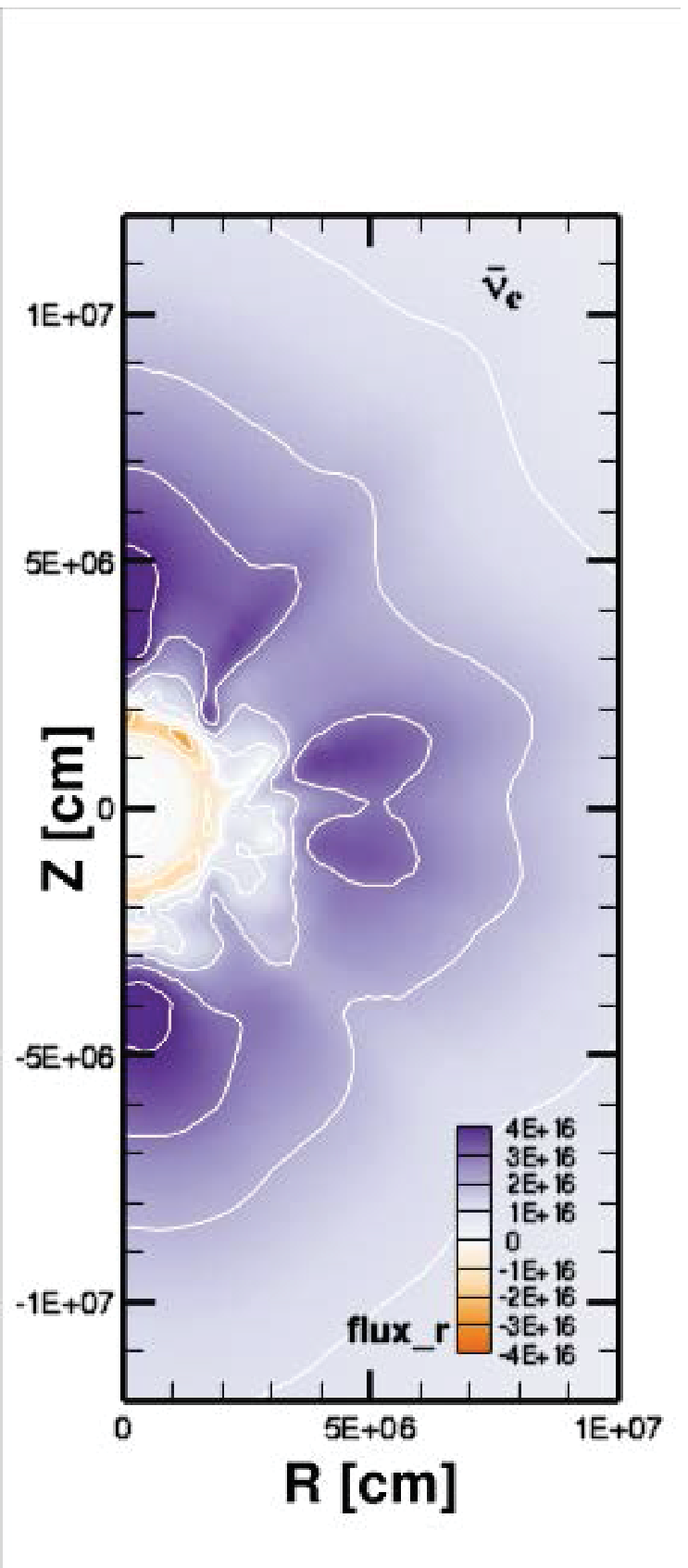}
\plotone{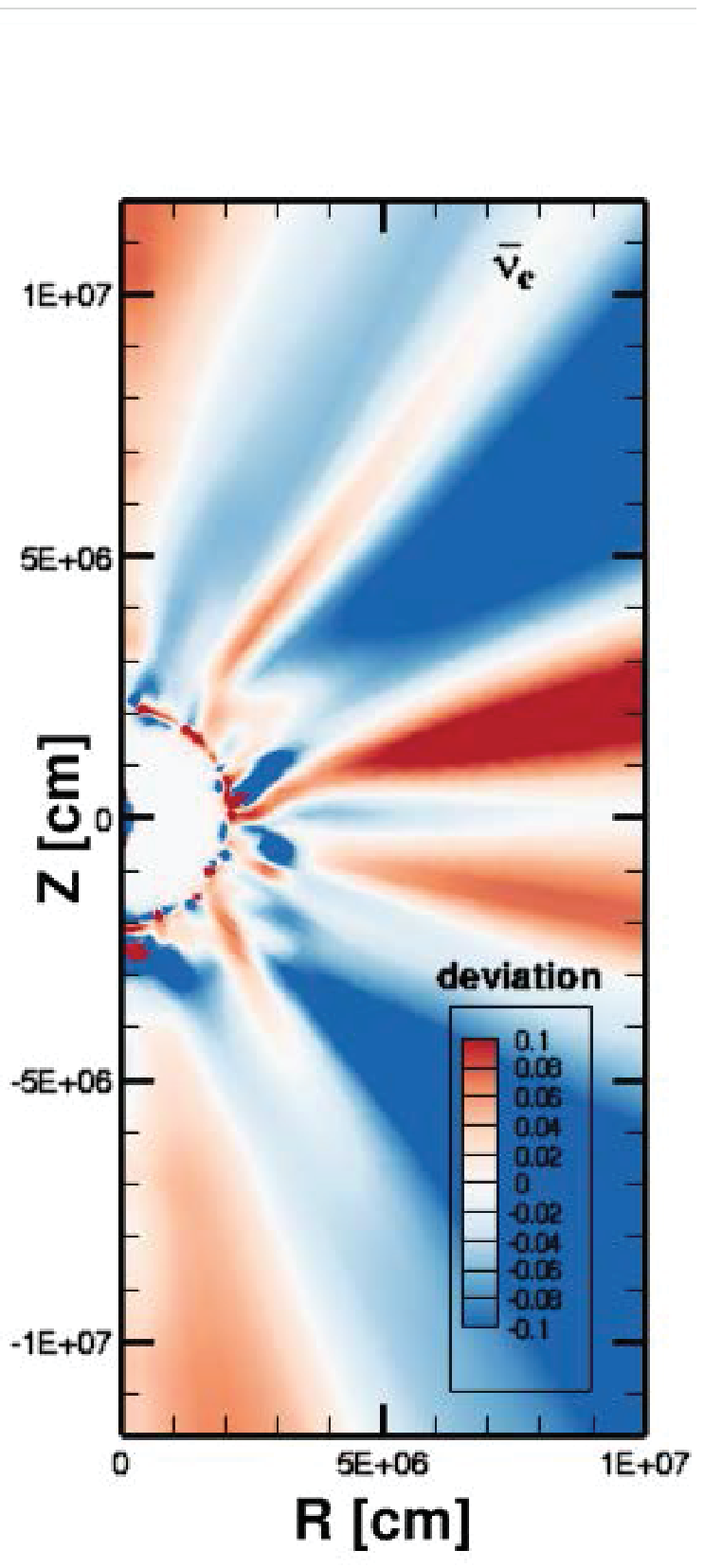}
\caption{Profiles of the number flux of $\nu_e$ (top) 
and $\bar{\nu}_e$ (bottom) 
in the ray-by-ray approximation 
and its relative differences with respect to the 6D Boltzmann evaluation 
on the meridian slice at $\phi$=51$^{\circ}$ 
are shown in the left and right panels, respectively, 
for the 11M model at 150 ms 
corresponding to Fig. \ref{fig:3db-flux.iph05.slice.inx}.  }
\label{fig:rbr.3db-flux.iph05.slice.inx}
\end{figure}

\begin{figure}
\epsscale{0.27}
\plotone{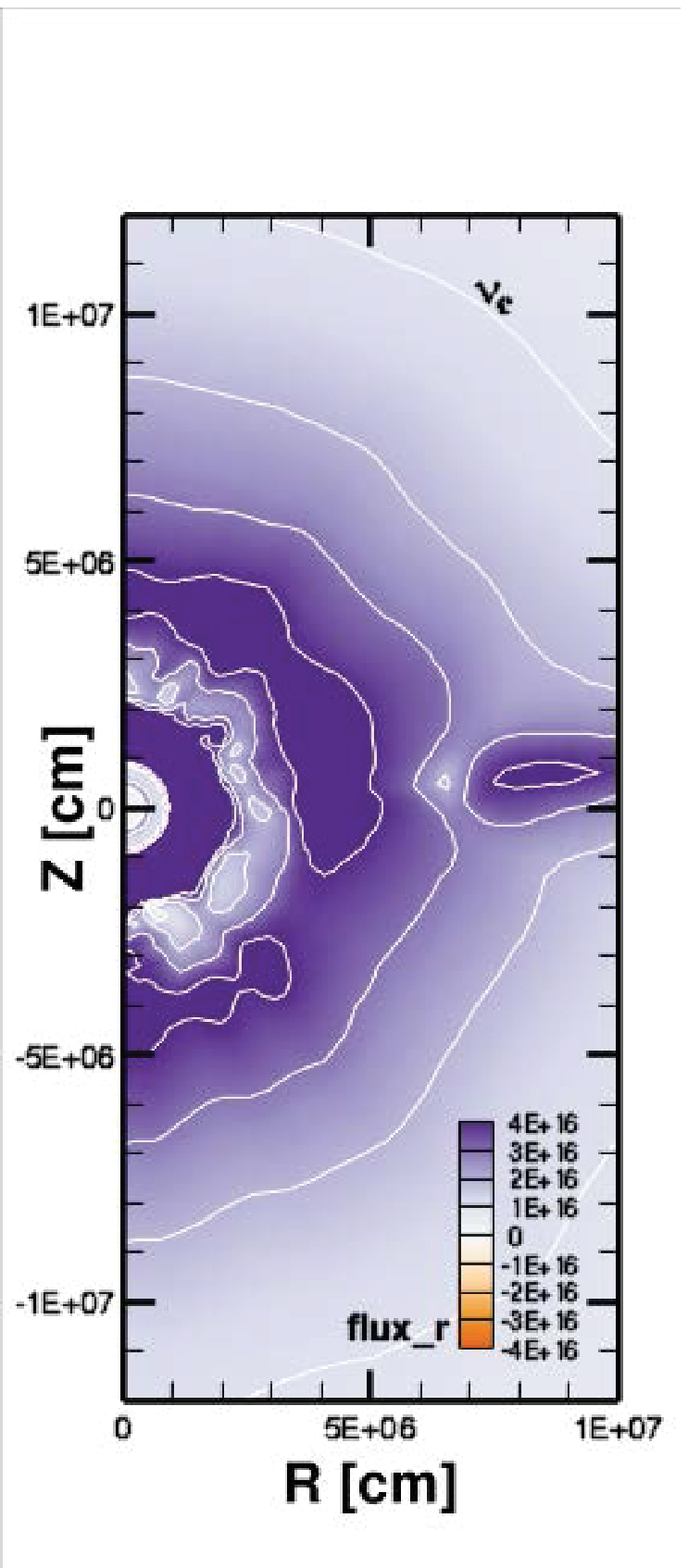}
\plotone{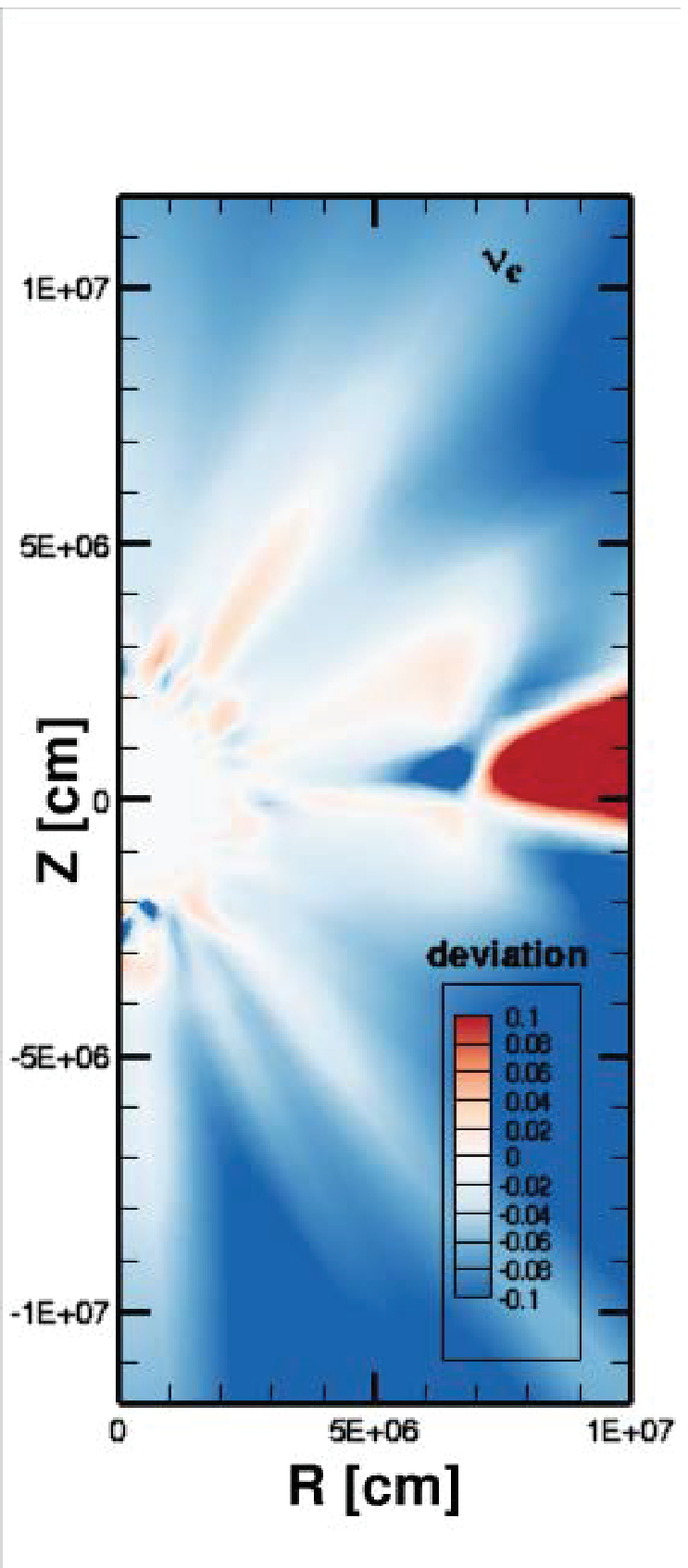}\\
\plotone{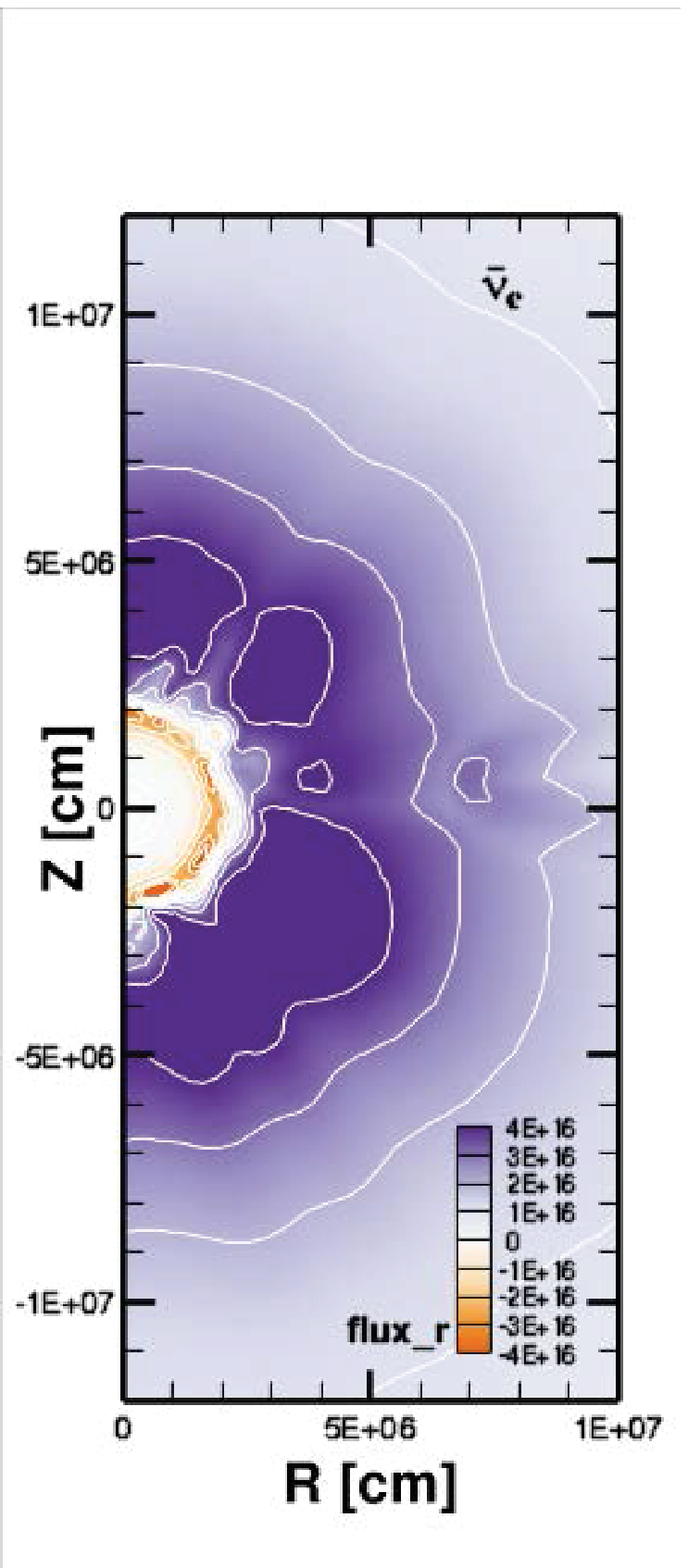}
\plotone{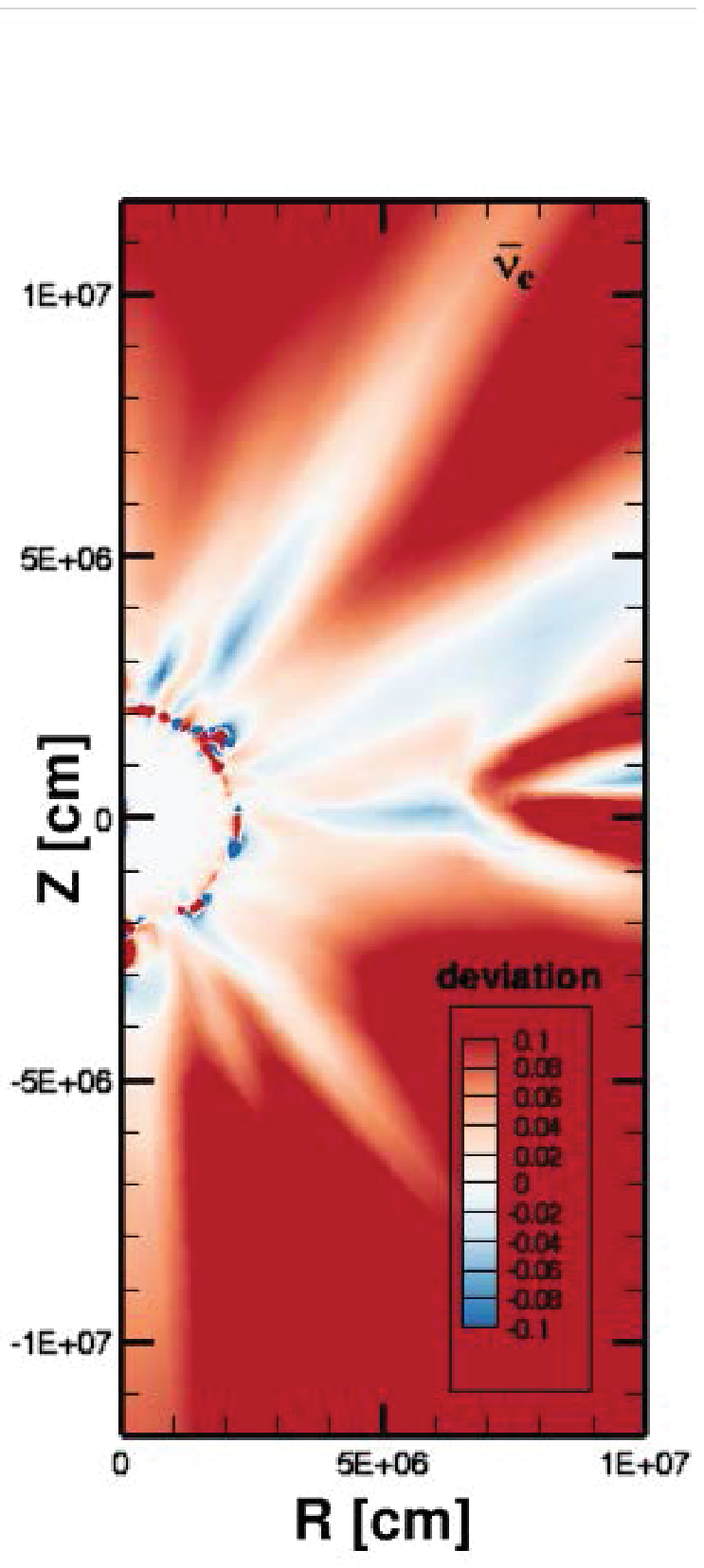}
\caption{Same as Figure \ref{fig:rbr.3db-flux.iph05.slice.inx} 
but on the meridian slice at $\phi$=141$^{\circ}$ 
corresponding to Fig. \ref{fig:3db-flux.iph13.slice.inx}.  }
\label{fig:rbr.3db-flux.iph13.slice.inx}
\end{figure}

\begin{figure}
\epsscale{0.32}
\plotone{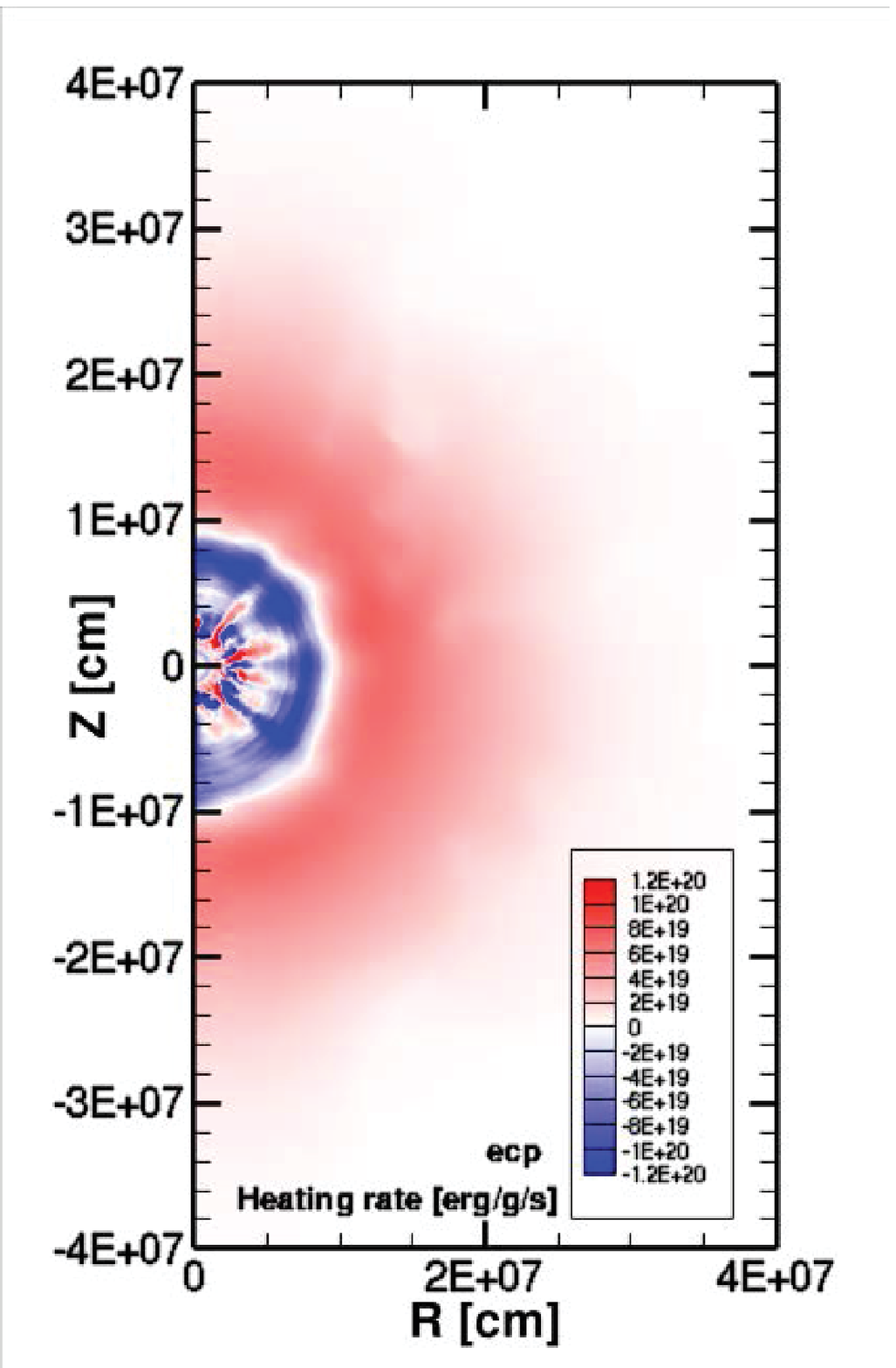}
\plotone{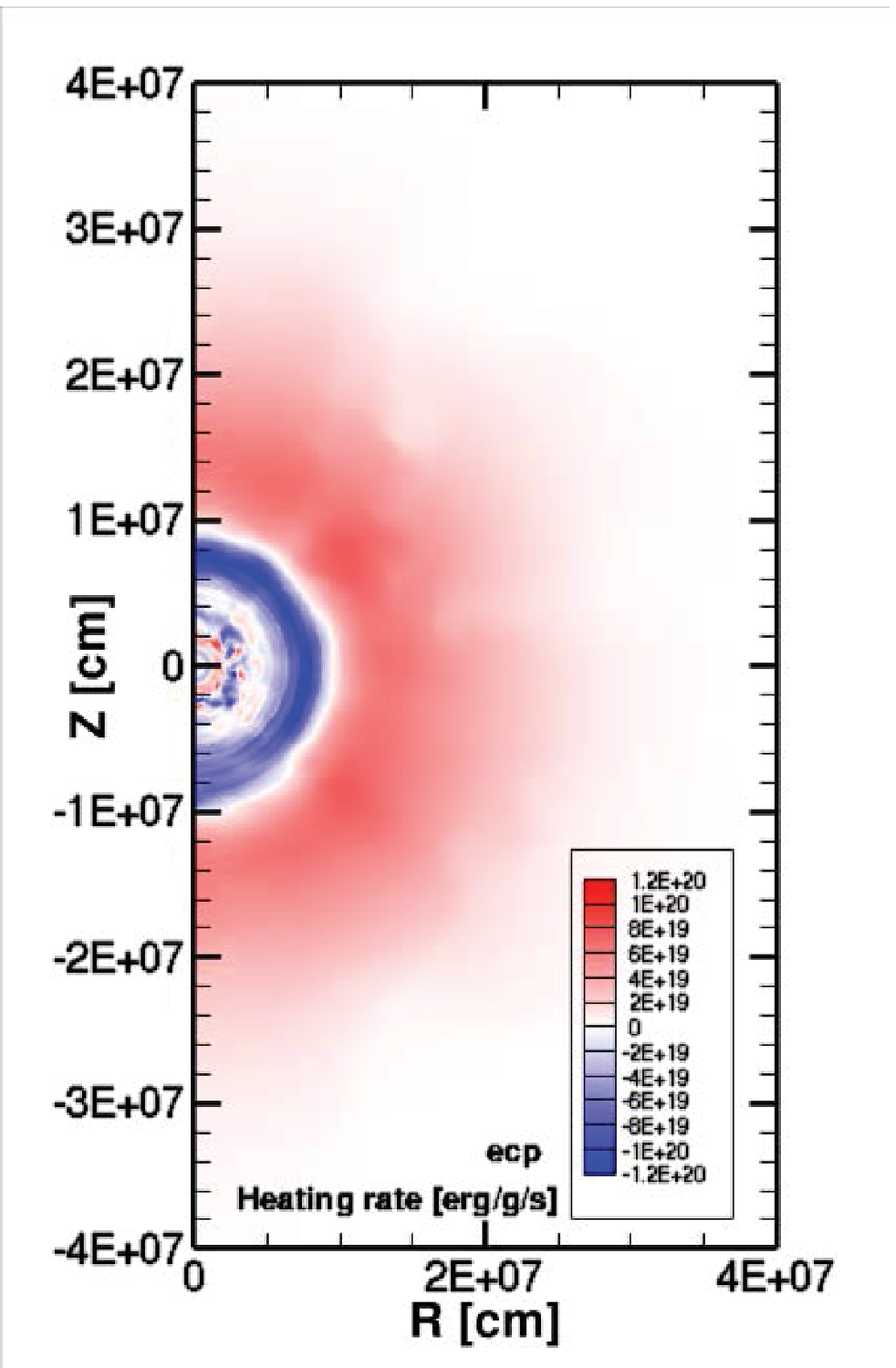}
\plotone{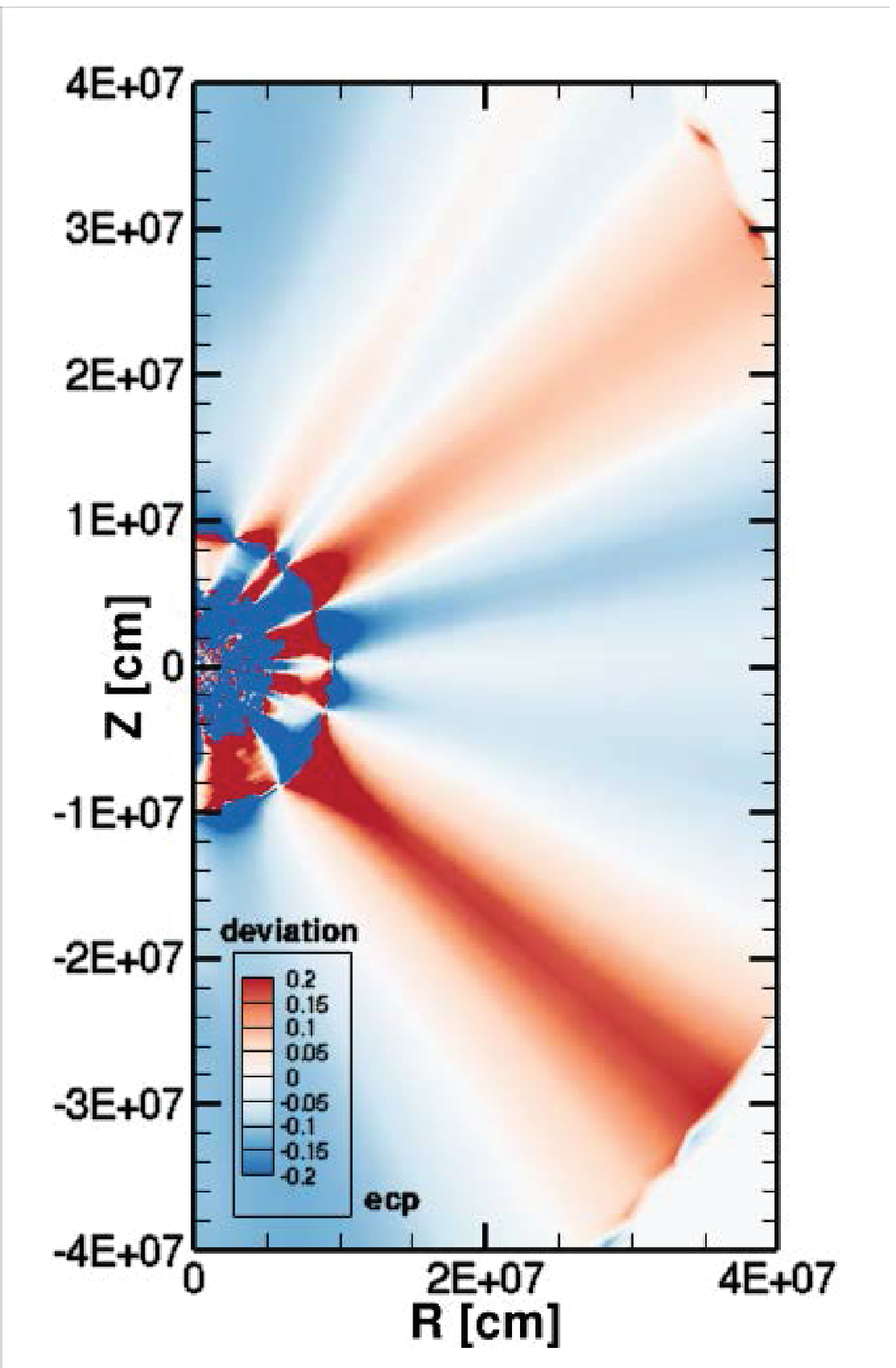}\\
\plotone{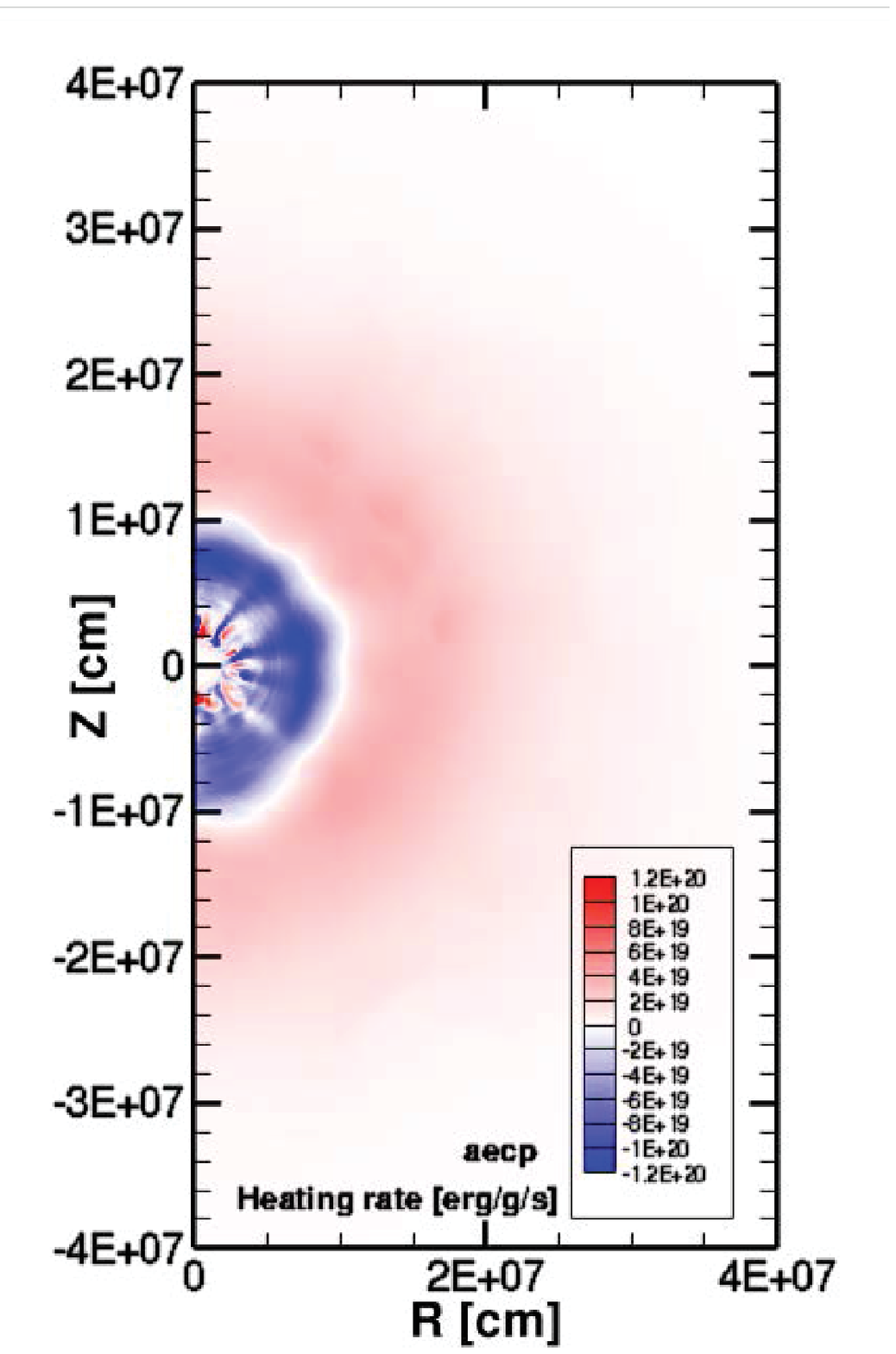}
\plotone{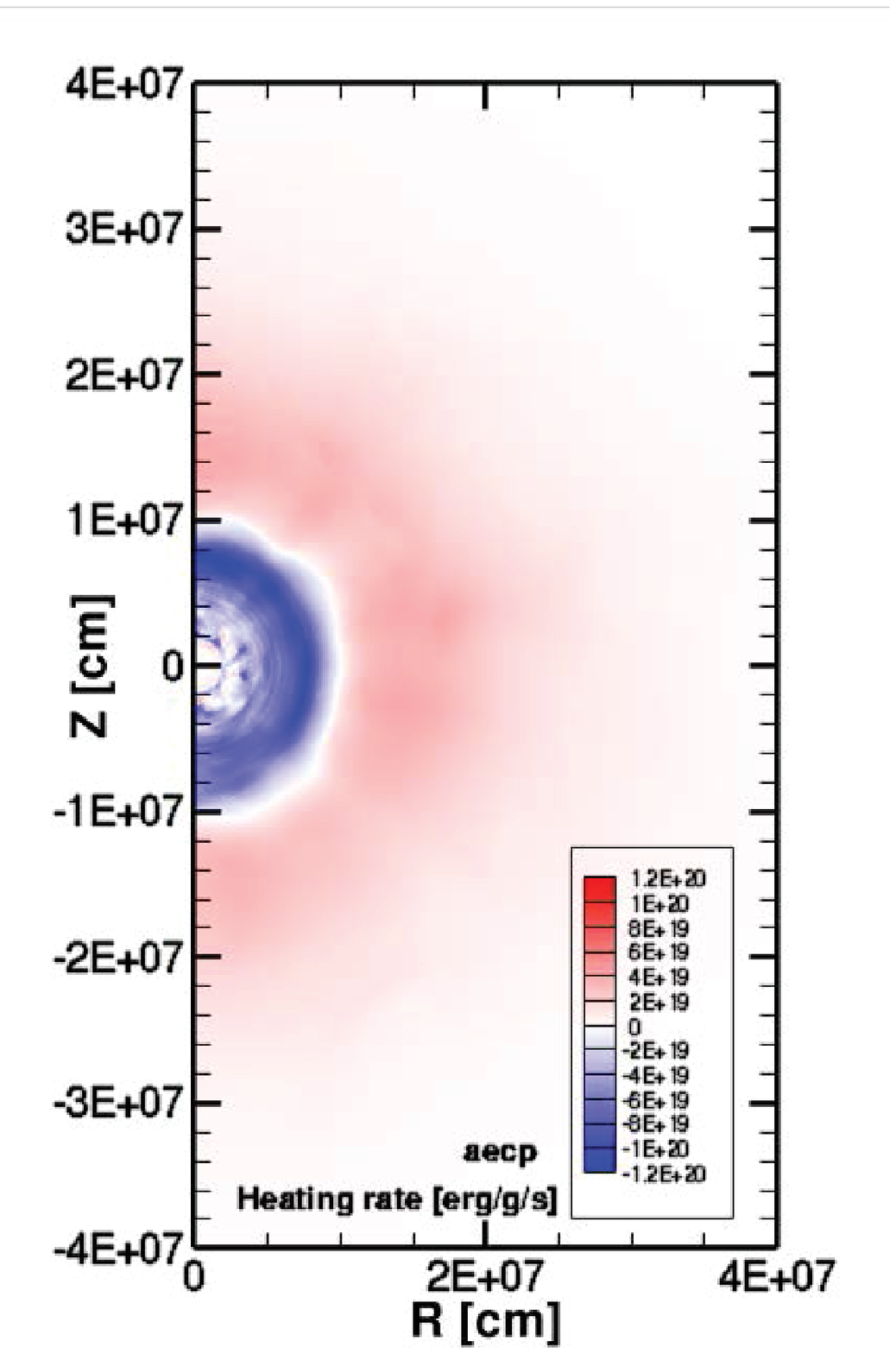}
\plotone{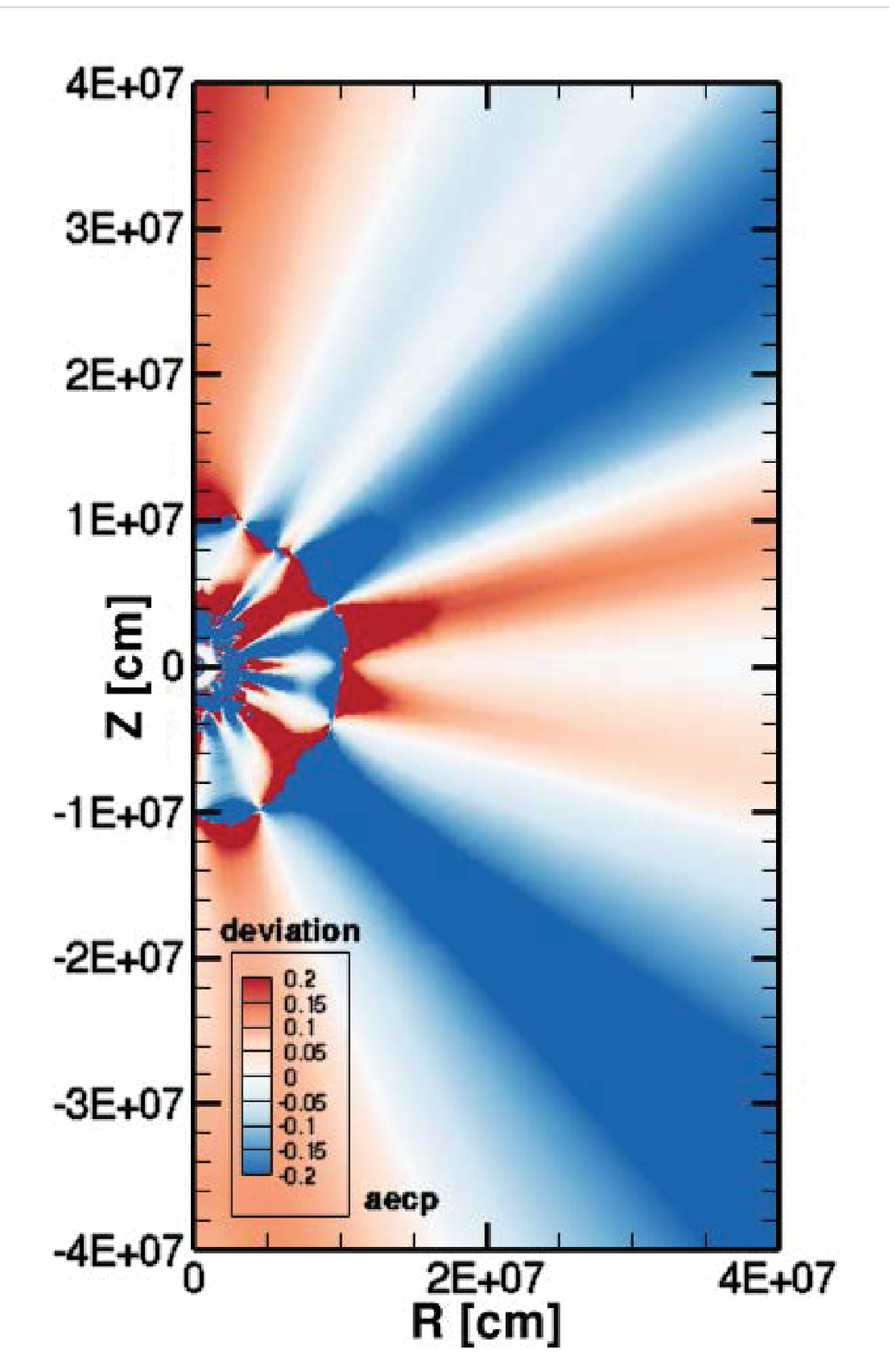}
\caption{Color maps for the heating rate 
via $\nu_e$ and $\bar{\nu}_e$ absorptions on nucleons 
on the meridian slice at $\phi$=51$^\circ$ 
are shown 
for the 11M model at 150 ms 
in the upper and lower panels, respectively.  
The left and middle panels display 
the 6D Boltzmann and ray-by-ray evaluations 
together with the relative difference of 
the ray-by-ray evaluation in the right panel.  }
\label{fig:rbr.2d.phslice-heating.3db.iph05.reactions}
\end{figure}

While the differences of the {\it total} heating rate appear 
in the limited directions, 
those of each heating rate for major reactions 
generally occur in the whole region.  
In Figure \ref{fig:rbr.2d.phslice-heating.3db.iph05.reactions}, 
we show the heating rates and their relative differences between the two methods 
for $\nu_e$ and $\bar{\nu}_e$ absorptions on nucleons 
on the meridian slice at $\phi$=51$^\circ$.  
One can see the pattern of enhancement or reduction of the individual heating rate 
just in the same way as those in the densities and fluxes.  
On the one hand, 
significant differences with an alternate sign 
appear along the radial ray 
for the $\nu_e$ absorption in the upper panel.  
This pattern arises from the differences of $\nu_e$ fluxes 
seen in Fig. \ref{fig:rbr.3db-flux.iph05.slice.inx}.  
On the other hand, 
differences of the similar size for the $\bar{\nu}_e$ absorption 
appear in the same pattern but with a different sign.  
This pattern reflects the corresponding pattern for $\bar{\nu}_e$ 
in Fig. \ref{fig:rbr.3db-flux.iph05.slice.inx}.  
Due to the production mechanism of the fluxes for two species, 
the overestimation and underestimation of the heating rates 
arise in the opposite way.  
Therefore, the total heating rates do not have 
large differences in most of the regions due to cancellation.  
This mechanism of the cancellation of the two heating rates 
is generally seen in our analyses of the models.  

The anisotropy of heating rates 
in the ray-by-ray approximation may have influences 
on the hydrodynamics in multi-dimensions.  
The enhancement of heating beamed along a certain direction 
may lead to the modification of shock dynamics.  
In contrast, the 6D Boltzmann evaluation provides 
mildly anisotropic distributions of the neutrino heating 
by the integration from various angles.  
We remark here that the anisotropy in the ray-by-ray approximation 
is found in the fixed hydrodynamics.  
It is necessary to examine whether these features persist 
for a long time scale in dynamical situations 
by neutrino-radiation hydrodynamics (See also \S \ref{section:h-total.time}).  

\subsection{Heating rates in other profiles}\label{section:h-boltz-ray.others}

We explore generic characteristics of the heating rates by examining the profiles 
at different timing and models.  
We demonstrate that the characteristic differences along the radial ray 
due to the ray-by-ray approximation 
occur also in the other cases.  
Their magnitude and sign, however, vary depending on the environment.  

\subsubsection{Different snapshots of 11M}\label{section:h-boltz-ray.others.11M}

\begin{figure}
\epsscale{0.4}
\plotone{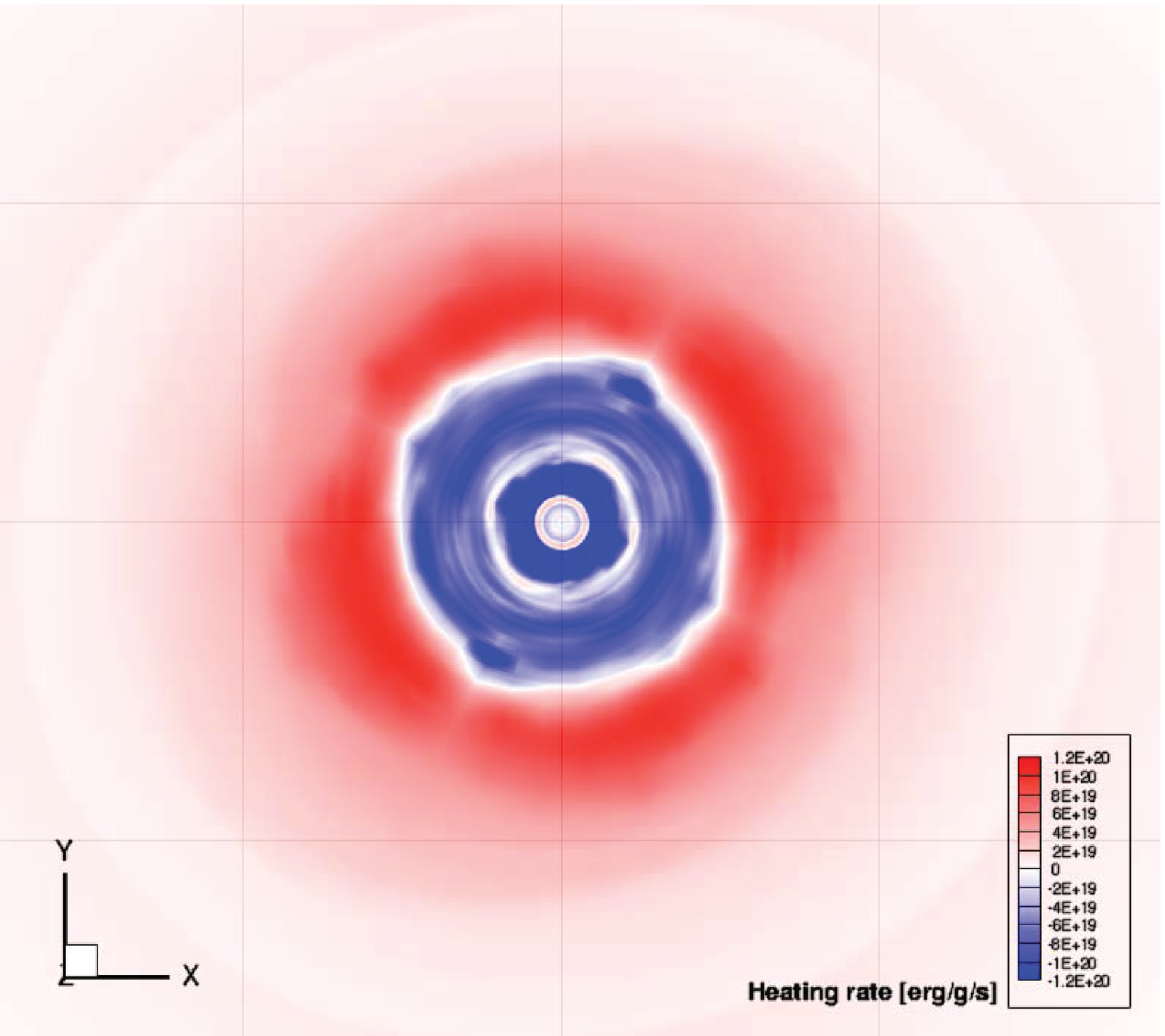}
\plotone{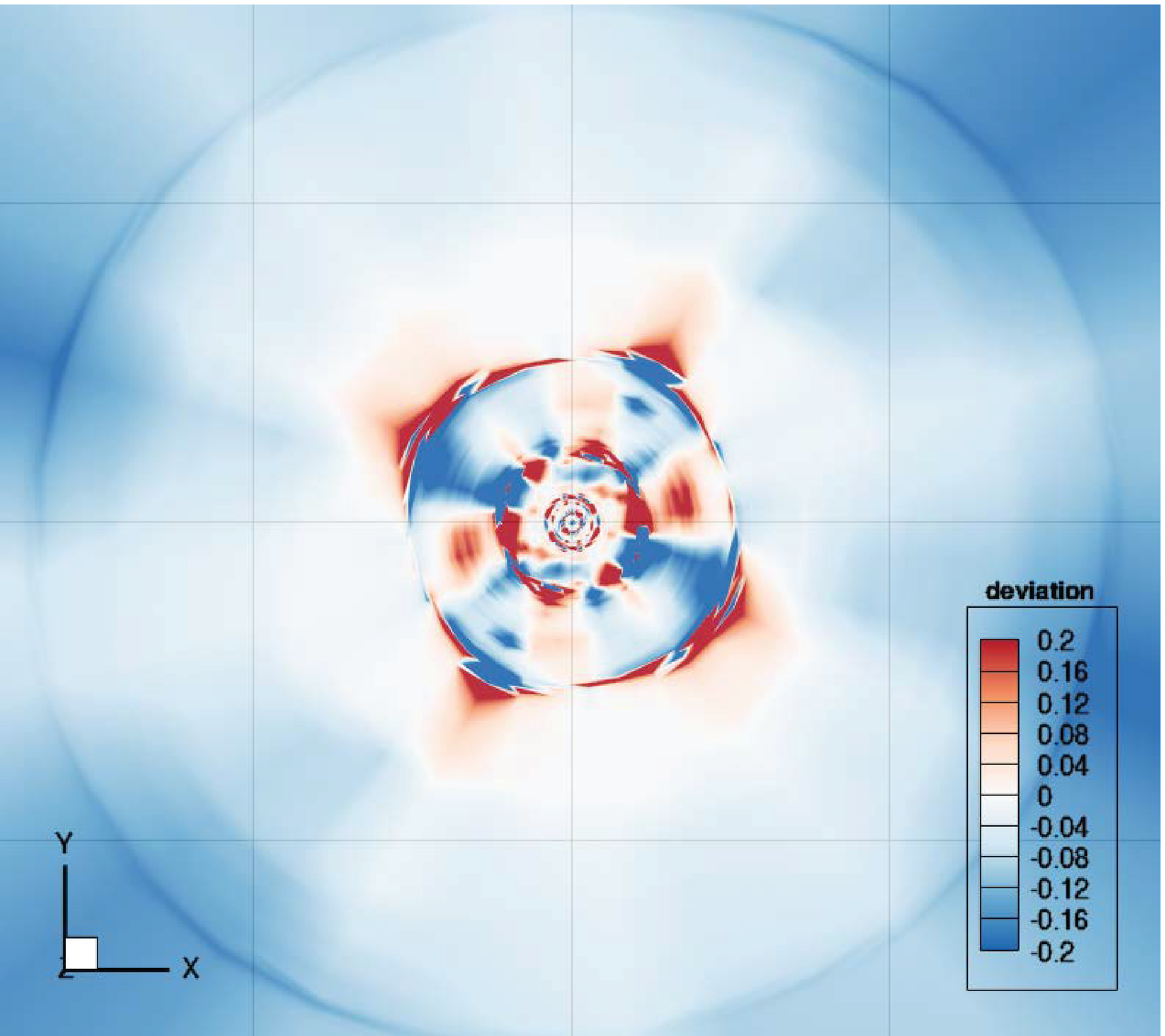}
\caption{Color maps of the heating rate 
obtained by the 6D Boltzmann solver 
for the 11M model at 100 ms after the bounce 
are shown in the left panel.  
The relative difference of the ray-by-ray evaluation 
is shown in the right panel.  
Grid lines with 200 km spacing are shown in the background.  }
\label{fig:2d.xyslice-heating.3da}
\end{figure}

\begin{figure}
\epsscale{0.32}
\plotone{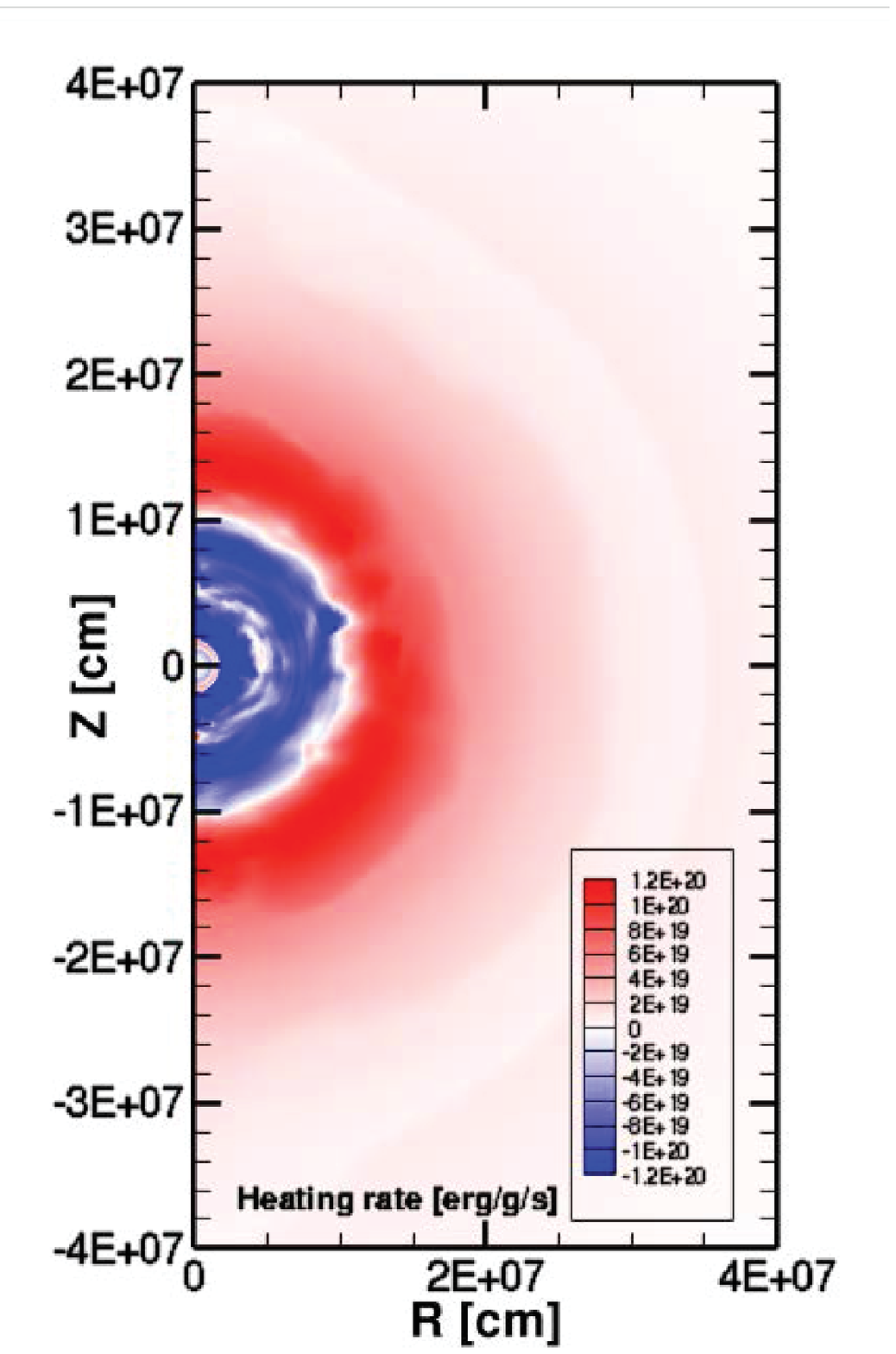}
\plotone{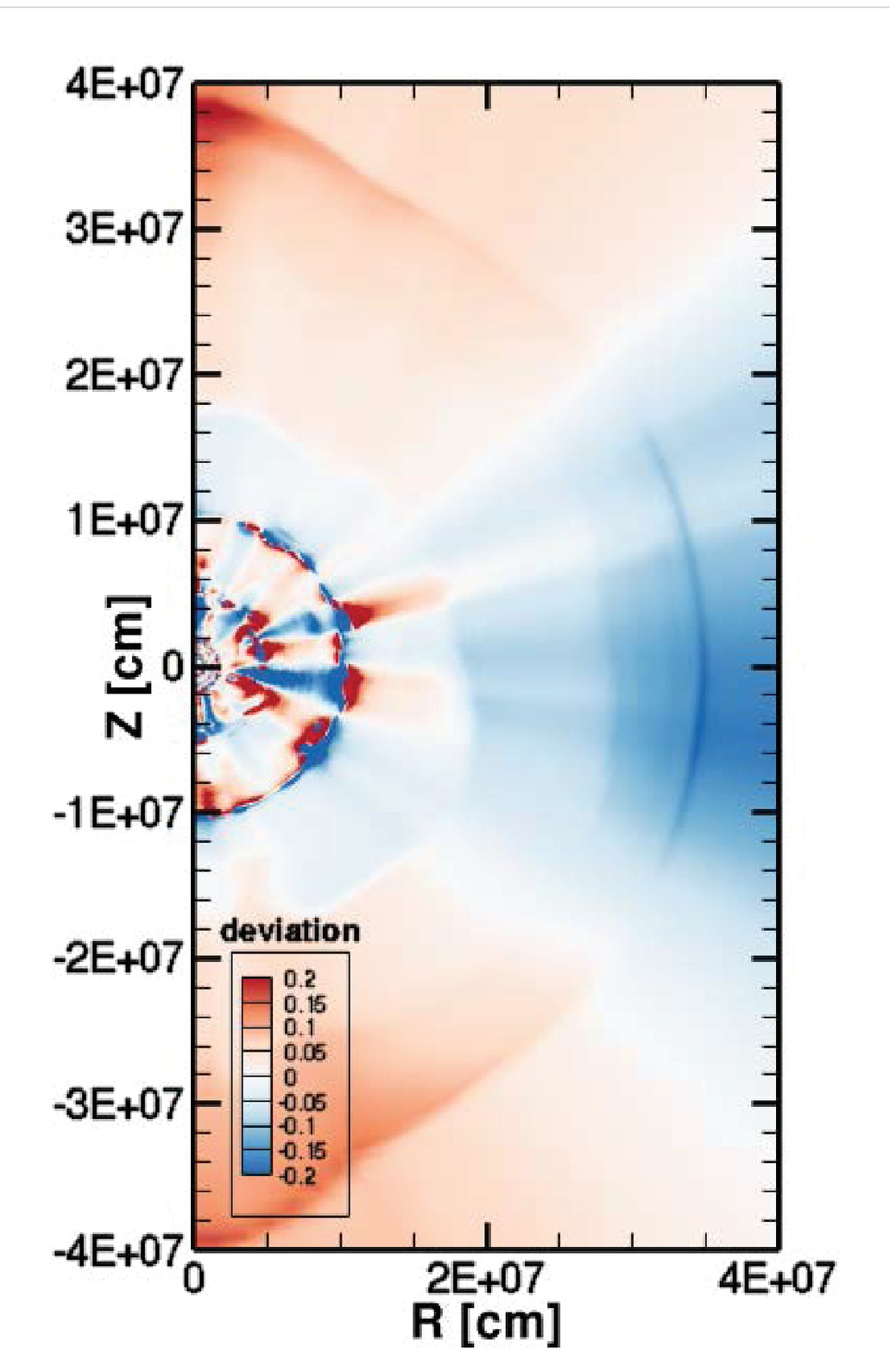}\\
\plotone{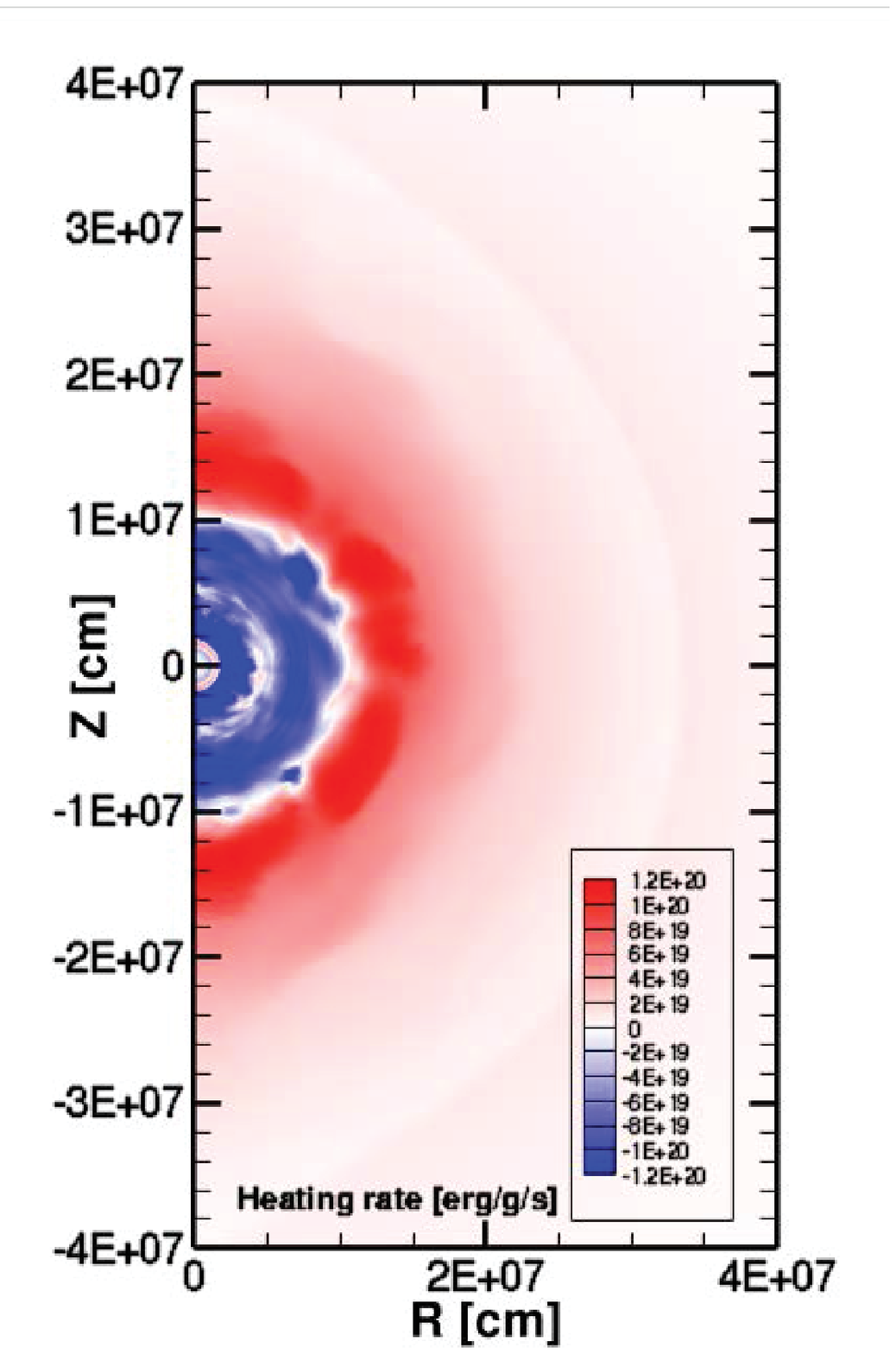}
\plotone{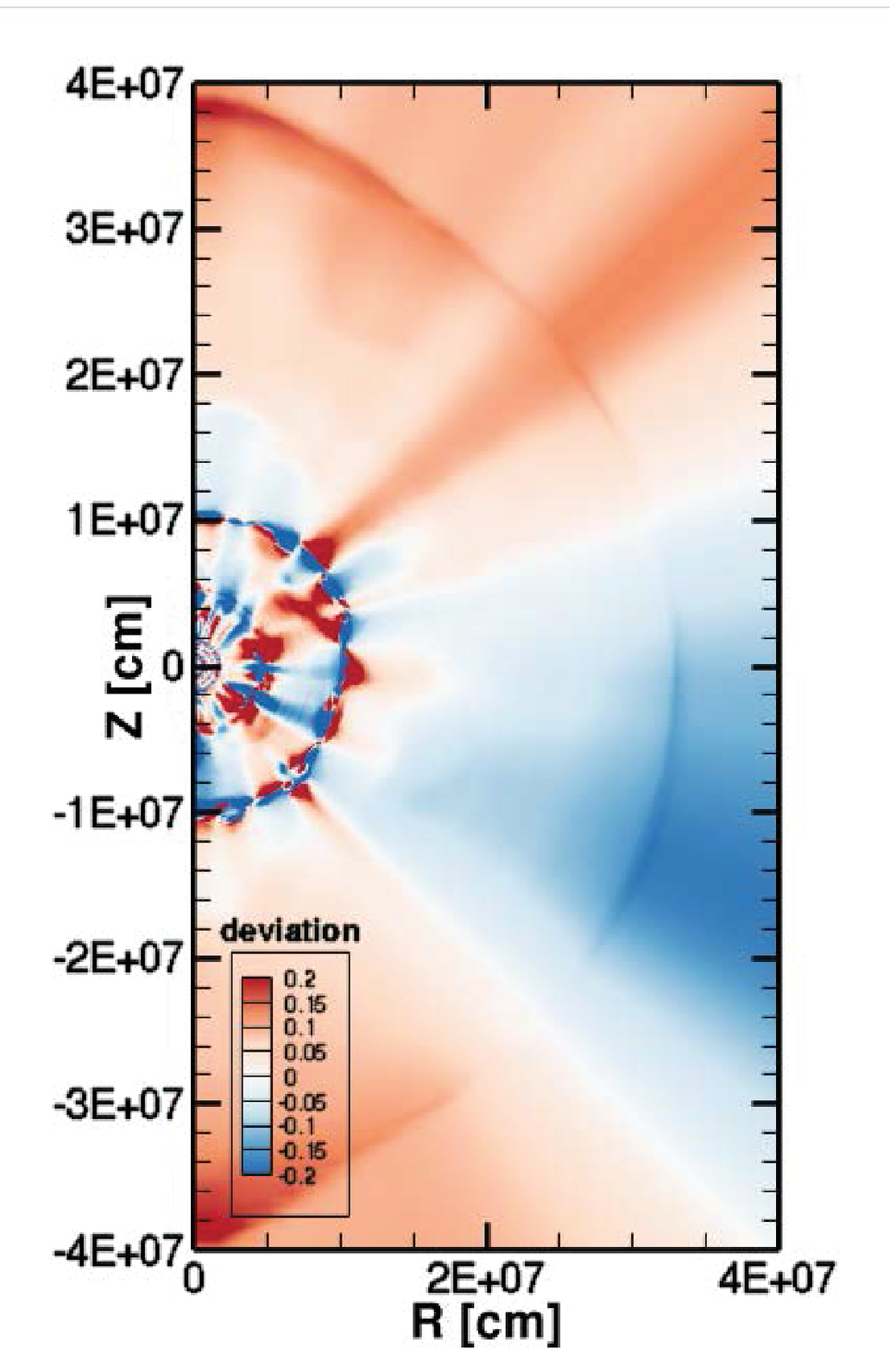}
\caption{Color maps of the heating rate 
by the 6D Boltzmann evaluation 
are shown in the left panels 
for the 11M model at 100 ms after the bounce.  
The corresponding profiles 
for the relative difference of the ray-by-ray evaluation 
are shown in the right panels.  
The top and bottom panels show the profiles 
on the meridian slices at $\phi$=51$^\circ$ and 141$^\circ$, respectively.  }
\label{fig:rbr.2d.phslice-heating.3da.iphx}
\end{figure}

The heating rates at 100 ms after the bounce by the 6D Boltzmann evaluation 
are shown in 
Figs. \ref{fig:2d.xyslice-heating.3da} 
and \ref{fig:rbr.2d.phslice-heating.3da.iphx} 
together with the relative differences of the ray-by-ray evaluation.  
One can see, in Fig. \ref{fig:2d.xyslice-heating.3da}, 
the deformed shape of the heating region and 
the underestimation of the heating rate 
in the ray-by-ray evaluation 
widespread on the xy-plane, 
being independent of $\phi$-direction, 
with a fluctuating size.  
In Fig. \ref{fig:rbr.2d.phslice-heating.3da.iphx}, 
the overestimation of the heating rate 
occurs around the z-axis within a wide polar angle 
while the underestimation is confined around the equator.  

\begin{figure}
\epsscale{0.4}
\plotone{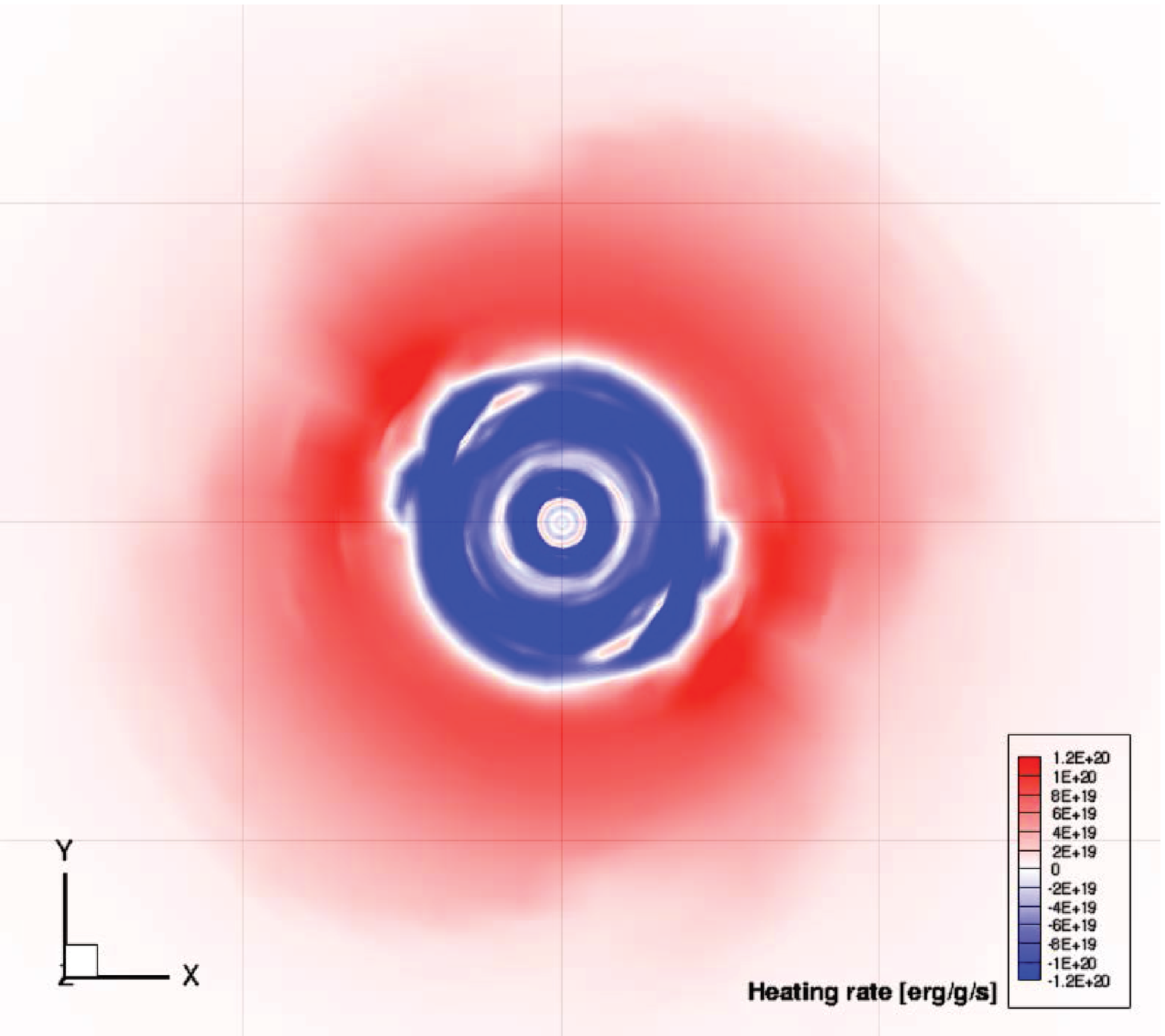}
\plotone{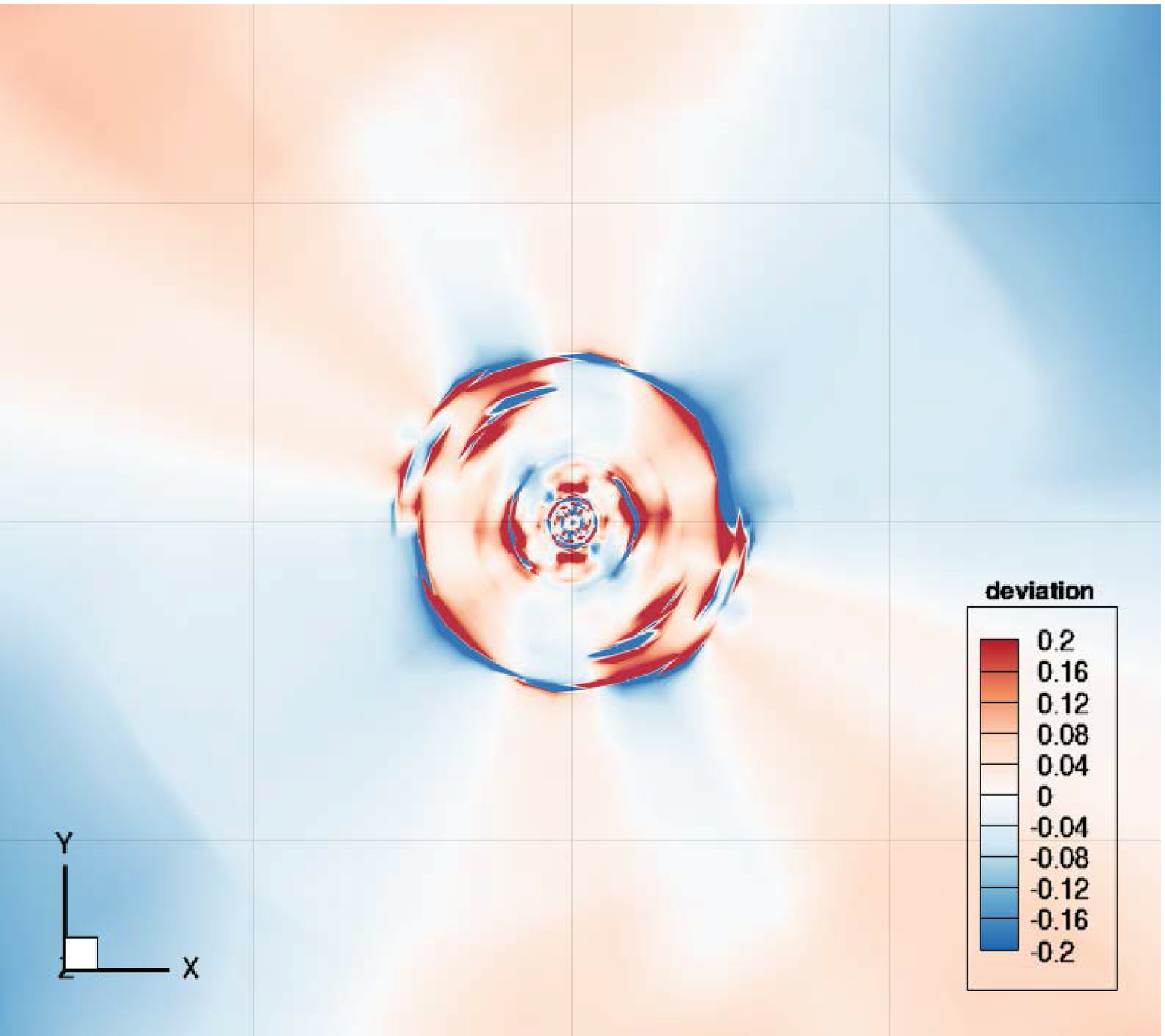}
\caption{Same as Fig. \ref{fig:2d.xyslice-heating.3da} but 
for the 11M model at 200 ms after the bounce.  }
\label{fig:2d.xyslice-heating.3dc}
\end{figure}

\begin{figure}
\epsscale{0.32}
\plotone{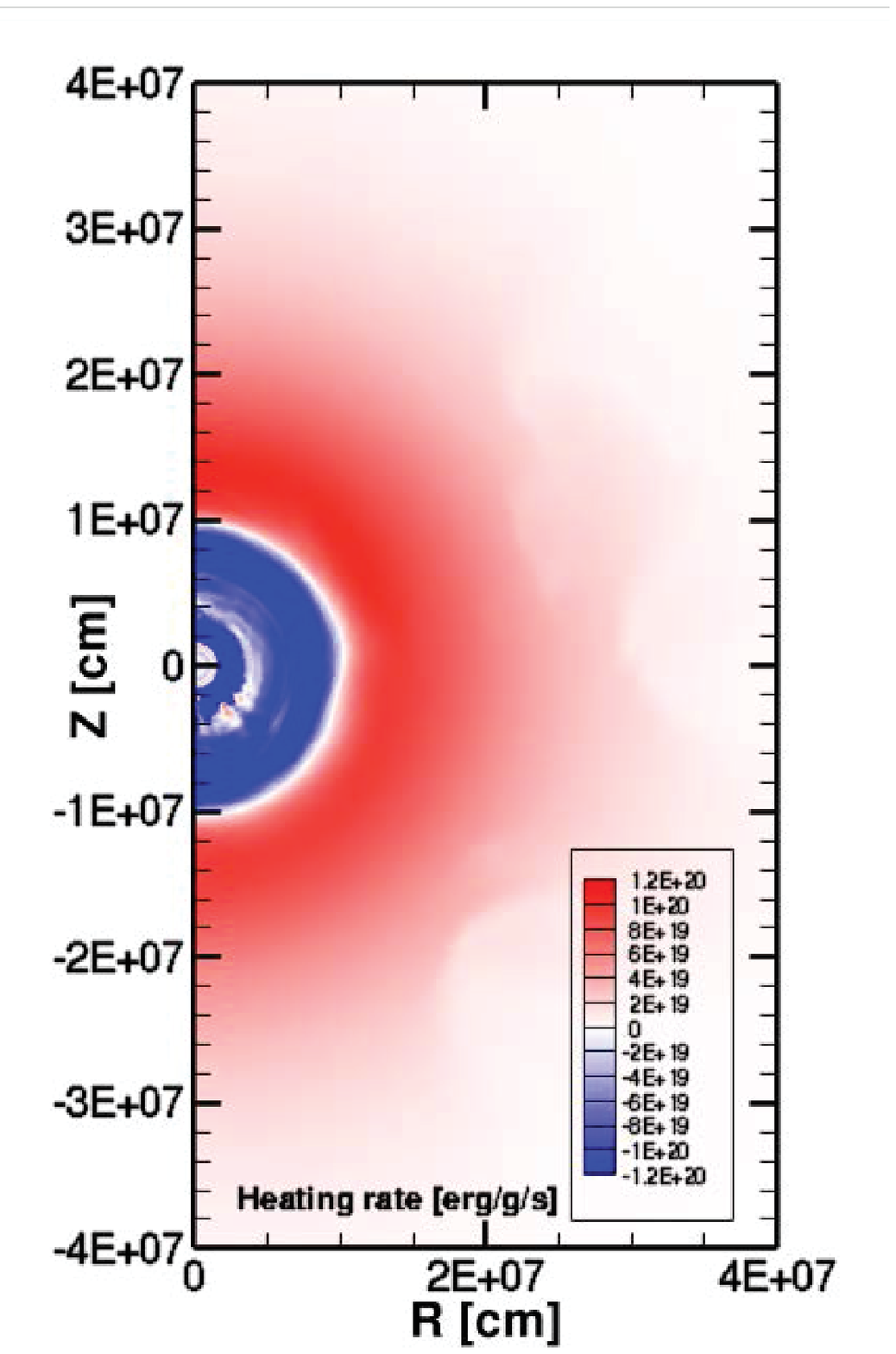}
\plotone{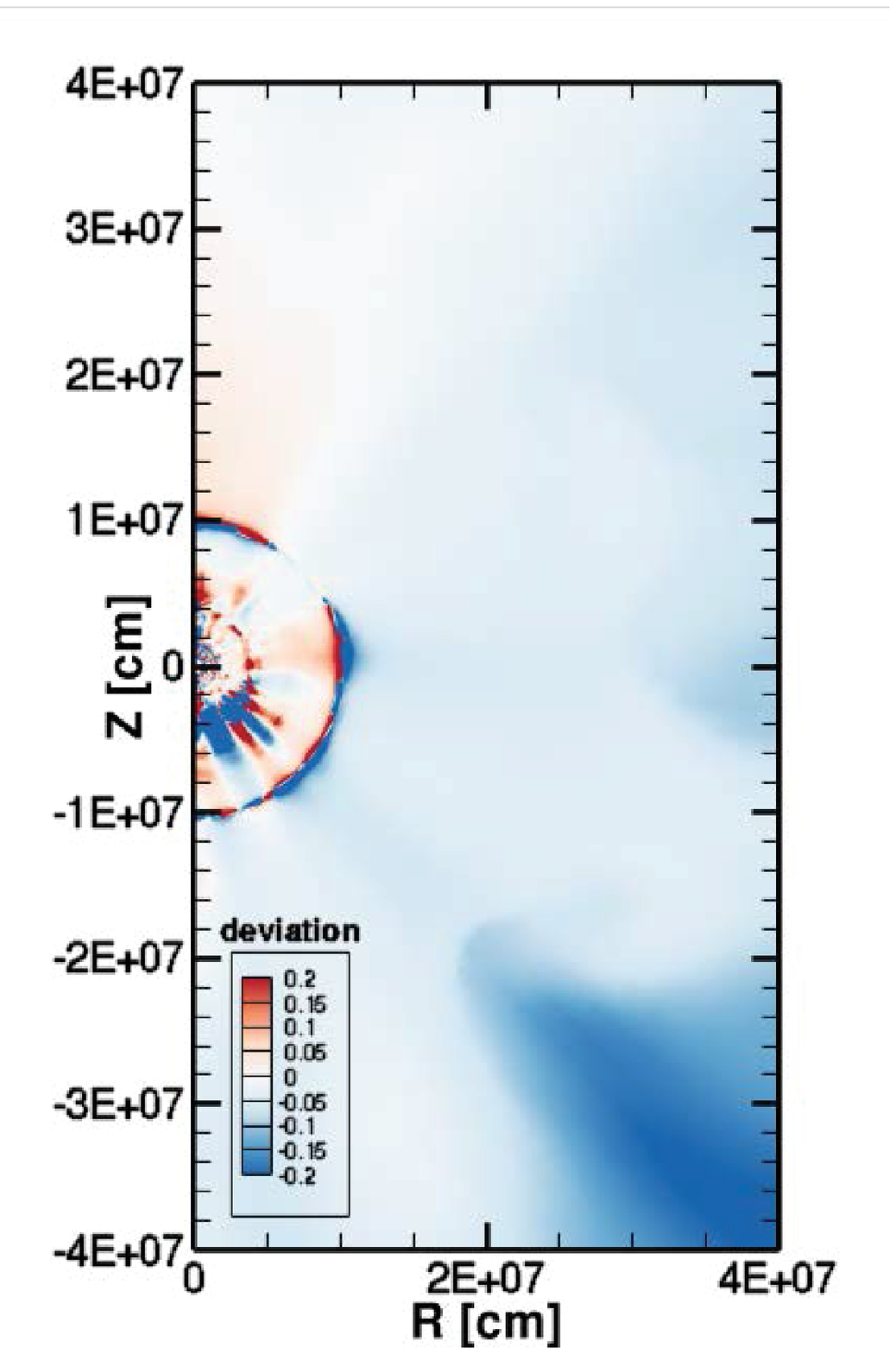}\\
\plotone{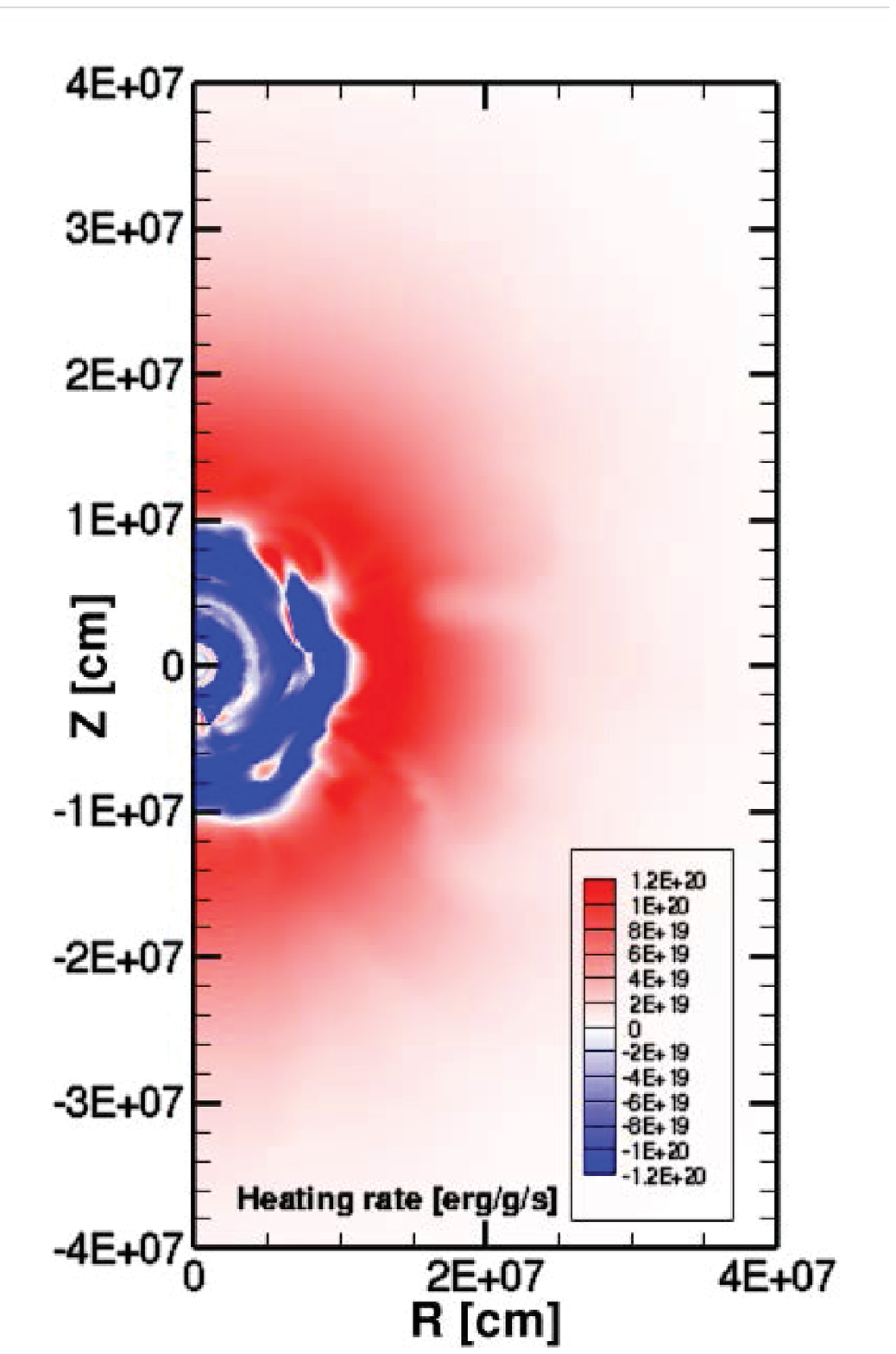}
\plotone{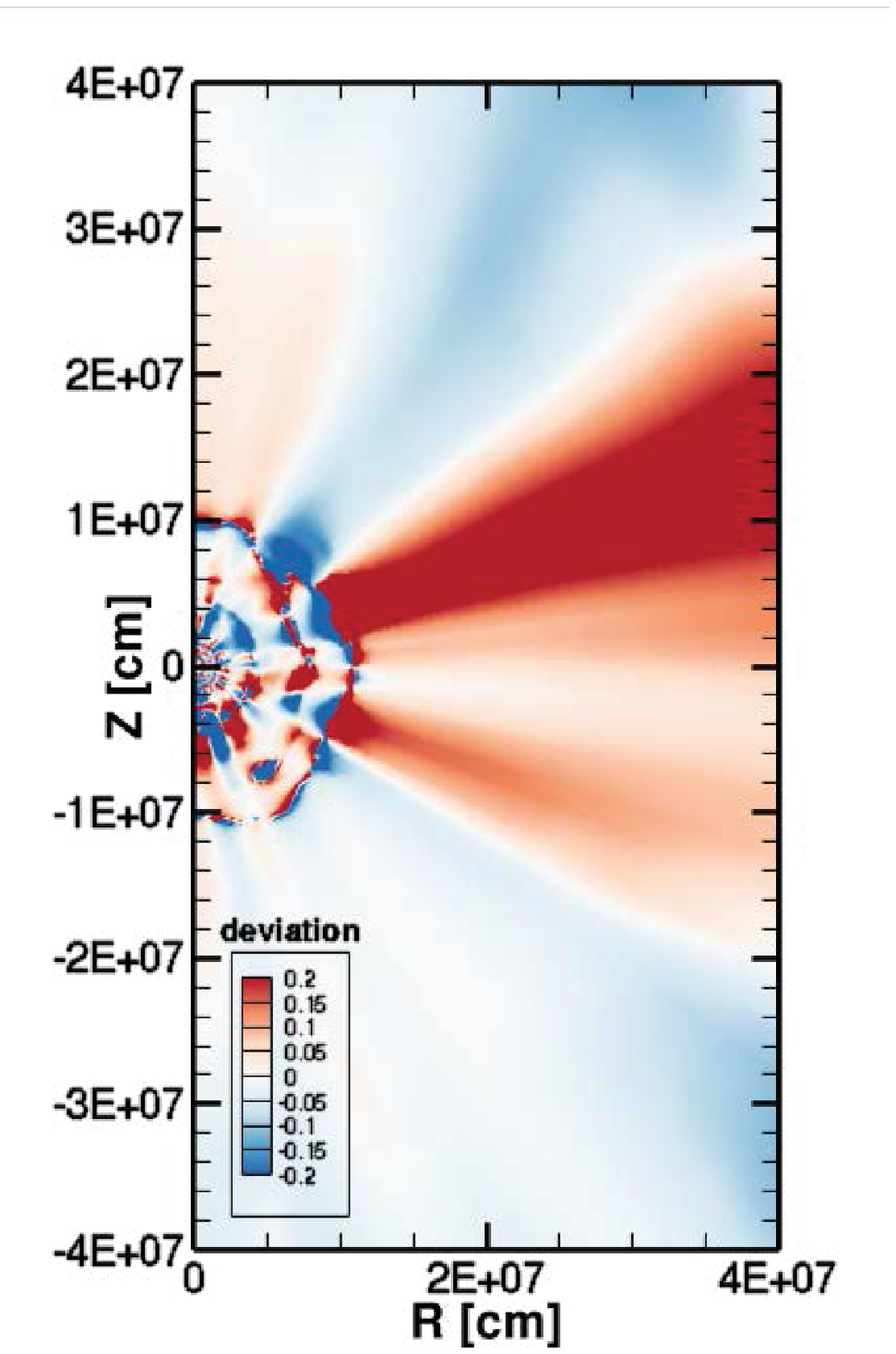}
\caption{Same as Fig. \ref{fig:rbr.2d.phslice-heating.3da.iphx} but 
for the 11M model at 200 ms after the bounce.  }
\label{fig:rbr.2d.phslice-heating.3dc.iphx}
\end{figure}

The pattern of differences looks differently, 
depending on the environment.  
We show the case 
at 200 ms in Figs. \ref{fig:2d.xyslice-heating.3dc} 
and \ref{fig:rbr.2d.phslice-heating.3dc.iphx}.  
The deviations alternately appear 
depending on the $\phi$-direction on the xy-plane 
as seen in Fig. \ref{fig:2d.xyslice-heating.3dc}.  
They depend on the polar angle and 
there are large variations on the meridian slice 
in Fig. \ref{fig:rbr.2d.phslice-heating.3dc.iphx}.  
While the underestimation with a large magnitude 
appears in the limited region along a certain direction 
in the case of $\phi$=51$^\circ$, 
significant overestimation appears 
at two polar directions 
on the meridian slice at $\phi$=141$^\circ$.  
These beam-like structures resemble the situation 
in the case of $\phi$=141$^\circ$ at 150 ms.  
In fact, the neutrino degeneracy around the neutrino-sphere 
is locally high at the two spots, which lead to high neutrino 
fluxes and, therefore, heating rates.  
These spots are located around a narrow channel 
between the two mushroom-like flows 
like the case for 150 ms.  






\subsubsection{Snapshots of 27M}\label{section:h-boltz-ray.others.27M}

We examine the heating rate for the 27M models 
in order to find the generic characteristics of the ray-by-ray approximation.  
In contrast to the 11M models considered so far, 
the shock wave in this model remains stagnated and 
does not show any hint of shock revival 
until the end of the simulation ($\sim$700 ms), 
which is consistent with other group's results \citep{han13,tam14}.  
This model is a representative of those models, 
which seem to be most difficult for the neutrino heating to produce explosions, 
and will hence serve our purpose here.  

\begin{figure}
\epsscale{0.4}
\plotone{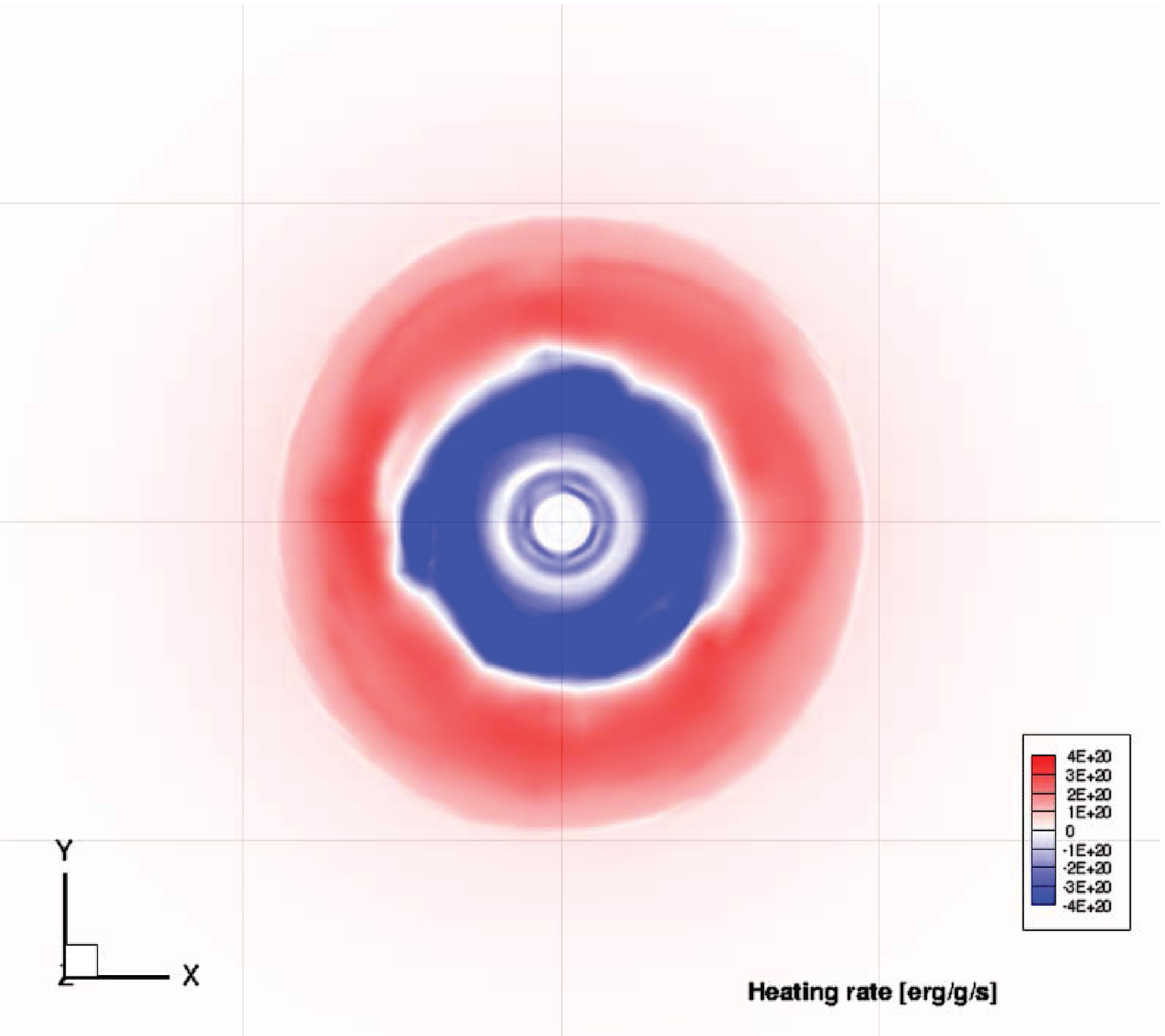}
\plotone{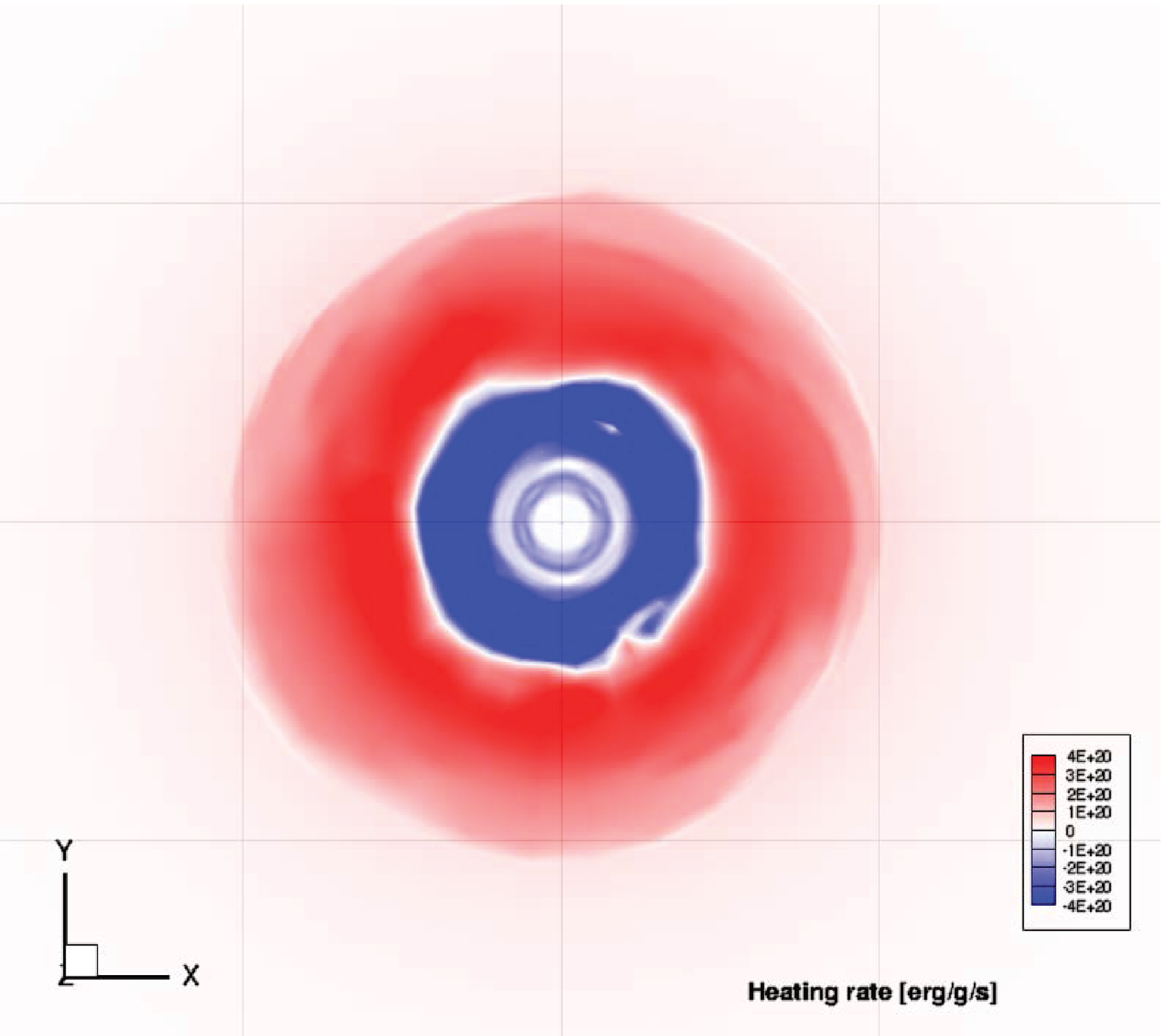}\\
\plotone{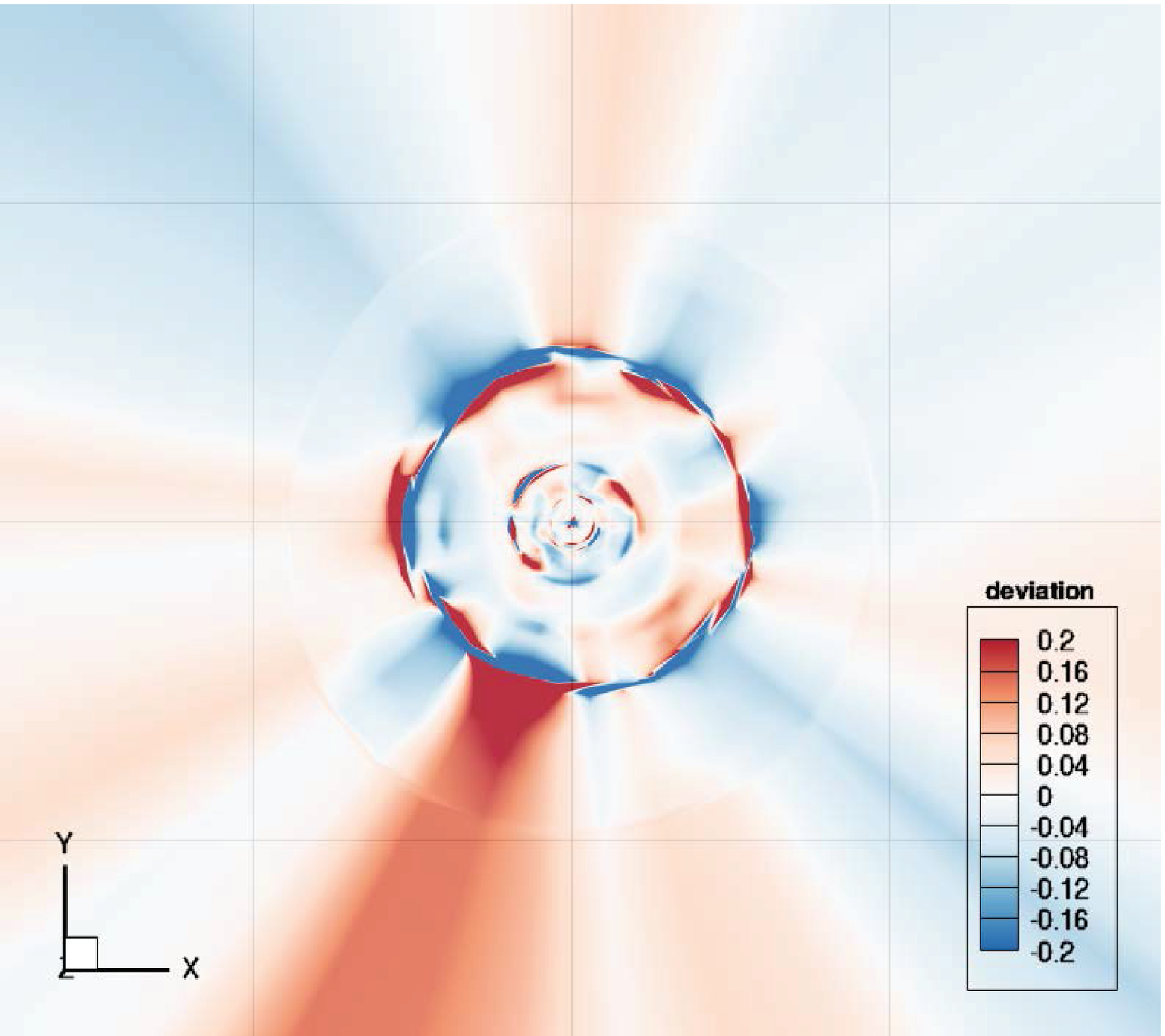}
\plotone{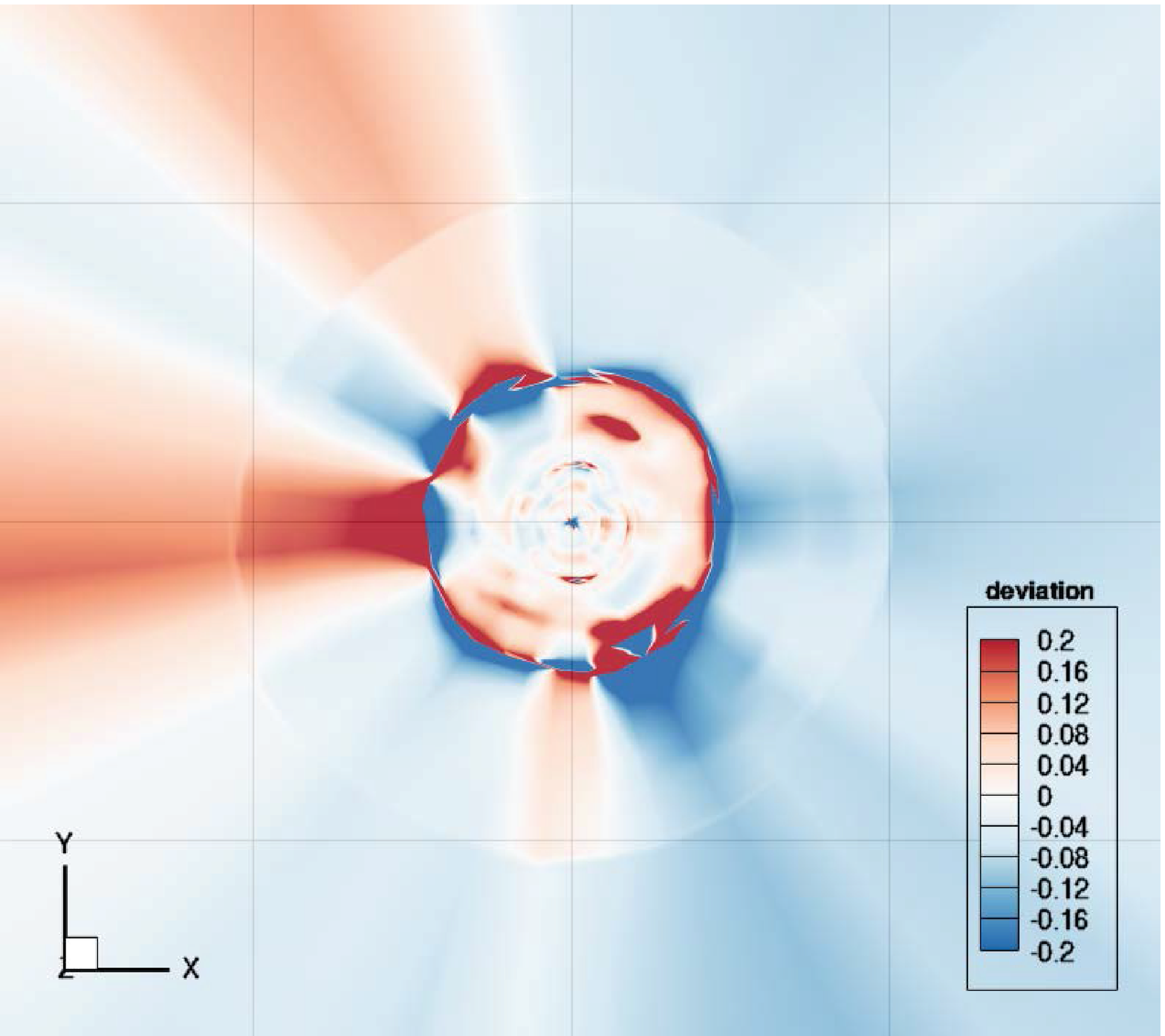}
\caption{
Color maps of the heating rate 
obtained by the 6D Boltzmann solver 
for the 27M model are shown in the upper panels.  
The relative difference of the ray-by-ray evaluation 
is shown in the lower panels. 
The left and right panels correspond to 
the snapshots at 150 and 200 ms after the bounce, respectively.  
Grid lines with 200 km spacing are shown in the background.  
}
\label{fig:2d.xyslice-heating.3d.s27}
\end{figure}

\begin{figure}
\epsscale{0.32}
\plotone{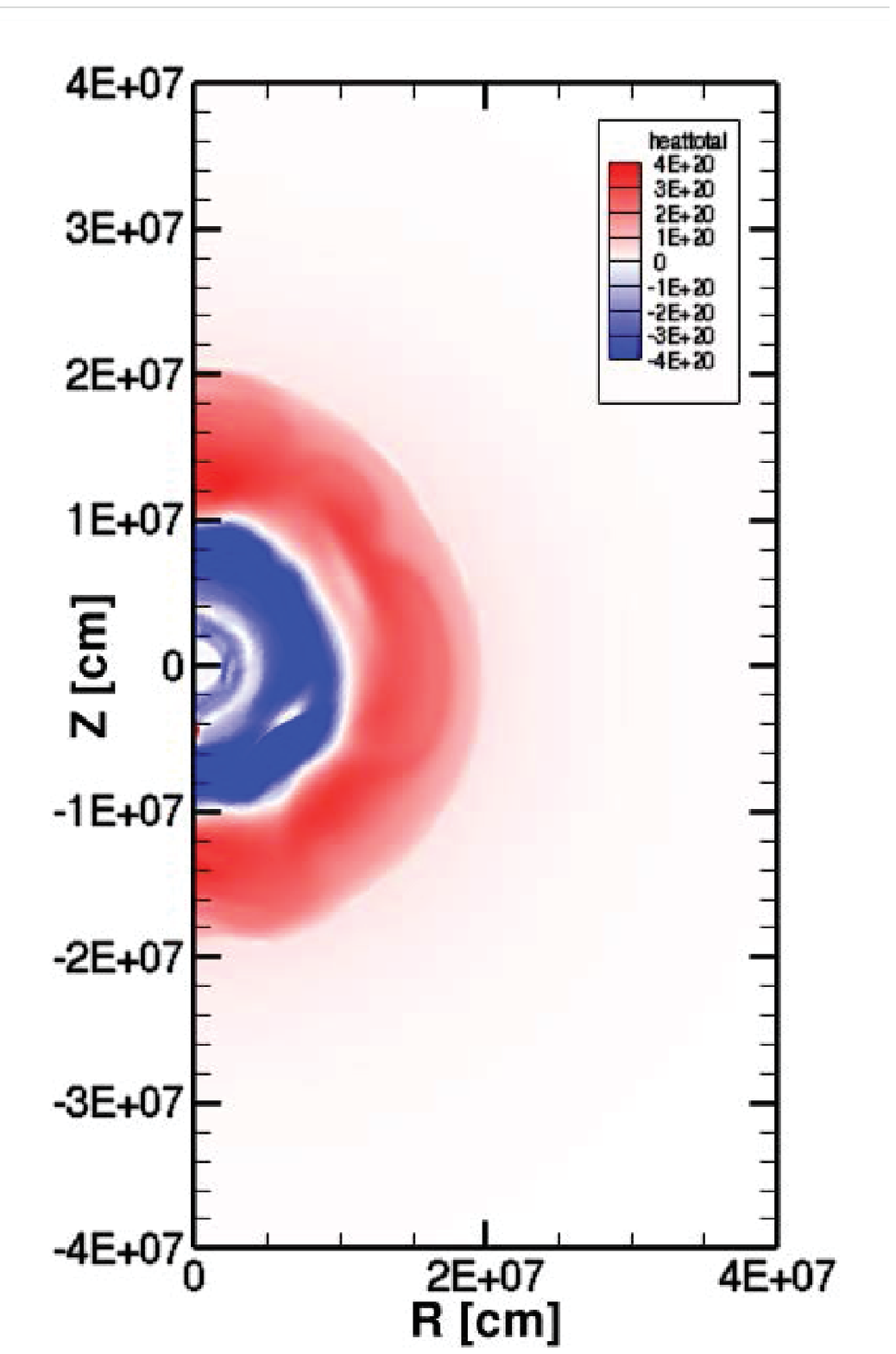}
\plotone{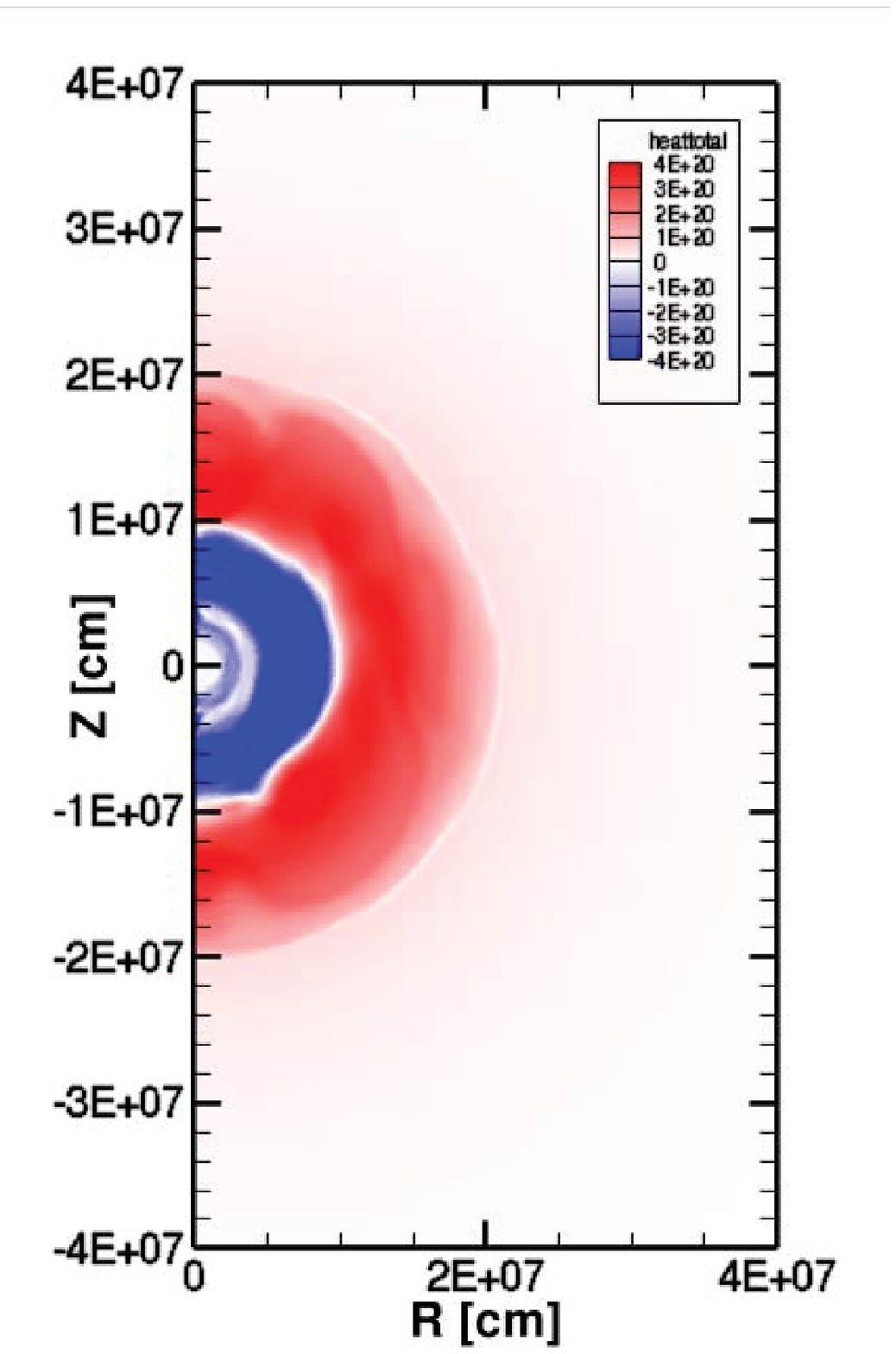}\\
\plotone{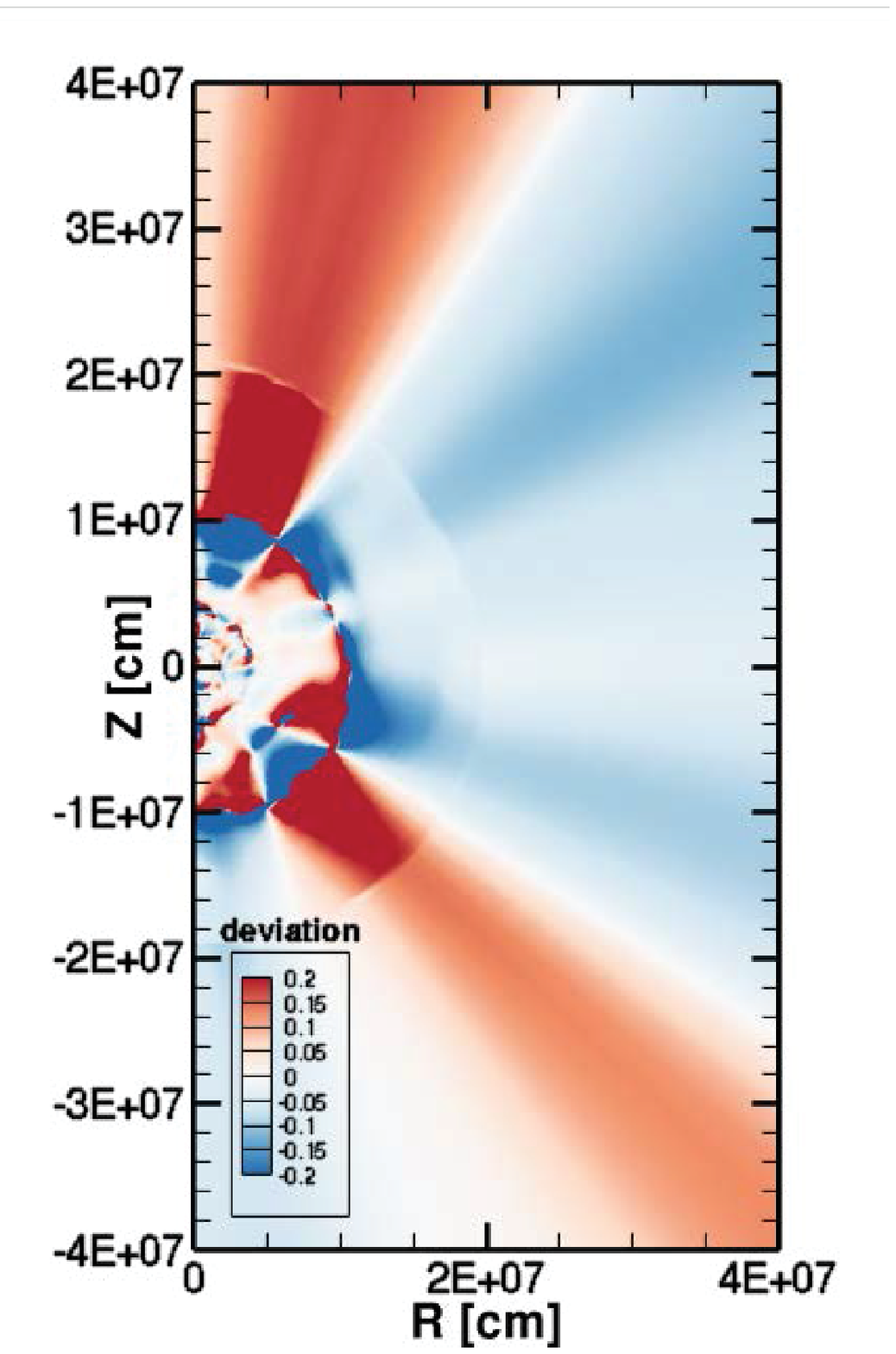}
\plotone{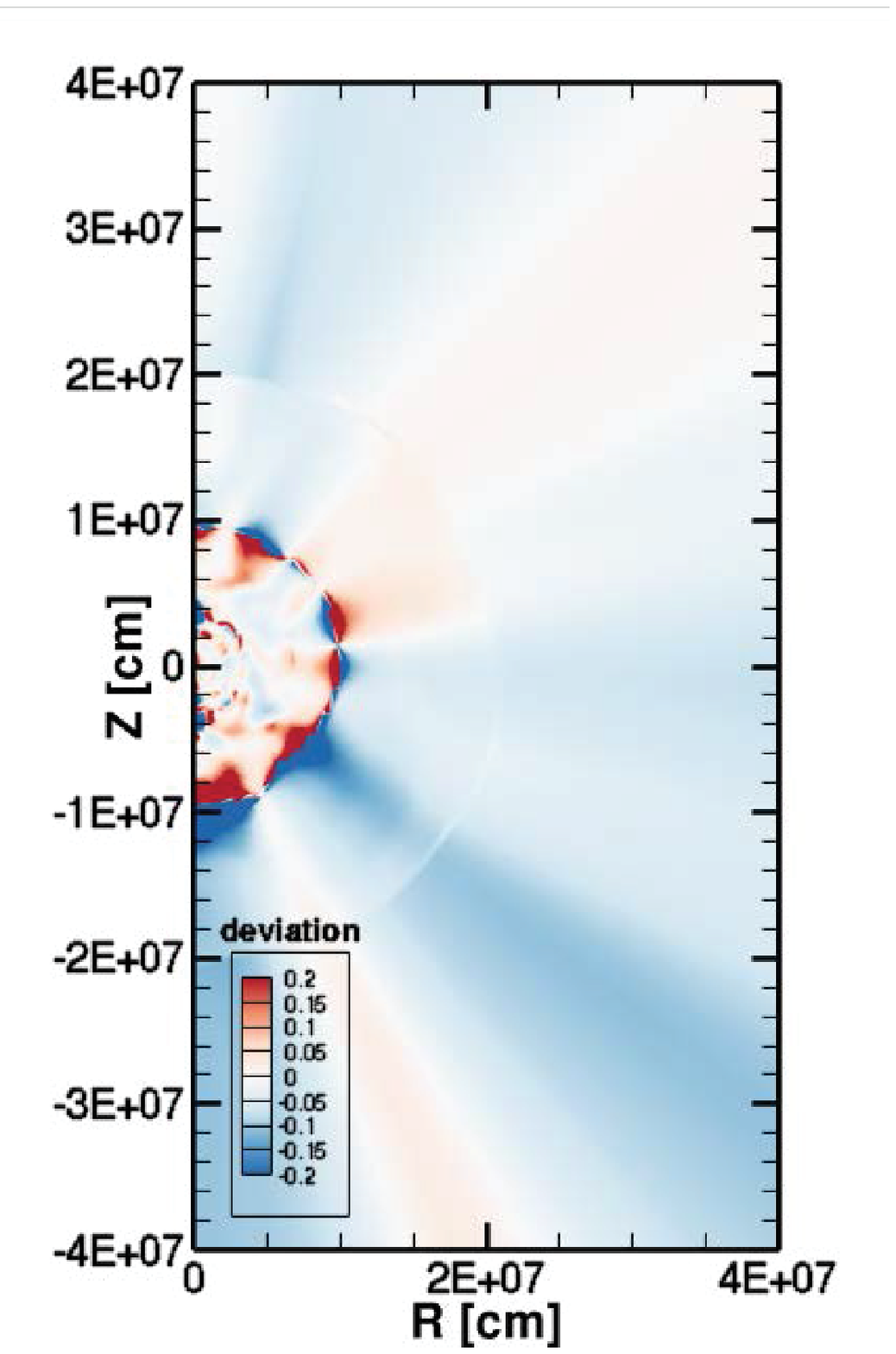}
\caption{
Color maps of the heating rate 
by the 6D Boltzmann evaluation 
for the 27M model on the meridian slice at $\phi$=51$^\circ$ 
are shown in the upper panels.  
The corresponding profiles 
for the relative difference of the ray-by-ray evaluation 
are shown in the lower panels.  
The left and right panels correspond to 
the snapshots at 150 and 200 ms after the bounce, respectively.  
}
\label{fig:rbr.2d.phslice-heating.3d.s27.iph05}
\end{figure}

We found that 
differences between the ray-by-ray approximation and 
the full Boltzmann treatment are no less significant 
in this case.  
We show the heating rates 
as well as the relative differences 
at 150 and 200 ms after the bounce 
in Figs. \ref{fig:2d.xyslice-heating.3d.s27} 
and \ref{fig:rbr.2d.phslice-heating.3d.s27.iph05}.  
The upper panels display the heating rates 
obtained by the 6D Boltzmann solver on the xy-plane 
and the meridian slice at $\phi$=51$^\circ$.  
The lower panels present 
the relative differences of the ray-by-ray evaluation.  
One can see 
that the differences are conserved along radial rays, 
i.e., over- and underestimations occur as radial beams 
just as in the 11M models.  
The deviations are substantial in the heating region, 
particularly near the gain radius, 
but remain beyond the shock wave located at $\sim$200 km.  
We confirm that this behavior 
is generic and is observed for other slices.  
When there is a hot spot on the neutrino-sphere that 
the down flow of proto-rich matter hits, then 
the ray-by-ray approximation tends to 
propagate the local enhancement in the neutrino density 
along the radial ray that emanates from the spot.  
These generic features of the ray-by-ray approximation 
may have non-negligible influences on the revival of the stalled shock wave 
and should be carefully assessed further.

\subsection{Volume-integrated heating rates}\label{section:h-total}

Since we found that the heating rates are locally different, 
it is of great importance to examine the total amount of 
the heating rate 
by the 6D Boltzmann solver and the ray-by-ray approximation.  
We evaluate the volume-integrated heating rates in the two evaluations 
to check this point.  
The specific heating rates, 
which we have discussed in the previous sections, 
are defined by
\begin{equation}
q_{\nu} = - \int d\Omega E_{\nu}^2 dE_{\nu}~E_{\nu} 
\left( \frac{\delta f}{\delta t} \right)_{collision} ~, 
\label{eq:spec.heat}
\end{equation}
with the collision term in the Boltzmann equation.  
The total amount of the heating rate is calculated by the integral 
over the volume in the heating region, 
\begin{equation}
Q_{\nu} = \int dV q_{\nu} \rho ~,
\label{eq:tot.heat}
\end{equation}
where $\rho$ is the mass density.  
The range of the integral covers the heating region, 
where the energy transfer is positive.  
We set the inner boundary of the heating region 
at the gain radius, where $q_{\nu}$ becomes zero along the radial coordinate.  
We define the outer boundary by the location of the shock wave 
when the entropy per baryon becomes 
5.5 in the unit of the Boltzmann constant, $k_B$.  

\subsubsection{Snapshots}\label{section:h-total.snapshots}

In Table \ref{table:heating}, 
we compare the volume-integrated heating rates 
for the 11M and 27M models 
in the two evaluations.  
The volume-integrated heating rates in the ray-by-ray approximation 
are slightly smaller than the values in the 6D Boltzmann evaluation.  
The difference is within $\sim$2$\%$ for 11M and less than 1$\%$ for 27M.  
One may wonder the reason why 
the volume-integrated heating rate by the ray-by-ray approximation 
is not significantly different from the one 
by the 6D Boltzmann solver.  
This is because the overestimation and underestimation of the heating rates 
in the local regions along the radial ray 
cancel out by the integral over the solid angle.  
The current analysis suggests that 
the local deviations in the ray-by-ray approximation 
are more influential than the total deviations.  
It should be noted, however, that 
even a slight difference in the heating rate 
can be crucial to the revival of the shock wave 
in the marginal situation \citep{jan96}.  
It is necessary to study this influence further 
by examining the hydrodynamical evolution.  

\subsubsection{Time variations}\label{section:h-total.time}

So far we have discussed the features of stationary neutrino distributions 
obtained for snapshots that are mutually separated in time by 50 ms.  
One may argue that rapid temporal changes in matter distributions might erase 
some of those features by averaging.  
In order to check this, we perform another series of simulations.  
We adopt consecutive snapshots from the period of 145 ms to 155 ms 
with an interval of 1 ms 
from the same supernova simulation of 11.2M$_{\odot}$ star; 
there are hence 11 snapshots including the one at 150 ms, 
which we have already discussed; 
we then follow the time evolution of neutrino distributions 
until they become stationary for each time slice and 
take the average of the results over these 11 simulations.  
In this way we can judge how sensitive our findings are 
to the short-time variations of the background.  

\begin{figure}
\epsscale{0.75}
\plotone{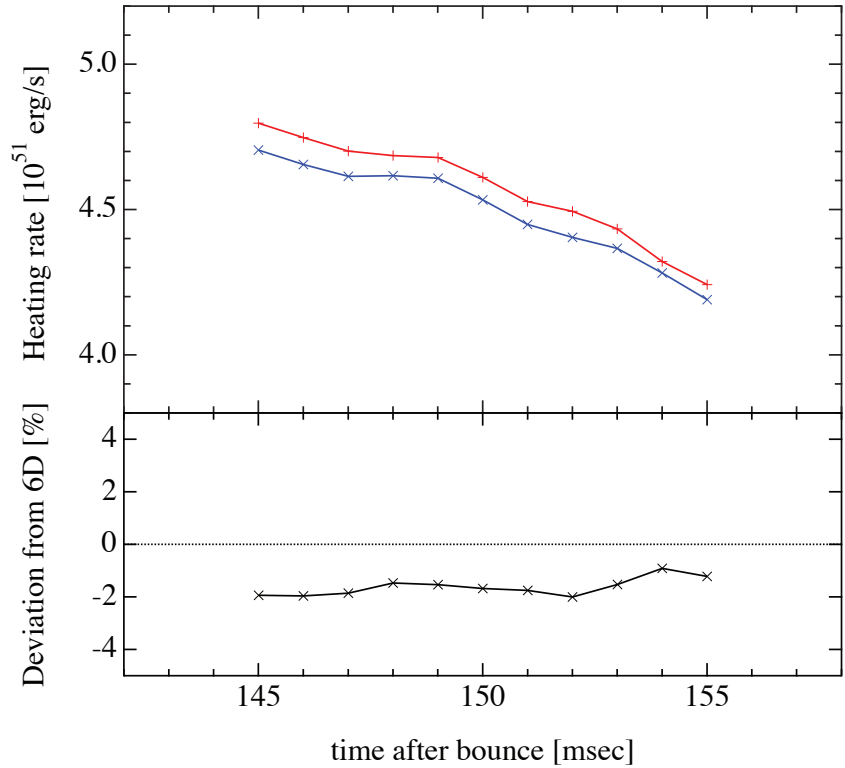}
\caption{
The volume-integrated heating rates 
are shown 
for the series of snapshots 
during the time period of 145--155 ms 
from the 3D supernova simulation (11M).  
The values at every 1 ms 
obtained by the 6D Boltzmann and ray-by-ray evaluations 
are shown by red and blue lines, respectively, 
with symbols in the upper panel.  
The relative difference of the ray-by-ray evaluation 
with respect to the 6D Boltzmann evaluation  
is shown in the lower panel.  
}
\label{fig:deviation-heating.tslice}
\end{figure}

We show the temporal evolution of the total heating rates 
during the 10 ms in the two evaluations 
in Fig. \ref{fig:deviation-heating.tslice}.  
In both cases, 
there is a gradual decrease owing to 
the change in hydrodynamical conditions.  
One can see that 
the ray-by-ray approximation provides systematically smaller rates 
than the 6D Boltzmann evaluation over this period.  
The deviation of the ray-by-ray evaluation 
stays around $-2\%$.  
This result suggests that 
the underestimation by the ray-by-ray approximation is 
not an artifact of the fixed-background models, 
but a generic feature originated from 
the nature of the approximation itself.  
In fact, we find that the local deviations over 20$\%$, 
which are seen in Figs. \ref{fig:rbr.2d.xyslice-heating.3db} 
and \ref{fig:rbr.2d.phslice-heating.3db.iphx}, 
are also observed in this case and last for $\sim$10 ms.  

\section{Angle moments}\label{section:moments}

\begin{figure}
\epsscale{0.72}
\plotone{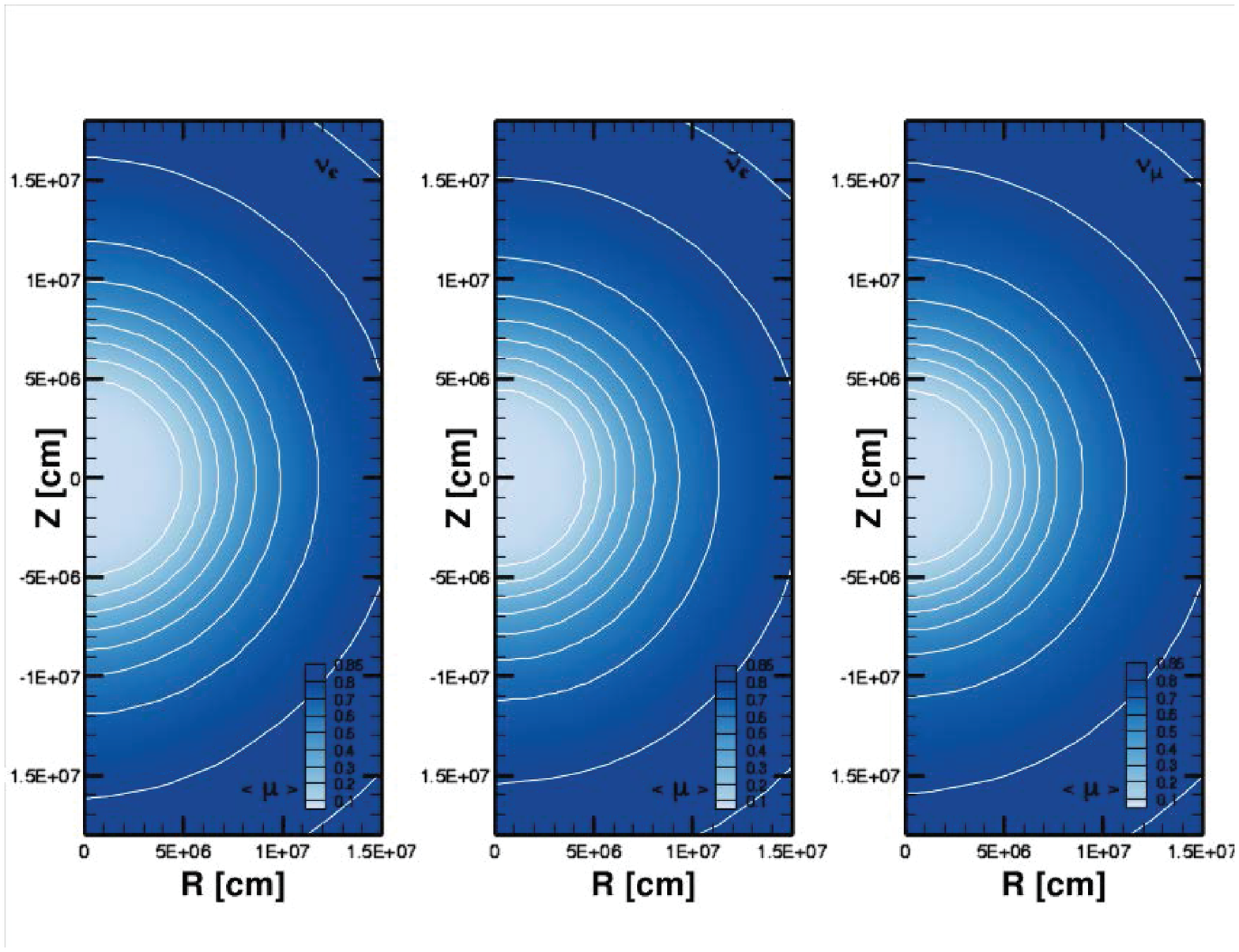}
\plotone{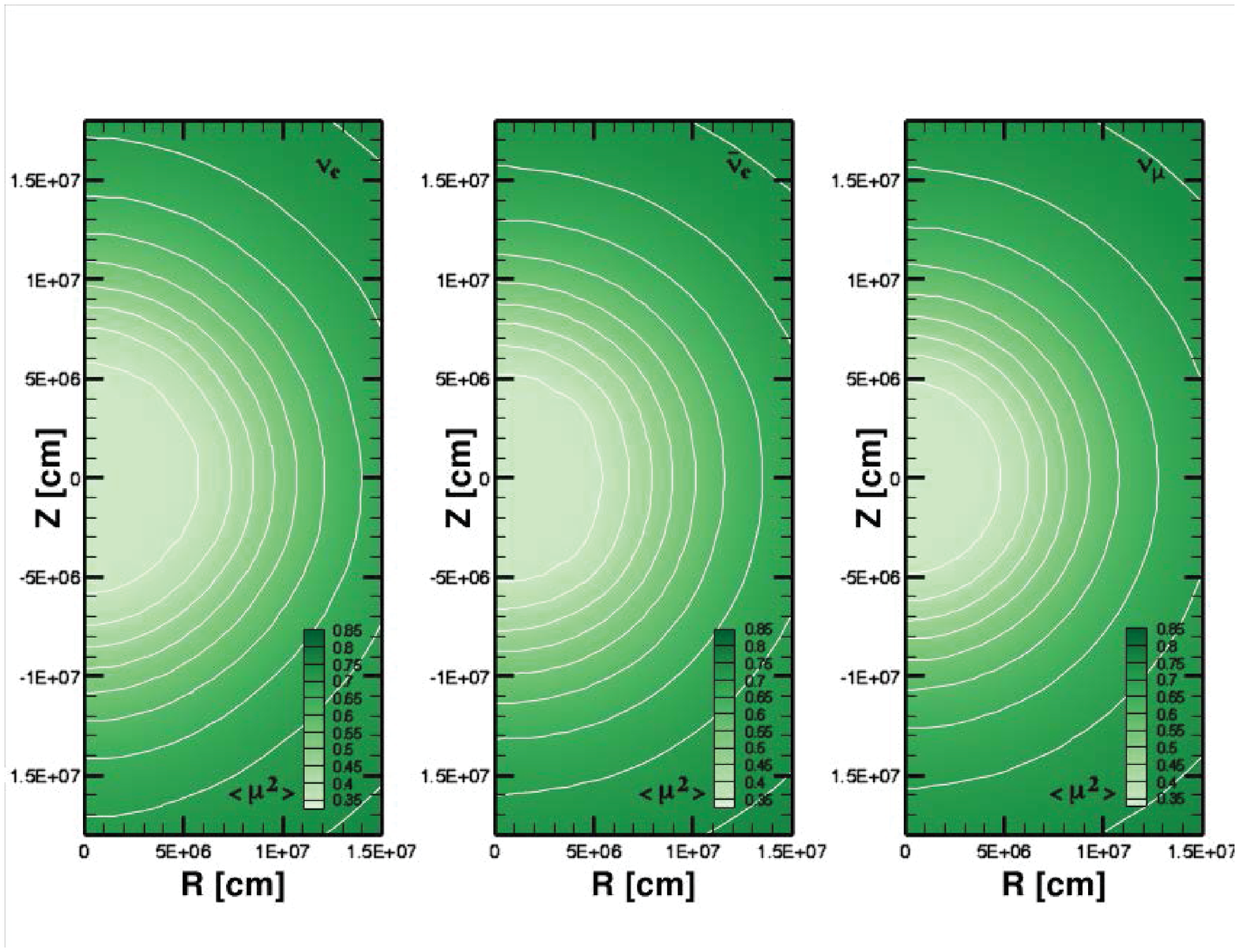}
\caption{Contour maps of the flux factor, $\langle \mu_{\nu} \rangle$, (top) 
and the second angle moment, $\langle \mu_{\nu}^2 \rangle$, (bottom) 
on the meridian slice at $\phi$=51$^{\circ}$ 
are shown for the 11M model at 150 ms after the bounce.  
The left, middle and right panels display 
profiles for three species ($\nu_e$, $\bar{\nu}_e$ and $\nu_{\mu}$), 
respectively.  }
\label{fig:3db-mom.mu.iph05.slice}
\end{figure}
\begin{figure}
\epsscale{0.72}
\plotone{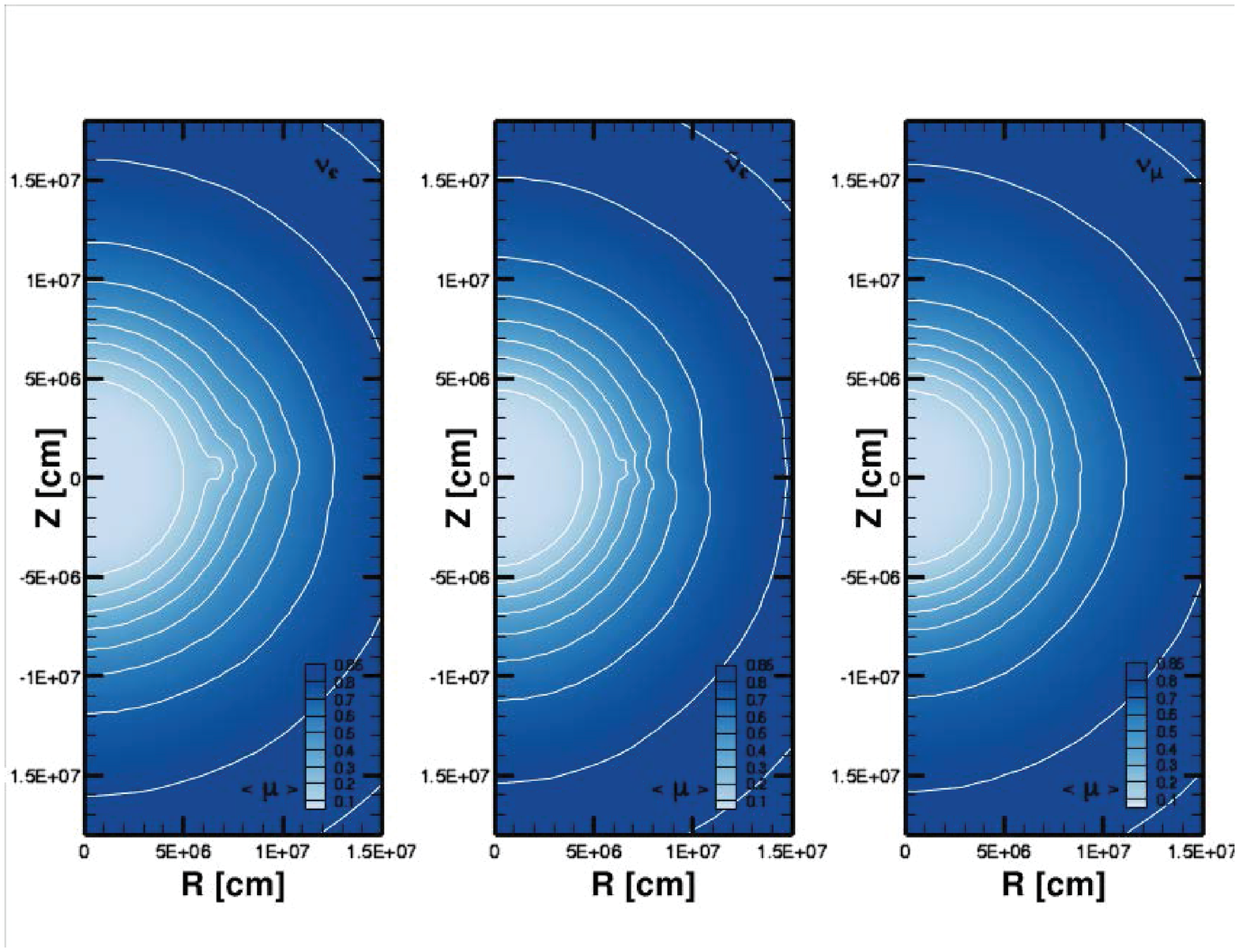}
\plotone{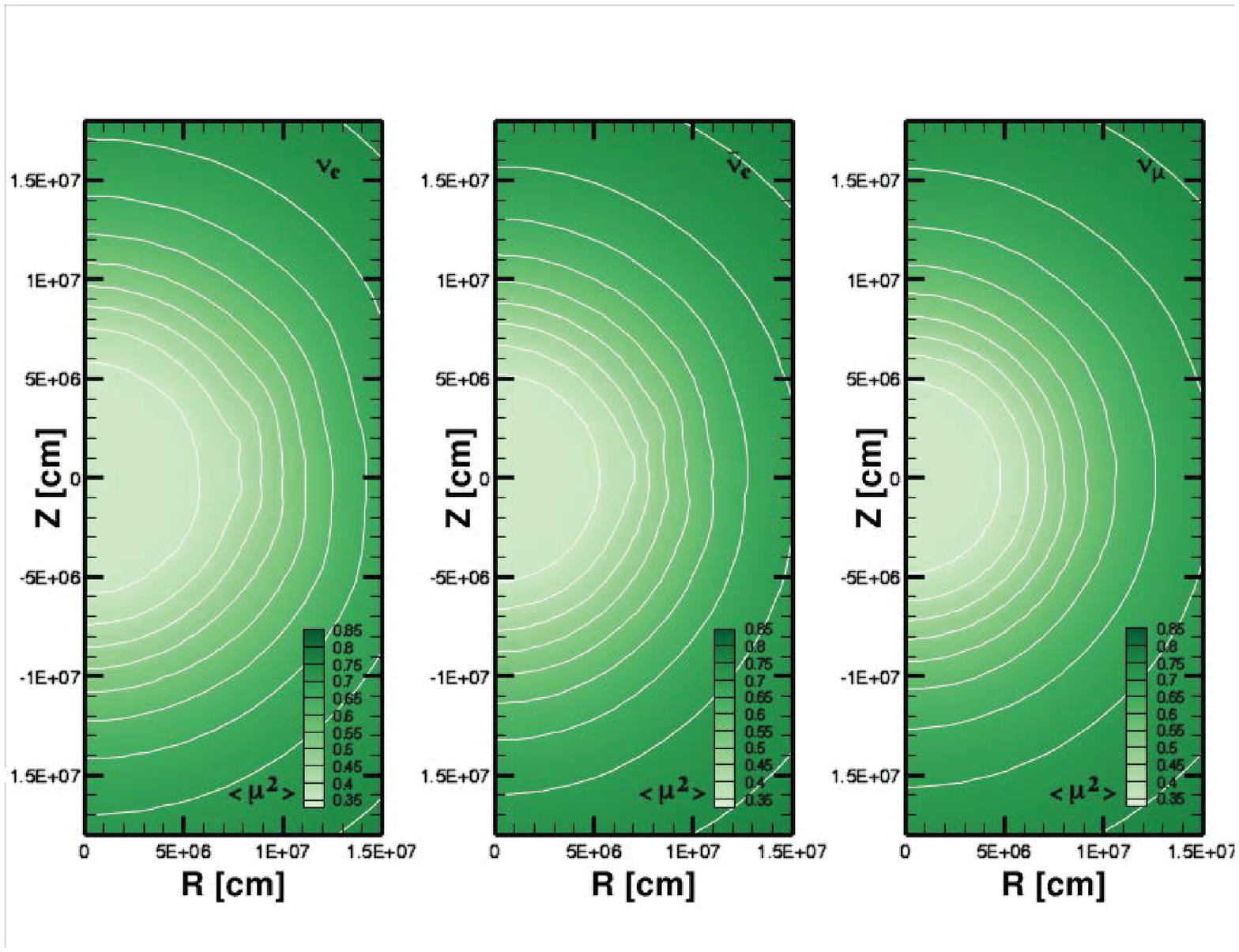}
\caption{Same as Figure \ref{fig:3db-mom.mu.iph05.slice} 
but on the meridian slice at $\phi$=141$^{\circ}$.  }
\label{fig:3db-mom.mu.iph13.slice}
\end{figure}

Various moments of the neutrino distribution function are important 
to reveal the 3D features of the neutrino transfer.  
They are also useful to develop new methods for the closure 
relation in the moment formalism (see \citet{shi11} for example).  
In principle, we can derive various moments with respect to angle and energy 
from the obtained neutrino distribution function in six dimensions.  
As an example, we show the contour maps of 
the angle moments for three species of neutrinos 
in Figs. \ref{fig:3db-mom.mu.iph05.slice} and \ref{fig:3db-mom.mu.iph13.slice} 
on the meridian slices at $\phi$=51$^\circ$ and $\phi$=141$^\circ$, respectively, 
for the 11M model at 150 ms after the bounce.  
The spatial distributions of both the first angle moment or 
the flux factor, $\langle \mu_{\nu} \rangle$, 
and the second moment, $\langle \mu_{\nu}^2 \rangle$, 
are almost spherical with a slight deformation in the central part.  
Some bumps are seen in the narrow region near the equator 
on the meridian slice at $\phi$=141$^{\circ}$.  
The asymptotic behavior of these angle moments at large radii 
is consistent with those obtained in 
the previous studies on the flux factor \citep{mes98,yam99} 
and on the basic validation of the 6D Boltzmann code \citep{sum12}.  

\begin{figure}
\epsscale{0.72}
\plotone{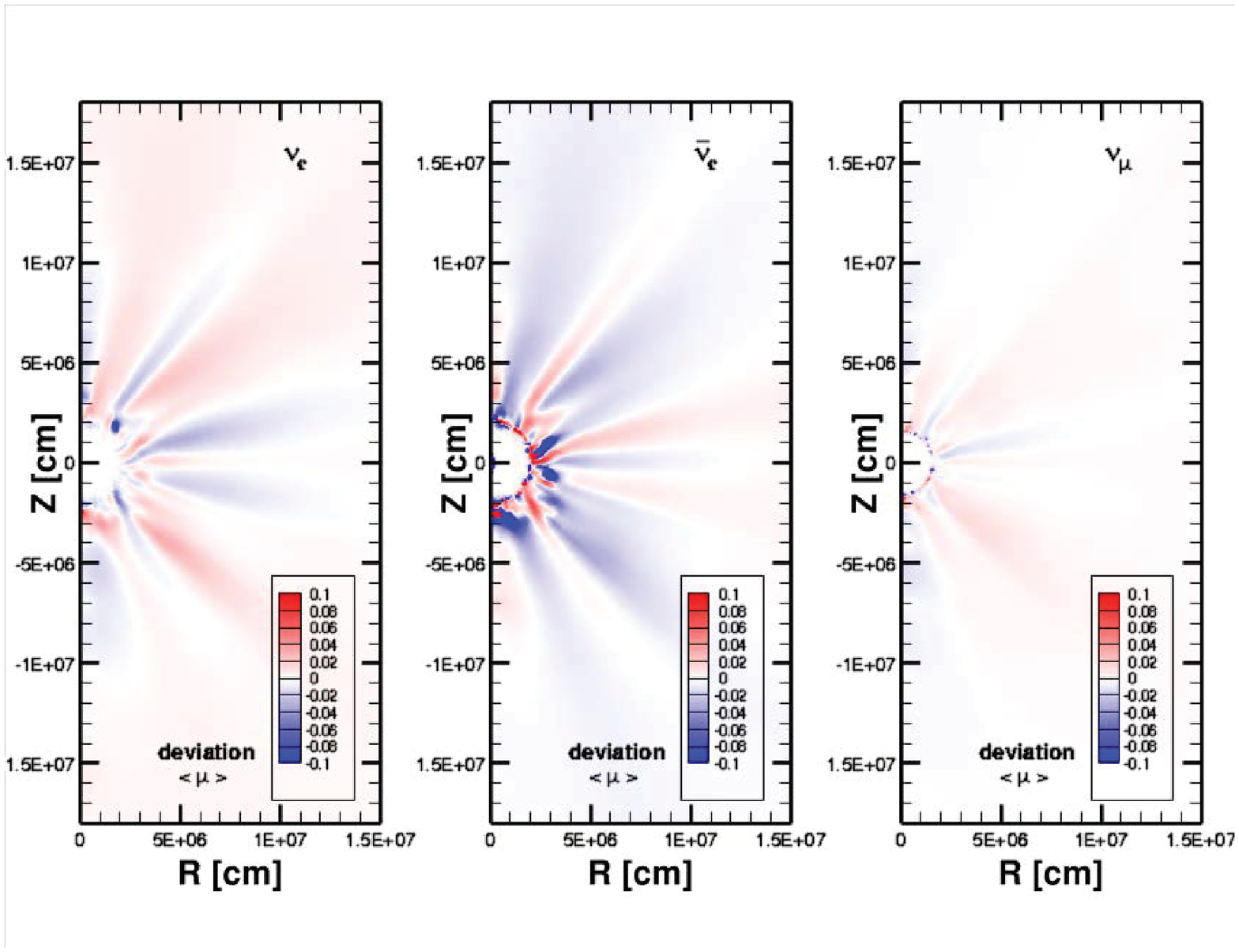}
\plotone{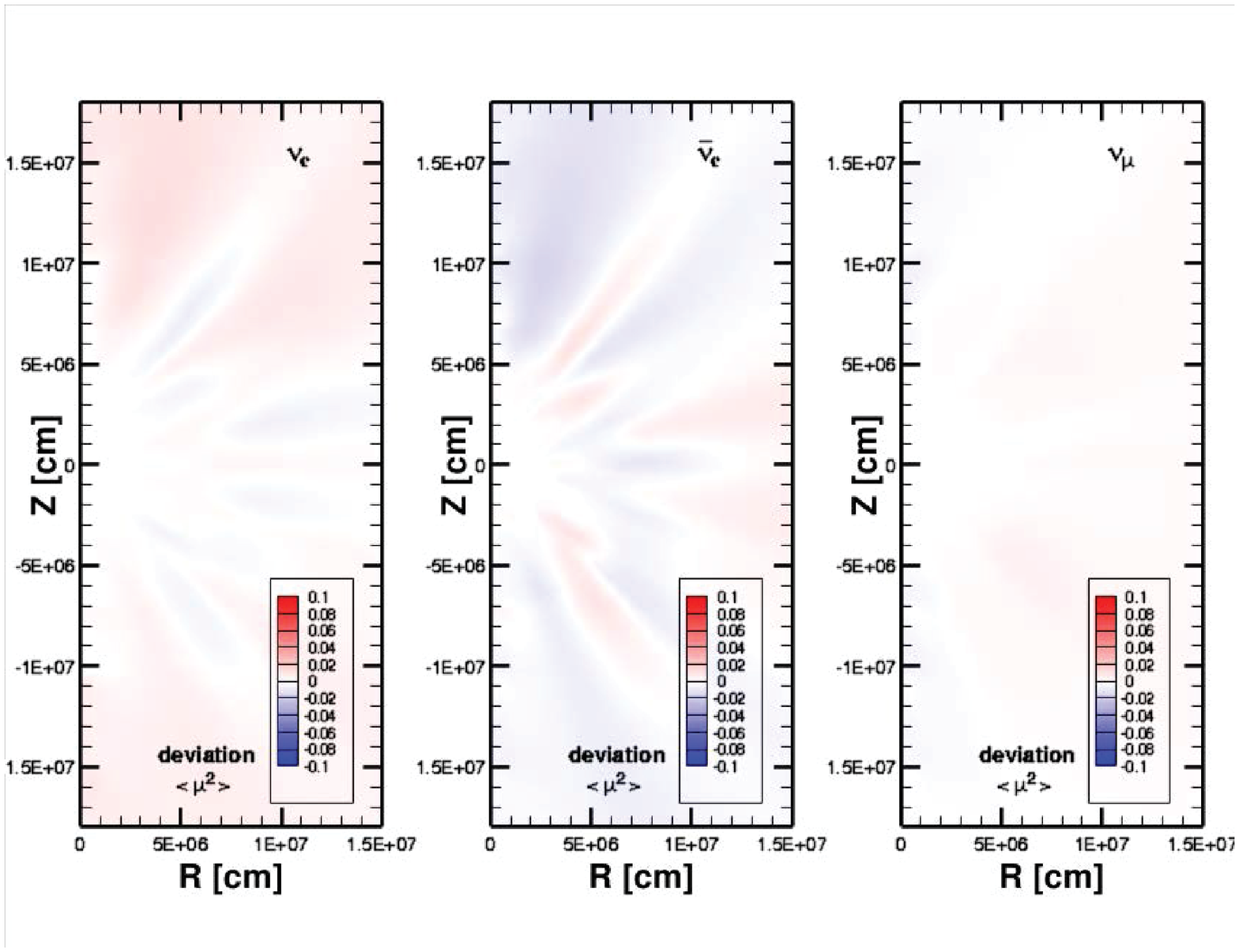}
\caption{Color maps of the relative differences  
of the ray-by-ray evaluation 
with respect to the 6D Boltzmann evaluation 
are shown for 
the angle moments $\langle \mu_{\nu} \rangle$ (top) 
and $\langle \mu_{\nu}^2 \rangle$ (bottom).  
The left, middle and right panels display profiles 
for three species ($\nu_e$, $\bar{\nu}_e$ and $\nu_{\mu}$), respectively, 
corresponding to those in Fig. \ref{fig:3db-mom.mu.iph05.slice}.  }
\label{fig:ratio.3db-mom.mu.iph05.slice}
\end{figure}
\begin{figure}
\epsscale{0.72}
\plotone{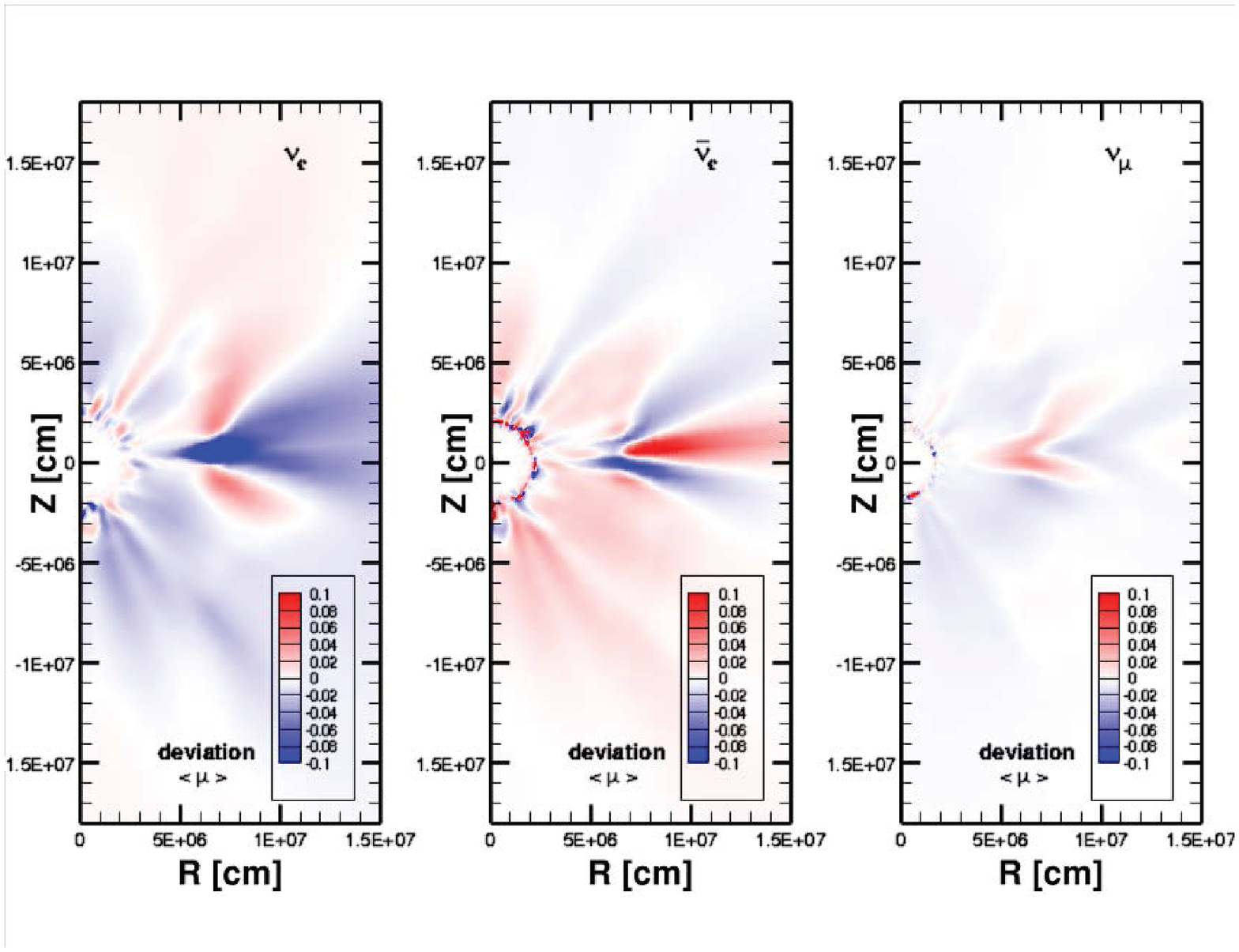}
\plotone{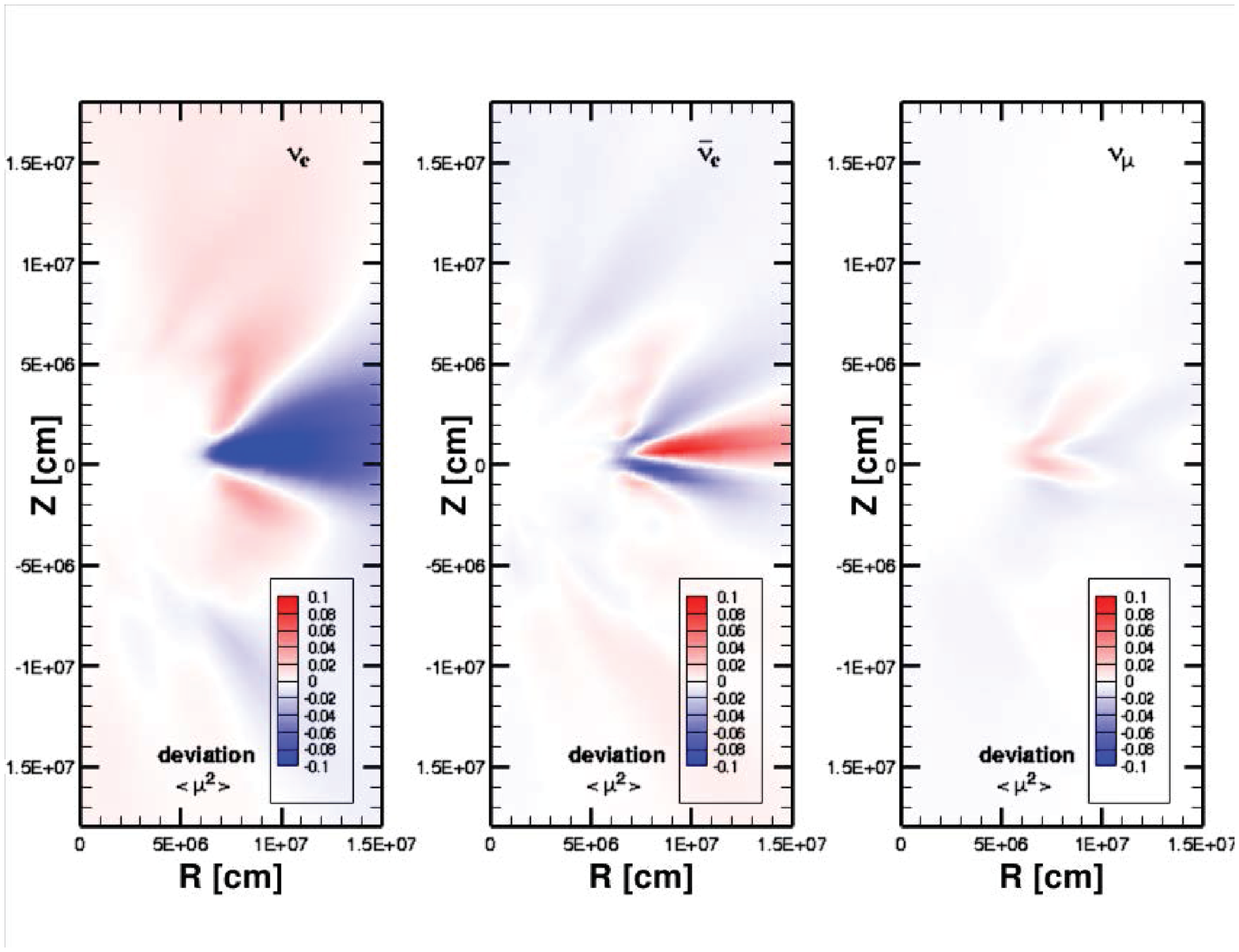}
\caption{Same as Figure \ref{fig:ratio.3db-mom.mu.iph05.slice} 
but on the meridian slice at $\phi$=141$^{\circ}$, 
corresponding to those in Fig. \ref{fig:3db-mom.mu.iph13.slice}.  }
\label{fig:ratio.3db-mom.mu.iph13.slice}
\end{figure}

We further examine the angle moments 
obtained by the ray-by-ray approximation 
through comparisons with the 6D Boltzmann evaluation.  
In Figs. \ref{fig:ratio.3db-mom.mu.iph05.slice} and 
\ref{fig:ratio.3db-mom.mu.iph13.slice}, 
we show the relative differences of the angle moments 
evaluated by the ray-by-ray approximation 
with respect to the 6D Boltzmann evaluation 
shown in Figs. \ref{fig:3db-mom.mu.iph05.slice} and \ref{fig:3db-mom.mu.iph13.slice}.  
Appreciable deviations of the angle moments arise in some regions 
where non-radial fluxes are large in the neutrino distributions.  
The deviations become large up to $\sim10\%$ and 
is comparable to those of the other quantities we have seen.  
The deviations are very small within $\sim3\%$ for $\langle \mu_{\nu}^2 \rangle$ 
on the meridian slice at $\phi$=51$^{\circ}$ where the distribution is almost spherical.  
On the meridian slice at $\phi$=141$^{\circ}$, 
the deviation is significant around the narrow channel near the equator 
due to the hot spot effect.  
Alternate deviations of underestimation and overestimation 
appear depending on the radial direction.  

\begin{figure}
\epsscale{0.72}
\plotone{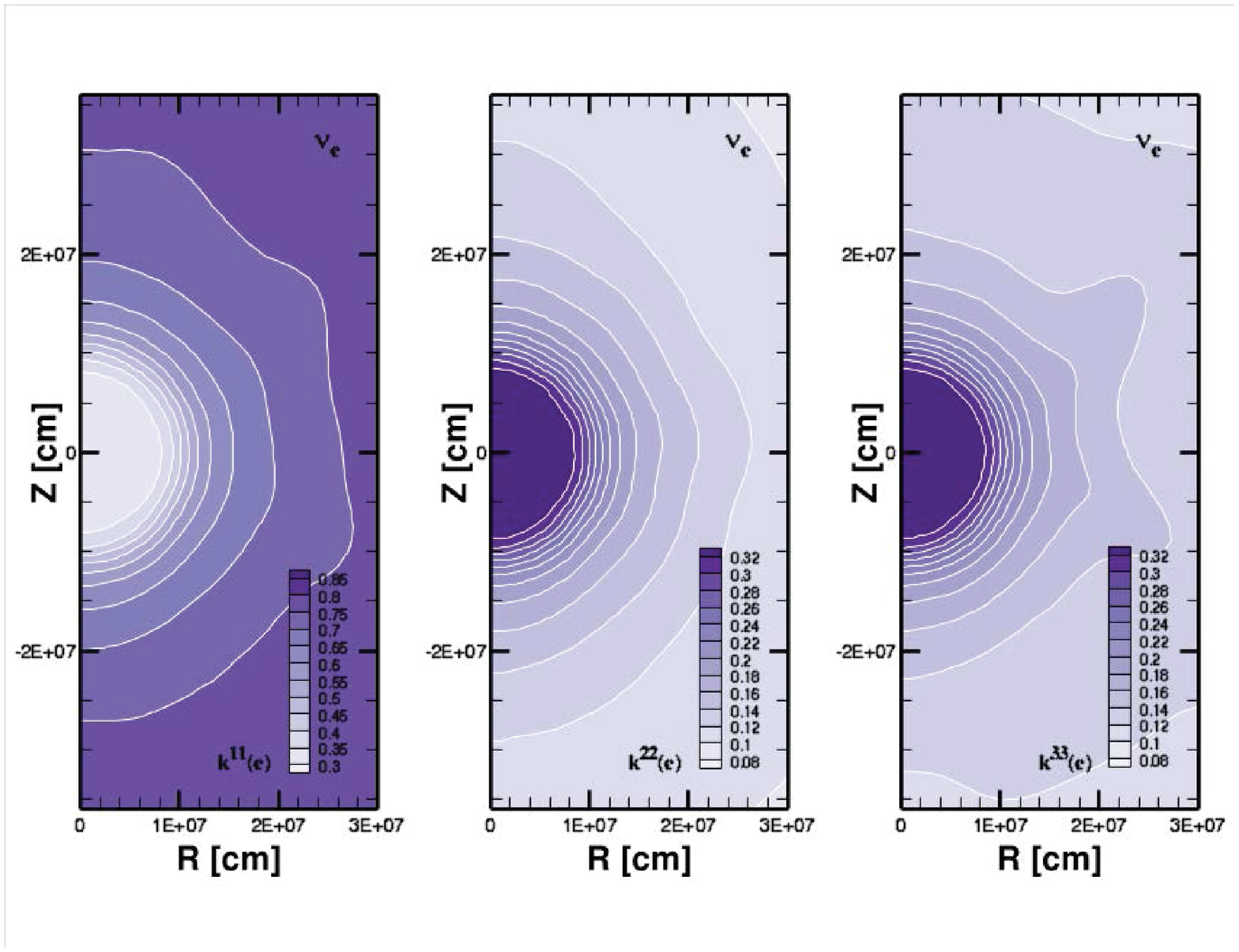}
\plotone{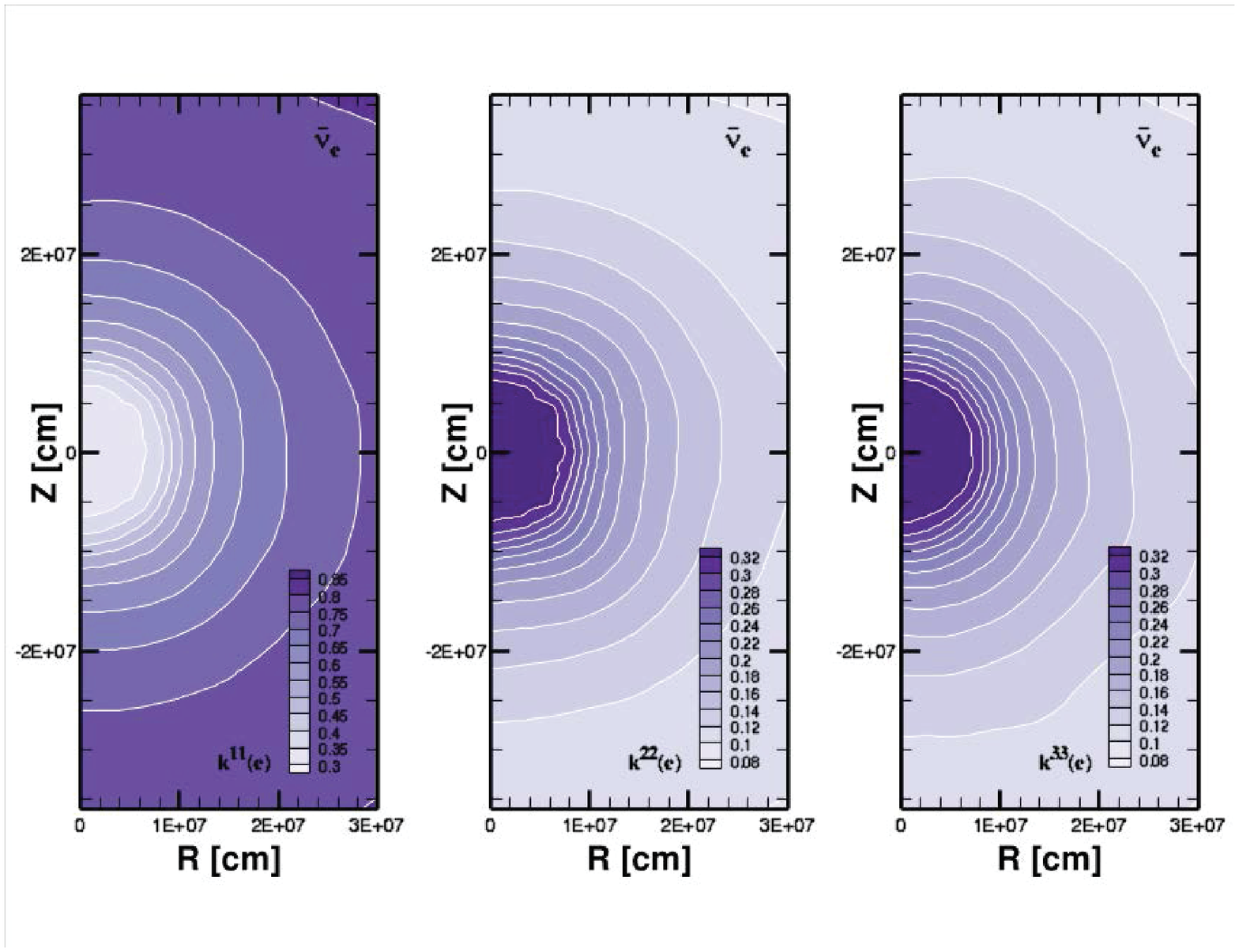}
\caption{Contour maps of the Eddington tensor 
on the meridian slice at $\phi$=51$^{\circ}$ are shown 
for the 11M model at 150 ms after the bounce.  
The diagonal elements for the energy bin ($E_{\nu}=$ 34 MeV) 
are shown in the left ($r r$), middle ($\theta \theta$) 
and right ($\phi \phi$) panels.  
The profiles for two species $\nu_e$ and $\bar{\nu}_e$ 
are displayed in the top and bottom panels, respectively.  } 
\label{fig:3db-eddi.iph05.slice.inx}
\end{figure}
\begin{figure}
\epsscale{0.72}
\plotone{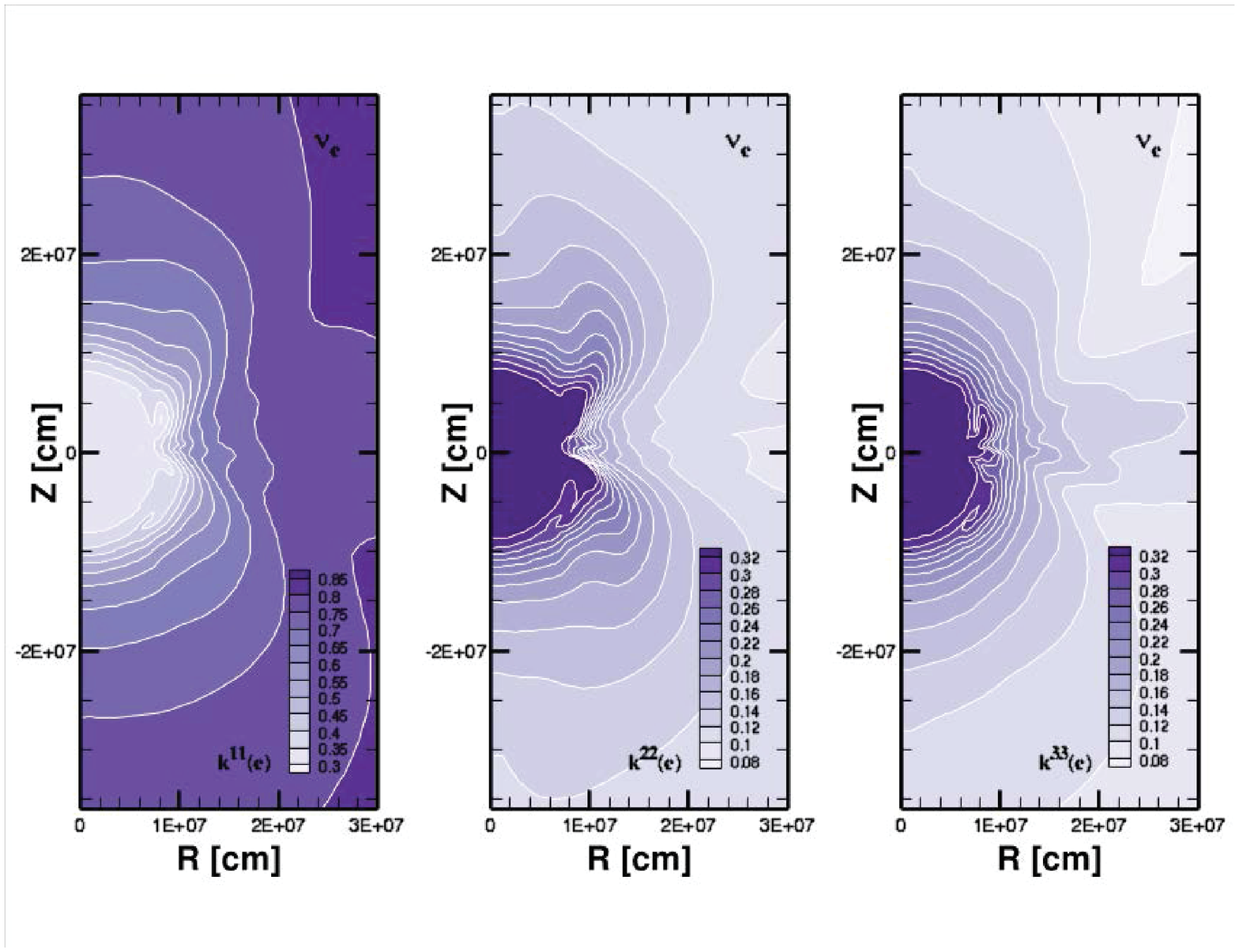}
\plotone{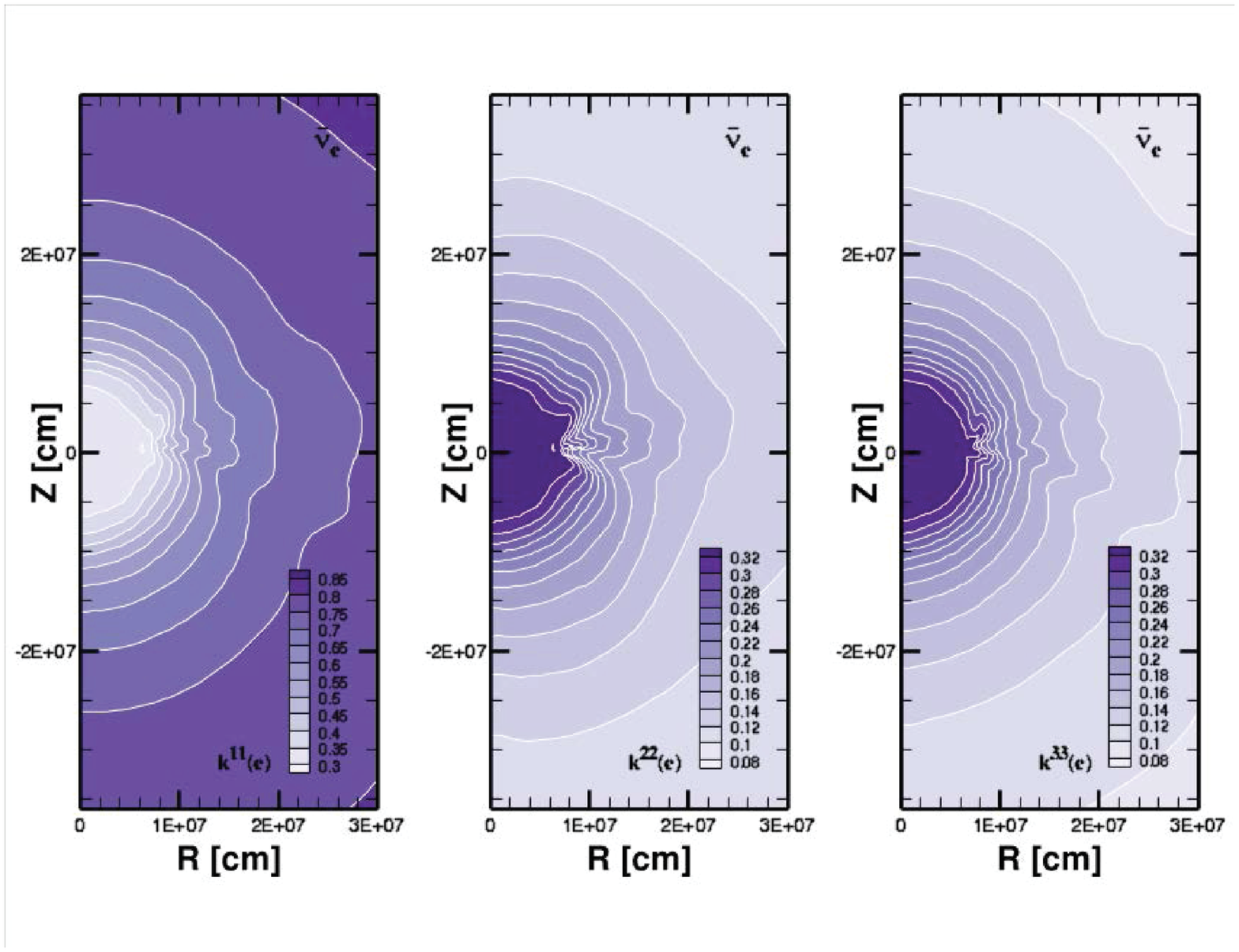}
\caption{Same as Figure \ref{fig:3db-eddi.iph05.slice.inx} 
but on the meridian slice at $\phi$=141$^{\circ}$.  }
\label{fig:3db-eddi.iph13.slice.inx}
\end{figure}

It is profitable to note that 
we can examine all (diagonal and non-diagonal) elements 
of the Eddington tensors by the 6D Boltzmann solver.  
Components of the Eddington tensor show 
interesting features with non-spherical shapes in some cases.  
The diagonal elements of the Eddington tensor 
for the energy bin ($E_{\nu}=$ 34 MeV) 
of two species of neutrinos ($\nu_e$ and $\bar{\nu}_e$) are shown 
in Figs. \ref{fig:3db-eddi.iph05.slice.inx} 
and \ref{fig:3db-eddi.iph13.slice.inx} 
on the two meridian slices 
for the 11M model at 150 ms.  
In the case of $\phi$=51$^{\circ}$, 
the distributions of the three components exhibit deformed shapes 
in the outer region ($\sim$250 km).  
In the central part, 
they have spherical shapes for $\nu_e$ 
and non-spherical ones for $\bar{\nu}_e$, 
reflecting their distributions.  
In the case of $\phi$=141$^{\circ}$, 
the non-spherical features of the distribution are prominent 
in a wide region for the two species.  
The bumpy structure reflects the non-uniform shape of the central core, 
which has the narrow channel between the two mushrooms.  

\begin{figure}
\epsscale{0.72}
\plotone{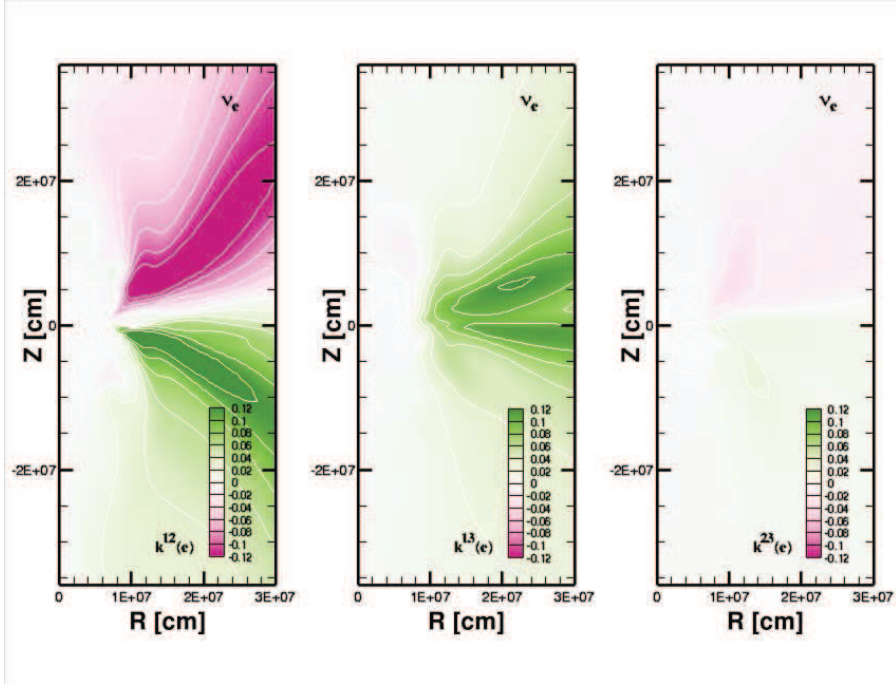}
\caption{Contour maps of the Eddington tensor 
on the meridian slice of $\phi$=141$^{\circ}$ are shown 
for the 11M model at 150 ms after the bounce.  
The non-diagonal elements for the energy bin ($E_{\nu}=$ 34 MeV) 
for $\nu_e$ 
are shown in the left ($r \theta$), middle ($r \phi$) 
and right ($\theta \phi$) panels.  }
\label{fig:3db-eddi.subd.iph13.slice.in1}
\end{figure}

In usual situations with nearly spherical shapes, 
only the diagonal elements are dominant 
with zero non-diagonal elements.  
We found that there are some regions 
where the non-diagonal elements are non-negligible.  
In Fig. \ref{fig:3db-eddi.subd.iph13.slice.in1}, 
we show, as an example, the color maps of the non-diagonal 
($r \theta$), ($r \phi$) and ($\theta \phi$)-components 
for the energy bin ($E_{\nu}=$ 34 MeV) of $\nu_e$ 
on the meridian slice at $\phi$=141$^{\circ}$ in the 11M model at 150 ms.  
Their magnitude amounts to $\sim$0.1 and is not negligible 
around the channel along the equator.  

These non-trivial properties of the Eddington tensors are 
valuable information to express the anisotropy of neutrino distribution functions 
in 3D supernova cores.  
Since the neutrino distribution function is symmetric 
around the radial coordinate in the ray-by-ray approximation, 
only the three diagonal components are non-zero and 
the $rr$-component determines 
the other two $\theta\theta$ and $\phi\phi$-components.  
Hence, the ray-by-ray approximation loses the information 
of the non-trivial structure of the Eddington tensor.  
Further studies on the detailed features of the Eddington tensors 
obtained by the 6D Boltzmann solver 
are underway to develop new closure relations for the 3D problems 
as a separate article.  






%
%
%




\section{Discussions}\label{section:discuss}

We discuss here some possible implications 
of our finding for 
the supernova mechanism and associated phenomena.  

The 6D Boltzmann solver provides 
a seamless description of the neutrino transfer 
from optically thick to thin regions 
in 3D space.  
Non-radial neutrino transfer occurs in general 
up to large distances from the center beyond the neutrino-sphere, 
which can have influences 
on the convective motions of neutrino-rich matter 
inside the proto-neutron star 
as well as the emission properties 
around the neutrino-sphere.  
These may in turn affect the neutrino heating 
in the gain region and ultimately the revival of shock wave.  

Although the diffusion approximation can describe 
the non-radial transfer of neutrinos in the optically thick regime, 
the transition to the transparent regime is handled 
by a prescribed flux limiter, 
which is not easy to justify \citep{jan92,mes98}.  
The non-radial neutrino flux 
at large radii in the semi-transparent and transparent regimes, 
which we have seen above, 
can not be properly described by the diffusion-based approximation, 
since the flux in these regimes is 
not determined by the local gradient of neutrino density 
but is obtained non-locally 
as a superposition of neutrinos flying in different directions 
from distant points.  
The 6D Boltzmann treatment is a unique solution 
to treat such circumstances rigorously.  
It is, of course, more desirable to conduct 
the neutrino-radiation hydrodynamics simulations, 
which will be a future work, though.  

A recent study by the multi-energy group flux-limited diffusion approximation in 2D 
reported non-explosions for 12--25M$_{\odot}$ stars \citep{dol14}, 
being different from the successful 2D simulations 
by other groups \citep{suw10,mul12b,bru13,knak14a,suw14b}.  
This example again casts a light on uncertainties 
due to different approximations in neutrino transfer, 
which might lead to totally different outcomes.   
The 6D Boltzmann code 
will be helpful to decipher the discrepancies 
among them.  
(See \citet{lie09} for the validation of IDSA in 1D.  )  
%

The comparison of the ray-by-ray approximation 
with the 6D Boltzmann evaluation 
is beneficial for the supernova research, 
since this approximation is routinely used 
for the currently state-of-the-art numerical simulations 
in 2D and 3D spatial dimensions 
but its validation is rather scarce.  
While the ray-by-ray approximation is an efficient way 
to handle the neutrino transfer in multi-dimensions, 
it inevitably neglects some of the non-radial neutrino transport 
(See \citet{bura06} for partial coupling) 
and, as a consequence, tends to enhance artificially 
the contrast among different radial rays.  
These features found by the 6D Boltzmann solver in this paper 
are in accord with the comparison in 2D 
between the Boltzmann transport and the flux-limited 
diffusion approximation 
by \citet{ott08} and \citet{bra11} 
(See also \citet{dol14} for the discussions with a formal solution).  
It remains to be seen if this effect persists 
in 3D dynamical simulations.  

If the artificial enhancement of neutrino anisotropy 
lasts for a long period, it may have impact 
on the hydrodynamics.  
We have actually seen that it lasts at least for $\sim$10 ms.  
A larger neutrino-heating rate 
along a certain radial ray will give a 
strong push to the stalled shock wave locally in that direction.  
The opposite situation 
can arise in another direction, 
where neutrino heating is suppressed erroneously.  
The ray-by-ray approximation, 
therefore, may enhance the deformation of shock wave 
and might induce stronger hydrodynamical instabilities.  
In the 6D Boltzmann treatment, such anisotropies tend to be 
smoothed by the non-radial transport.  
It is advisable to pay an appropriate attention 
to this effect when one studies 
the anisotropy of neutrino distributions produced by 
local matter inhomogeneities.  

It remains to be seen whether the improvement by the 6D Boltzmann treatment 
indeed changes 
the behavior of hydrodynamical instabilities 
such as SASI \citep{blo07a,iwa08,nor10,han12,tak12,han13}.  
Very recently \citet{tam14} reported with the ray-by-ray approximation 
a new instability named 
Lepton-number Emission Self-suitained Asymmetry (LESA), 
which produces self-sustained, globally dipolar configurations.  
Note, however, that \citet{dol14} have found no such instability 
with the flux-limited diffusion approximation.  
LESA may hence be an artifact of the approximation. 
In fact, since it is claimed that the global coupling of anisotropy 
between the neutrino-sphere and shock wave is the key to the instability, 
accurate evaluations of the neutrino anisotropy may be crucial.  
It is hence interesting to see what the Boltzmann transport would produce.  
It may be that eventually 
only by performing full neutrino-radiation hydrodynamics 
simulations with Boltzmann transport can we disentangle this new twist.  

Our results suggest that the errors by 
the ray-by-ray approximation may be moderate 
for most of the cases.  
This will be a good news 
for the supernova research in 3D, 
since the current supercomputing resources are limited 
and allow only the 3D simulations with an approximate neutrino transfer.  
In future, however, it is necessary to remove 
even such moderate uncertainties.  
In fact, even a $\sim10\%$ error in the heating rate 
may have an impact on the revival of shock wave particularly 
when it is marginal.  

The neutrino distributions 
in the transparent regions 
are determined 
not locally but globally as a superposition of neutrinos 
flying in different directions from various distant points.  
The Boltzmann solver can fully take this into account 
in sharp contrast to other approximate methods.  
The neutrino spectra 
are also affected by such non-local features in neutrino transfer, 
reflecting thermodynamical states not just at a single point 
but in a wide region on the neutrino-sphere.  
It may hence be possible that the Boltzmann solver reveals 
new features in the neutrino spectra.  
It would then be also interesting to examine 
its influences 
on the nucleosynthesis of heavy elements 
via neutrino processes.  

This will be particularly the case when 
the proto-neutron star is strongly deformed 
due to rapid rotation.  
Since the core is globally asymmetric, 
the resulting fluxes are intrinsically non-radial.  
Such an effect was pointed out by \citet{ott08} 
in the case of a rapidly rotating 2D supernova core.  
Its observational consequences were studied \citep{bra11}.  
In the case of 3D 
the neutrino-sphere may be globally non-axisymmetric 
if the core rotates very rapidly.  
It would be interesting then to study 
its possible observational implications with 
the 6D Boltzmann transport.  

The main limitation of the current Boltzmann code is 
the angular resolution 
in describing the forward-peaked neutrino distribution functions 
in the optically thin region.  
A large number of grid points would be required 
if one were to fully resolve such configurations, 
which is simply unaffordable. 
Fortunately, the area, which we 
are most concerned with and where the Boltzmann solver is most needed, 
is the region from beneath the neutrino-sphere 
up to the stalled shock wave and 
the neutrino distribution function is not so forward-peaked yet there; 
the moderate angular resolution we can afford 
will be acceptable particularly 
for the evaluations of heating and cooling rates 
although detailed convergence tests are certainly necessary.  
If one wants to obtain the neutrino luminosities and spectra 
at different viewing angles quantitatively, 
one may need a finer angular mesh, which may 
require prohibitive computing resources 
in the current approach.  
(See, however, \S 4 in \citet{kot12} for the scaling of computational load.  ) 
It would then be necessary to develop alternative techniques 
such as the variable angle mesh \citep{yam99} 
or 
the Monte Carlo method \citep{abd12}.  


Although 
the current study employs only 
the snapshots from the 3D supernova simulations 
with the ray-by-ray IDSA \citep{tak12,tak14,hor14,tak14b}, 
it is desirable to investigate 
other background models, since 
the neutrino transfer 
may be somewhat affected 
indirectly by the approximation employed in the supernova simulations.  
It will be also preferable to 
study the influences of the numerical grid employed 
(ex. spherical-polar vs 3D-cartesian as well as its resolution) 
and progenitor models.  
World wide collaborations will be indispensable 
in order to fully understand the 3D neutrino transfer.  
It may become possible then 
to choose suitable tool for a given problem  
(See \citet{ili06} for the cosmological radiative transfer codes, 
for example).  
The validation of the approximate methods currently used 
in 3D supernova simulations will be 
a necessary first step in that direction.  

The Boltzmann solver can be combined with the magneto-hydrodynamical code 
with no problem to study magneto-hydrodynamics scenarios of core-collapse supernovae 
in more realistic settings. 
The application will not be limited to the supernova.  
The neutrino transfer in the collapsar, 
which originates from more massive stars, for example, will be 
important for the jet formation in the gamma ray burst.  
The Boltzmann transport is particularly suitable 
to study such highly anisotropic and non-local problem.  
In particular information  
on the angle moments of neutrino distribution functions 
will be very helpful to develop a closure relation 
for the moment formalism, another approximate method.  
These issues are now under investigation 
with the 6D Boltzmann solver and will be reported elsewhere.  

Improvements of 
the current 6D Boltzmann code 
are certainly needed and may somewhat change the results.  
Among other things, the implementation of 
the modern reaction rates 
instead of the conventional set is mandatory
to investigate the post-bounce dynamics quantitatively further 
as demonstrated in the previous papers \citep{bura06,len12a}.  
The modifications of the emission and absorption rates 
affect directly the cooling and heating rates.  
Addition of inelastic scatterings of neutrinos will 
modify their energy spectra and, as a result, anisotropy as well. 
Even the electron captures on nuclei 
affect the post-bounce dynamics by setting the initial stage.  

The proper account of relativistic corrections is another issue 
toward the ultimate realism \citep{lie04,sum05,bura06,len12b}.  
Implementations of the special relativistic effects 
such as the Doppler shift and aberration 
have been actually done recently \citep{nag14}.  
Combined with a Newtonian hydrodynamics code, 
it is now applied to the 2D supernova simulation and 
the results will be reported elsewhere.  
The incorporation of general relativistic corrections to the present code 
is also currently underway (See also \citet{car13b,shi14}).  
Its coupling to a general relativistic hydrodynamics code and 
a solver of the Einstein equation in 3D will be a grand challenge.  

\section{Summary}\label{section:summary}

We have explored the 3D properties of the neutrino transfer 
in supernova cores by solving the Boltzmann equation 
in 6 dimensions (3 in space and 3 in momentum space).  
We have performed the numerical simulations to obtain 
the stationary state of neutrino distribution functions 
in 6D phase space 
for the fixed backgrounds of the 3D supernova core.  
We used the typical profiles of the supernova core 
taken from the 3D supernova simulations of 
11.2M$_{\odot}$ and 27.0M$_{\odot}$ stars 
by \citet{tak12,hor14,tak14b}.  
The numerical code to solve the Boltzmann equation in 6D 
handles the neutrino transfer 
in multi-energy and multi-angle group for three neutrino species.  
The three dimensional propagation of neutrinos is described 
by the S$_n$ method.  
The standard set of neutrino reactions is implemented 
in the collision term together with the table of equation of state.  

This is the application of 
the 6D Boltzmann solver to the realistic profiles 
of the 3D supernova cores for the first time.  
The 6D Boltzmann equation directly 
provides the energy and angle distributions 
of neutrinos in the whole region of the 3D supernova cores.  
Most importantly, the transport of neutrinos 
in the multi-angle description 
provides the information of the non-radial 
(polar and azimuthal) directions as well as the radial one.  
We have demonstrated that the 6D Boltzmann solver describes 
the 3D features of neutrino transfer, 
removing the approximations often used 
in the currently state-of-the-art simulations.  
The numerical results show that the non-radial fluxes 
are generally seen inside the proto-neutron star 
and can extend over the wide region beyond the optically thick region.  

We find that the ray-by-ray approximation, which is a popular 
approximation in the current supernova simulations, provides 
enhanced directional variations in neutrino distributions.  
We have computed the 3D neutrino transfer in the ray-by-ray approximation 
by using our numerical code, 
dropping off the non-radial transport terms.  
We have examined the characteristics of the ray-by-ray approximation 
for the same 3D profiles 
through the comparison with the 6D Boltzmann evaluation.  
In the ray-by-ray approximation, 
the non-spherical profiles of neutrino densities and fluxes 
tend to prevail from center to large distances.  
This is in marked contrast to the 6D Boltzmann evaluation 
in which the integral from many directions provides 
the nearly spherical distribution 
outside the proto-neutron star.  

Our analysis has revealed that the ray-by-ray evaluation 
provides noticeable overestimation or underestimation 
of densities and fluxes along the radial ray.  
By examining the relative differences of the ray-by-ray evaluation 
with respect to the 6D Boltzmann evaluation, 
the deviations of $\sim20\%$ was found to 
develop depending on the environment.  
Once the neutrino emission is enhanced due to a local 
hot spot of the neutrino chemical potential, 
this modification continues along the radial ray.  
The reduced neutrino emission is similarly maintained.  
Therefore, we find alternate distributions of 
the rays with underestimation and overestimation 
outside the proto-neutron star.  
This behavior is seen commonly 
for $\nu_e$ and $\bar{\nu}_e$ distributions, 
which reflect the non-uniform distribution 
of the neutrino degeneracy.  
The sign of the deviations for the two species 
is opposite each other 
because of the opposite effects on the emission 
through the degeneracy.  

We have also found that 
the heating rates evaluated by the ray-by-ray approximation 
have deviations along the radial ray.  
The relative differences from the 6D Boltzmann evaluation amount to 
$\sim20\%$ in some regions but smaller than 
the case of the densities and fluxes in wide regions.  
This is because the overestimation and underestimation 
of the heating rates by $\nu_e$ and $\bar{\nu}_e$ 
cancel each other due to their opposite deviations 
in many cases.  
Significant deviations occur in the regions 
where the neutrino degeneracy 
is prominent from the average distribution, for example, 
in the narrow channels between the two mushroom-like outflows.  
This hot spot provides the enhanced heating rate 
along the particular radial ray.  
The hot spot effect may last for $\sim$10 ms 
as found in our study using the time series of the profile.  
We have found that 
the total amount of heating in the whole heating region 
in the ray-by-ray evaluation 
is slightly smaller than that in the 6D Boltzmann evaluation.  
These features of the ray-by-ray characteristics are 
common for the adopted profiles from 
the 11.2M$_{\odot}$ and 27.0M$_{\odot}$ stars, 
which represent the exploding and stalled shock situations.  
Our study calls for careful assessment of the angle-dependent 
heating effects which may affect the hydrodynamics.  
We remark that these features are found 
in the fixed background profiles and 
further studies in neutrino-radiation hydrodynamics 
are necessary.  

We have examined the basic features of the angle moments 
and the Eddington tensor in the 3D supernova cores.  
We show that the angle moments of the neutrino 
distribution functions obtained by the 6D Boltzmann solver 
have rather spherical profiles.  
The relative differences of the ray-by-ray evaluation 
appear in some regions, but limited within the inner part.  
The components of the Eddington tensor 
exhibit deformed distributions in more wide regions 
than the case of angle moments.  
Since we have the 6D neutrino distribution functions, 
we can compute all components of the Eddington tensor 
and clarify the behavior of the diagonal and non-diagonal elements.  
We have demonstrated that 
there are regions 
with appreciable non-diagonal components 
in one of the snapshots as an example.  
Studies of 
these basic quantities will be helpful to develop 
new closure relations for the moment formalism 
in largely deformed profiles such as collapsars.  

The current study is a step toward the complete treatment 
of neutrino-radiation hydrodynamics 
implementing the full list of neutrino reactions 
in the general relativistic framework.  
It is of great interest how the 6D Boltzmann treatment 
affects the neutrino heating mechanism 
in hydrodynamical instabilities.  
The coupling of the 6D Boltzmann solver 
with the hydrodynamics is made recently \citep{nag14} 
and numerical studies 
to explore the dynamical outcome for explosions 
are currently underway.  


\section*{Acknowledgments}

We are grateful to K. Kotake, H. Suzuki 
for the collaborations on supernova simulations.  
K. S. thanks to H. Nagakura, W. Iwakami, 
K. Nakazato, S. Furusawa, Y. Sekiguchi, K. Kiuchi and N. Ohnishi 
for collaborative works and profitable discussions.  
K. S. expresses his thanks to T.-H. Janka, C. Ott and A. Mezzacappa 
for valuable comments 
on the numerical methods of neutrino transfer in 2D and 3D.  
We thank A. Imakura and T. Sakurai 
for the project of computational science 
for sparse matrix solvers 
and the supporting team of HITACHI 
for tuning the numerical code on SR16000.  

The numerical computations in this work were mainly performed 
on the supercomputers 
at High Energy Accelerator Research Organization (KEK) 
under the support of its Large-scale Simulation Program
(Nos.12/13-05, 13/14-10, 14/15-17).  
K. S. acknowledges also the extensive usage of the supercomputing resources at 
Yukawa Institute for Theoretical Physics (YITP) in Kyoto University, 
The University of Tokyo, K-computer 
and 
Research Center for Nuclear Physics (RCNP) in Osaka University.  
This research is supported by the computational resources of the K computer 
provided by the RIKEN Advanced Institute for Computational Science 
through the HPCI System Research project (Project ID:hp130025, hp140211).  
Large-scale storage of numerical data 
is supported by the JLDG constructed over the SINET4 of NII.  

This work is partially supported by 
the Grant-in-Aid for Scientific Research on Innovative Areas (Nos. 20105004, 20105005, 24103006) 
and 
the Grant-in-Aid for the Scientific Research (Nos. 19104006, 21540281, 22540296, 24244036, 26870823) 
from the Ministry of Education, Culture, Sports, Science and Technology (MEXT) in Japan.  

This numerical study on core-collapse supernovae using the supercomputer facilities 
is supported by the HPCI Strategic Program of MEXT, Japan.  





\bibliographystyle{apj}                       
\bibliography{apj-jour,sumi}    

\clearpage


\begin{table}
\caption{Volume-integrated heating rates 
in the heating region 
are listed in the unit of [erg/s] 
for the 6D Boltzmann and ray-by-ray evaluations.  
The total amount of the heating rate, $Q_{\nu}$ [erg/s], 
is calculated by the volume-integration 
of the specific heating rate, $q_{\nu}$ [erg/g/s], 
times the mass density, $\rho$ [g/cm$^3$], 
where $q_{\nu}$ is positive.  
(See the definition in \S \ref{section:h-total}.  )
Relative differences of the ray-by-ray evaluation with respect to 
6D Boltzmann evaluation are also listed.  }
\label{table:totalheating}
\begin{center}
\begin{tabular}{ccrrr} \hline \hline
Model & Timing & Ray-by-ray        & 6D Boltzmann      & Relative difference \\
      &        & $Q_{\nu}$~[erg/s] & $Q_{\nu}$~[erg/s] &  [$\%$]   \\
\hline
11M   & 100 ms & $6.84 \times 10^{51}$  & $6.93 \times 10^{51}$ & -1.4 \\
11M   & 150 ms & $4.53 \times 10^{51}$  & $4.61 \times 10^{51}$ & -1.7 \\
11M   & 200 ms & $2.65 \times 10^{51}$  & $2.68 \times 10^{51}$ & -1.1 \\
\hline
27M   & 150 ms & $1.25 \times 10^{52}$  & $1.26 \times 10^{52}$ & -0.8 \\
27M   & 200 ms & $1.16 \times 10^{52}$  & $1.16 \times 10^{52}$ & -0.3 \\
%
\hline \hline
\end{tabular}
\end{center}
\label{table:heating}
\end{table}%



\newpage

\newpage

\newpage

\newpage

\newpage

\newpage


\clearpage

\newpage

\newpage

\newpage

\newpage

\newpage

\newpage

\clearpage

\newpage

\newpage

\newpage

\newpage

\newpage

\newpage

\newpage

\clearpage

\newpage

\newpage

\newpage

\newpage

\newpage

\newpage

\newpage

\newpage


\newpage


\clearpage

\newpage

\newpage

\newpage

\newpage

\newpage

\newpage

\end{document}